\documentclass[12pt,a4paper,twoside,openleft]{report}

\usepackage{amsthm,latexsym,multibox,amssymb}
\usepackage{graphicx,lscape,fancyhdr,array}
\usepackage[arrow,matrix,curve]{xy}\SilentMatrices
\def\xyma{\xymatrix@M.7em}
\def\xymas{\xymatrix@M.1em}

\usepackage{color}

\definecolor{rouge}{rgb}{1,0,0}
\definecolor{bleu}{rgb}{0,0,1}
\definecolor{vert}{rgb}{0,1,0}
\definecolor{magenta}{cmyk}{0,1,0,0}
\definecolor{orange}{cmyk}{0,0.6,0.8,0}
\definecolor{jaune}{cmyk}{0,0,1,0}
\definecolor{navy}{rgb}{0,0,0.5}
\definecolor{rougef}{rgb}{0.56,0,0}
\definecolor{vertf}{rgb}{0,0.5,0}
\definecolor{brun}{rgb}{0.5,0.1,0.1}
\definecolor{noir}{gray}{0}

\newcommand{\be}{\begin{equation}}
\newcommand{\ee}{\end{equation}}
\newcommand{\ba}{\begin{eqnarray}}
\newcommand{\ea}{\end{eqnarray}}

\newtheorem{corollary}{Corollary}[chapter]
\newtheorem{proposition}{Proposition}[chapter]
\newtheorem{lemma}{Lemma}[chapter]
\newlength{\mylength}
\newcommand{\cadre}[3]
{\ \\[2ex]
  \setlength{\fboxsep}{1ex}
  \setlength{\mylength}{\linewidth}
  \addtolength{\mylength}{-2\fboxsep}
  \addtolength{\mylength}{-2\fboxrule}
  \framebox[\linewidth]{
    \begin{minipage}{\mylength}
      \vspace{1ex}
      \begin{#1}\label{#2}
        {\rm #3}
      \end{#1}
    \end{minipage}
  }\ \\[2ex]}

\def\nn{\nonumber}

\def\a{\alpha}
\def\b{\beta}
\def\g{\gamma}

\def\d{\delta}
\def\D{\Delta}
\def\e{\epsilon}
\def\ve{\varepsilon}

\def\f{\phi}
\def\F{\Phi}
\def\p{\psi}
\def\P{\Psi}

\def\k{\kappa}
\def\l{\lambda}
\def\L{\Lambda}
\def\m{\mu}
\def\n{\nu}

\def\r{\rho}
\def\s{\sigma}
\def\S{\Sigma}

\def\O{\Omega}
\def\g{\gamma}

\def\cD{{\mathcal D}}
\def\cF{{\mathcal F}}

\def\cL{\epsilon}
\def\cH{{\mathcal H}}
\def\cl{{\cal L}}

\def\IR{\relax{\rm I\kern-.18em R}}
\def\ZZ{\relax{\hbox{\cmss Z\kern-.4em Z}}}

\def\su{\sqrt{-u^2}}
\def\half{\frac{1}{2}}
\newcommand{\dH}[2]{H^{(-)#1}_{#2}}
\newcommand{\uH}[2]{H^{(-)#1#2}}
\def\ra{\rightarrow}
\def\dslash{\hbox{\ooalign{$\displaystyle\partial$\cr$/$}}}
\csname @addtoreset\endcsname{equation}{section}

\newlength{\blength}
\renewcommand{\proof}[1]{\vspace{-.05cm}
\begin{list}{\bf Proof:}
{\listparindent=\parindent\parsep=0pt \labelwidth=-0.5cm
\labelsep=\parindent \addtolength{\labelsep}{-\blength}
\addtolength{\labelsep}{1.5cm}
\itemindent=-\blength
\addtolength{\itemindent}{\parindent} \leftmargin=1.0cm}
\item
#1~$\qedsymbol$\end{list}
\vspace{.0cm}}

\begin{document}

%%%%%%%%%%%%%%%%%%%%%%%%%%%%%%%%%%%%%%%%%%%%%%%%%%%%%%%%%%%%

\begin{titlepage}
\begin{center}
{\large UNIVERSIT\'E LIBRE DE BRUXELLES \\
Facult\'e des Sciences \\\vspace{1mm} Service de Physique
Th\'eorique et Math\'ematique}

\vspace{6cm}

{\LARGE\bf Issues in electric-magnetic duality}

\vspace{3cm}

Dissertation pr\'esent\'ee en vue de l'obtention \\
du grade de Docteur en Sciences \\

\vspace{3cm}
Xavier Bekaert\\
\vspace{3cm}
Ann\'ee acad\'emique 2001--2002 \\
\end{center}

\end{titlepage}

\thispagestyle{empty}

%%%%%%%%%%%%%%%%%%%%%%%%%%%Dédié%%%%%%%%%%%%%%%%%%%%%%%%%%%%%%%%%
\newpage \thispagestyle{empty}

\cleardoublepage

\thispagestyle{empty}

\vspace*{6cm}

\begin{flushright}

{\large\textit{Aux Mauvaises Herbes.}}

\end{flushright}

%\vspace{1cm}
%\textit{En esp\'erant qu'elles continuent encore longtemps à pousser en libert\'e le long des chemins mal fr\'equent\'es.}

\newpage \thispagestyle{empty}

\cleardoublepage
%%%%%%%%%%%%%%%% Remerciements %%%%%%%%%%%%%%%%%%%%%%%%%

\thispagestyle{empty}

\vspace*{4cm}

{\Large \bf Remerciements}
\\
\\

Avant toute chose, je constate qu'il est difficile de rendre
justice aux innombrables personnes qui ont pu m'aider et
m'influencer durant ces quatre ans de th\`ese.

Il va sans dire que les travaux pr\'esent\'es ici n'ont pas
\'et\'e accomplis de mani\`ere isol\'ee. Ils trouvent tous leurs
sources dans diverses discussions et dans de nombreuses
collaborations. Je voudrais commencer par remercier toutes les
personnes avec qui j'ai eu la joie de travailler directement. Ma
th\`ese a d\'ebut\'e par une collaboration avec Christiane
Schomblond et Bernard Knaepen. \`A l'instar de nombreuses discussions
avec Glenn Barnich, cette collaboration m'a familiaris\'ee avec les
techniques de la cohomologie BRST, ce qui m'a \'et\'e
extr\^emement utile par la suite. Elle a \'et\'e suivie par une
collaboration avec Marc Henneaux et Alexander Sevrin sur les
d\'eformations des th\'eories de champs de jauges chiraux. J'ai eu
ensuite le plaisir de travailler avec Odile Saliu et Constantin
Bizdadea en Roumanie sur un projet que nous n'avons malheureusement
pas r\'eussi \`a conclure. Cependant j'ai tir\'e grand
b\'en\'efice de cette rencontre puisqu'elle m'a amen\'e \`a
apprendre concr\`etement la quantification BRST Hamiltonienne. La
rencontre avec Sorin Cucu et les travaux qui en suivirent ont \'et\'e une
longue exp\'erience tr\`es enrichissante. Les visites d'Andres
Gomberoff \`a Bruxelles m'ont donn\'e l'opportunit\'e de travailler
sur la non-conservation de la charge \'electrique en supergravit\'e.
Derni\`erement, j'ai effectu\'e une br\`eve incursion dans la
ph\'enom\'enologie des branes avec Nicolas Boulanger et Justin
Vazquez-Poritz. Cet doctorat s'ach\`eve par une
collaboration avec Nicolas Boulanger et Marc Henneaux sur la
dualisation de la gravitation lin\'earis\'ee.

%Les rencontres furent trop nombreux et vari\'es que pour en faire
%le compte-rendu d\'etaill\'e ici.
Parmi les personnes avec qui j'ai convers\'e pendant
la dur\'ee de cette th\`ese, je remercie particuli\`ement toutes
celles qui ont particip\'e aux divers groupes d'\'etudes et
s\'eminaires informels qui ont jalonn\'e ces quatres ann\'ees.
J'ai particuli\`erement appr\'eci\'e les multiples \'echanges que
nous avons pu avoir entre membres du service de physique
th\'eorique et du groupe de physique math\'ematique des
interactions fondamentales de l'ULB. J'ai \'egalement profit\'e
des rencontres, plus rares mais tout aussi enrichissantes, avec
les
% - alternativement - invit\'es et h\^otes
membres du Centro de Estudios
Cient\'ificos de Chile.

J'ai partag\'e un bureau avec Nicolas Boulanger pendant trois ans.
Ce dernier a pu faire connaissance avec mes humeurs changeantes et
je le f\'elicite pour sa patience. Cette cohabitation prolong\'ee
m'a beaucoup apport\'e sur les plans de la physique et de
l'amiti\'e.

C'est avec plaisir que je remercie \'egalement la chaleureuse
hospitalit\'e roumaine dont je n'ai jamais rencontr\'e dans nul
autre pays d'analogue aussi frappant. En particulier, un grand
merci \`a tous les amis roumains de l'Universit\'e de Craiova.

Je suis tr\`es reconnaissant \`a Glenn Barnich, Bernard Knaepen,
et Olivier Debliquy pour leur aide r\'ep\'et\'ee et patiente en
informatique. Cette th\`ese n'aurait probablement jamais pu voir
le jour sans les milles coups de main qu'ils m'ont donn\'es dans
ce domaine qui conservera toujours pour moi sa part d'absurde.

Je suis \'egalement tr\`es reconnaissant \`a Fabienne Deneyn pour
son aide indispensable dans bon nombre de questions
administratives.

Je remercie chaleureusement Christiane Schomblond pour sa
disponibilit\'e, son aide scientifique et son soutien moral durant
ces ann\'ees d'\'etudes.

Je remercie vivement mon promoteur Marc Henneaux pour ses id\'ees
stimulantes, ses remarques, ses encouragements et ses conseils
opportuns lors de tous les moments d\'ecisifs du long parcours
scientifique (et initiatique) qu'est le doctorat.

Une fois de plus, je remercie Christiane Schomblond, Marc Henneaux
et Nicolas Boulanger; cette fois ci, pour leur lecture attentive
de cette th\`ese et leurs commentaires constructifs. Pour ces
m\^emes raisons, je remercie \'egalement les membres du jury
Laurent Houart, Alexander Sevrin, Mario Tonin et Michel Tytgat.

Je n'ai pas tant d'amis que pour ne pouvoir les citer tous ici
mais, de peur d'en omettre un seul, je pr\'ef\`ere les remercier
tous ensemble, d'un seul \'elan. Chacun d'entre eux s'y
reconnaitra ais\'ement car ils savent combien leur amiti\'e m'est
chaque jour indispensable.

Pour terminer, mes remerciements vont \`a mes parents pour toutes
ces innombrables choses, petites ou grandes, sans lesquelles ce
travail n'aurait certainement pas pu voir le jour.

\thispagestyle{empty}

\newpage \thispagestyle{empty}

\cleardoublepage
%%%%%%%%%%%%%%%%%%%%Table of content%%%%%%%%%%%%%%%%%%%%%%%%%

\pagenumbering{roman}

\tableofcontents

%%%%%%%%%%%%%%%%%%%%%%%%%%%%%%%%%%%%%%%%%%%%%%%%%%%%%%%%%%%%

\chapter{Introduction to electric-magnetic duality}

\setcounter{page}{0}

\pagenumbering{arabic}

\par

Electric-magnetic duality symmetry has a rather old history, going
back to the birth of Maxwell's equations:
\begin{eqnarray}
{\bf \nabla} \cdot {\bf E} &=& \rho\,,\label{G}
\\
{\bf \nabla} \cdot {\bf B} &=&0\,,\label{M}
\\
{\bf \nabla} \times {\bf B} &=& {\bf j} \,+\,  \frac{\partial {\bf
E}}{\partial t}\,,\label{A}
\\
{\bf \nabla} \times {\bf E} &=& -\,  \frac{\partial {\bf
B}}{\partial t}\;\;.\label{F}
\end{eqnarray}
It seems likely that the symmetric role played by the electric and
the magnetic fields in the laws of electromagnetism, has been one
of the motivations of Maxwell to introduce in the Amp\`ere law,
the ``displacement current" $\frac{\partial {\bf E}}{\partial t}$,
which was not sanctioned by experiment at that time\footnote{The
displacement current also ensures electric charge conservation,
which was another reason to introduce the former.}. Significantly,
this led him to the correct equation (\ref{A}). When there is no
electric current (${\bf j}=0$) this equation takes indeed the same
form as the Faraday law (\ref{F}) up to a change of sign. Maxwell
was known to have an aesthetic appreciation for mathematical
structures: {\it``I always regarded mathematics as the method of
obtaining the best shapes and dimensions of things; and this meant
not only the most useful and economical, but chiefly the most
harmonious and the most beautiful".}\footnote{Quotation of
Maxwell's letter to Galton \cite{Maxwell}.} In this light, it is
reasonable to accept the conclusion of Roger Penrose: {\it``It
would seem that the symmetry of these equations and the aesthetic
appeal that this symmetry generated must have played an important
role for Maxwell in his completion of these equations"}
\cite{Penrose}.

%%%%%%%%%%%%%%%%%%%%%%%%%%%%%%%%%%%%%%%%%%%%%%%%%%%%%%%%%%%%%%

\section{Modifying Maxwell's equations with Dirac}

The above-mentioned symmetry can be more precisely expressed as
the invariance of sourceless ($\rho={\bf j}=0$) Maxwell's
equations under a ``rotation" of the electric and magnetic fields
\begin{eqnarray}
{\bf E}&\rightarrow&\,\, \cos\alpha\, {\bf E} - \sin\alpha\,{\bf
B},\nonumber\\ {\bf B}&\rightarrow& + \sin\alpha \,{\bf E}
+\cos\alpha\,{\bf B}.\label{rotation}
\end{eqnarray}
A physical important point is that the energy-momentum tensor is
also left invariant under the transformation (\ref{rotation}).
This can be easily checked for the energy and momentum densities
of the electromagnetic field: \be {\cal E}=\frac12\,({\bf
E}^2+{\bf B}^2),\quad {\bf{\cal P}}={\bf E\times B}\,, \ee as
well as for the stress tensor $T_{ij}=-E_i\,E_j-B_i\,B_j+{\cal
E}\,\d_{ij}$.

\subsection{Magnetic monopoles}\label{Magnmonop}

Basically, the equation (\ref{M}) states that if one breaks a
magnet bar in two pieces in its middle, one always obtains two
smaller magnets and never a North pole in one hand and a South
pole in the other hand, as it could occur for an electric dipole.
The absence of isolated magnetic charge in Maxwell's equations
(\ref{G})-(\ref{F}) spoils the possibility of the symmetry
(\ref{rotation}) in the presence of sources. Since it appears
deceiving that Nature is deprived from such a beautiful symmetry,
it would be satisfactory to find an explanation of its absence.
The first natural step is to assume the existence of magnetic pole
and investigate whether this leads to any contradiction with known
principles. This is what Dirac did in 1931 \cite{Dirac:1931}.

To establish the symmetry between the electric and magnetic fields
in the presence of sources, we have to assume the existence of
magnetic densities of charge and current $\rho _{m}$ and ${\bf
j}_{m}$, in addition to the usual electric charge density and
current $\rho _{e}=\rho$ and $j_{e}={\bf j}$. The Maxwell
equations then take the form
\begin{eqnarray}
\mathbf{\nabla }\cdot\bf{E} &=& \rho _{e} \,,\nonumber\\
-\bf{\nabla }\times \bf{E}
&=&\bf{j}_{m}+\frac{\partial \bf{B}}{\partial t}\,,\nonumber \\
\bf{\nabla }\cdot\bf{B} &=& \rho _{m}\,,\nonumber \\
\bf{\nabla }\times \bf{B} &=&\bf{j}_{e}+\frac{\partial
\bf{E}}{\partial t}\,.\label{modified}
\end{eqnarray}

The EM symmetry can easily be seen by rewriting the modified
Maxwell equations as
\begin{eqnarray}
\bf{\nabla }\cdot(\bf{E}+i\,\bf{B}) &=& \rho_e+i\,\rho_m\,,\nonumber \\
\bf{\nabla }\times (\bf{E}+i\,\bf{B})
&=&i\,\left[\left(\bf{j}_{e}+i\,\bf{j}_{m}\right)+\frac{\partial}{\partial
t}\left(\bf{E}+i\,\bf{B}\right)\right]\,.
\end{eqnarray}
In terms of the fields, the rotation (\ref{rotation}) reads :
\begin{eqnarray}
\bf{E}+i\,\bf{B} &\rightarrow& e^{i\,\alpha }\left(
\bf{E}+i\,\bf{B}\right)\,,\nonumber\\
\rho_e+i\,\rho_m &\rightarrow& e^{i\,\alpha }\left(
\rho_e+i\,\rho_m\right)\,,\nonumber\\
\bf{j}_{e}+i\,\bf{j}_{m}&\rightarrow& e^{i\,\alpha }
\left(\bf{j}_{e}+i\,\bf{j}_{m}\right)\,.\label{crotation}
\end{eqnarray}
The electromagnetic energy density reads \be{\cal
E}=\frac12\,|\bf{E}+i\,\bf{B}|^2\,,\ee which is manifestly
invariant under (\ref{crotation}).

As noticed in \cite{Jackson}, it is a matter of convention to
speak of a single electrically charged but not magnetically
charged particle : ``{\it The only meaningful question is whether
or not \underline{all} particles have the same ratio of magnetic
to electric charge.}" If they do, we can choose an appropriate
duality rotation which makes the magnetic charge and current to
vanish, and we recover the ordinary Maxwell equations.

\subsection{Charge quantization}

In 1931 Dirac showed that if one wants the associated quantum
theory to be consistent in the presence of an electric charge $e$
and a magnetic monopole of charge $g$, they have to satisfy the
following quantization relation \cite{Dirac:1931}
\begin{equation}
e\,g=n\,h\quad\quad n\in\mathbb N\,.\label{Diracqcond}
\end{equation}
Therefore the existence of a single magnetic charge could explain
the quantization of electric charge in Nature. This is how Dirac
left the subject in 1931. He came back to it in 1948, to clarify
their dynamical behaviour \cite{Dirac:1948}. However many
questions were left unresolved, including the puzzling remark that
the quantization condition (\ref{Diracqcond}) does not respect the
proposed rotational symmetry (\ref{rotation}). Several of these
questions will be addressed in the sequel; the chapter
\ref{chargeq} is devoted to the particular topic of the charge
quantization condition.

The known unit of pure electric charge is known to be small:
$e^2/4\pi=\a\ll 1$. We point out that (\ref{Diracqcond}) implies
that the magnetic charge of a monopole is expected to be large in
natural units: $g\gg 1$ since $e\ll 1$.

The 1931 and 1948 Dirac's arguments for charge quantization were
essentially topological in nature\footnote{The appendix
\ref{topology} will cover some useful tools in algebraic topology
to follow Dirac's arguments in a modern fashion.}. It seems to be
the first notable appearance of topology in twentieth century
physics\footnote{There exist a pleasant historical presentation by
C. Nash of the major fruitful interactions between topology and
physics since the seminal work of Gauss where one finds the modern
definition of the linking number in its close integral form, which
was found during his researches on electromagnetism
\cite{Nash:1997}.}. It is surprising that one had to wait until
the 1959 paper of Aharonov and Bohm (almost thirty years after
Dirac's monopole !) to pinpoint a significant instance of an
interaction between topology and physics. As is well known, it has
been followed by an amazing number of other fruitful appearances
in many areas of theoretical physics during the past four decades.

\subsection{Dirac's aesthetic}

In some sense, Dirac pursued Maxwell's quest of mathematical
beauty by modifying the electromagnetism equations on aesthetic
grounds. Dirac might be the physicist who insisted the most on the
role played by the inherent mathematical beauty of physics
equations. At the advanced age of 78, he described his main
attitude regarding mathematics and physics: ``{\it A good deal of
my research work in physics has consisted in not setting out to
solve some particular problem, but simply examining mathematical
quantities of a kind that physicists used and trying to fit them
together in an interesting way, regardless of any application that
the work may have. It is simply a search for pretty mathematics.
It may turn out later that the work does have an application. Then
one has good luck.}" \cite{Dirac}.

Nevertheless, since the existence of the magnetic monopole is
still not experimentally established, it is not yet clear if
Dirac's monopole will follow the same fate as Maxwell's
displacement current or Dirac's positron. Anyway, the
investigation of the magnetic monopole is, according to David I.
Olive, ``{\it perhaps the single contribution that best
illustrates Dirac's fearlessness}"\footnote{Quoted from the very
nice non-technical introduction to the subject given by Olive in a
lecture celebrating Dirac's life and work \cite{Dirac}.} because
``{\it it certainly required courage to initiate a theory of an
undetected particle. Because of what Dirac found, that theory has
continued to intrigue researchers and continues to develop (...)
The story of magnetic monopole is still far from complete and
indeed promises more revelations}". The role played by dualities
in the emergence of the M-theory conjecture and the better
understanding of EM duality obtained in this framework extensively
confirms the fertility of Dirac's seminal idea as well as Dirac's
``good luck" in discovering new areas in physics.

%%%%%%%%%%%%%%%%%%%%%%%%%%%%%%%%%%%%%%%%%%%%%%%%%%%%%%%%%%%%%%%%%%%%

\section{Preliminary remarks}

Let us stress an important restriction in the present thesis: the
absence of fermions. Despite the fact that supersymmetry and, more
recently, supergravity together with superstring theory gave birth
to numerous developments in the subject, we will restrict
ourselves to purely bosonic theories (or to the bosonic sectors of
supersymmetric ones). This is legitimated by the fact that
fermions are inert under EM duality rotation in all known cases,
therefore they will play no role in the topics discussed
here.\footnote{There exist a huge number of reviews on
electric-magnetic duality in supersymmetric gauge theories. The
reader interested in this topic could for instance have a look at
the introductions \cite{Harvey:1996,Kiritsis:1999}.}

In order to separate the original results obtained during this
thesis from the works on which they are based, we reserve the
label ``theorem" to the main original mathematical material
presented here. Results imported from outside will be referred to
as lemmas or propositions, without respect of their scientific
importance. Multiple citations are given in the chronological
order of publication. Mathematical definitions are written in bold
type. Physical definitions are italicized.

We use a unit system for which $\hbar=c=1$.

%%%%%%%%%%%%%%%%%%%%%%%%%%%%%%%%%%%%%%%%%%%%%%%%%%%%%%%%%%%%%%%%%

\section{Overview of the thesis}

\hspace{.5cm} In the {\bf chapter \ref{journey}}, our discussion
will mainly focus on the non-manifest classical duality symmetry
of the equations of motion (derived from standard actions) of
various kind of bosonic brane electrodynamics. Our theorem
\ref{t:analyticity} is given and its relevance for
four-dimensional non-linear electrodynamics is explained
\cite{Bekaert:2001}. A new discussion of open brane charge in the
language of relative (co)homology is presented in section
\ref{surgery}.

Possible generalizations of EM duality for linearized gravity are
discussed in the {\bf chapter \ref{spintwo}}. We make some
comments on the differences with electrodynamics. The main result
of this chapter is our theorem \ref{GenPoin} which generalizes the
Poincar\'e lemma to tensors in arbitrary representations of the
general linear group $GL(D,\mathbb R)$ \cite{Bekaert:2002}. This
theorem allows a systematic discussion of duality properties of
such tensors that is done in section \ref{arbitraryYoung}.

The {\bf chapter \ref{manifestd}} is devoted to manifest duality
symmetry. Different ways to implement the duality symmetry at the
level of the action are reviewed. In the section
\ref{D-branelectro}, we introduce a topological derivation of the
D-brane-boundary rule from the Wess-Zumino coupling in the
duality-symmetric formulation of IIA supergravity.

We perform the Batalin-Vilkovisky quantization of Maxwell's theory
in its manifestly Lorentz and duality symmetric formulation in the
{\bf chapter \ref{BRSTquantization}} \cite{Bekaert:2000}. The
Hamiltonian BRST quantization of Maxwell's theory is reviewed.

In the {\bf chapter \ref{chargeq}}, we first review some
quantization conditions that generalizes Dirac's work and then we
derive a quantization condition of the Chern-Simons coefficient of
eleven-dimensional supergravity in section \ref{CCS}
\cite{Bekaert:2002i}.

The {\bf chapter \ref{deformation}} is a discussion on consistent
deformations of self-dual gauge fields in the M-theory
perspective. Our no-go theorem \ref{Noway} on local consistent
deformations is presented
\cite{Bekaert:1999,Bekaert:2000i,Bekaert:2001}. More specifically,
in the section \ref{s:self} we also compute all consistent,
continuous, local and Lorentz invariant deformations of a system
of Abelian self-dual vector gauge fields \cite{Bekaert:2001}. Our
no-go theorem \ref{nogothree} on deformations of a mixed symmetry
type gauge field of interest in gravity duality properties is also
given \cite{Bekaert:2002ii}.

The first four {\bf appendices \ref{preliminaries}-\ref{Young}}
review mathematical definitions (that the reader might be familiar
with) which are used throughout the thesis in order to be as much
self-contained as possible (and to provide some reference map for
the author himself not to get completely lost in this mathematical
jungle !). Some choices of conventions and notations are also
presented there. All this material is placed at the end of the
thesis to separate the physical exposure from the necessary
mathematical machinery exposition.

The full proofs of the three theorems
\ref{t:analyticity}-\ref{Noway} are given in the last three {\bf
appendices \ref{l:analyticity}-\ref{pmain}} since they are
technical and lengthy.

%%%%%%%%%%%%%%%%%%%%%%%%%%%%%%%%%%%%%%%%%%%%%%%%%%%%%%%%%%%%%%%%%%%%%
%%%%%%%%%%%%%%%%%%%%%%%%%%%%%%%%%%%%%%%%%%%%%%%%%%%%%%%%%%%%%%%%%%%%%

\chapter{Classical duality in bosonic brane electrodynamics}\label{journey}

\par

This introductory chapter is aimed to give an overview (far from
being complete) of Maxwell's electrodynamics generalizations which
appear in string/M-theory, and of several possible extensions of
the electric-magnetic (EM) duality symmetry (\ref{rotation}) in
this context.

In the first section we generalize the concept of EM duality to
higher dimensional sources and spacetime. The electrokinematics
section \ref{electrok} tries to motivate geometrically the use of
$p$-form gauge fields in generalizing Maxwell's theory. It may be
skipped by the reader without inconvenience. The section
\ref{Diracproc} introduces the mathematical tools chosen to deal
with extended sources. In section \ref{linelectro}, linear
electrodynamics for extended objects is presented and its
properties are discussed in details. The section \ref{non-linear}
gives up the linearity assumption and generalizes the duality
transformation in the (Abelian) self-interacting context. In the
short section \ref{sl2r}, we briefly discuss some consequences of
the addition of a theta term in the Lagrangian in dimensions
multiple of $4$. A reformulation of charge non-conservation issues
in the presence of a Chern-Simons (CS) coupling is presented in
the language of de Rham currents in section \ref{surgery}. We end
up this chapter by applying the previous discussion to two
specific electrodynamics with CS coupling in the last section
\ref{CScoupling}.

%%%%%%%%%%%%%%%%%%%%%%%%%%%%%%%%%%%%%%%%%%%%%%%%%%%%%%%%%%%%%%%%%%%%%%

\section{Let's use differential forms}\label{difff}

In order to render higher dimensional generalizations more
transparent in the next section, we will reformulate the modified
Maxwell equations in terms of differential forms\footnote{For
fundamental definitions and conventions, see section
\ref{difform}.}. Their use is very convenient because their
minimal coupling with a background metric is automatically
implemented. To proceed, we define the electric and magnetic
fields as components of the field strength two-form $F$: the
electric and magnetic fields are the components of the field
strength: $E^i\equiv F^{0i}$, $B^i\equiv -(*F)^{0i}=\frac12
\epsilon^{ijk}F_{jk}$. The charge and current densities are,
respectively, the time and spatial components of a one-form $J$.
Equations (\ref{modified}) are rewritten as
\begin{equation}
d \left( \begin{array}{c} F\\ *F
\end{array}
\right) = \left( \begin{array}{c} *J_m\\ *J_e
\end{array}
\right)\,.\label{Max}
\end{equation}

In the absence of magnetic monopole (i.e. $J_m=0$), the Poincar\'e
lemma is used to derive from the Bianchi identity ($dF=0$) the
existence of a potential vector $A$ such that $F=dA$. The field
equation $d*F=*J_e$ arises from the Maxwell action
\begin{equation}
S_M[A,J_e]=-\frac{1}{2}\int F\wedge *F+\int A\wedge
*J_e.\label{Maxy}
\end{equation}
When $J_m\neq 0$, one introduces a Dirac string $G$ such that
$d*G=*J_m$; therefore the modified Bianchi $dF=*J_m$ gives
$F=dA+*G$. The field equation is derived from the same action with
$F$ replaced by $dA+*G$. In subsection \ref{Diracproc} we will
explain in more details the precise definition of the Dirac
strings and their use to deal with magnetic monopoles in an action
principle. Electric charges and magnetic monopoles are not the
only possibilities. Particles carrying both electric and magnetic
charges also make sense, they have been christened as \emph{dyons}
by Schwinger \cite{Schwinger:1969}.

\subsection{Duality rotation}

The duality rotation (\ref{crotation}) takes the form
\begin{eqnarray}
&&\left( \begin{array}{c} F\\ $*$F
\end{array}
\right) \rightarrow R\left( \begin{array}{c} F\\ $*$F
\end{array}
\right)
 \,,\nonumber\\&&
\left( \begin{array}{c} J_m\\ J_e
\end{array}
\right) \rightarrow R\left( \begin{array}{c} J_m\\ J_e
\end{array}
\right)
 \, ,\label{transfo}
\end{eqnarray}
where $R\in SO(2)$.

Let us stress two points that arise if the fieldstrength is
expressed in terms of a gauge field $A$ (which is the fundamental
field in the variational principle). The first point is that the
duality rotation is only well-defined on-shell. In the chapter
\ref{double}, a formulation for which duality rotation is an
off-shell symmetry is presented. For an infinitesimal rotation of
angle $\delta \a$ one has $\d F=*F\,\d\a$, therefore the variation
of the Bianchi identity requires $d*F=0$ because $d\,\d F=\d
\,dF=0$. The second point is that the duality rotation becomes a
non-local map when expressed in terms of the gauge field $A$.
Indeed, for an infinitesimal rotation of angle $\delta \a$ we have
$\delta A=d^{-1}(*F)\,\delta\alpha$, where $d^{-1}$ stands for the
non-local operator which is the inverse of the differential $d$.

\subsection{Energy-momentum tensor}

Let $S=\int {\cal L}(g_{\m\n})$ be a matter action in a metric
background $g_{\m\n}$. The \emph{energy-momentum tensor} is
defined as
\begin{equation} T_{\mu \nu
}*1=-2\,\frac{\partial \,{\cal L}}{\partial g^{\mu \nu
}},\label{tmunu}\end{equation} where $*1$ is the volume form on
$\cal M$. The Maxwell energy-momentum tensor is defined by using
the differential form ${\cal L}_{Maxwell}=-\frac12\, *F\wedge F$
in (\ref{tmunu}) to get \be T^{Maxwell}_{\mu \nu }*1=
F_{\m}\,F_{\n}\, +\, g_{\m\n}\,{\cal L}_{Maxwell} \,,\ee where
$F_\m$ is a one-form defined by (\ref{am}). We made use of the
identity (\ref{para*b}).

Let us introduce the $(D-3)$-form $(*F)_\m$ defined by (\ref{am}).
The relation \be *(*F)_{\m}\wedge(*F)_{\n}- *F_{\m}\wedge F_{\n}=2
g_{\m\n}\,{\cal L}_{Maxwell}\,,\ee follows from the identity
(\ref{par*a*b}). So, one can rewrite the energy-momentum tensor in
a more convenient form \cite{Deser:1997}\be 2\,T^{Maxwell}_{\mu
\nu }*1= *F_{\m }\wedge F_{\n}\, +\,
*(*F)_{\m}\wedge(*F)_{\n}\,.\label{tmunuMax}\ee It is now
straightforward to check the invariance of the energy-momentum
$T_{\mu\nu}$ since $*^2F=-F$.

%%%%%%%%%%%%%%%%%%%%%%%%%%%%%%%%%%%%%%%%%%%%%%%%%%%%%%%%%%%%%%%%%%%%

\section{Brane electrokinematics}\label{electrok}

Having string theory in mind, it is natural to try to extend the
usual Maxwell point particles electrodynamics to incorporate
extended objects. The objects we will consider are the so-called
$p$-branes. They are ($p+1$)-dimensional submanifolds of the
$D$-dimensional spacetime manifold $\cal M$ (called the
\emph{target space} in this context) with the induced metric of
signature $(-,+,\ldots,+)$. One has $0\leq p+1\leq D$.

The limiting case $p=-1$ corresponds to \emph{instantons} which
are localized both in space \emph{and} time. The $0$-branes are
the standard relativistic particles while the $1$-branes are
called \emph{strings} and the $2$-branes have been christened
\emph{membranes}. This is summarized in the table \ref{branelexic}
\begin{table}[ht]
\begin{center}
\begin{tabular}{|c|c|c|}
  \hline
  $p$ & $p$-brane & worldhistory \\
  \hline
$-1$  & instanton & point\\
$0$  & particle & worldline\\
$1$  & string & worldsheet\\
$2$  & membrane & worldvolume\\
$3$  & $3$-brane & worldvolume\\
$\vdots$  & $\vdots$ & $\vdots$ \\
  \hline
\end{tabular}
\caption{$p$-brane dictionary \label{branelexic}}
\end{center}
\end{table}

In 1986, Teitelboim and Nepomechie already studied generalized
electrodynamics and the Dirac quantization condition in the
presence of magnetic extended sources
\cite{Teitelboim:1986i,Nepomechie:1985}.

A $p$-brane worldvolume is a manifold $\in\Lambda_{p+1}({\cal M})$
(= the space of all $(p+1)$-dimensional smooth orientable
submanifolds in the manifold ${\cal M}$) defined by the parametric
equations $x^\mu=x^\mu(\tau,\sigma^1,\dots,\sigma^p)$. We assume
the target space and the worldvolume to be smooth orientable
manifolds throughout this thesis. The index $\mu$ is the spacetime
index ($\mu=0,\dots,D-1$), $\tau$ is the time coordinate along the
brane worldvolume and $\sigma=(\sigma^1,\dots,\sigma^p)$ denote
the set of spatial coordinates.

The remaining of this section tries to motivate and to discuss on
geometrical grounds the introduction of $(p+1)$-form gauge fields
when considering $p$-branes. It is heuristic in spirit and is not
essential for understanding the following ``brane
electrodynamics".

\subsection{Gauge field theory of branes}

In a naive gauge field theory of branes \`a la Schr\"odinger, we
would consider ``wave-functionals" $\Psi[x(\sigma)]$ defined on
the configuration space $\Lambda_{p}({\cal M})$.
%We removed the time dependence by analogy with standard field theory defined on $\cal M$ which is interpreted as a theory of particles in a first quantization scheme.
%A second quantization would bring back the time evolution in the Feynman interpretation of processes.

We define a $p$-derivative $\frac{\delta}{\delta x^\mu(\sigma)}$
which generalizes the usual derivative $\frac{\partial}{\partial
x^\mu}$ and the loop-derivative $\frac{\delta}{\delta x^\mu(s)}$
\cite{Chan:1986}. It can be seen (without genuine mathematical
rigor) in the following way: Given $x^\nu(\sigma')$, we can make a
$\delta$-function variation in the direction $\mu$ at any point of
coordinates $\sigma$. This variation gives
$x^{'\nu}(\sigma')=x^\nu(\sigma')+\Delta\delta^\nu_\mu\delta(\sigma'-\sigma)$.
The derivative $\frac{\delta}{\delta x^\mu(\sigma)}$ of the
functional $\Psi[x]$, defined on $\Lambda_p({\cal M})$, is defined
by
\begin{equation}
\frac{\delta \Psi[x]}{\delta x^\mu(\sigma)}=\lim_{\Delta
\rightarrow 0} \frac{\Psi[x']-\Psi[x]}{\Delta}.
\end{equation}
The set of all $p$-derivatives $\{\frac{\delta}{\delta
x^\mu(\sigma)}\}$ provides a basis of the tangent space to
$\Lambda_p({\cal M})$.

Formally, we can now write a Dirac equation for a
``wave-functional" $\Psi[x(\sigma)]$ (invariant under Lorentz
transformation of the embedding Minkowski space) by extending
naturally the usual Dirac equation. Let us assume that this
``first quantized" theory possess a global invariance under an
internal Lie group $G$ (which, for generality, will be assumed to
be semi-simple) acting as
\begin{equation}
\Psi[x(\sigma)]\rightarrow \exp (ig\alpha)\,\Psi[x(\sigma)]
\label{intern}
\end{equation}
where $\alpha$ is an element of the matrix representation of the
Lie algebra ${\cal G}$ tangent to the group $G$. We would extend
(\ref{intern}) into a local gauge invariance, i.e. replace
$\alpha$ by a $x(\sigma)$-dependent matrix $\alpha[x(\sigma)]$ in
(\ref{intern}). In order to preserve the invariance of the ``Dirac
equation", we would replace the derivatives $\frac{\delta}{\delta
x^\mu(\sigma)}$ by covariant derivatives $\frac{\Delta}{\Delta
x^\mu(\sigma)}$.

To do so, let us introduce a ${\cal G}$-Lie-algebra-valued
connection 1-form $A[x(\sigma)]$ defined on $\Lambda_p(M)$. The
parallel transport of the ``wave-function" along a path
$x_u^\mu(\sigma)$ in $\Lambda_{p+1}(M)$, enters the theory through
the equation
\begin{equation}
\delta_\parallel
\Psi[x_u(\sigma)]=igA[x_u(\sigma)]\Psi[x_u(\sigma)]du,
\label{parallequ}
\end{equation}
where $g$ is a coupling constant. This equation defines the
infinitesimal parallel transport of the wave-function from
$x_u(\sigma)$ to $x_{u+du}(\sigma)$ as
\begin{equation}
\Psi_\parallel[x_{u+du}(\sigma)]=\Psi[x_{u+du}(\sigma)]+\delta_\parallel\Psi[x_u(\sigma)].
\end{equation}
Hence, the covariant derivative of the wave-function is given by
\begin{equation}
\frac{\Delta}{\Delta
x^\mu(\sigma)}\Psi[x_u(\sigma)]=\left(\frac{\delta}{\delta
x^\mu(\sigma)}+igA[x_u(\sigma]\right)\Psi[x_u(\sigma)]
\end{equation}
The searched-for covariantized ``Dirac equation" will be invariant
under the gauge transformation of the wave-function
\begin{equation}
\Psi[x(\sigma)]\rightarrow  U[x(\sigma)]\Psi[x(\sigma)],
\end{equation}
where $U[x(\sigma)]=\exp (ig\alpha[x(\sigma)])$ if, at the same
time, $A$ transforms as a connection
\begin{equation}
A[x(\sigma)]\rightarrow\left(\frac{i}{g}\,\delta U[x(\sigma)] +
U[x(\sigma)]A[x(\sigma)]\right)U^{-1}[x(\sigma)],
\end{equation}
where $\delta$ is the exterior differential on $\Lambda_p({\cal
M})$. As usual, this is the patching condition of the $A$ seen as
a local connection on the principal bundle $P(\Lambda_p({\cal
M}),G)$ of basis $\Lambda_p({\cal M})$ and of structure group $G$.

We can now parallel-transport the wave-function from $x_0(\sigma)$
to $x_1(\sigma)$ along the $p+1$-dimensional path $\Gamma_{p+1}$.
The solution of (\ref{parallequ}) is then
\begin{equation}
\Psi[x_1(\sigma)]=P_u \exp
\left[ig\int_{\Gamma_{p+1}}A\right]\Psi[x_0(\sigma)].
\label{transport}
\end{equation}
The symbol $P_u$ denotes path ordering in the $u$-parameter.

Up to now, we have no guarantee that the previous objects are
well-defined and that nothing prevents the existence of the
connection $A$. To address more rigorously this issue, we add a
requirement necessary to make contact with standard local field
theory: the requirement of \emph{locality}. As shown in
\cite{Teitelboim:1986,Henneaux:1986} this will impose strong
restrictions on the gauge group for $p\geq 1$.

\subsection{Antisymmetric tensor fields as Abelian
connections}\label{connection}

The locality requirement states that along the path
$\Gamma_{p+1}$, the connection $A$ should be given by
\cite{Henneaux:1986}
\begin{equation}
A[x_u(\sigma)]\,du=\frac{1}{(p+1)!}\int_{\Gamma^u_{p}}A_{\mu_1\dots\mu_{p+1}}(x_u)\,dx_u^{\mu_1}\wedge\dots\wedge
dx_u^{\mu_{p+1}}
\end{equation}
where $A_{\mu_1\dots\mu_{p+1}}$ are the components of a
$(p+1)$-form on the spacetime manifold $\cal M$, and the integral
is extended over the $p$-dimensional manifold $\Gamma^u_{p}\equiv
x=x_u(\sigma)$ ($u $ is fixed). The exterior product
$dx^{\mu_1}\wedge\dots\wedge dx^{\mu_{p+1}}$ is the tangent to the
path $\Gamma_{p+1}$, and is equal to
\begin{equation}
dx^{\mu_1}\wedge\dots\wedge dx^{\mu_{p+1}} =\frac{\partial
x^{[\mu_1}}{\partial \sigma^1}\dots\frac{\partial
x^{\mu_{p}}}{\partial \sigma^p}\frac{\partial
x^{\mu_{p+1}]}}{\partial u} d\sigma^1\wedge\dots\wedge
d\sigma^p\wedge du.
\end{equation}

An important constraint is that: in order to be well-defined, the
parallel transport operator in (\ref{transport}) must be
reparametrisation invariant, that is, it must remain invariant
under general coordinate transformations of the coordinates
$(u,\sigma^1,\dots,\sigma^p)\equiv(\xi^0,\dots,\xi^p)$ on
$\Gamma_{p+1}$. Teitelboim proved by geometrical arguments that
this is impossible for $p\geq 1$, except if the gauge group is
Abelian \cite{Teitelboim:1986}. In the Abelian case, the path
ordering on $u$ is no longer necessary and the ``integral" in the
exponential of the parallel transport in (\ref{transport}) becomes
the usual volume integral
$\int_{\Gamma_{p+1}}A_{\mu_1\dots\mu_{p+1}}(x)\,dx^{\mu_1}\wedge\dots\wedge
dx^{\mu_{p+1}}$, the reparametrisation invariance of which is well
known.

In other words, \emph{Abelian $(p+1)$-forms do not allow for
simple local non-Abelian extensions when $p\geq 1$}. The
possibility for $p=0$ is obvious, because it is nothing else than
the Yang-Mills theory. It can even be proved that, under certain
assumptions, the Yang-Mills theory is the unique local
consistent\footnote{In the chapter \ref{deformation} we will
explain in more details what we mean by ``consistent". Roughly it
says that the (deformed) theory should be free of: negative-energy
(ghost) propagating excitations, algebraic inconsistencies among
field equations, discontinuities in the degree-of-freedom content,
etc.} non-trivial deformation of the free Abelian theory
\cite{Barnich:1994}. In the papers
\cite{Henneaux:1997,Henneaux:1997i,Henneaux:1998,Henneaux:1999},
general local and continuous deformations of free Abelian forms of
rank higher than one have been classified at first order in the
coupling constant, using the powerful algebraic tools of homology
(reviewed in chapter \ref{deformation}), thereby dropping any
geometric prejudice. Though both already known and novel
deformations were discovered, none of them had the required
property that the gauge algebra becomes ``genuinely" non-Abelian
(i.e. at first order in the coupling constant $g$, as in the
Yang-Mills construction). The possibility of non-local (or
non-perturbative) extensions of antisymmetric tensor gauge fields
of course remains, but such a theory (if it exists) is still
lacking\footnote{The author of \cite{Nepomechie:1983} tried to
construct such a theory, but as he explaines, this attempt was
unsuccessful. The reason might be the small loop limit that this
author eventually takes. Indeed, in that limit all (possibly
unavoidable) non-local features of non-Abelian $p$-forms are
lost.}. Furthermore, it is not granted that such a theory could be
really interpreted as a field theory of a connection for extended
objects, as in the Abelian case explained here. Anyway, M-theory
seems to require a non-Abelian extension of chiral $2$-forms (see
chapter \ref{deformation}), hence there still exists some clues of
the existence of a field theory (maybe ``exotic") of non-Abelian
$(p+1)$-forms.

To conclude this subsection, let us briefly remark that a more
precise mathematical formulation of antisymmetric tensors as
connections for extended objects is provided by Abelian gerbes
\cite{Zunger:2000}. The Abelian $p$-gerbes are the generalization
of bundles for $(p+2)$-field strength. For $p=0$, they correspond
to the usual principal bundle formulation of point particle gauge
theory. Recently, a global approach to the study of duality
transformations has been undertaken in the geometrical setting of
gerbes \cite{Caicedo:2002}.

%%%%%%%%%%%%%%%%%%%%%%%%%%%%%%%%%%%%%%%%%%%%%%%%%%%%%%%%%%%%%%%%%%%

\section{de Rham currents and Dirac branes}\label{Diracproc}

We start by introducing a convenient tool to deal with charged
extended objects: the \emph{de Rham currents}\footnote{It may be
useful for the reader to first read the review given in the
introduction of appendix \ref{current} before going on, since this
formulation is not so frequent in the physics literature. Some
notations and definitions used in this section are presented in
the appendix \ref{current}. We will refer to them when they will
be needed.}.

\subsection{Extended sources as de Rham currents}

Let ${\cal W}_{p+1}$ be the worldvolume of a $p$-brane with charge
density $q$. The image of ${\cal W}_{p+1}$ by the Poincar\'e dual
map (\ref{Pdualmap}) is the differential form $P({\cal
W}_{p+1})=*\a_{p+1}$ of form degree $D-p-1$. Its current $J_p$ is
defined to be equal to $q\a_{p+1}$. Explicitly, one has
\begin{equation}
J^{\mu_1\ldots\mu_{p+1}}(x) = q\int_{{\cal W}_{p+1}}
\,\delta^{(D)}\left(x-W(\xi)\right) dW^{\mu_1} \wedge \ldots\wedge
dW^{\mu_{p+1}} \,,
\end{equation}
if $x^\mu=W^\mu(\xi)$ is the parametric equation of the $p$-brane
worldvolume ${\cal W}_{p+1}$.

For several, say $n$, branes of worldvolumes ${\cal
W}^{(i)}_{p+1}$ and charge $q_i$ ($i=1,\ldots,n$), we define the
manifold ${\cal W}_{p+1}$ to be the sum ${\cal
W}_{p+1}\equiv\sum_i {\cal W}^{(i)}_{p+1}$ (which is well defined
in $\Omega_*({\cal M};\mathbb Z)$). Without loss of generality the
submanifolds ${\cal W}^{(i)}_{p+1}$ are assumed not to intersect
with each other, ${\cal W}^{(i)}_{p+1} \cap{\cal
W}^{(j)}_{p+1}=\emptyset$ for $i\neq j$. We define the current
$J_p$ as the weighted sum $\sum_i q_i \a_{p+1}^{(i)}\equiv J_p$
with $P({\cal W}^{(i)}_{p+1})=*\a_{p+1}^{(i)}$. The Poincar\'e
dual forms $*\a_{p+1}^{(i)}$ are integer forms (This kind of
property will be extremely useful in the chapter \ref{chargeq}).

\subsection{Dirac branes}

From now on, we assume that \emph{the spacetime manifold has no
boundary, $\partial{\cal M}=0$}. Let us further assume that the
brane charge is conserved\footnote{The general case will be
considered the in subsection \ref{surgery}.}, that is $d*J_p=0$.
The Poincar\'e dual statement is that the worldvolume has no
boundary $\partial{\cal W}^{(i)}_{p+1}=0$. If the $(p+1)$-th
homology group of the spacetime manifold $\cal M$ is trivial, i.e.
$H_{p+1}({\cal M})=0$, then one finds that ${\cal W}^{(i)}_{p+1}$
is the boundary of some manifold ${\cal V}^{(i)}_{p+2}$ (see
appendix \ref{shomology}), called a \emph{Dirac $(p+1)$-brane}
worldvolume, ${\cal W}^{(i)}_{p+1}=\partial{\cal V}^{(i)}_{p+2}$.
The Dirac brane current $G$ is naturally taken to be $G\equiv
\sum_i q_i \b_{p+2}^{(i)}$ with $P({\cal
V}^{(i)}_{p+2})=(-)^{D-p-1}*\b_{p+2}^{(i)}$ such that $d*G=*J$ due
to the formula (\ref{Pmorphism}).

\subsection{Charge conservation}

A first application of algebraic topology considerations arises in
the definition of charge (it might be useful for the reader to
look at the subsection \ref{linking}). Let us take a
$(D-p-1)$-dimensional (spacelike) submanifold $\Sigma_{D-p-1}
\subset \cal M$ that intersects the manifold ${\cal W}_{p+1}$ at a
finite number of points. From the definition (\ref{intersection}),
\be \int_{\Sigma_{D-p-1}}*J=\sum_i \,q_i\, I(\Sigma_{D-p-1},{\cal
W}_{p+1}^{(i)})\,.\ee where $I(\Sigma_{D-p-1},{\cal
W}_{p+1}^{(i)})$ is the intersection between $\Sigma_{D-p-1}$ and
${\cal W}_{p+1}^{(i)}$. Therefore if $\Sigma_{D-p-1}^{(i)}$
\begin{description}
  \item[(i)] intersects the $p$-brane worldvolume ${\cal W}^{(i)}_{p+1}$ only
and
  \item[(ii)] is such that $\partial \Sigma_{D-p-1}^{(i)}$ wraps once
around ${\cal W}^{(i)}_{p+1}$,
\end{description}
then \be q_i=\int_{\Sigma_{D-p-1}^{(i)}}*J\,,\label{chargedef.}\ee
since from the definition (\ref{linkingn}) the linking number
$L(\partial\Sigma_{D-p-1}^{(i)},{\cal W}^{(i)}_{p+1})$ is equal to
one. More precisely, the charge (\ref{chargedef.}) is defined by
the linking homology class\footnote{For more information, see the
definition (\ref{equivr}) of linking homology class.}
$[\Sigma_{D-p-1}^{(i)}]$ with respect to ${\cal W}^{(i)}_{p+1}$,
whose linking number is equal to one. In general, the boundary
$\partial{\cal W}^{(i)}_{D-p-1}$ has the topology of a
$(D-p-2)$-sphere $S_{D-p-2}$ wrapping around the brane
worldvolume.

Let us assume that the spacetime can be foliated by time as ${\cal
M}={\mathbb R}\times \Sigma$ with $\Sigma(t)$ a spacelike slice at
time $t$ ($\Sigma_t\subset\Sigma$). As is standard, we say that
the charge $q_i$ is conserved because the charge $q_i(t)$ at time
$t$ is defined by a homology class $[\Sigma_{D-p-1}^{(i)}(t)]$
such that $\Sigma_{D-p-1}^{(i)}(t)\subset\Sigma(t)$ for all $t$.
The charge variation between two times $t_1$ and $t_0$, \be
q_i(t_1)-q_i(t_0)=\int_{\Sigma_{D-p-1}^{(i)}(t_1)}*J-\int_{\Sigma_{D-p-1}^{(i)}(t_0)}*J\,,\ee
is equal to minus the integral of the $p$-brane current on
$[t_0,t_1]\times
\partial \Sigma_{D-p-1}^{(i)}$, which is the volume covered by the
boundary of the manifolds $\Sigma_{D-p-1}^{(i)}(t)$ over the time
interval $[t_0,t_1]$. Since the manifolds
$\Sigma_{D-p-1}^{(i)}(t)$ are assumed to belong to a well-defined
linking homology class the charge variation vanishes. Indeed they
never intersect the $p$-brane, the variation vanishes and the
charge (\ref{chargedef.}) is indeed conserved over time.

%%%%%%%%%%%%%%%%%%%%%%%%%%%%%%%%%%%%%%%%%%%%%%%%%%%%%%%%%%%%%%%%%%%

\section{Linear electrodynamics}\label{linelectro}

The Maxwell equations describes the electrodynamics of charged
particles in $3+1$ dimensional spacetimes. It is easy to
generalize it to electrodynamics of higher dimensional objects
embedded in a spacetime of dimension other than $4$.

\subsection{Electrodynamics equations}

The electromagnetic field strength $F$ ruling the dynamics of
electric $p$-branes is expected to be a $(p+2)$-form. A direct
generalization of point particle electrodynamics provides the
following set of equation
\begin{equation}
d \left( \begin{array}{c} F\\ *F
\end{array}
\right) = \left( \begin{array}{c} *J_m\\ *J_e
\end{array}
\right)\,.\label{Max+}
\end{equation}
The electric and magnetic ``currents", $J_e$ and $J_m$ are
conserved since the action of the operator $d$ on both sides of
(\ref{Max+}) gives
\begin{equation}
d *\left( \begin{array}{c} J_m\\ J_e
\end{array}
\right) = 0 \,.
\end{equation}
The elementary extended objects are electric $p$-branes and
magnetic $\tilde{p}$-branes. A simple counting of form degrees
indicates that $\tilde{p}(D,p)=D-p-4$. The table \ref{pdual}
summarizes the first possibilities that arise.

\begin{table}[ht]
\begin{center}
\begin{tabular}{|c|c|c|c|c|c|}
  \hline
  $p$ & $\tilde{p}(D=2)$ & $\tilde{p}(D=3)$ & $\tilde{p}(D=4)$ & $\tilde{p}(D=5)$ & $\tilde{p}(D=6)$ \\
  \hline
  instanton & instanton & particle & string & membrane & $3$-brane \\
  particle & $-$ & instanton & particle & string & membrane \\
  string & $-$ & $-$ & instanton & particle & string \\
  membrane & $-$ & $-$ & $-$ & instanton & particle \\
  $3$-brane & $-$ & $-$ & $-$ & $-$ & instanton \\
  \hline
\end{tabular}
\caption{Electric $p$-branes and their respective magnetic
$\tilde{p}$-branes in $D$ dimensions. \label{pdual}}
\end{center}
\end{table}

Following \cite{Lechner:1999,Lechner:2000} one describes charged
sources in the appropriate mathematical scheme of de Rham currents
(see appendix \ref{current}). This is the task of next subsection.

\subsection{Electric and magnetic sources}

For simplicity, one will consider a single brane (For several
branes one introduces the appropriate sum over the branes.). In
that case ${\cal M}_e$ (${\cal M}_m$) denotes the worldvolume of
an electric (magnetic) brane of dimension $p+1$ (resp. $D-p-3$)
and charge density $e$ (resp. $g$). The image of ${\cal M}_e$
(${\cal M}_m$) by the Poincar\'e dual map (\ref{Pdualmap}) is
written $*\a_e$ (resp. $*\a_m$), and is of form degree $D-p-1$
($p+3$). The electric (magnetic) current $J_e$ ($J_m$) is equal to
$e\a_e$ ($g\a_m$).

To be more explicit, let the electric brane worldvolume ${\cal
M}_e$ be defined by the equation $X^\mu=X^\mu(\xi^a)$
($a=0,1,\ldots,p$). Then the electric current reads
\begin{equation}
J_e^{\mu_1\ldots\mu_{p+1}}(z) = e\int_{{\cal M}_e}
\,\delta^{(D)}\left(z-X(\xi)\right) dX^{\mu_1} \wedge \ldots\wedge
dX^{\mu_{p+1}} \,,\label{je}
\end{equation}
In the same way, if the parametric equation $x^\mu=Y^\mu(\zeta^m)$
($m=0,1,\ldots,D-p-4$) defines the magnetic brane worldvolume
${\cal M}_e$, the magnetic current explicitly reads
\begin{equation}
J_m^{\mu_1\ldots\mu_{D-p-3}}(z) = g\int_{{\cal M}_m}
\,\delta^{(D)}\left(z-Y(\zeta)\right) dY^{\mu_1} \wedge
\ldots\wedge dY^{\mu_{D-p-3}} \,.\label{jm}
\end{equation}

Let $\Sigma_{D-p-1}$ be a spacelike manifold such that $\partial
\Sigma_{D-p-1}^{(i)}$ wraps once around ${\cal M}_e$, then one can
use (\ref{Max+}) to express the electric charge as an electric
flux through the boundary of $\Sigma_{D-p-1}^{(i)}$ \be
e=\int\limits_{\Sigma_{D-p-1}}*J_e=\int\limits_{\partial\Sigma_{D-p-1}}*F\,.\ee
The analogue construction for the magnetic charge gives \be
g=\int\limits_{\Sigma_{p+3}}*J_e=\int\limits_{\partial\Sigma_{p+3}}F\,.\ee

As one can see, the de Rham current scheme indeed provides an
appropriate mathematical formulation of the standard definitions
(\ref{je}) and (\ref{jm}). Its main interest for the present
purposes is that the Poincar\'e duality provides a simple
translator of the brane \emph{geometrical} properties in terms of
\emph{algebraic} relations of their currents (that practically one
works with). As one will see, it also replaces tedious tensorial
manipulations by mere algebraic ones.

\subsection{Dirac procedure}

As previously seen in the section \ref{difff}, in order to find an
action principle for (\ref{Max+}) one should first solve $dF=*J_m$
in terms of a potential $A$. To that end, one follows the
remarkable idea of the seminal paper \cite{Dirac:1948} by
introducing the Dirac branes for magnetic charges.

For simplicity, one will assume from now on that ${\cal M}$ is
such that \emph{the homology group $H_*(\partial,{\cal M})$ is
trivial} (which is equivalent to ask that $H^*(d,{\cal M})$ is
trivial). In that case one can simply define a Dirac brane ${\cal
M}_D$ from the condition $\partial {\cal M}_D={\cal M}_m$ coming
from the fact that the magnetic worldvolume $\partial {\cal M}_m$
is closed. The Dirac brane current $G$ is equal to $*G=(-)^{p+1}
gP({\cal M}_D)$. More explicitly,
\begin{equation}
G^{\mu_1\ldots\mu_{D-p-2}}(z) = (-)^{p+1}g\int_{{\cal M}_D}
\,\delta^{(D)}\left(z-Z(\varsigma)\right) dZ^{\mu_1} \wedge
\ldots\wedge dZ^{\mu_{D-p-2}} \,.
\end{equation}
with $x^\m=Z^\m(\varsigma^r)$ ($r=0,\ldots,D-p-3$) the parametric
equation for the Dirac brane worldvolume ${\cal M}_D$.

Since the Poincar\'e dual map is a morphism of complexes by
(\ref{Pmorphism}), the equation $\partial {\cal M}_D={\cal M}_m$
becomes in the language of forms $d*G=*J_m$. This is very helpful
since now one has the equation $d(F-*G)=0$ which allows the
fieldstrength to derive from a potential $A$: $F=dA+*G$. Even if
$F$ is smooth in a neighborhood of the Dirac brane, there is a
singularity in $A$ at the location of the Dirac brane. This point
can be seen in explicit monopole solutions\footnote{At the
magnetic source is located an ``essential" singularity in the
explicit solution while only a ``coordinate" singularity takes
place at the Dirac brane location.}.

For Maxwell's electrodynamics ($D=4$, $p=0$) the Dirac brane is
the famous \emph{Dirac string} \cite{Dirac:1948}. In that case,
$-G$ can be physically interpreted as the magnetic field produced
by a semi-infinite extremely thin solenoid surrounding the Dirac
string which ends at the magnetic monopole location. The solenoid
brings from infinity the magnetic flux that evades from the
magnetic monopole. Thus, in the magnetic field configuration
corresponding to $F-*G$ the magnetic flux vanishes around the
magnetic monopole. So we are back in standard Maxwell's theory
where the existence of scalar and vector potentials results from
(\ref{M}) and (\ref{F}).

\subsection{Gauge freedom}

To end up with the ingenious procedure of \cite{Dirac:1948}, one
should remember that a Dirac brane is unphysical and that its
location must therefore be unobservable. Indeed, one has some
freedom in the choice of a Dirac brane because the equation
$\partial {\cal M}_D={\cal M}_m$ only defines the homology class
$[{\cal M}_D]$ of ${\cal M}_D$ since the homology is trivial.
Otherwise, one has the following freedom in the choice of the
Dirac brane: ${\cal M}_D\rightarrow{\cal M}_D+\partial{\cal V}$.
In terms of forms, one gets the new gauge freedom
\be*G\rightarrow*G+d*V\,,\label{gtransfo1}\ee where $*V=-gP({\cal
V})$. But the physics must remain unchanged, i.e. $F$ must be left
unaffected; that requires a simultaneous transformation of $A$ \be
A\rightarrow A+d\Lambda-*V\,.\label{gtransfo2}\ee The exact term
$d\Lambda$ is the usual ($p+1$))-form gauge field transformation.
In the literature, the third term is sometimes interpreted as a
(singular) gauge transformation. This comes from the fact that
outside $\cal V$, the differential form $*V$ is closed, and is
therefore locally exact.

\subsection{Action principle}

Let $A$ be an Abelian $(p+1)$-form gauge field (i.e. a local
Abelian connection of $p$-branes, or the connection of an Abelian
$p$-gerbe) the dynamics of which is given by the action \ba
&&S[A_{\mu_1\ldots\mu_{p+1}},G^{\mu_1\ldots\mu_{D-p-2}},J_e^{\mu_1\ldots\mu_{p+1}}]=\nn\\
&&\quad\quad=-\frac12\int *F\wedge F+(-)^{D-p-1} \int *J_e\wedge A
+ I_K,\label{linear} \ea where $I_K$ is a sum of kinetic terms for
the branes. The field equation obtained by varying $A$ is
precisely $d*F=*J_e$.

The energy momentum tensor is given by \be 2\,T_{\mu \nu }*1=
F_{\m }\,F_{\n}\, +\, (*F)_{\m}\,(*F)_{\n}\,,\label{tmn}\ee which
generalizes (\ref{tmunuMax}). For later purpose, one notes that
the trace of (\ref{tmn}) is equal to\be
T*1=\big(D-2(p+2)\big)\,*F\wedge F\,.\label{trace}\ee To get this,
(\ref{a*b}) and (\ref{niceid}) are used.

The variation of the action under a finite gauge transformation
(\ref{gtransfo1})-(\ref{gtransfo2}) is equal to \be \Delta
S=(-)^{D-p-1} \int *J_e\wedge *V=(-)^{D-p} \,eg\,L({\cal
M}_e,\Delta {\cal M}_D)\,,\ee with $L({\cal M}_e,\Delta {\cal
M}_D)$ the linking number between the electric brane worldvolume
${\cal M}_e$ and the Dirac brane worldvolume variation $\Delta
{\cal M}_D$ in the spacetime $\cal M$ (see subsection
\ref{linking}). Since this is a topological quantity, the action
is strictly invariant under infinitesimal gauge transformations.
Anyway, the e.o.m. do not depend on the choice of the Dirac
string, therefore the physics is independent of the location of
the string, as it should be. Therefore everything seems to be all
right; one succeeds to derive the electrodynamics equation
(\ref{Max+}) from an action principle. Still, one should care not
to have introduced dynamical inconsistencies with the Dirac brane.

Let us vary independently from $A$ the Dirac brane position
$Z^\m(\varsigma^r)$ without varying the magnetic brane
worldvolume, in geometrical terms $\d {\cal M}_m=0$. Hence $\d
{\cal M}_D=\partial {\cal V}$, where ${\cal V}$ is an
infinitesimal band, one border of which is the Dirac brane. One
defines the current $*V\equiv -gP({\cal V})$. Explicitly, \ba
&&V^{\mu_1\ldots\mu_{D-p-2}}(z) =\nn\\&& -g\int_{{\cal M}_D}
\,\delta^{(D)}\left(z-Z(\varsigma)\right) \frac{\partial
}{\partial\varsigma^0}\,Z^{[\mu_1}\ldots\frac{\partial}{\partial\varsigma^{D-p-3}}\,Z^{\mu_{D-p-2}}\,\delta
Z^{\mu_{D-p-3]}}\,. \nn\ea The variation of the action then
reads\be\d_Z S=(-)^{D-p}\int_{\cal M} d*F\wedge
*V=g(-)^{(D-p)p+1}\int_{\cal V}d*F\,.\ee since $\d *G=d*V$. A
sufficient condition to ensure that the equations of motion
obtained from varying $A$ and $G$ are consistent with each other
is the \emph{Dirac veto}: \underline{a Dirac brane must never
touch an electric brane}. Better, the Dirac brane e.o.m. is then a
consequence of the gauge field e.o.m., as one expects since the
Dirac brane is unphysical and should not introduce any new
dynamics.

The Abelian form $A$ is of mechanical dimension $L^{-D/2+p+2}$.
The gauge field $A$ is minimally coupled to the $p$-branes of
electric current $J_e$ by the term \be \int_{{\cal M}} *J_e\wedge
A =\sum\limits_i e_i\int_{{\cal M}^{(i)}_e} A\, \ee where the
coupling constants $e_i$ are the electric charge densities of
mechanical dimension $L^{D/2-p-2}$. The gauge field couples
non-minimally to magnetic $(D-p-4)$-brane current $J_m$ by the
Dirac $(D-p-3)$-brane of current $G$ which enters in the
combination $F=dA+*G$.

The brane kinetic term is taken to be the sum of the Nambu-Gotto
(NG) actions for the electric and magnetic branes, \be
I_K=S_{NG}[J_e^{\mu_1\ldots\mu_{p+1}}]+S_{NG}[J_m^{\mu_1\ldots\mu_{D-p-3}}]\,.\ee
The NG action, for a $p$-brane of worldvolume ${\cal
W}_{p+1}\equiv X^\mu=X^\mu(\xi^a)$ ($a=0,\ldots,p$), is
proportional to the ``proper" volume of ${\cal W}_{p+1}$
(Lorentzian signature). More precisely \ba
S_{NG}[J^{\mu_1\ldots\mu_{p+1}}]&=&-T_p\int\limits_{{\cal
W}_{p+1}} (*1)\nonumber\\
&=&-T_p\int\sqrt{-\det \left(g_{ab}
\right)}\,\,d^{p+1}\xi\,,\label{NGact}\ea where $T_p$ is the
tension (units $[L]^{-(p+1)}=[M][L]^{-p}$), $*1$ is the volume
form on the worldvolume ${\cal W}_{p+1}$ and $g_{ab}\equiv
G_{\mu\nu}\partial_a X^\mu
\partial_b X^\nu$ is the pulled back of the bulk metric $G_{\mu\nu}$.

With a kinetic term for the branes, they now become dynamical. The
position fluctuations of the electric $p$-brane satisfies \ba
&&T_p\,\frac{\partial}{\partial\xi^c}\left(\sqrt{-\det
\left(g_{ab} \right)}\,g^{cd}\,G_{\mu\nu}(X)\, \frac{\partial
X^\nu}{\partial\xi^d}
\right)+\nonumber\\&&+\frac{e}{(p+1)!}\,F_{\nu_1\ldots\nu_{p+1}\mu}(X)\,\frac{\partial
X^{\nu_1}}{\partial\xi^0}\,\ldots\frac{\partial
X^{\nu_{p+1}}}{\partial\xi^p}=0\,,\ea which is a generalized
``Lorentz force" formula. The magnetic $(D-p-4)$-brane position
$Y^\mu(\zeta^m)$ obeys \ba
&&T_{D-p-4}\,\frac{\partial}{\partial\zeta^q}\left(\sqrt{-\det
\left(g_{mn} \right)}\,g^{qr}\,G_{\mu\nu}(Y)\, \frac{\partial
Y^\nu}{\partial\zeta^r}
\right)+\nn\\&&+(-)^{D(p+1)}\frac{g}{(D-p-2)!}\,(*F)_{\nu_1\ldots\nu_{D-p-2}\,\mu}(Y)\,\frac{\partial
Y^{\nu_1}}{\partial\zeta^0}\,\ldots\frac{\partial
Y^{\nu_{D-p-2}}}{\partial\zeta^{D-p-3}}=0\,.\nn\ea

\subsection{Counting physical degrees of freedom}\label{degreesf}

Let $A_{[q]}$ be an Abelian $q$-form gauge field in a
$D$-dimensional spacetime. It undergoes a gauge transformation
$A_{[q]}\rightarrow A_{[q]}+d\Lambda_{[q-1]}$. This gauge
transformation is of reducibility order $q-1$ due to the chain of
reducibility identities \ba \delta A_{[q]}=0\,\quad &\mbox{when}&
\Lambda_{[q-1]}=d\Lambda_{[q-2]}\,,\nonumber\\
\Lambda_{[q-i]}=0\,\quad &\mbox{when}&
\Lambda_{[q-i]}=d\Lambda_{[q-i-1]}\,\quad
(i=2,\ldots,q-1)\,.\nonumber\ea As a remark, let us stress that
this chain of gauge and reducibility parameters finds a
geometrical interpretation in terms of transition functions
between patches for an Abelian $(q-1)$-gerbe \cite{Zunger:2000}.

The number of physical degrees of freedom of an Abelian $q$-form
gauge field living in $D$ dimensions is given by \be \left(
\begin{array}{c}D-2\\ q
\end{array}
\right)\equiv C^q_{D-2}\,.\ee A completely systematic proof
amounts to apply the Hamiltonian procedure of a system with
constraints, that is: take into account all primary and secondary
constraints that appear, compute their algebra and then, separate
them between first and second class constraints. Here, due to its
simplicity, we shortcut the detailed analysis and give the set of
Hamiltonian (all first class) and reducibility constraints \ba
&\Phi_{(0)}^{i_1\ldots
i_{q-1}}\equiv\pi^{0\,i_1\ldots i_{q-1}}=0\,,&\\
&\Phi_{(1)}^{i_1\ldots i_{q-1}}\equiv\partial_j\pi^{j \,i_1\ldots
i_{q-1}}=0&\\
&\Phi_{(2)}^{i_1\ldots i_{q-2}}\equiv\partial_j\Phi_{(1)}^{j\,
i_1\ldots i_{q-2}}=0\,,&\\
&\Phi_{(3)}^{i_1\ldots i_{q-3}}\equiv\partial_j\Phi_{(2)}^{j
\,i_1\ldots i_{q-3}}=0\,,& \\
&\vdots&\\
&\Phi_{(q)}\equiv\partial_j\Phi_{(q-1)}^j=0\,,& \ea where
$\pi^{\m_1\ldots\m_q}$ is the momentum conjugate to
$A_{\m_1\ldots\m_q}$. The set of constraints $\Phi_{(0)}$ is
irreducible: hence they count for $C^{q-1}_{D-1}$. The set of
constraints $\Phi_{(1)}$ (the ``Gauss law") is of order $q-1$.
This set is equivalent to a set of $C^{q-1}_{D-2}$ independent
constraints since \be
C^{q-1}_{D-1}-C^{q-2}_{D-1}+C^{q-3}_{D-1}-\ldots+(-)^{q+1}=C^{q-1}_{D-2}\,.\ee
The gauge field $A_{[q]}$ has $C^q_{D}$ independent components. To
end up, we subtract the number of first class constraints and
obtain the correct number of physical degrees of freedom:
$C^q_D-C^{q-1}_{D-1}-C^{q-1}_{D-2}=C^q_{D-2}$.

\subsection{Duality symmetry group}\label{dualitygroup}

For an electric-magnetic duality rotation to be meaningful, a
necessary condition is that the electric and magnetic fields
should have the same rank. This happens when \be
D=2p+4\,,\quad\quad (p=-1,0,1,2,\ldots)\label{dimension}\ee Then
the possibility of branes both electrically \emph{ and}
magnetically charged arises (as can be checked on table
\ref{pdual}). Such dyonic branes carrying both types of charge can
only exist in spacetimes of the appropriate dimensionality
(\ref{dimension}). Another property suggesting duality symmetry in
these dimensions is the matching of physical degrees of freedom
since \be \left(
\begin{array}{c}D-2\\ p+1
\end{array}
\right)= \left(
\begin{array}{c}D-2\\ D-p-3
\end{array}
\right)\,.\ee

These dimensions are also of particular interest because precisely
in these dimensions (\ref{dimension}) the coupling constant is
dimensionless. Another noteworthy related property is that
$F\wedge *F$ is invariant under pointwise conformal rescaling of
the metric as is reflected in the identical tracelessness of the
energy momentum tensor (Take a look at (\ref{trace})).

To find the duality symmetry group one follows the straightforward
derivation of \cite{Deser:1997}. At first sight, the system
(\ref{Max+}) of equations is invariant under any linear
transformation
\begin{eqnarray} &&\left(
\begin{array}{c} F\\ $*$F
\end{array}
\right) \rightarrow A\left( \begin{array}{c} F\\ $*$F
\end{array}
\right)
 \,,\\&&
\left( \begin{array}{c} J_m\\ J_e
\end{array}
\right) \rightarrow A\left( \begin{array}{c} J_m\\ J_e
\end{array}
\right)
 \,,\label{GLsource}
\end{eqnarray} with \be A=\left( \begin{array}{c c} a & b\\
c & d
\end{array}
\right)\,\in\,GL(2,{\mathbb R}) \,.\ee But one should remember
that $*^2 F=(-)^{p+1}F$ in our specific case; that implies\be
a=d\,,\quad b=(-)^{p+1}c\,.\label{requi}\ee Hence the
\emph{duality symmetry group of $p$-brane electrodynamics field
equations} is
\begin{itemize}
  \item ${\mathbb R}^{+}\times SO(2)$ for $p$ even, and
  \item ${\mathbb R}^{+}\times SO(1,1)\times {\mathbb Z}_2$ for $p$
  odd,
\end{itemize}
where the factor ${\mathbb Z}_2$ for $p$ odd corresponds to the
exchange $F\leftrightarrow*F$. The ${\mathbb R}^{+}$ factor
corresponds to global scale transformation.

The last requirement is that the duality symmetry should leave the
energy momentum tensor invariant . The transformation of
(\ref{tmn}) is \ba T_{\m\n}*1\,\rightarrow\,
(a^2+b^2)\,T_{\m\n}*1\,+\frac12\,ab\,\left(1+(-)^{p+1}\right)\left[*F_{\m}\wedge(*F)_{\n}\,
+\, *(*F)_{\m}\wedge F_{\n} \right]\,.\nn\ea This leads to the
\emph{duality symmetry group of $p$-brane electrodynamics}, which
is
\begin{itemize}
  \item $SO(2)$ for $p$ even, and
  \item ${\mathbb Z}_2\times {\mathbb Z}_2$ for $p$
  odd.
\end{itemize}
It was expected that the global scale transformation would be
eliminated. The more subtle issue is that hyperbolic rotations
{\emph do not} preserve the energy-momentum tensor. The extra
factor ${\mathbb Z}_2$ for $p$ odd that survived corresponds to
the trivial sign change $F\rightarrow -F$.

%%%%%%%%%%%%%%%%%%%%%%%%%%%%%%%%%%%%%%%%%%%%%%%%%%%%%%%%%%%%%%%%%%%

\section{Non-linear electrodynamics}\label{non-linear}

The investigation of non-linear electrodynamics can be traced back
to the years 1932-1934 with the birth of the Born-Infeld (BI)
action \cite{Born:1934}. The initial motivations for its
introduction were two-fold: 1) to have a unitary\footnote{In the
sense that Maxwell's electrodynamics theory is ``dual" because it
requires two independent ingredients: the electromagnetic field
and the charged sources. Born and Infeld failed in their attempt
to construct a theory of electric charges solely from the
electromagnetic field because of their misunderstanding of the
Dirac delta.} formulation of electromagnetism and 2) to find a
finite value fot the proper energy of an electron. A nice (but
somewhat old-fashioned) review of non-linear electrodynamics may
be found in \cite{Born:1939}.

The dual Lagrangian has been explicitly worked out by Born and
Infeld themselves in their original paper \cite{Born:1934} but
they did not insist on the symmetry property. The $SO(2)$ duality
symmetry of BI theory has really been pointed out by Schr\"odinger
one year after when he formulated the BI theory in terms of
complex electromagnetic fields \cite{Schrodinger}.

After an analysis of duality invariance of any non-linear
electrodynamics by Gaillard and Zumino \cite{Gaillard:1981},
Gibbons and Rasheed found the necessary and sufficient condition
to possess EM duality invariance \cite{Gibbons:1995}. Due to its
elegance and simplicity, we will briefly reproduce their main
derivations here.

\subsection{Action principle}

Let us start with the action
\begin{equation}
S[A_{\mu_1\ldots\mu_{p+1}},G_{\mu_1\ldots\mu_{D-p-2}}]=\int {\cal
L} (F_{\mu_1\ldots\mu_{p+2}})+(-)^{D-p-1}\int *J_e\wedge A.
\end{equation}The charged sources are here assumed to be fixed because their
dynamics plays no role. One defines the antisymmetric tensor
\begin{equation}
E^{\mu_1\ldots\mu_{p+2}}=-\frac{\partial L}{\partial
F_{\mu_1\ldots\mu_{p+2}}},\label{constitution}
\end{equation}
where we take the derivatives not treating the
$F_{\mu_1\ldots\mu_{p+2}}$ with permuted indices as independent
variables and ${\cal L}(F)=L(F)\,*1$. For ${\cal L}=-\frac12\,
*F\wedge F$ one has $E=F$. In general, a variation of $F$ gives
$\delta {\cal L}= -*E\wedge \delta F$ hence the non-linear
electrodynamics equations are
\begin{equation}
d \left( \begin{array}{c} F\\ *E
\end{array}
\right) =* \left( \begin{array}{c} J_m\\ J_e
\end{array}
\right)\,.\label{MaxII}
\end{equation} The energy-momentum tensor is found to be
\be T_{\mu \nu }*1= *E_{(\m}\wedge F_{\n)}\, +\, g_{\m\n}\,{\cal
L} \,.\ee

\subsection{Four dimensional nonlinear electrodynamics}

In four dimensions there is a simple physical interpretation of
the tensor field $E_{\m\n}$. Let us define the electric induction
vector $\bf D$ and the magnetic intensity $\bf H$ as \be D^{i}
=E^{0i} \,,\quad H^{i} =\frac{1}{2}\varepsilon
^{ijk}E_{jk}\label{definitionnn} \ee The equation (\ref{MaxII})
rewrites as
\begin{eqnarray*}
\bf{\nabla }.\bf{D} &=&\rho _{e} \\
-\bf{\nabla }\times \bf{E}
&=&\bf{j}_{m}+\frac{\partial \bf{B}}{\partial t}
\\
\bf{\nabla }.\bf{B} &=&\rho _{m} \\
\bf{\nabla }\times \bf{H} &=&\bf{j}_{e}+\frac{\partial
\bf{D}}{\partial t}
\end{eqnarray*}
These equations have exactly the form of the phenomenological
Maxwell's equations in the presence of exterior sources. The
definition (\ref{constitution}) now reads
\begin{equation}
\bf{D}=\frac{\partial L}{\partial \bf{E}}\ \quad\bf{H}=-%
\frac{\partial L}{\partial \bf{B}} \label{constitution'}
\end{equation}
which suggests to call (\ref{constitution}) the \emph{constitution
relations}. The energy density is given by\be {\cal
E}=\sqrt{|g|}\,T_{00}\,,\quad T_{00}=\bf{D\cdot E}-L\,.\ee

\subsection{Duality condition}

In order to speak about duality symmetry we have to restrict the
discussion to branes with $p=2k-2$ embedded in spacetimes of
dimension $D=4k$. Looking at (\ref{MaxII}) the duality rotation is
easily generalized to
\begin{eqnarray}
&&\left( \begin{array}{c} F\\ $*$E
\end{array}
\right) \rightarrow R\left( \begin{array}{c} F\\ $*$E
\end{array}
\right)
 \,,\nonumber\\&&
\left( \begin{array}{c} J_m\\ J_e
\end{array}
\right) \rightarrow R\left( \begin{array}{c} J_m\\ J_e
\end{array}
\right)
 \, ,\label{transfoII}
\end{eqnarray}
where $R\in SO(2)$ is a rotation of angle $\a$. The equation
(\ref{MaxII}) is formally invariant under (\ref{transfoII}).
Though we should not forget that $F$ and $E$ are not independent
fields. As a result it is necessary and sufficient to preserve the
constitution relation (\ref{constitution}) in order for the
duality rotation to be a symmetry of the non-linear
electrodynamics. For an infinitesimal duality rotation this
requirement leads to \cite{Gibbons:1995}
\begin{equation}
F\wedge F=E\wedge E + 2C*1\,, \label{dualitycond}\end{equation}
where $C$ is a constant of integration. We will refer to the
equation (\ref{dualitycond}) as the \emph{duality condition}. If
the Lagrangian goes to the quadratic Lagrangian in the weak field
limit: $C=0$. In four dimensions, the duality condition with $C=0$
reads
\begin{equation}
\bf{E\cdot B}=\bf{D\cdot H}
\end{equation}
due to (\ref{definitionnn}).

Furthermore the variation of the stress-energy-tensor under an
infinitesimal duality rotation of parameter $\delta\a$ is equal to
\cite{Gibbons:1995}
\begin{equation}
\delta T_{\m\n}=C\,g_{\m\n}.
\end{equation} Therefore \emph{the
vanishing of the constant $C$} in the duality condition is a
\emph{necessary and sufficient condition to get the
energy-momentum tensor invariant} under duality rotations.

As an example, non-linear Lagrangians reducing to the free theory
in the weak field limit and leading to duality invariant
electrodynamics equations have a duality invariant energy-momentum
tensor.

\subsection{Courant-Hilbert equation}

A nice feature in four dimensions is that only two independent
Lorentz invariants can be built from the field strength
$F_{\m\n}$. In that case, the condition (\ref{dualitycond})
translates into a non-linear partial differential equation in two
variables \cite{Gibbons:1995} which has later been re-obtained
from many different approaches in different (but related) contexts
\cite{Perry:1997,Schwarz:1997,Deser:1998,Bekaert:1998}.

We start from the manifestly gauge and Lorentz invariant action
$S[A_\m]=\int d^4x\, L(x,y).$ The function $L$ depends only on the
two independent Lorentz scalars constructed from the curvature
$F_{\m\n}$, namely \be x =
-\frac{1}{4}F_{\m\n}F^{\m\n}=\frac{1}{2}(\bf{E}^{2}-\bf{B}^{2})\,,\ee
and \be y =-\frac{1}{64}(F_{\m\n}*F^{\m\n})^2=-\frac{1}{4}\left(
\bf{E\cdot B}\right)^{2}\,.\ee An important physical
requirement is to recover Maxwell's theory in the weak field
limit. For this, we require $L(x,y)$ to be \emph{analytic} in the
neighborhood of $x=y=0$ and
\begin{equation}\label{e:weakfield}
L(x,y)=x+O(x^2,y).
\end{equation}
It turns out to be most convenient to pass to a Lorentz frame in
which both $\bf E$ and $\bf B$ are parallel. This is always
possible when ${\bf E\cdot B}\neq 0$. In such a case the two
invariants can be taken to be the norms of the electric and
magnetic field in this frame, $E=|{\bf E}|$ and $B=|{\bf B}|$. If
we ask for (*) duality invariance of the equations of motion
together with (**) the weak field limit (\ref{e:weakfield}), we
find the equation
\[
\frac{\partial L}{\partial E}\frac{\partial L}{\partial B}=-EB
\] where we simply used
the constitution relation (\ref{constitution'}).

We make the change of variable
\begin{equation}\label{e:xy}
x = u_+ + u_- \,,\;\; y = u_+ u_-.
\end{equation}
and obtain for the function \be f(u_+,u_-):=L(x,y)\ee the
remarkably simple first-order differential equation
\cite{Gibbons:1995,Perry:1997,Deser:1998,Bekaert:1998}
\begin{equation}\label{duality}
f_+ f_-=1.
\end{equation}
We will refer to this equation enforcing duality symmetry as the
\emph{Courant-Hilbert equation}. We just mention that,
surprisingly, this equation also appears as a condition on the
Hamiltonian for a self-interacting massless scalar field in four
dimensions to be Lorentz covariant \cite{Deser:1998}. The
Courant-Hilbert equation will reappear several times in this
thesis.

For instance the Courant-Hilbert equation has also been obtained
for a chiral two-form in six dimensions, with a choice of the
fifth direction in \cite{Perry:1997}, or the time direction in
\cite{Bekaert:1998}. This comes from the fact that dimensional
reduction of a chiral two-form from six to four dimensions gives
duality-symmetric electrodynamics, at the linear
\cite{Verlinde:1995} and non-linear \cite{Berman:1997} level. This
will be explained in more details in the chapter \ref{manifestd}.

Extending the analysis of \cite{Perry:1997}, we determine the
necessary and sufficient set of conditions for the analyticity (in
the weak field limit) of solutions to this differential equation
in next subsection.

\subsection{Analytic solutions to the Courant-Hilbert \\
equation}\label{CHsols}

We have seen that the Courant-Hilbert equation is central in the
study of EM duality symmetric systems in 4 dimensions, and that it
reappears in many different approaches. This provides a strong
motivation to study in detail the solutions of this simple (but
highly non-linear) differential equation.

As pointed out in \cite{Perry:1997}, the general solution of
(\ref{duality}) has been given by Courant and Hilbert
\cite{courant}.\footnote{Of course, this is the reason why we
refer it as the Courant-Hilbert equation. Let us mention that the
authors of \cite{Hatsuda:1999} gave an interesting alternative
form of the general solutions.} But the general solution is given
only implicitly in terms of an arbitrary function $z(t)$:
\be\left\{\begin{array}{c} f = {2u_+\over \dot z(t)} + z(t)\,, \\
u_- = {u_+\over (\dot z(t))^2} + t.
\end{array}\right.\label{e:general}\ee
The dot means the derivative of the function with respect of its
argument. In principle, the second equation determines $t$ in
terms of $u_+$ and $u_-$, which can then be substituted into the
first one to give $f$ in terms of $u_+$ and $u_-$. Unfortunately,
in practice this method for generating solutions is not tractable
for arbitrary $z(t)$.

Furthermore, we should not forget that $L(x,y)$ is required (i) to
be analytic at the origin and (ii) to obey (\ref{e:weakfield}).
The following theorem states that these two requirements can be
equivalently translated into a precise condition on the generating
function $z(t)$ \cite{Perry:1997,Bekaert:2001}.

\cadre{theorem}{t:analyticity}{Let $f(u_+,u_-)$ be a solution of:
$f_+f_-=1$. The function $L(x,y)\equiv
f\left(u_+(x,y),u_-(x,y)\right)$
\begin{itemize}
\item is analytic near $(x,y)=(0,0)$ and
\item satisfies $L(x,y)=x+O(x^2,y)$,
\end{itemize}
{\bf if and only if} the boundary condition $L(t,0) \equiv z(t)$
is such that the function $\Psi(t)\equiv -t\dot{z}^2(t)$
\begin{itemize}
\item[(i)] is equal to its inverse: $\Psi\left(\Psi(t)\right)=t$,
\item[(ii)] is distinct from the identity: $\Psi(t)\neq t$,
\item[(iii)] is analytic near the origin $t=0$ and
\item[(iv)] vanishes at the origin: $\Psi(0)=0$.
\end{itemize}}

\proof{\noindent
\begin{description}
\item[$\underline{\,\bf\Rightarrow:}$] It has been shown by
Perry and Schwarz that $(i)-(iv)$ were necessary
\cite{Perry:1997}. We will not reproduce their proof here.
\item[$\underline{\,\bf\Leftarrow:}$] The proof that it is also sufficient was given in
\cite{Bekaert:2001}. It amounts to glue together the lemmas
\ref{l:analytic}-\ref{l:z} with the last remark of appendix
\ref{l:analyticity}. The proof is a bit lengthy because we make
use of $f(u_+,u_-)$ as an intermediate function to propagate an
equivalence between the analyticity conditions on Lagrangian
density $L(x,y)$ that generates duality-invariant equations of
motion and the conditions (i)-(iv) on the generating function
$z(t)=L(t,0)$.
\end{description}}
One of the main physical interests of the theorem
\ref{t:analyticity} is to allow the proof of the
\begin{corollary}\label{infiniteclass}There exists an infinite class of
four-dimensional electromagnetism Lagrangians which
\begin{itemize}
  \item are gauge and Lorentz invariant,
  \item approach Maxwell's Lagrangian in the weak-field regime,
  \item are analytic functions in the weak-field regime, and
  \item have duality-symmetric equations of motion.
\end{itemize}\end{corollary} Although only a single explicit example is known (the BI
Lagrangian) there exists an \emph{infinite} class of physically
relevant duality-symmetric theories. To see this, we use the
procedure given by Perry and Schwarz \cite{Perry:1997} to generate
a large class of solutions for $\Psi\left(\Psi(t)\right)=t$. We
prove the
\begin{proposition}\label{implicite}
Let $F(s,t)$ be an analytic function near the origin,
\begin{itemize}
\item symmetric in its two arguments $F(s,t)=F(t,s)$ and
\item such that $F(s,t)=s+t+O(s^2,t^2,st)$.
\end{itemize}
Then the implicit equation \ba F(s,t)=0\quad\Leftrightarrow\quad
s=\Psi(t) \nonumber\ea defines a function $\Psi(t)$ satisfying the
conditions $(i)-(iv)$ of theorem \ref{t:analyticity}.
\end{proposition}
The proof is straightforward. The property (i) comes from the
symmetry property of $F$. Indeed $F(s,t)=F(t,s)=0$ implies
$t=\Psi(s)=\Psi\left(\Psi(t)\right)$, that is, the implicit
function $\Psi(t)$ is equal to its inverse. The analyticity near
the origin of the function $\Psi(t)=-t+O(t^2)$ follows as a mere
application of the implicit function theorem \cite{Dieu}.

\subsection{Born-Infeld electrodynamics}

Let us consider the simplest non-trivial example of functions
satisfying the assumption of proposition \ref{t:analyticity} to
generate analytic solutions: $F(s,t) = s + t + \alpha s t$. It
generates the BI electrodynamics at the end of the whole procedure
($\a=0$ corresponds to Maxwell's theory, which is not considered
here as a distinct example of solution). Explicitly, \be f
=-\frac{2}{\alpha }\left( 1-\sqrt{\left( 1+\alpha u_{+}\right)
\left( 1+\alpha u_{-}\right) }\right) \label{BIuseful}\ee or \be
L=-\frac{2}{\alpha }\left( 1-\sqrt{1+\alpha x+\alpha ^{2}y}\right)
\ee Let us denote $\alpha =-2b^{2}$, one finds
\[
L=\frac{1}{b^{2}}\left\{ 1-\sqrt{1+b^{2}\left( \bf{B}^{2}-\bf{E}%
\right) ^{2}-b^{4}\left( \bf{E\cdot B}\right) ^{2}}\right\}
\]
which is the BI Lagrangian in its original form. The constant
$1/|b|$ was called \emph{absolute field}.

Unfortunately, this procedure for generating solutions becomes
rapidly cumbersome and no other explicit example of
duality-symmetric theory is known. Anyway, our theorem shows that
duality invariance together with analyticity is not enough to
single out uniquely BI theory\footnote{The authors of
\cite{Deser:1998i} imposed both duality and shock-free
propagation. These two requirements single out BI, without even
requiring the solutions to reduce to Maxwell in the weak field
limit (something that is used to select BI in the derivations of
each separate demand).}, contrary to what might have been
conjectured from the fact that only one explicit example is known.
The theorem \ref{t:analyticity} ensures that we can generate
implicitly an infinite class of analytic solutions at the origin.

It is standard to express the BI action in a determinant form. We
take a coordinate system such that the metric $G_{\m\n}$ at the
point $x$ is equal to $\eta_{\m\n}$, hence ${\cal
L}(x)\,=\,L(F)\,d^{D}x$. Moreover one has still enough freedom to
take the coordinate system such that ${\bf E}(x)$ and ${\bf B}(x)$
are parallel, we take the two invariants to be the norms of the
electric and magnetic field in this frame: $E$ and $B$. In this
coordinate, the field strength at the point $x$ takes the form
\[
F_{\mu \nu }(x)=\left(
\begin{array}{cccc}
$0$ & $-E$ & $0$ & $0$ \\
$E$ & $0$ & $0$ & $0$ \\
$0$ & $0$ & $0$ & $B$ \\
$0$ & $0$ & $-B$ & $0$
\end{array}
\right)
\] if one takes the $x^1$-axis to be parallel to ${\bf E}(x)$. One
computes the determinant \[ \det (\eta _{\mu \nu }+bF_{\mu \nu
})=\det \left(
\begin{array}{cccc}
$-1$ & $-bE$ & $0$ & $0$ \\
$bE$ & $1$ & $0$ & $0$ \\
$0$ & $0$ & $1$ & $bB$ \\
$0$ & $0$ & $-bB$ & $1$
\end{array}
\right) =-1+b^2(E^{2}-B^{2})+b^4(EB)^{2}\,.
\]
Therefore the BI action can be written as
\begin{equation} S_{BI}
=\int d^{4}x\frac{1}{b^{2}}\left\{ 1-\sqrt{-\det (G_{\mu \nu
}+bF_{\mu \nu })}\right\} \,  \label{B-I 0}
\end{equation}

Due to its square root form, the BI theory is roughly speaking to
the Maxwell theory what relativistic particles are to Newtonian
particles. In fact, this analogy was the guiding principle that
led Mie, in 1912, to write a first version of non-linear
electrostatics, followed by 1932 Born's version for a non-linear
electrodynamics (without the $b^4$ term). It was the
electromagnetic analogue to the replacement of the Newtonian
Lagrangian $\frac12 m{\bf v}^2$ by the relativistic expression
$mc^2(1-\sqrt{1-{\bf v}^2}/c^2)$ (with $v^i=\frac{dx^i}{dx^0}$).

The relativistic particle velocity and the BI electric field share
a common ``limiting principle". The velocity is limited by the
speed of light, $|{\bf v}|\leq c$. For the BI theory, the electric
field norm $E$ is bounded by the absolute field, $E\leq
\frac{1}{|b|}$. To see this, let $u_+$ and $u_-$ be the two roots
of the second order polynomial in $u$
\begin{equation}
R(u)\equiv u^2-x\,u+y=(u-u_+)(u-u_-)\,.\label{roots}
\end{equation}
This definition of $u_{\pm}$ is equivalent to (\ref{e:xy}). Now we
notice that \be\a^2R(-1/\a^2)=\left( 1+\alpha u_{+}\right) \left(
1+\alpha u_{-}\right)\,,\ee which should be positive since it is
under the square root in (\ref{BIuseful}). Moreover, \be R(u)\geq
0\quad \Leftrightarrow\quad u\leq u_- \quad\mbox{or}\quad u_+\leq
u\,.\ee The limiting value for the electric field follows.

Above all, this analogy is crowned by the ``unification" of the NG
action (that generalizes the relativistic particle action for
extended objects) and the BI action in the Dirac-Born-Infeld
action describing the fluctuations of the D-brane.

%%%%%%%%%%%%%%%%%%%%%%%%%%%%%%%%%%%%%%%%%%%%%%%%%%%%%%%%%%%%%%%%%%%

\section{$SL(2,\IR)$ symmetry}\label{sl2r}

The EM duality symmetry can be generalized further in the presence
of scalars, as in many supergravity theories. For instance, in 4
dimensions the bosonic sector of ${\cal N}=4$ supergravity and
string theory compactified on a six-torus may be described, at
lowest order, by the following Lagrangian \cite{Cremmer:1978} \be
L = R - \frac12\,\left(\nabla\phi\right)^2 - \frac12
e^{2\phi}\left(\nabla a\right)^2 +
\frac14\,a\,F_{\mu\nu}(*F)^{\mu\nu} -
\frac{1}{4}\,e^{-\phi}F_{\mu\nu}F^{\mu\nu} \label{LowEn} \ee where
for simplicity we consider only a single $U(1)$ gauge field.

The resulting theory admits an $SL(2,{\mathbb R})$ duality
symmetry which mixes the electromagnetic field equations with the
Bianchi identities and also transforms the axion $a$ and dilaton
$\phi$.

\subsection{Theta angle and Witten effect}

If we restrict ourselves to bosonic $2k$-brane electrodynamics in
dimension $D=4(k+1)$, one can add a parity-breaking gauge
invariant term proportional to $F\wedge F$ in the Lagrangian, the
so-called $\theta$-term.\footnote{It is also possible to add
$\theta$-like terms in dimensions $2$ modulo $4$ if there are
several Abelian forms \cite{Deser:1998ii}.}

For later convenience and in order to make contact with standard
conventions in this context, we make a standard redefinition of
the fields $F\rightarrow 1/e\,F$. Therefore the charges underwent
$e\rightarrow 1$ and $g\rightarrow g/e$. This provides the
following linear action of $4(k+1)$-dimensional $2k$-brane
electrodynamics \be
S[A_{\mu_1\ldots\mu_{2k+1}},G^{\mu_1\ldots\mu_{2(k+1)}}]=-\frac{1}{2\,e^2}
\int *F\wedge F+\frac{\theta}{4\pi} \int F\wedge
F-\int\limits_{{\cal M}_e} A\label{sl2action} \ee where $e$ and
$\theta$ are dimensionless coupling constants. The constant
$\theta$ is known as the \emph{vacuum angle}. The minimal coupling
term is gauge invariant up to a boundary term. As in the previous
section, the charged sources are assumed to be fixed (This
assumption will also be done in the following sections). In four
dimensions this action is obtained from (\ref{LowEn}) with given
expectation values of the axion and dilaton.

If there is no magnetic source the curvature form $F$ is exact and
the equations of motion are not modified by the presence of the
$\theta$-term since $\int dA\wedge dA$ is a boundary term. If
there are magnetic sources such that their Dirac branes do not
intersect, then then the identity $*G\wedge *G\equiv 0$ holds. We
can rewrite the action as
\begin{equation}
S=-\frac{1}{2e^2} \int *F\wedge F +\int
A\wedge(*J_e+\frac{\theta}{2\pi}\, *J_m)\,.
\end{equation}
The electrodynamics equation derived from this action are now
\begin{equation}
d \left( \begin{array}{c} F\\ \frac{1}{e^2}\,*F
\end{array}
\right) = \left( \begin{array}{c} *J_m\\
*J_e+\frac{\theta}{2\pi}\, *J_m
\end{array}
\right)\,.\label{MaxIII}
\end{equation}
We notice that in the case of a dyonic brane of magnetic charge
$g$, its electric charge is shifted by an amount $\frac{\theta
g}{2\pi}$. Pointed out for the first time in \cite{Witten:1979},
this charge shift is the by now celebrated \emph{Witten effect}.

\subsection{M\"obius transformation}

It is standard to combine the coupling constants into the complex
modulus
$$ \tau =\tau_1+i\tau_2= \frac{\theta}{2\pi} + {i\over e^2}\,.$$
Since one works with complex numbers, one introduces the complex
fieldstrength $$\overline{F}\equiv F+i*F\,.$$ Then the
electrodynamics equation (\ref{MaxIII}) can be rewritten as
\begin{equation}
d \,\,\mbox{Re}\left( \begin{array}{c} \overline{F}\\
-\tau\,\overline{F}
\end{array}
\right) = *\,\left( \begin{array}{c} J_m\\
J_e
\end{array}
\right)\,.\label{MaxIIII}
\end{equation}
It is convenient to define a complex 2-component vector $\psi$ by
\be \psi = \left({1\atop -\tau}\right). \ee The electrodynamics
equation are left invariant by the (formal) duality transformation
defined by $\psi \rightarrow A\,\psi$ with $A\in GL(2,{\mathbb
R})$ if the sources transform as in (\ref{GLsource}).

One can check that the energy momentum tensor is unchanged under
the action of $SL(2,{\mathbb R})$ \cite{Gibbons:1996}. If \be A =
\left(
\begin{array}{cc}
p & q \\
r & s
\end{array}
\right) \quad\mbox{where  }ps-qr=1, \ee then the induced
transformations of the coupling constants are given by a M\"obius
transformation of $\tau$ \be \tau \rightarrow {p\,\tau+q\over
r\,\tau+s}. \label{axdiltransf} \ee

%%%%%%%%%%%%%%%%%%%%%%%%%%%%%%%%%%%%%%%%%%%%%%%%%%%%%%%%%%%%%%%%%%%

\section{``Brane surgery"}\label{surgery}

The presence of CS terms in the dynamics of gauge fields is one of
the unusual properties of supergravity theories. A dual property
is that the Bianchi identities are ``modified".\footnote{The
section title is the one of a Townsend's paper
\cite{Townsend:1997} from which the subsequent discussion is
partly inspired.} Such CS terms complicate our understanding of
charge\footnote{Three distinct type of charge definitions that
appeared in the strings/M physics literature were identified in
the reference \cite{Marolf:2000}.} since they allow an ``opening
of brane". This section attempts to clarify this issue by
generalizing the discussion made in section \ref{Diracproc}.

Let the submanifold ${\cal W}_{p+1}\subset{\cal M}$ be defined by
the equation $W^\mu=W^\mu(\xi^a)$ ($a=0,1,\ldots,p$). The
submanifold ${\cal W}_{p+1}$ is the worldvolume of an arbitrary
$p$-brane with charge density $q_p$. If $P({\cal
W}_{p+1})=*\a_{p+1}$, then the current $J_p$ is defined to be
equal to $J_p=q_p\a_{p+1}$. Explicitly,
\begin{equation}
J_p^{\mu_1\ldots\mu_{p+1}}(x) = q_p\int_{{\cal W}_{p+1}}
\,\delta^{(D)}\left(x-W(\xi)\right) dW^{\mu_1} \wedge \ldots\wedge
dW^{\mu_{p+1}} \,.
\end{equation}

For simplicity, we explicitly cover the case of a single brane of
each type at a time because it is straightforward to generalize
the procedure for an arbitrary configuration.

\subsection{Closed brane}

This case is the simplest one since charge is always conserved. We
already covered it with many details in the section
\ref{Diracproc}. Physically speaking, this case corresponds to a
brane electrodynamics with no CS coupling.

\subsection{Open brane}

Let $\overline{\cal W}_p$ be the boundary of the (so-assumed) open
$p$-brane worldvolume, i.e. $\partial{\cal W}_{p+1}=\overline{\cal
W}_p$. By definition the $(p-1)$-brane current $\bar{J}_{p-1}$
associated to the worldvolume $\overline{\cal W}_p$ is set equal
to $\bar{J}_{p-1}=\overline{q}_{p-1}\overline{\a}_p$ where
$P(\overline{\cal W}_p)=*\overline{\a}_p$. Let us parametrize the
worldvolume $\overline{\cal W}_p$ of the closed $(p-1)$-brane by
$w^\mu=w^\mu(\zeta^\a)$ ($\a=0,1,\ldots,p-1$). The current
explicit expression reads\begin{equation}
\bar{J}_{p-1}^{\mu_1\ldots\mu_p}(x) =
\overline{q}_{p-1}\int_{\overline{\cal W}_p}
\,\delta^{(D)}\left(x-w(\zeta)\right) dw^{\mu_1} \wedge
\ldots\wedge dw^{\mu_p} \,.
\end{equation}

The $p$-brane charge non-conservation and the $(p-1)$-brane charge
conservation read\be d*J_p=\a *\bar{J}_{p-1}\,,\quad\quad
d*\bar{J}_{p-1}=0\,,\label{openbranecons}\ee The proportionality
coefficient $\a$ will be called the \emph{Chern-Simons
coefficient}. Equations like (\ref{openbranecons}) appear for
$p$-brane electrodynamics with linear Chern-Simons coupling
considered in section \ref{qCS}. In this example, the magnetic
$(p-1)$-branes live at the boundary of electric $p$-brane
worldvolumes.

Putting everything together we find that the charge densities must
satisfy the relation\be q_p=(-)^{D-p}\,\a\,
\overline{q}_{p-1}\,.\ee

This discussion can be formulated in the setting of relative
(co)homology, as we do below. Nevertheless this mathematical
reformulation is not mandatory for further understanding and can
be skipped by the reader in a first reading.

\subsubsection{Relative (co)cycle condition}

Mathematically speaking, the set of two equations in
(\ref{openbranecons}) states that the relative $(D-p)$-form $(\a
*\bar{J}_{p-1},*J_p)$ is a cocycle of the cohomology
$CH^{D-p}({\cal M})$ of the de Rham complex cone
$C\Omega^{D-p}({\cal M})$ since\footnote{For more details
concerning relative cohomology we refer to sections
\ref{complexes} and \ref{toutestrelatif}.} \be d(\a
*\bar{J}_{p-1},*J_p):=(\a\,
d*\bar{J}_{p-1}\,,\,\a*\bar{J}_{p-1}-d*J_p)=(0,0)\,.\ee In fact,
this is the translation by Poincar\'e of the fact that
$(\overline{\cal W}_p,{\cal W}_{p+1})$ is a $(p+1)$-cycle of the
$CH_{p+1}({\cal M})$ homology group of the cone
$C\Omega_{p+1}({\cal M})$, that is \be
\partial(\overline{\cal W}_p,{\cal
W}_{p+1}):=(\partial\overline{\cal W}_p,\overline{\cal
W}_p-\partial{\cal W}_{p+1})=(0,0)\,.\label{suggests}\ee

\subsection{Brane ending on brane}

More specifically we consider the case of an open $p$-brane ending
on a closed $q$-brane. The (absolute) boundary $(p-1)$-brane
worldvolume $\overline{\cal W}_p$ is a submanifold of a
$(q+1)$-dimensional closed brane worldvolume ${\cal V}_{q+1}$
\cite{Strominger:1996}, that is $\overline{\cal W}_p \subset {\cal
V}_{q+1}$ with $\partial{\cal V}_{q+1}=0$. We consider $p\leq q$
since the case $p=q+1$ is the previous case, called ``open brane"
case.

An example of ``brane ending on brane" case is the $p$-brane
electrodynamics in $(3p+5)$ dimensions with quadratic $CS$
coupling that will be considered in section \ref{cCS}. The $EM$
duals to electric $p$-branes are magnetic $(2p+1)$-branes, on
which electric branes can end. At the boundary of the electric
brane, lives a dyonic $(p-1)$-brane.

Obviously the $q$-brane current $J_q$ is defined to be equal to
$J_q=q_q\a_{q+1}$ with $P({\cal V}_{q+1})=*\a_{q+1}$. Let
$P_{{\cal V}_{q+1}}:\Omega_p( {\cal V}_{q+1})\rightarrow
\Omega^{q-p+1}({\cal V}_{q+1})$ be the Poincar\'e dual map on
${\cal V}_{q+1}\in H_q({\cal M})$ and let
$\overline{*}:\Omega^p({\cal V}_{q+1})\rightarrow
\Omega^{q-p+1}({\cal V}_{q+1})$ be the Hodge dual operator on
${\cal V}_{q+1}$. Since the $(p-1)$-brane is confined on the
$q$-brane, it is natural to define the $(p-1)$-brane current as
$\bar{J}_{p-1}=\overline{q}_{p-1}\overline{\a}_p$, where $P_{{\cal
V}_{q+1}}(\overline{\cal W}_p)=\overline{*}\,\overline{\a}_p$.
Since the worldvolumes ${\cal V}_{q+1}$ and $\overline{\cal W}_p$
have no boundary, one has \be d*J_q=0\,,\quad
d\bar{*}\bar{J}_{p-1}=0\label{brol}\,.\ee The Chern-Simons
coefficient is defined by the relation
\begin{equation}q_p=(-)^{D-p}\,\a\,q_q\,\overline{q}_{p-1}\,,\label{CScoeff}
\end{equation} the choice of which is such that \be \mbox{\begin{tabular}{|c|}
  \hline\\
  $d*J_p=\a *J_q\wedge
\bar{*}\bar{J}_{p-1}\,,$ \\\\
  \hline
\end{tabular}}\label{comparedwith}
\end{equation} since $(d \circ P)({\cal
W}_{p+1})=(-)^{D-p}P(\overline{\cal W}_p)$ and $P({\cal
V}_{q+1})\wedge P_{{\cal V}_{q+1}}=P$ from the properties
(\ref{Pmorphism}) and (\ref{wedgeP}).

\subsubsection{Surgical graft}

There exists a surgical operation that cures the charge
non-conservation illness forbidding a naive definition of the
$p$-brane charge as in subsection \ref{Diracproc}. Let
$\overline{\cal V}_{p+1}\subset {\cal V}_{q+1}$ a Dirac $p$-brane
worldvolume attached to the $(p-1)$-brane closed worldvolume
$\overline{\cal W}_p$. The Dirac $p$-brane worldvolume (embedded
in the $q$-brane worldvolume) is seen as the necessary organ to
graft on to the $p$-brane worldvolume so that the cured brane
${\cal W}_{p+1}-\overline{\cal V}_{p+1}$ has no
boundary\footnote{This surgical operation was inspired by a
comment made by Sorokin at the end of \cite{Sorokin:1998} in the
context of M-branes couplings.} because this surgical operation
allows a well-defined definition of $q_p$.

The Dirac $p$-brane current $\bar{G}$ is naturally taken to be
$\bar{G}\equiv \overline{q}_{p-1} \overline{\b}_{p+1}$ with
$P(\overline{\cal V}_{p+1})=(-)^{q-p+1}*\b_{p+1}$ such that
$d\bar{*}\,\bar{G}=\bar{*}\bar{J}_{p-1}$. From the definitions, we
have \be *J_p+(-)^{D-q}\a*J_q\wedge \bar{*}\,\bar{G}=q_p\, P({\cal
W}_{p+1}-\overline{\cal V}_{p+1})\,.\label{krakoukas}\ee

Let us take a spacelike $(D-p-1)$-dimensional manifold
$\Sigma_{D-p-1}\subset \cal M$ in the linking number one homology
class of ${\cal W}_{p+1}-\overline{\cal V}_{p+1}$, then a
definition of the $p$-brane charge analogous to (\ref{chargedef.})
is \ba
q_p&=&\int\limits_{\Sigma_{D-p-1}}(*J_p+(-)^{D-q}\a*J_q\wedge
\bar{*}\,\bar{G})\nn\\&=&\int\limits_{\Sigma_{D-p-1}}*J_p\,+\,(-)^{D-q}\,\a
q_q\int\limits_{{\Sigma_{D-p-1}}\cap {\cal V}_{q+1}}
\bar{*}\,\bar{G}\,.\label{qp}\ea Basically, this definition works
because \be d(*J_p+(-)^{D-q}\a*J_q\wedge \bar{*}\,\bar{G})=0\,.\ee
Of course, physically speaking, this definition is a bit odd since
the Dirac $p$-brane is unphysical, however this definition seems
unavoidable if one wants a ``topological" definition of the open
$p$-brane charge.

As before this discussion can be formulated in the setting of
relative currents. We repeat that it can be skipped by the reader
in a first reading.

\subsubsection{Relative currents}

The study of the previous case suggests\footnote{Let us mention
that, besides relative (co)homology, there exist alternative
mathematical tool for brane surgery, e.g. Thom classes
\cite{Papadopoulos:1999}.} that we can generically describe an
open $p$-brane ending on a $q$-brane\footnote{If ${\cal V}_{q+1}$
was not assumed to be closed, one should presumably consider
instead a higher order relative homology group (i.e. ``relative of
relative").} by a relative $(p+1)$-cycle $(\overline{\cal
W}_p,{\cal W}_{p+1})\in H_{p+1}({\cal M},{\cal V}_{q+1})$, due to
the equation (\ref{suggests}). The Poincar\'e duality map
(\ref{relativeP}) relative to ${\cal V}_{q+1}$ applied on this
$(p+1)$-cycle gives \be P(\overline{\cal W}_p,{\cal W}_{p+1})=
\left(\,-P(\overline{\cal W}_p)\,,\,(-)^{D-p-1}P({\cal
W}_{p+1})\,\right)\,.\ee The relative Poincar\'e duality map $P$
is a morphism, thus the \emph{relative current} \be (\a *J_q
\wedge \bar{*}\bar{J}_{p-1}\,,\,*J_p)\equiv (-)^{D-p-1}q_p\,
P(\overline{\cal W}_p,{\cal W}_{p+1})\,,\label{invim}\ee is a
cocycle of $CH^{D-p}({\cal M})$, i.e. \be d(\a *J_q \wedge
\bar{*}\bar{J}_{p-1}\,,\,*J_p)=\left(\a\, d(*J_q\wedge
\bar{*}\bar{J}_{p-1})\,,\,\a *J_q\wedge
\bar{*}\bar{J}_{p-1}-d*J_p\right)=(0,0)\,.\ee A natural object to
consider that uses the Poincar\'e dual (\ref{definPoin}) is the
cocycle of $H^{q-p+1}({\cal V}_{q+1},{\cal M})$ \be (a\,\a
\bar{*}\bar{J}_{p-1}\,,\,*J_p)= (-)^{D-p-1}q_p \,P_{{\cal
V}_{q+1},{\cal M}}(\overline{\cal W}_p,{\cal W}_{p+1})\,,\ee which
is the inverse image of (\ref{invim}) by the map (\ref{ext}) with
$a$ a mere sign equal to $a=(-)^{(D-q-1)(q+1-p)}$.

We introduce the Dirac membrane $\underline{\Sigma}_{D-p}$
corresponding to the representative $\Sigma_{D-p-1}\in
H_{D-p-1}({\cal M})$ introduced above equation (\ref{qp}), i.e.
$\partial\underline{\Sigma}_{D-p}=\Sigma_{D-p-1}$. From
(\ref{krakoukas}) we find that the topological charge definition
(\ref{qp}) is proportional to \ba
\int\limits_{\Sigma_{D-p-1}}P({\cal W}_{p+1}-\overline{\cal
V}_{p+1})&=&\int\limits_{\Sigma_{D-p-1}}P({\cal
W}_{p+1})+(-)^{D-p-1}\int\limits_{\underline{\Sigma}_{D-p}}P(\overline{\cal
W}_p)\nn\\&=&-\int\limits_{(\underline{\Sigma}_{D-p},\Sigma_{D-p-1})}P(\overline{\cal
W}_p,{\cal W}_{p+1})\nn\\\nn\\ &=&
I\left[(\underline{\Sigma}_{D-p},\Sigma_{D-p-1}),(\overline{\cal
W}_p,{\cal W}_{p+1})\right]\,,\label{welldef}\ea where $I$ is the
relative intersection number defined by (\ref{relint}) which is
well-defined in (\ref{welldef}) because
$(\underline{\Sigma}_{D-p},\Sigma_{D-p-1})$ and $(\overline{\cal
W}_p,{\cal W}_{p+1})$ are cocycles of the relative homology
$H_*({\cal M},{\cal V}_{q+1})$.

One of the main point of this last discussion was to point out
that while the $p$-brane current $*J_p$ \emph{is not} conserved,
the relative current $(a\,\a \bar{*}\bar{J}_{p-1}\,,\,*J_p)$
\emph{is} conserved, in the sense that it is a cocycle of
$H^{q-p+1}({\cal V}_{q+1},{\cal M})$. The interests of relative
currents is that if one replaces everywhere the ``absolute"
objects by ``relative" ones, the standard discussion of section
\ref{Diracproc} can be repeated exactly without any change. It is
also useful for Dirac brane construction purposes to know from
proposition \ref{coniso} that all the relative (co)homology groups
are trivial if the homologies $H_*({\cal W}_{q+1})$ and $H_*({\cal
M})$ are trivial.

%%%%%%%%%%%%%%%%%%%%%%%%%%%%%%%%%%%%%%%%%%%%%%%%%%%%%%%%%%%%%%%%%%%

\section{Chern-Simons coupling}\label{CScoupling}

A CS coupling was first introduced in 1982 by Deser, Jackiw and
Templeton \cite{Deser:1982} by considering three-dimensional
Yang-Mills (YM) theory augmented by a topological term which
preserves local gauge invariance. They pointed out that the vector
field becomes massive, despite the gauge invariance. CS terms,
cubic in the Abelian gauge fields, appear in the bosonic sector of
many supergravity theories. Example of such a theory is given by
the celebrated eleven-dimensional supergravity
\cite{Cremmer:1978}, describing the low energy effective action of
M-theory. The five-dimensional simple supergravity also contains
such a CS coefficient. This theory is known to resemble $D=11$
supergravity in many respects
\cite{Chamseddine:1980,Mizoguchi:1998} and can be used as a toy
model to test various ideas of $M$-theory in a simpler setting.
Anyway, this similarity is understood from the fact that $D=5$
simple supergravity can be realized as a specific truncation of a
Calabi-Yau compactification of $D=11$ supergravity
\cite{Cadavid:1995,Ferrara:1996}.

\subsection{Topologically massive electrodynamics}\label{qCS}

It is possible to add a quadratic CS term $A\wedge dA$ in the
action in dimensions $D=2p+3$. This term is a boundary term (i.e.
$A\wedge dA=1/2 d(A\wedge A)$) when $p$ is odd. Therefore we will
focus on $2k$-brane electrodynamics in spacetimes of dimension
$D=4k+3$.

As usual, we assume for simplicity the existence of a single
electric $2k$-brane of worldvolume ${\cal M}_e$ and a single
magnetic $(2k-1)$-brane the worldvolume of which is ${\cal M}_m$.
Their respective currents are $*J_e=eP({\cal M}_e)$ and
$*J_m=gP({\cal M}_m)$. Following the procedure sketched in
subsection \ref{Diracproc}, we graft a Dirac $2k$-brane of
worldvolume ${\cal M}_D$ on to the magnetic $(2k-1)$-brane
worldvolume ${\cal M}_m$, that is $\partial {\cal M}_D={\cal
M}_m$. The Dirac current is taken to be $*G=-gP({\cal M}_D)$ such
that $d*G=*J_m$. The Dirac brane position is pure gauge: ${\cal
M}_D\rightarrow{\cal M}_D+\partial{\cal V}$. In terms of forms, we
get the gauge freedom \be*G\rightarrow*G+d*V\,,\quad A\rightarrow
A+d\Lambda-*V\,.\label{gaugetr}\ee where $*V=-gP({\cal V})$.

The action of topologically massive electrodynamics is
\cite{Henneaux:1986i}
\begin{equation}
S[A_{\mu_1\ldots\mu_{2k+1}},G^{\mu_1\ldots\mu_{2k+1}}]=-\frac12\int
*F\wedge F- \frac{\a}{2}\int A\wedge dA-\a\int *G\wedge A+\int
*J_e\wedge A \,.
\end{equation} The electric charge has the unit $L^{-1/2}$, the
magnetic $L^{1/2}$ and the Chern-Simons coefficient $L^{-1}$.

The e.o.m. obtained from varying the gauge field $A$ together with
the ``Bianchi identity" is the system \be d*F+\a F=*J_e\,,\quad
dF=*J_m \,.\label{brrr}\ee Applying the differential $d$ on these
equations, we get \be d*J_e=\a *J_m\,,\quad
d*J_m=0\,,\label{nonconser}\ee The gauge field e.o.m. are
therefore consistent with the assumption that the magnetic charge
is conserved, necessary to define the Dirac brane according to the
procedure exposed in subsection \ref{Diracproc}.

The set of equation (\ref{nonconser}) can be formulated as the
(relative) cocycle condition of $(\a *J_m,*J_e)$. Furthermore, a
property specific to topologically massive electrodynamics
($\a\neq 0$) is that its field equations (\ref{brrr}) are
equivalent to a single condition on $(\a F,-*F)$: \be d(\a
F,-*F)=(\a *J_m,*J_e)\,.\ee

An infinitesimal variation $\d {\cal M}_D=\partial {\cal V}$ of
the Dirac brane worlvolume location leads to the following
variation of the action \be\d S=-\int_{\cal M} (d*F+\a\, dA)\wedge
*V\,.\label{varact}\ee where we defined the current $*V= -gP({\cal
V})$ such that $\d *G=d*V$. It can be proved that $*G\wedge
*V\equiv 0$, thus the variation of the action (\ref{varact}) can
be rewritten as \be\d S=g\int_{\cal V}(d*F+\a F)\,.\ee Hence, the
Dirac veto is still a sufficient condition to ensure that the
Dirac brane is unphysical and does not introduce any new dynamics.

The first equation of (\ref{nonconser}) states that the the
magnetic $(2k-1)$-brane must live on the boundary of the electric
$2k$-brane worldvolume. For instance, in the limiting three
dimensional case a magnetic instanton either creates an electric
particle or is the endpoint of an electric worldline. Consistency
requires the following relation between the electric and magnetic
charge \be e+\a g=0\,.\ee This kind relation allows us to observe
the success of the surgical transplant that grafted the Dirac
brane on to the magnetic charge. In fact, the Dirac brane can be
considered as the continuation of the electric brane in spacetime,
since the manifold ${\cal M}_e -{\cal M}_D$ has no boundary.
Accordingly, we rewrite the action as
\begin{equation}
S=-\frac12\int\limits_{\cal M} *F\wedge F-
\frac{\a}{2}\int\limits_{\cal M} A\wedge
dA\,+\,e\int\limits_{{\cal M}_e -{\cal M}_D} A \,.
\end{equation}
A large gauge transformation (\ref{gaugetr}) leads to the
following variation of the action
\begin{equation}
\Delta S=e\int\limits_{{\cal M}_e -{\cal M}_D}(d\Lambda-
*V)=-egL({\cal M}_e -{\cal M}_D,\Delta{\cal M}_D) \,.
\end{equation}

\subsection{Cubic Chern-Simons term}\label{cCS}

A non-vanishing quadratic coupling in the e.o.m. of $p$-brane
electrodynamics exists only in $(3p+5)$-dimensional spacetimes
with $p$ even. For non-vanishing CS coefficient, electric
$p$-branes can end on magnetic $(2p+1)$-branes.

\subsubsection{Five-dimensional toy model}

We start with the $5$-dimensional theory with only pointlike
electric sources\footnote{By analogy with eleven-dimensional
supergravity, we could call them M$0$-branes while the magnetic
strings could be christened M$1$-branes.} as charged object:
\begin{equation}
S = -\frac{1}{2} \int\limits_{{\cal M}_5} *F\wedge F -
\frac{\alpha}{6}\int\limits_{{\cal M}_5} F \wedge F \wedge A +
\int\limits_{{\cal M}_5} *J_e\wedge A \ ,
\end{equation}
where $A$ is the gauge field, $F=dA$ its curvature, $J_e$ is the
electric current. The equations of motion and Bianchi identity are
\begin{eqnarray}
\left\{
\begin{array}{c}
d* F +\frac{\alpha}{2} F\wedge F = *J_e \\
dF = 0\,.
\end{array}
\right.\nonumber
\end{eqnarray}
The action is gauge invariant if $J_e$ is a conserved current,
i.e. if  $d*J_e=0$. This requirement is consistent with the
equations of motion, as we can see if one takes the exterior
derivative of the first one, taking into account the second. Let
be an electric particle of charge $e$ and of worldline ${\cal
M}_1$. Since the electric charge is conserved, ${\cal M}_1$ is a
closed submanifold of ${\cal M}_5$. Its associated current is
defined as $*J_e=e\,P({\cal M}_1)$, where $P({\cal M}_1)$ is the
Poincar\'e dual of ${\cal M}_1$ with the whole worldline as
support.

We would now like to introduce magnetic sources. In 5D these will
correspond to one-dimensional extended objects (because $*F$ is a
3--form), that we will call magnetic strings. The magnetic string
current will be described by a 2--form, $J_m$, and we expect the
equations to take the form \be \left\{
\begin{array}{c}
d* F +\frac{\alpha}{2} F\wedge F = *J_e
\\
dF = *J_m  \, .
\end{array}
\right.\label{m1} \ee

These equations are gauge invariant, but the electric current
$J_e$ cannot be conserved. In fact, taking the exterior derivative
of the first equation of (\ref{m1}) we get
\begin{equation}
d*J_e =\a *J_m \wedge F \, .\label{nc}
\end{equation}
This last equation is telling us that electric current does not
need to be conserved on the worldsheet of a magnetic string. Let
be one magnetic string of two-dimensional closed worldsheet ${\cal
M}_2$ since the magnetic current $J_m$ is conserved, we can also
safely assume it to be defined by : $*J_m=g\,P({\cal M}_2)$, where
$g$ is a constant measuring the charge of the magnetic source.

We now construct an action principle giving rise to the equations
(\ref{m1}). First of all, we consider the modified Bianchi
identity. Following the standard procedure, we introduce a Dirac
membrane ${\cal M}_3$ with the magnetic string as boundary: ${\cal
M}_2=\partial{\cal M}_3$. This is translated in terms of the
Poincar\'e duals into $*J_m=d*G$ with $*G=-gP({\cal M}_3)$. As
usual one sets $F= dA + *G$. The ``gauge freedom" associated to
the position of the Dirac brane alone is translated into
$*G\rightarrow*G+d*V$, where $*V=-gP({\cal V}_4)$ with ${\cal
V}_4$ the manifold described by the displacement of the Dirac
brane $\delta {\cal M}_3=\partial{{\cal V}_4}$.

Consider the following action in $5$ spacetime dimensions:
\begin{eqnarray}
I &=& -\frac{1}{2}\int *F\wedge F - \frac{\alpha}{6}\int dA \wedge
dA \wedge A
-\frac{\alpha}{2}\int * G\wedge dA \wedge A \nonumber \\
& &+ \int  *J_e \wedge A +  \frac{\alpha}{2} \int *J_m\wedge\left(
f\wedge A
 + \frac{v}{2}\, \bar{*}f \wedge f + \Phi  \bar{*}j
 \right)
   +  I_K\,.\nn\\
\label{I}
\end{eqnarray}
Here $f= d\Phi - i^*A +\bar{*}g$ where $i^*A$ denotes the pullback
of $A$ on ${\cal M}_2$, and $\bar{*}$ is the Hodge star on the
worldvolume ${\cal M}_2$. $\Phi$ is a scalar field living on the
string, $v$ some real parameter and $I_K$ is the sum of the
kinetic terms for the three present branes. We will assume for
simplicity that electric source $J_e$ is external and therefore
not to be varied in the action principle.

Finally, $g$ defines a Dirac worldline which originates from the
instanton on the magnetic string. The case of one instanton of
strength $\nu$ located at the point ${\cal M}_0$ (denoted in that
way to make higher dimensional generalization straightforward) is
described by the instanton ``current" $\bar{*}j=\nu\,P_{{\cal N
}_2}({\cal M}_0)$, a two-form proportional to the Poincar\'e dual
to ${\cal M}_0$ in ${\cal M}_2$. The Dirac point is defined by a
worldine ${\cal N}_1$ ending or originating at the instanton, in
such a way that $\bar{*}j= d\bar{*}g$ if $\bar{*}g=\nu\,P_{{\cal N
}_2}({\cal N}_1)$. The set of manifolds and their corresponding
brane is summarized in the table \ref{branelex}
\begin{table}[ht]
\begin{center}
\begin{tabular}{|c|c|}
  \hline
  manifold & brane worldhistory \\
  \hline
${\cal M}_0$  & dyonic instanton location\\
${\cal M}_1$  & electric particle worldline\\
${\cal M}_2$  & magnetic string worldsheet\\
${\cal M}_3$  & Dirac membrane worldvolume\\
${\cal N}_1$  & Dirac string worldline\\
  \hline
\end{tabular}
\caption{Worldhistory notations \label{branelex}}
\end{center}
\end{table}

Using the definition of the Poincar\'e dual, we can rewrite the
action as
\begin{eqnarray}
I &=& -\frac{1}{2}\int\limits_{{\cal M}_5} *F\wedge F -
\frac{\alpha}{6}\int\limits_{{\cal M}_5}  dA \wedge dA \wedge A
+\frac{\alpha g}{2}\int\limits_{{\cal M}_3}dA \wedge A \nonumber
\\\nn & &+ \,e\int\limits_{{\cal M}_1}  A  + \frac{\alpha g}{2}
\int\limits_{{\cal M}_2} f\wedge A + \frac{\alpha g v}{4}
\int\limits_{{\cal M}_2} \bar{*}f \wedge f + \frac{\alpha g\nu}{2}
\int\limits_{{\cal M}_0} \Phi  +  I_k\,.
\end{eqnarray}

Note that, to simplify notations, we sometimes make use of $d\Phi$
and $\bar{*}g$ in the action as forms of $\Omega({\cal M}_5)$.
This should be understood as extensions whose pull-back gives the
corresponding differential form of $\Omega({\cal M}_2)$ on the
string worldsheet.

The equation of motion coming from varying $A$ is
\begin{equation}
d* F +\frac{\alpha}{2} F\wedge F + \frac{\alpha}{2}*J_m\wedge
(v\bar{*}f-f)  = *J_e \,, \label{eom}
\end{equation}
where one used the fact that the highly singular product vanishes,
\begin{equation}
* G \wedge *G = 0  \,,\label{singular}
\end{equation} when \emph{the Dirac membranes never touch
each other}\footnote{This follows from general arguments of
appendix \ref{PD}.}, which will be assumed throughout this
discussion. Varying $\Phi$ we obtain,
\begin{equation}
(df + v \,d \bar{*}f) \wedge *J_m =0 \label{eomphi}
\end{equation}
Taking the exterior derivative of (\ref{eom}) we get
\begin{equation}
\alpha\,\, *J_m\wedge \bar{*}j= d*J_e \,. \label{nc2}
\end{equation}
We made use of (\ref{eomphi}) and $*J_m\wedge *G\equiv 0$, which
is a consequence of (\ref{singular}). The equation (\ref{nc2})
means that we can specify external sources (contrarily to
(\ref{nc})), but they must satisfy the above condition. Also note
that in the absence of instantons on the worldsheet of the string,
we get conservation of the electric charge. Electric charge can
end on magnetic string and create an instanton on it. The eleven
dimensional analogue is the well known fact that M$2$-branes can
end on M$5$-brane with a string as intersection. In that case, a
spacelike picture is possible because there is ``more room" in
eleven dimensions.

Since ${\cal M}_1$ is not closed, the linking number is ill
defined (see appendix \ref{linking}). In the same vein, the
electric charge is not simply equal to the integral of $*J_e$
because the electric current is not conserved. But the problem can
be solved if we rewrite the electric charge non-conservation law
as
\begin{equation}
d(*J_e+\alpha *J_m\wedge \bar{*}g)=0
\end{equation}
Let $S^3$ be a hypersphere wrapped around the electric worldline
continued in the magnetic worldvolume by the Dirac worldine of the
instanton. We can define the electric charge as
\begin{equation}
\int\limits_{D^4}(*J_e+\alpha *J_m\wedge \bar{*}g)
\end{equation}
with $\partial D^4=S^3$. It is a well defined number because it
only depends on the homology class $D^4\sim D^4+\partial B^5$.
Then, let us take two different representatives, $D^4_{(1)}$ and
$D^4_{(2)}$. The first one does not intersect the magnetic
worldsheet while the second one has no intersection with the
worldline. If we evaluate the tension with $D^4_{(1)}$ we get
\begin{eqnarray}
\int\limits_{D^4_{(1)}}*J_e=e\nonumber
\end{eqnarray}
and with $D^4_{(2)}$
\begin{eqnarray}
\alpha\int\limits_{D^4_{(2)}} *J_m\wedge \bar{*}g=\alpha
g\int\limits_{S^1}\bar{*}g=\alpha
g\int\limits_{B^2}\bar{*}j=\alpha g\nu\nonumber
\end{eqnarray}
where $D^4_{(2)} \cap {\cal M}_2 = S^1 =\partial B^2$. In
conclusion, the charges are related by
\begin{equation}
e= \alpha g \nu.\label{charges}
\end{equation}

Note that (\ref{eom}) is similar to but not identical with the
first equation in (\ref{m1}). They differ from each other on the
worldsheet of the string. In order to get the equations we want,
we have to impose one further condition: self--duality of $f$ on
the string worldsheet
\begin{equation}
f = \bar{*}f \, .\label{selfdu}
\end{equation}
Obviously, this makes the instanton dyonic. Now equation
(\ref{eom}) takes the form
\begin{equation}
d* F +\frac{\alpha}{2} F\wedge F + \frac{\alpha(v-1)}{2}*J_m\wedge
f  = *J_e \ . \label{eom2}
\end{equation}
If the action (\ref{I}) aims to represent the bosonic sector of
simple D=5 supergravity, we should take $v=-1$. In this paper, for
simplicity, the self-duality condition is imposed by hand at the
level of the equations of motion but it is possible to write an
action similar to (\ref{I}) for which (\ref{selfdu}) arises as a
consequence of the variation of the action. It has been done in
$D=11$ (and $v=-1$) in \cite{Bandos:1998}. A regularized version
has been proposed recently in \cite{Lechner:2001i}.
%\AV{Witten argued}

\begin{figure}[ht!]
     \centerline{\includegraphics[width=0.3\linewidth]{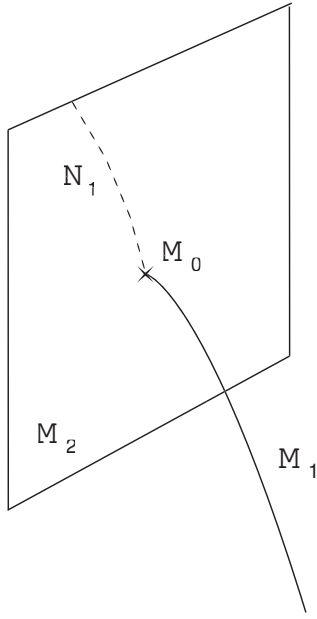}}
\caption{An electric worldline ${\cal M}_1$ ends on a magnetic
worldsheet ${\cal M}_2$ and produces a magnetic instanton ${\cal
M}_0$.}
\end{figure}

With all this in mind, we can now rewrite the charge definition.
For $v=-1$, we have the explicit expression
\begin{equation}
e=\int\limits_{S^3}(*F+\frac{\alpha}{2}A\wedge dA)-\alpha g
\int\limits_{S^1} \bar{*}f\,.
\end{equation}
where $S^1=S^3\cap{\cal M}_2$.

The Dirac membrane should be classically unobservable. This means
that the action (\ref{I}) should possess the gauge invariance
\begin{eqnarray*}
A &\longrightarrow& A + d\Lambda -*V \\
\Phi &\longrightarrow& \Phi + i^*\Lambda \\
 *G &\longrightarrow& *G +
d*V\,,
\end{eqnarray*}
where $A$ (and $\Phi$) transforms in such a way that the field
strength $F$ (and $f$) is invariant. The pull-back of $*V$ can be
shown to vanish and $\Phi$ can be left invariant. The action
(\ref{I}) is invariant under small displacements of the Dirac
brane. To check this, the identity $*G\wedge *V= 0$ is needed
(This identity follows from taking the variation of the identity
(\ref{singular}) and using $*V\wedge *V=0$).

The same remark applies to the Dirac point living on the magnetic
string. The corresponding gauge transformation is
\begin{eqnarray}
\bar{*}g\rightarrow \bar{*}g + d\bar{*}v\,,\, \Phi \rightarrow
\Phi-\bar{*}v.\label{worldshit}
\end{eqnarray}
The scalar $\bar{*}v$ is proportional to the Poincar\'e dual of
the surface described by the Dirac worldline in the string
worldsheet. The action is obviously invariant under infinitesimal
transformations (\ref{worldshit}).

\subsubsection{M-theoretic example}

Due to its modern relevance, we briefly mention the case $p=2$,
that is bosonic gauge sector of M-theory in the weak field limit
and fixed gravitational background.

The fundamental objects living in the eleven dimensional bulk
${\cal M}_{11}$ are
\begin{eqnarray}
&& \mbox{\textbf{M2-brane} of worldsheet }{\cal M}_3\,\mbox{and
electric charge density }e\nn
\\
&& \mbox{\textbf{M5-brane} of worldsheet }{\cal M}_6\,\mbox{and
magnetic charge density }g\nn
\end{eqnarray}
On the M$5$-brane lives a \ba \mbox{\textbf{Dyonic string} of
worldsheet }{\cal N}_2\,\mbox{and self-dual charge }\nu \nn\ea

%%%%%%%%%%%%%%%%%%%%%%%%%%%%%%%%%%%%%%%%%%%%%%%%%%%%%%%%%%%%%%%%%%%%%%%
%%%%%%%%%%%%%%%%%%%%%%%%%%%%%%%%%%%%%%%%%%%%%%%%%%%%%%%%%%%%%%%%%%%%%%%

\chapter{Massless spin two gauge theory}\label{spintwo}

It is well known that the gravity field equations in
four-dimensional spacetime are formally invariant under a duality
rotation. As usual the Bianchi identities (I and II) get exchanged
with Einstein field equations. But a problem (present also in
Yang-Mills theories) seems to prevent this duality rotation from
being a true symmetry of gravity (This question is addressed in
more details in section \ref{nogo} for Yang-Mills gauge theories).
However, linearized gravity does not present this problem
 and the duality rotation is a true symmetry. Massless spin two \emph{free}
fields are
similar to Abelian form gauge fields in many aspects. This chapter
provides a reformulation of known results in a systematic unified
mathematical framework. Complete proofs are also presented.

The section \ref{Lingravi} is a review of massless spin two gauge
fields. In section \ref{MixedY}, we review the Hodge duality for
linearized gravity and the free dual gauge fields obtained, which
are in the representation of $GL(D,\mathbb R)$ corresponding to
Young diagrams with one row of two columns and all other rows of
length one. The sections \ref{N-c}, \ref{symt} and \ref{rectd}
provides a review of linearized gravity gauge structure in the
language of $N$-complexes. Multiforms and some operations useful
later on, are introduced in section \ref{Hodged}. Linearized
gravity field equations and their duality properties are discussed
in \ref{fieldequs} in the introduced mathematical framework. The
section \ref{GenPl} presents our theorem \ref{GenPoin}, which
generalizes the standard Poincar\'e lemma to arbitrary finite
dimensional representations of $GL(D,\mathbb R)$. Some useful corollaries
are also given. All this mathematical machinery is used in the
section \ref{arbitraryYoung} to provide a systematic treatment of
tensor gauge fields in arbitrary representations of $GL(D,\mathbb
R)$.

%%%%%%%%%%%%%%%%%%%%%%%%%%%%%%%%%%%%%%%%%%%%%%%%%%%%%%%%%%%%%%%%%%%%%%%%%%%%%%%%%

\section{Linearized gravity}\label{Lingravi}

\subsection{Pauli-Fierz action}

A free symmetric tensor gauge field $h_{\mu\nu}$ in $D$ dimensions
has the gauge symmetry \be \delta h_{\mu\nu}=2 \partial
_{(\mu}\xi_{\nu)} .\label{2gauge}\ee The linearized Riemann tensor
for this field is \be R_{\mu\nu\,
\sigma\tau}\equiv\frac12(\partial_{\mu}
\partial_{\sigma}h_{\nu\tau}+\ldots)=-2\partial_{[\mu}h_{\nu][\sigma,\tau]}
\label{Riemannlin}
\ee It satisfies \be R_{\mu\nu\,
\sigma\tau}=R_{\sigma\tau\,\mu\nu}\label{symR}\ee together with
the first Bianchi identity \be R_{[\mu\nu\, \sigma]\tau} =0
\label{BianchiI}\ee and the second Bianchi identity
\be\partial_{[\rho}R_{\mu\nu]\, \sigma\tau}=0 \label{BianchiII}\ee

It has been shown by Pauli and Fierz \cite{Fierz:1939} that there
is a unique, consistent action describing a pure massless spin-two
field. This action happens to be the Einstein action linearized
around a Minkowski background\footnote{Note that, the way back to
full gravity is quite constrained. It has been shown that there is
no local consistent coupling, with at most two derivatives of the
fields, that can mix various gravitons \cite{Boulanger:2000}. In
other words, there are no Yang-Mills-like spin-2 theories (like
for antisymmetric tensor gauge theories, as reviewed in subsection
\ref{connection}). The only possible deformations are given by a
sum of individual Einstein-Hilbert actions. Therefore in the case
of one graviton, \cite{Boulanger:2000} provides a strong proof of
the uniqueness of Einstein's theory.}\be S_{EH}[g_{\mu
\nu}]=\frac{2}{\kappa^2}\int d^D x \,\sqrt{-g}\,R_{full}\,,\quad
g_{\m\n}=\eta_{\m\n}+\kappa h_{\m\n}\,,\ee with $R_{full}$ the
scalar curvature for the metric $g_{\m\n}$. The constant $\kappa$
has mechanical dimensions $L^{D/2-1}$. The term of order
$1/\kappa^2$ in the expansion of $S_{EH}$ vanishes since the
background is flat. The term of order $1/\kappa$ is equal to zero
because it is proportional to the (sourceless) Einstein equations
evaluated at the Minkowski metric. The next order term in the
expansion in $\kappa$ is (up to a boundary term) the action for a
massless spin-2 field in spacetime dimension $D$
\cite{Fierz:1939}\ba S_{PF}[h_{\mu \nu}] &=& \int d^D x \left[
-\frac{1}{2}\left(\partial_{\mu}{h}_{\nu\rho}\right)\left(\partial^{\mu}{h}^{\n\r}\right)
+\left(\partial_{\m}{h}^{\m}_{~\n}\right)\left(\partial_{\r}{h}^{\r\n}\right)\right.\nonumber\\
&&\left.-\left(\partial_{\n}{h}^{\m}_{~\m}\right)\left(\partial_{\r}{h}^{\r\n}\right)
+\frac{1}{2}\left(\partial_{\m}{h}^{\n}_{~\n}\right)
\left(\partial^{\m}{h}^{\r}_{~\r}\right)\right] \, . \label{PF}\ea
Since we linearize around a flat background, spacetime indices are
raised and lowered with the flat Minkowskian metric $\eta_{\mu
\nu}$. For $D=3$ the Lagrangian is a total derivative so we will
assume $D\geq 3$. The (vacuum) equations of motion are the natural
free field equations \be R^\sigma{}_{\mu\, \sigma\nu}=0
\label{Einstein}\ee which are the linearized Einstein equations.
Together with (\ref{Riemannlin}) the Einstein equations imply
\be\partial^\mu R_{\mu\nu\, \sigma\tau}=0\,.\label{d*R}\ee

\subsection{Minimal coupling}

The Euler-Lagrange variation of the Pauli-Fierz action is \be
\frac{\d S_{PF}}{\d h_{\m\n}}=R^{\s}_{\,\m\s\n}-\frac12\,
\eta_{\m\n}\,R^{\s\tau}_{\quad\s\tau}\,.\ee It can be shown that
the second Bianchi identities imply\be \partial^\m
R_{\m\s\n\r}+\partial_\r R_{\n\m\s}^{\quad\m}-\partial_\n
R_{\r\m\s}^{\quad\m}=0\,, \ee and taking the trace again leads to
\be
\partial^\m R^{\s}_{\,\m\s\n}-\frac12\,
\partial_\n R^{\s\tau}_{\quad\s\tau}=0\,.\label{conservedR}\ee
From another perspective, the equations (\ref{conservedR}) are the
Noether identities corresponding to the gauge transformations
(\ref{2gauge}).

Let us introduce a source $T_{\m\n}$ which couples minimally to
$h_{\m\n}$ through the term \be S_{minimal}=-\kappa\int d^D x \,
h^{\m\n}T_{\m\n}\,.\ee We add this term to the Pauli-Fierz action
(\ref{PF}) together with a kinetic term $S_K$ for the sources to
get the action\be S=S_{PF}+S_K+S_{minimal}\,.\ee The field
equations for the symmetric gauge field $h_{\m\n}$ are the
linearized Einstein equations \be R^{\s}_{\,\m\s\n}-\frac12\,
\eta_{\m\n}\,R^{\s\tau}_{\quad\s\tau}= \kappa
T_{\m\n}\,.\label{Einsteinwsource}\ee Consistency with
(\ref{conservedR}) imply that the linearized energy-momentum
tensor is conserved $\partial^\m T_{\m\n}=0$.

The simplest example of source is the one of a free particle of
mass $m$ following a worldline $x^\m(s)$ with $s$ the proper time
along the worldline. The Polyakov action for the massive particle
reads \be S_{Polyakov}[x^\m(s)]= -m\int ds\,
g_{\m\n}\frac{dx}{ds}^\m \frac{dx}{ds}^\n \,.\ee It results as the
sum of the two actions \ba S_K&=&-m\int ds\,
\eta_{\m\n} \frac{dx}{ds}^\m \frac{dx}{ds}^\n \,,\label{kineticc}\\
S_{minimal}&=&-m\kappa \int ds \,h_{\m\n} \frac{dx}{ds}^\m
\frac{dx}{ds}^\n\,,\label{minimalc}\ea from which one infers that
the (matter) source $T_{\m\n}$ for a massive particle is equal
to\be T^{\m\n}(x)=m\int ds \,\d^{D}\left(x-x(s)\right)\,
\frac{dx}{ds}^\m \frac{dx}{ds}^\n\,.\ee It is conserved if and
only if $\frac{d^2x}{ds^2}^\m=0$, which means that the test
particle follows a straight worldline. In general, if one
considers a free massless spin-two theory coupled with matter,
then the latter has to be constrained in order to be consistent
with linearized energy-momentum tensor conservation\footnote{This
should not be too surprising since it is well known that the
Einstein equations simultaneously determine the gravity field
\emph{and} the motion of matter.}. At first sight, it is however
inconsistent with the natural expectation that matter reacts to
the gravitational field. Anyway, the condition
$\frac{d^2x}{ds^2}^\m=0$ is mathematically inconsistent with the
e.o.m. obtained from varying (\ref{kineticc}) and (\ref{minimalc})
with respect to the worldline $x^\m(s)$ which constrains the
massive particle to follow a geodesic for $g_{\m\n}$ (and
\emph{not} a straight line). In fact, if one wants the matter to
respond to the gravity field, we must add a source $\kappa
T^{self}_{\m\n}$ for the gravitational field itself, in such a way
that the sum $T_{\m\n}+T^{self}_{\m\n}$ is conserved if the matter
obeys its own equation to first order in $\kappa$ and if the
gravity field obeys (\ref{Einsteinwsource}). This gravitational
self-energy must come from a first order (in $\kappa$) deformation
of the Pauli-Fierz action. This modification has been the starting
point of Feynman in the late fifities when he derived the Einstein
gravity action by deforming the Pauli-Fierz action consistently
with matter back reaction\footnote{In 1962, Feynman presented this
derivation in his sixth Caltech lecture on gravitation
\cite{Feynman}. In their foreword of Feynman's lectures, J.
Preskill and K.S. Thorne give an interesting historical account of
the field-theoretic approach to Einstein's theory as a
self-interacting massless spin-two field theory (\cite{Feynman},
pp. x-xv, and references therein). One of the intriguing features
of this viewpoint is that the initial flat background is no longer
observable in the full theory. In the same vein, the fact that the
self-interacting theory has a geometric interpretation is
``\emph{not readily explainable - it is just marvelous}", as
Feynman expressed.}. At the end of the perturbative procedure, one
obviously obtains that the free-falling particle must follow a
geodesic for consistency with the (full) Einstein equations.

\section{Mixed Young arrays}\label{MixedY}

\subsection{Duality in linearized gravity}

Let us mention for further purpose that the last equation
(\ref{d*R}) can be directly deduced from the equations
(\ref{BianchiI})-(\ref{BianchiII})-(\ref{Einstein}) for the
linearized Riemann tensor without using its explicit expression
(\ref{Riemannlin}). To simplify the proof and prepare the ground
for a discussion about duality properties, let us introduce the
tensor \be
(*R)_{\mu_1\ldots\mu_{D-2}\,|\,\rho\sigma}=\frac12\,\varepsilon_{\mu_1\,\ldots\,\mu_D}\,R^{\mu_{D-1}\mu_D}_{\quad\quad\quad
\rho\sigma}\label{*R}\ee The linearized second Bianchi identity
and the Einstein equations rewrite in terms of this new tensor
respectively as
\be\partial^\mu\left[(*R)_{\mu\,\ldots\nu\,|\,\rho\sigma}\right]=0\label{EinsteinII}\ee
and
\be(*R)_{[\mu\,\ldots\nu\,|\,\rho]\sigma}=0.\label{BianchiIV}\ee
If one takes the divergence of (\ref{BianchiIV}) with respect to
the first index $\mu$, and uses (\ref{EinsteinII}), one gets
\be\partial^\mu\left[(*R)_{\rho\,\ldots\nu\,|\,\mu\sigma}\right]=0\ee
which is equivalent to \be \partial^\mu R_{\a\b\,
\mu\sigma}=0\label{p*R}\ee as follows from the definition
(\ref{*R}). Using the symmetry property (\ref{symR}) of the
Riemann tensor we recover (\ref{d*R}).

Later on (in the corollary \ref{coro}), we will prove that the
equations \be R^\sigma{}_{\mu\, \sigma\nu}=0\,,\quad\partial^\mu
R_{\mu\nu\,\sigma\tau}=\partial^\sigma
R_{\mu\nu\,\sigma\tau}=0\label{d*R'}\ee are (locally) equivalent
to the following equation \cite{Hull:2001}
\be(*R)_{\mu_1\ldots\mu_{D-2}\,|\,\rho\sigma}=\partial_{[\mu_1}\tilde{h}_{\mu_2\ldots\mu_{D-2}]\,|\,[\rho,\sigma]}\,,\label{dualcurva}\ee
which defines the tensor field
$\tilde{h}_{[\mu_1\ldots\mu_{D-3}]\,|\,\rho}$ called the
\emph{dual gauge field} of $h_{\m\n}$ and which is said to have a
\emph{mixed symmetry} because it is neither (completely)
antisymmetric nor symmetric. In fact, it obeys the identity\be
\tilde{h}_{[\mu_1\ldots\mu_{D-3}\,|\,\rho]}\equiv 0\,.\ee Still,
for $D=4$ the dual gauge field is a symmetric tensor
$\tilde{h}_{\mu\nu}$, which signals a possible duality symmetry.
The curvature dual (\ref{dualcurva}) is left invariant by the
transformations
\be\delta\tilde{h}_{\mu_1\ldots\mu_{D-3}\,|\,\rho}=
\partial_{[\mu_1}S_{\mu_2\ldots\mu_{D-3}]\,|\,\rho}
+\partial_{\rho}A_{\mu_2\ldots\mu_{D-3}\mu_1}
+A_{\rho[\mu_2\ldots\mu_{D-3},\mu_1]}\label{gtransfoht}\ee where
the complete antisymmetrization of the gauge parameter
$S_{[\mu_1\ldots\mu_{D-4}]\,|\,\mu_{D-3}}$ vanishes while the
other gauge parameter $A_{\mu_1\ldots\mu_{D-3}}$ is completely
antisymmetric.

\subsection{Mixed symmetry type gauge fields}\label{mixed}

More generally, let us consider massless gauge fields
$M_{\m_1\m_2\ldots\m_n\,|\,\m_{n+1}}$ having the same symmetries
than the above-mentioned dual gauge field
$\tilde{h}_{\mu_1\ldots\mu_{D-3}\,|\,\rho}$. It is represented by
the Young diagram \be \footnotesize
\begin{picture}(30,30)(0,0)
\multiframe(0,10)(10.5,0){1}(10,10){$1$}
\multiframe(10.5,10)(10.5,0){1}(17,10){$n\!\!+\!\!1$}
\multiframe(0,-0.5)(10.5,0){1}(10,10){$2$} \put(5,-12){$\vdots$}
\multiframe(0,-24)(10.5,0){1}(10,10){$n$}
\end{picture}\normalsize\,,
\label{YS}\ee \\
\\
which implies that the field obeys the identity \be
M_{[\m_1\m_2\ldots\m_n\,|\,\m_{n+1}]}\equiv 0. \ee Such tensors
gauge fields have been studied two decades ago by the authors of
\cite{Curtright:1985,Aulakh:1986,Labastida:1986} and appear in the
bosonic sector of some odd-dimensional CS supergravities
\cite{Troncoso:1998,Troncoso:1998i,Banados:2001}. Here, $n$ will
denote the number of antisymmetric indices carried by the field
$M_{\m_1\m_2\ldots\m_n\,|\,\m_{n+1}}$. This is also the number of
boxes in the first column of the corresponding Young array. The
tensors $M_{\m_1\m_2\ldots\m_n\,|\,\m_{n+1}}$ have
$\frac{n(D+1)!}{(n+1)!(D-n)!}$ components in $D$ dimensions.

The action of the free theory is \be
S_0[M_{\m_1\m_2\ldots\m_n\,|\,\m_{n+1}}]=\int d^D x \,\cl
\label{startingpoint} \ee where the Lagrangian is
\cite{Aulakh:1986} \ba {\cal L}&=&M_{\m_1\ldots\m_n\,|\,\m_{n+1}}
\partial^2 M^{\m_1\ldots\m_n\,|\,\m_{n+1}}
 -2n M_{\m_1\ldots\m_n\,|\,\m_{n+1}}\partial^{\m_1}\partial^{\l}
M_{\l}^{~\m_2\ldots\m_n\,|\,\m_{n+1}}
\nonumber \\
&-&n M^{\m_1}_{~~\m_2\ldots\m_n\,|\,\m_1}\partial^{2}
M^{\m_1\ldots\m_n}_{~~~~~~~\m_1} +n(n-1)
M_{\m_1\m_2\ldots\m_n\,|\,\m}
\partial^{\m_1}\partial^{\n} M_{\n}^{~\m_2\ldots\m_{n-1}\m\,|\,\m_n}
\nonumber \\
&+&n(n-1) M_{\b\g\m_3\ldots\m_n}^{~~~~~~~~~\b}
\partial^{\g}\partial^{\m} M_{\n\m}^{~~\m_3\ldots\m_n\,|\,\n}
\nonumber \\
 &+&2n  M^{\m_1}_{~~\m_2\ldots\m_n\,|\,\m_1}\partial^{\m}\partial_{\n}
M_{\m}^{~\m_2\ldots\m_n\,|\,\n}. \label{Lagrangian} \ea The field
equations derived from (\ref{Lagrangian}) are equivalent to\be
\eta^{\m_1\n_1} K_{\m_1\m_2\ldots\m_{n+1}\,|\,\n_1\n_2}=0\,\ee
where \be K_{\m_1\m_2\ldots\m_{n+1}\,|\,\n_1\n_2}\equiv
\partial_{[\m_1}M_{\m_2\ldots\m_{n+1}]\,|\,\,[\n_1\,,\,\n_2]}\,,\label{curvM}\ee is the
curvature; it obeys the algebraic identity \be
K_{[\m_1\ldots\m_{n+1}\,|\,\n_1]\n_2}=0\,.\ee Notice that for
$n=1$ one recovers the linearized gravity theory.

The action (\ref{startingpoint}) and the curvature (\ref{curvM})
are invariant under the following gauge transformations
\begin{equation}
\d_{S,A}M_{\m_1\ldots\m_{n+1}}=
\partial_{[\m_1}S_{\m_2\ldots\m_n]\,|\,\m_{n+1}}
+\partial_{[\m_1}A_{\m_2\ldots\m_n]\m_{n+1}}
+\partial_{\m_{n+1}}A_{\m_2\ldots\m_n\m_1} \label{invdejauge}
\end{equation}
where the gauge parameters $S_{\m_2\ldots\m_n\,|\,\m_{n+1}}$ and
$A_{\m_2\ldots\m_{n+1}}$ have the symmetries \footnotesize
\begin{picture}(40,30)(0,0)
\multiframe(0,10)(10.5,0){1}(10,10){$2$}
\multiframe(10.5,10)(10.5,0){1}(17,10){$n\!\!+\!\!1$}
\multiframe(0,-0.5)(10.5,0){1}(10,10){$3$} \put(5,-12){$\vdots$}
\multiframe(0,-24)(10.5,0){1}(10,10){$n$}
\end{picture}\normalsize
and $~~~$ \footnotesize \begin{picture}(20,30)(0,0)
\multiframe(0,10)(10.5,0){1}(10,10){$2$}
\multiframe(0,-0.5)(10.5,0){1}(10,10){$3$} \put(5,-12){$\vdots$}
\multiframe(0,-24)(10.5,0){1}(15,10){$n\!${\scriptsize{$+$}}$\!1$}
\end{picture}\normalsize
\vspace{1cm} , respectively. These gauge transformations come with
a chain of reducibilities (actually $n-1$ of them) on the gauge
parameters. These reducibilities read, with $1\leq i\leq n$ \ba
\stackrel{(i)}{S}_{\m_1\ldots\m_{n-i}\,|\,\m_{n-i+1}}&=&
\partial_{[\m_1}\stackrel{(i+1)}{S}_{\m_2\ldots\m_{n-i}]\,|\,\m_{n-i+1}}
+\frac{(n+1)}{(n-i+1)}\,\partial_{[\m_1}\stackrel{(i+1)}{A}_{\m_2\ldots\m_{n-i}]\m_{n-i+1}}+
\nonumber\\
&+&\frac{(n+1)}{(n-i+1)}\,\partial_{\m_{n-i+1}}\stackrel{(i+1)}{A}_{\m_2\ldots\m_{n-i}\m_1}\,,\label{red1}
\\
\stackrel{(i)}{A}_{\m_1\ldots\m_{n-i+1}}&=&\partial_{[\m_1}\stackrel{(i+1)}{A}_{\m_2\ldots\m_{n-i+1}]}\label{red2}
\ea with the conventions that \ba
\stackrel{(1)}{S}_{\m_1\ldots\m_{n-1}\,|\,\m_n}&=&S_{\m_1\ldots\m_{n-1}\,|\,\m_n}\,,\quad
\stackrel{(n)}{S}_\m\,=\, 0\,,\nn\\
\stackrel{(1)}{A}_{\m_1\ldots\m_n}&=&A_{\m_1\ldots\m_n}\,.\nn\ea
The reducibility parameters at reducibility level $i$ have the
symmetry
\\
$\stackrel{(i+1)}{S}_{\m_1\ldots\m_{n-i-1}\,\mid\,\m_{n-i}}\simeq$
\footnotesize
\begin{picture}(40,30)(0,0)
\multiframe(0,10)(10.5,0){1}(10,10){$1$}
\multiframe(10.5,10)(10.5,0){1}(17,10){$n\!${\scriptsize{$-$}}$\!i$}
\multiframe(0,-0.5)(10.5,0){1}(10,10){$2$} \put(5,-12){$\vdots$}
\multiframe(0,-24)(10.5,0){1}(20,10)
{\scriptsize{$n\!${\scriptsize{$-$}}$\!i${\scriptsize{$-$}}$\!\!$\scriptsize
{$1$}}}
\end{picture}\normalsize
and $\stackrel{(i+1)}{A}_{\m_1\ldots\m_{n-i}}\simeq$ \footnotesize
\begin{picture}(20,30)(0,0)
\multiframe(0,10)(10.5,0){1}(10,10){$1$}
\multiframe(0,-0.5)(10.5,0){1}(10,10){$2$} \put(5,-12){$\vdots$}
\multiframe(0,-24)(10.5,0){1}(15,10){$n\!${\scriptsize{$-$}}$\!i$}
\end{picture}\normalsize
\vspace{1cm}.
\\
Notice that $\stackrel{(n-1)}{S}_{\m\n}\,\simeq$ \footnotesize
\begin{picture}(30,20)(0,0)
\multiframe(0,0)(10.5,0){2}(10,10){$\m$}{$\n$}
\end{picture}\normalsize.
These gauge transformations and reducibilities were introduced and
discussed in \cite{Curtright:1985,Aulakh:1986,Labastida:1986}. Our
theorem \ref{GenPoin} will provide a more systematic tool for the
investigation of mixed symmetry type gauge field theories.

Following closely the derivation of section \ref{degreesf}, one
obtains the number of physical degrees of freedom, which is equal
to \be \frac{(D-2)!\,D\,(D-n-2)\,n}{(D-n-1)!\,(n+1)!}\,.\ee This
number is manifestly invariant under the exchange
$n\leftrightarrow D-n-2$ which corresponds to a Hodge duality
transformation. This confirms that the dimension for which the
theory is dual to a symmetric tensor is equal to $D=n+3$; this is
also the critical dimension for the theory to have local physical
degrees of freedom.

%%%%%%%%%%%%%%%%%%%%%%%%%%%%%%%%%%%%%%%%%%%%%%%%%%%%%%%%%%%%%%%%%%%%%%

\section{$N$-complexes}\label{N-c}

The idea of
\cite{Olver:1983,Dubois-Violette:1999,Dubois-Violette:2001} is to
construct complexes for irreducible tensor fields of mixed Young
symmetry type generalizing thereby to some extent the calculus of
differential forms. This tool provides an elegant formulation of
symmetric tensor gauge fields and their Hodge duals (like
differential form notation within electrodynamics). It is strongly
advised for the reader to take a look at the appendix \ref{Young}.

Let $(Y)=(Y_p)_{p\in \mathbb N}$ be a given sequence of Young
diagrams such that the number of cells of $Y_p$ is $p$, $\forall
p\in \mathbb N$. For each $p$, there is a single shape $Y_p$) and
$Y_p\subset Y_q$ for $p<q$. We define $\Omega^p_{(Y)}({\cal M})$
to be the vector space of smooth covariant tensor fields of rank
$p$ on ${\cal M}$ which have the Young symmetry type $Y_p$ (i.e.
their components $T(x)$ belong to the Schur module associated to
$Y_p$). More precisely they obey the identity ${\bf
Y}_p\,T(x)=T(x)$, $\forall x\in\cal M$, with ${\bf Y}_p$ the
Young symmetrizer on tensor of rank $p$ associated to the Young
symmetry $Y_p$. Let $\Omega_{(Y)}({\cal M})$ be the graded vector
space $\oplus_p\Omega^p_{(Y)}({\cal M})$ of irreducible tensor
fields on $\cal M$. One can define a product in
$\Omega_{(Y)}({\cal M})$ but it is not associative
\cite{Dubois-Violette:1999,Dubois-Violette:2001}.

One then generalizes the exterior differential by setting
\cite{Dubois-Violette:1999,Dubois-Violette:2001}
\begin{equation}
d\equiv{\bf Y}_{p+1}\circ \partial :\Omega^p_{(Y)}({\cal
M})\rightarrow \Omega^{p+1}_{(Y)}({\cal M})\,,\label{defd}
\end{equation}
that is to say one first takes the partial derivative of the
tensor $T\in \Omega^p_{(Y)}({\cal M})$ and one acts with the Young
symmetrizer ${\bf Y}_{p+1}$ to obtain a tensor in
$\Omega^{p+1}_{(Y)}({\cal M})$. Examples are provided in the next
section. Let us briefly mention that there are no $dx^\m$ in this
definition of the operator $d$. The operator $d$ is not nilpotent
in general, therefore $d$ does not always endow
$\Omega_{(Y)}({\cal M})$ with a structure of differential complex.

If we want to generalize the calculus of differential forms, we
have to use the extension of the structure of differential complex
with higher order of nilpotency. Anyway, let us stress that the
operator $d$ is not really a differential because, in general, $d$
is neither nilpotent nor a derivative (even if one defines a
product in $\Omega_{(Y)}({\cal M})$ the non-trivial projections
destroy the Leibnitz rule).

A sufficient condition for $d$ to endow $\Omega_{(Y)}({\cal M})$
with a structure of $N$-complex is that the number of columns of
any Young diagram be strictly smaller than $N$.
\begin{lemma} Let $S$ be a non-vanishing integer and assume that the sequence $(Y)$ is
such that the number of columns of the Young diagram $Y_p$ is
strictly smaller than $S+1$ (i.e. $\leq S$) for any $p\in \mathbb
N$. Then the space $\Omega_{(Y)}({\cal M})$ endowed with the
operator $d$ is a $(S+1)$-complex.
\end{lemma}
\noindent Indeed, one has $d^{S+1}=0$ since the indices in one
column are antisymmetrized and $d^{S+1}\omega$ necessarily
involves at least two partial derivatives $\partial$ in one of the
columns (there are $S+1$ partial derivatives involved and at most
$S$ columns).

%%%%%%%%%%%%%%%%%%%%%%%%%%%%%%%%%%%%%%%%%%%%%%%%%%%%%%%%%%%%%%%%%%%%%%%%

\section{Symmetric gauge tensors and maximal
sequences}\label{symt}

A Young diagrams sequence of interest in theories of spin $S\geq
1$ is the maximal sequence $Y^S=(Y^S_p)_{p\in\mathbb N}$
\cite{Dubois-Violette:1999,Dubois-Violette:2001}. This sequence is
defined as the naturally ordered sequence of diagrams with
maximally filled rows.

\noindent\,\,{\bf Notation:} In order to simplify the notation, we
shall denote $\Omega^p_{(Y^S)}({\cal M})$ by $\Omega^p_S({\cal
M})$ and $\Omega_{(Y^S)}({\cal M})$ by $\Omega_S({\cal M})$.

\noindent If $D$ is the dimension of the manifold ${\cal M}$ then
the subcomplexes $\Omega^p_{S}({\cal M})$ with $p>SD$ are trivial
since, for these values of $p$, the Young diagrams $Y^S_p$ have at
least one column with more than $D$ cells.

\subsection{Massless spin one gauge field}

It is clear that $\Omega_1({\cal M})$ with the differential $d$ is
the usual complex $\Omega({\cal M})$ of differential forms on
${\cal M}$. The connection between the complex of differential
forms on ${\cal M}$ and the theory of classical $q$-form gauge
fields is well known. Indeed the subcomplex
\begin{equation}
\Omega^0({\cal M})\stackrel{d_0}{\rightarrow}\Omega^1({\cal
M})\stackrel{d_1}{\rightarrow}\ldots
\stackrel{d_{q-1}}{\rightarrow}\Omega^q({\cal
M})\stackrel{d_q}{\rightarrow}\Omega^{q+1}({\cal M})
\stackrel{d_{q+1}}{\rightarrow}\Omega^{q+2}({\cal
M})\label{complex}\end{equation} with $d_p\equiv
d:\Omega^p\rightarrow \Omega^{p+1}$, has the following
interpretation in terms of $q$-form gauge field $A_{[q]}$ theory
(see subsection \ref{degreesf}). The space $\Omega^{q+1}({\cal
M})$ is the space which the field strength $F_{[q+1]}$ lives in.
The space $\Omega^{q+2}({\cal M})$ is the space of Hodge duals to
magnetic sources $*J_m$ (at least if we extend the space of
``smooth" $(q+2)$-forms to de Rham currents) since
$dF_{[q+1]}=(*J_m)_{[q+2]}$. If there is no magnetic source, the
fieldstrength belongs to the kernel of $d_{q+1}$. The Abelian
gauge field $A_{[q]}$ belongs to $\Omega^q({\cal M})$. The
subspace Ker$d_q$ of $\Omega^q({\cal M})$ is the space of pure
gauge configurations (which are physically irrelevant). The space
$\Omega^{q-1}({\cal M})$ is the space of infinitesimal gauge
parameters $\Lambda_{[q-1]}$ and $\Omega^{q-2}({\cal M})$ is the
space of first reducibility parameters $\Lambda_{[q-2]}$. The
story continues for higher order reducibility parameters. If the
manifold ${\cal M}$ has the topology of $\mathbb R^D$ then
(\ref{complex}) is an exact sequence.

\subsection{Massless spin two gauge field}

As another example, we exhibit the correspondence between some
spaces Young diagrams in the maximal sequence with at most two
columns and their corresponding spaces in the differential
$3$-complex $\Omega_2({\cal M})$

\begin{table}[ht]
\begin{center}\begin{tabular}{|c|c|c|c|}
  \hline
  Young diagram & Vector space & Example & Components \\
  \hline
  \mbox{\footnotesize
\begin{picture}(38,15)(0,0)
\multiframe(10,0)(10.5,0){1}(10,10){}
\end{picture}\normalsize} & $\Omega^1_2({\cal M})$ & lin. diffeomorphism parameter & $\xi_\m$ \\
&&&\\
  \mbox{\footnotesize
\begin{picture}(38,15)(0,0)
\multiframe(10,-1)(10.5,0){2}(10,10){}{}
\end{picture}\normalsize} & $\Omega^2_2({\cal M}) $ & graviton & $h_{\m\n}$ \\
&&&\\
  \mbox{\footnotesize
\begin{picture}(38,15)(0,0)
\multiframe(10,5)(10.5,0){2}(10,10){}{}
\multiframe(10,-5.5)(10.5,0){1}(10,10){}
\end{picture}\normalsize
}  & $\Omega^3_2({\cal M}) $ & mixed symmetry type field & $M_{\m\n\,|\,\r}$ \\
&&&\\
  \mbox{\footnotesize
\begin{picture}(38,15)(0,0)
\multiframe(10,6)(10.5,0){2}(10,10){}{}
\multiframe(10,-4.5)(10.5,0){2}(10,10){}{}
\end{picture}\normalsize} & $\Omega^4_2({\cal M}) $ & Riemann tensor & $R_{\mu\nu\,\rho\sigma}$ \\
&&&\\
  \mbox{\footnotesize
\begin{picture}(38,15)(0,0)
\multiframe(10,9.5)(10.5,0){2}(10,10){}{}
\multiframe(10,-1)(10.5,0){2}(10,10){}{}
\multiframe(10,-11.5)(10.5,0){1}(10,10){}
\end{picture}\normalsize} & $\Omega^5_2({\cal M})$ & Bianchi identity & $\partial_{[\l}R_{\mu\nu]\,\rho\sigma}$ \\
&&&\\    \hline
\end{tabular}
\caption{Two-column maximal sequence and its physical relevance.}
\end{center}
\end{table}

The interest of $\Omega_2({\cal M})$ is its direct applicability
in free spin two gauge theory. Indeed in this case, the analog of
the sequence (\ref{complex}) is
\begin{equation}
\Omega^1_2({\cal M})\stackrel{d}{\rightarrow} \Omega^2_2({\cal
M})\stackrel{d^2}{\rightarrow} \Omega^4_2({\cal
M})\stackrel{d}{\rightarrow}\Omega^5_2({\cal M})\label{complex2}
\end{equation} since $\Omega^1_2({\cal M})$ is
the space of covariant vector fields ($\xi_\mu$) on ${\cal M}$,
$\Omega^2_2({\cal M})$ the space of covariant rank $2$ symmetric
tensor fields ($h_{\mu\nu}$) on ${\cal M}$, $\Omega^4_2({\cal M})$
the space of covariant tensor fields of degree 4
($R_{\mu\nu\,\rho\sigma}$) on ${\cal M}$ having the symmetries of
the Riemann curvature tensor and $\Omega^5_2({\cal M})$ the space
of covariant tensor fields of degree 5 having the symmetries of
the left-hand side of the Bianchi II identity. The action of the
$3$-differential writes explicitly in terms of components
\begin{eqnarray}
(d\xi)_{\mu\nu}&=&\frac12(\partial_\mu \xi_\nu+\partial_\nu \xi_\mu) \\
(d^2h)_{\lambda\mu\rho\nu}&=&\frac14(\partial_\lambda\partial_\rho
h_{\mu\nu} +\partial_\mu\partial_\nu h_{\lambda\rho}-\partial_\mu
\partial_\rho h_{\lambda\nu} -\partial_\lambda\partial_\nu
h_{\mu\rho})\label{d2h}\\
(dR)_{\lambda\mu\nu\alpha\beta}&=&\frac13(\partial_\lambda
R_{\mu\nu\,\alpha\beta}+\partial_\mu
R_{\nu\lambda\,\alpha\beta}+\partial_\nu
R_{\lambda\mu\,\alpha\beta}).\label{dR}
\end{eqnarray}

The generalized $3$-complex $\Omega_2({\cal M})$ can be pictured
as the commutative diagram \ba\xymas{&&&&\cdots\\
&&&\Omega^6_2({\cal M})\ar[r]_{d}\ar[ur]^{d^2}&\cdots\\
&&\Omega^4_2({\cal M})\ar[r]_{d}\ar[ur]^{d^2}& \Omega^5_2({\cal M})\ar[u]_{d}\\
&\Omega^2_2({\cal M}) \ar[r]_{d}\ar[ur]^{d^2}& \Omega^3_2({\cal M})\ar[u]_{d}\\
\Omega^0_2({\cal M}) \ar[r]_{d}\ar[ur]^{d^2}& \Omega^1_2({\cal
M})\ar[u]_{d} }\label{maxseq}\ea In terms of this diagram, the
higher order nilpotency $d^3=0$ translates into the fact that (i)
if one takes a vertical arrow followed by a diagonal arrow, or
(ii) if a diagonal arrow is followed by a horizontal arrow, one is
always mapped to zero.

%%%%%%%%%%%%%%%%%%%%%%%%%%%%%%%%%%%%%%%%%%%%%%%%%%%%%%%%%%%%%%%%%%

\section{Rectangular diagrams}\label{rectd}

The generalized cohomology \cite{Dubois-Violette:2000} of the
$N$-complex $\Omega_{N-1}({\cal M})$ is the family of graded
vector spaces $H_{(k)}(d)$ with $1\leq k \leq N-1$ defined by
$H_{(k)}(d)=\mbox{Ker}(d^k)/\mbox{Im}(d^{N-k})$. In general the
cohomology groups $H^p_{(k)}(d)$ are not empty, even when ${\cal
M}$ has a trivial topology. Nevertheless there exists a
generalization of the Poincar\'e lemma for $N$-complexes of
interest in gauge theories.

Let $Y^{S}$ be a maximal sequence of Young diagrams. The
(generalized) Poincar\'e lemma says that for ${\cal M}$ with the
topology of $\mathbb R^D$ the generalized
cohomology\footnote{Strictly speaking, the generalized Poincar\'e
lemma for rectangular diagrams was proved in
\cite{Dubois-Violette:1999,Dubois-Violette:2001} with an other
choice of convention for the Young symmetrizer $\overline{\bf
Y}$ where one first antisymmetrizes the columns. This other Young
symmetrizer is more convenient to proof the theorem in
\cite{Dubois-Violette:2001} but is inappropriate for considering
Hodge dualization properties. This explains our choice of
convention; still, as we will show later, the generalized
Poincar\'e lemma for rectangular diagrams remains true with the
definition (\ref{defd}).} of $d$ on tensors represented by
rectangular diagrams is empty in the space of maximal tensors
\cite{Olver:1983,Dubois-Violette:1999,Dubois-Violette:2001}. We
have the
\begin{proposition}(Generalized Poincar\'e lemma for rectangular
diagrams)\label{GenPoinrect}

\begin{itemize}
\item $H^0_{(k)}\left(\Omega_S(\mathbb R^D)\right)$ is the space of real
polynomial functions on $\mathbb R^D$ of degree strictly less than
$k$ ($1\leq k \leq N-1$) and
\item $H^{nS}_{(k)}\left(\Omega_S(\mathbb R^D)\right)=0$ $\forall n$ such that $1\leq n\leq D-1$.
\end{itemize}
\end{proposition}

This is the first theorem of \cite{Dubois-Violette:2001}, the
proof of which is given therein. This theorem strengthens the
analogy between the two complexes (\ref{complex}) and
(\ref{complex2}) since the generalized Poincar\'e lemma implies
that (\ref{complex2}) is also an exact sequence when ${\cal M}$
has a trivial topology. Exactness at $\Omega^2_2({\cal M})$ means
$H^2_{(2)}\left(\Omega_2(\mathbb R^D)\right)=0$ and exactness at
$\Omega^4_2({\cal M})$ means $H^4_{(1)}\left(\Omega_2(\mathbb
R^D)\right)=0$. These properties have a physical interpretation in
terms of the linearized Bianchi identity II and gauge
transformations. Let $R_{\mu\nu\rho\sigma}$ be a tensor
antisymmetric in its two pairs of indices
$R_{\mu\nu\rho\sigma}=-R_{\nu\mu\rho\sigma}=-R_{\mu\nu\sigma\rho}$,
namely it has the symmetry of the Young diagram \footnotesize
\begin{picture}(60,25)(0,0)
\multiframe(0,10)(10.5,0){1}(10,10){}
\multiframe(0,-0.5)(10.5,0){1}(10,10){}\put(20,3){$\bigotimes$}
\multiframe(40,10)(10.5,0){1}(10,10){}
\multiframe(40,-0.5)(10.5,0){1}(10,10){}\end{picture}\normalsize .
This one decomposes according to \ba\footnotesize
\begin{picture}(106,25)(0,0)
\multiframe(-80,10)(10.5,0){1}(10,10){}
\multiframe(-80,-0.5)(10.5,0){1}(10,10){}\put(-60,3){$\bigotimes$}
\multiframe(-38,10)(10.5,0){1}(10,10){}
\multiframe(-38,-0.5)(10.5,0){1}(10,10){} \put(-20,3){$=$}
\multiframe(0,10)(10.5,0){2}(10,10){}{}
\multiframe(0,-0.5)(10.5,0){2}(10,10){}{} \put(30,3){$\bigoplus$}
\multiframe(50,10)(10.5,0){2}(10,10){}{}
\multiframe(50,-0.5)(10.5,0){1}(10,10){}
\multiframe(50,-11)(10.5,0){1}(10,10){} \put(80,3){$\bigoplus$}
\multiframe(100,10)(10.5,0){1}(10,10){}
\multiframe(100,-0.5)(10.5,0){1}(10,10){}
\multiframe(100,-11)(10.5,0){1}(10,10){}
\multiframe(100,-21.5)(10.5,0){1}(10,10){}
\end{picture}\normalsize
\label{Yougdec}\\\nonumber\ea If we ask $R$ to obey the first
Bianchi identity (\ref{BianchiI}), we kill the last two terms in
its decomposition (\ref{Yougdec}) hence the tensor $R$ has the
symmetries of the Riemann tensor and belongs to $\Omega^4_2({\cal
M})$. Furthermore, from (\ref{dR}) it is obvious that the
linearized second Bianchi identity (\ref{BianchiII}) for $R$ reads
$dR=0$. Since the Riemann tensor has the symmetries of a
rectangular diagram, we get from the exactness of the sequence
(\ref{complex2}) that $R=d^2h$ with $h\in\Omega^2_2({\cal M})$.
This means that $R$ is effectively the linearized Riemann tensor
associated to the spin two field $h$, as can be directly seen from
(\ref{d2h}). However, the definition of the metric fluctuation $h$
is not unique : the gauge field $h+\delta h$ is physically
equivalent to $h$ if it leaves the physical linearized Riemann
tensor invariant, i.e. $d^2(\delta h)=0$. Since the sequence
(\ref{complex}) is exact we find : $\delta h=d\xi$ with
$\xi\in\Omega^1_2({\cal M})$. As a result we recover the standard
gauge transformations (\ref{2gauge}).

%%%%%%%%%%%%%%%%%%%%%%%%%%%%%%%%%%%%%%%%%%%%%%%%%%%%%%%%%%%%%%%%%%

\section{Multiforms and Hodge duality}\label{Hodged}

For practical purpose, we will consider the set ${\mathbb
Y}^{(S)}$ of all Young diagrams $Y_{(l_1,l_2,\ldots,l_S)}^{(S)}$
with at most $S$ columns of respective length $0\leq l_S\leq
l_{S-1}\leq\ldots\leq l_2\leq l_1\leq D-1$.

\noindent\,\,{\bf Notation:} The space
$\Omega_{\left(Y^{(S)}\right)}({\cal M})$ is a $(S+1)$-complex
that we denote $\Omega_{(S)}({\cal M})$. The subcomplex
$\Omega_{Y_{(l_1,l_2,\ldots,l_S)}^{(S)}}({\cal M})$ is denoted by
$\Omega^{(l_1,l_2,\ldots,l_S)}_{(S)}({\cal M})$.

This generalized complex $\Omega_{(S)}({\cal M})$ is the
generalization of the differential forms complex $\Omega({\cal
M})=\Omega_{(1)}({\cal M})$ we are looking for, in which each
proper space is invariant under the action of $GL(D,\mathbb R)$.
As an example, the previously mentioned mixed symmetry type gauge
field $M$ belongs to $\Omega^{(n,1)}_{(2)}({\cal M})$ by
(\ref{YS}).

A good mathematical understanding of the gauge structure of free
symmetric tensor gauge field theories is provided by the maximal
sequence and the vanishing of the rectangular diagrams cohomology.
Though, several new mathematical ingredients are needed as well as
an extension of the proposition \ref{GenPoinrect} to capture their
dynamics on-shell. A useful new ingredient is the obvious
generalization of Hodge's duality for $\Omega_S(\mathbb R^D)$,
which is obtained by contractions of the columns with the epsilon
tensor $\varepsilon^{\mu_1\dots\mu_D}$ of ${\cal M}$ and lowering
the indices with the Minkowskian metric. For rank $S$ symmetric
tensor gauge theories there are $S$ different Hodge operations
since the corresponding diagrams may contain up to $S$ columns. A
simple but important point to notice is the following: generically
the Hodge duality is not an internal operation in the space
$\Omega_S({\cal M})$. For that reason we must come back to the
complex $\Omega_{(S)}({\cal M})$ (defined at the end of the
subsection \ref{N-c}) for which the Hodge operation is internal
(up to some column reordering, that have to be understood in what
follows).

\subsection{Multiforms}

Another ingredient is the graded tensor product of $C^\infty({\cal
M})$ with $S$ copies of the exterior algebra $\Lambda({\mathbb
R}^{D*})$ where ${\mathbb R}^{D*}$ is the dual space of basis $d_i
x^\mu$ ($1\leq i\leq S$, $\m=0,1,\ldots,D$). Elements of this
space will be referred to as {\bf multiforms}
\cite{Dubois-Violette:2001}. They are sums of products of the
generators $d_ix^\mu$ with smooth functions on $\cal M$. The
components of a multiform define a tensor with the symmetry
properties of the product of $S$ columns
\vspace{0.3cm}\begin{center}\footnotesize
\begin{picture}(0,37)(0,-24)
\multiframe(-40,10)(10.5,0){1}(10,10){}
\multiframe(-40,-0.5)(10.5,0){1}(10,10){}
\put(-35.5,-12){$\vdots$}\put(-35.5,-23.5){$\vdots$}
\multiframe(-40,-35)(10.5,0){1}(10,10){} \put(-20,7){$\bigotimes$}
\multiframe(0,10)(10.5,0){1}(10,10){}
\multiframe(0,-0.5)(10.5,0){1}(10,10){} \put(4.5,-12){$\vdots$}
\multiframe(0,-25)(10.5,0){1}(10,10){} \put(20,7){$\bigotimes$}
\put(40,7){$\ldots$} \put(60,7){$\bigotimes$}
\multiframe(80,10)(10.5,0){1}(10,10){}\put(85,-1.5){$\vdots$}
\multiframe(80,-15)(10.5,0){1}(10,10){}
\end{picture}\normalsize\end{center}\vspace{0.3cm}

\noindent\,\,{\bf Notation:} We shall denote this complex by
$\Omega_{[S]}({\cal M})$. The subspace
$\Omega^{l_1,l_2,\ldots,l_S}_{[S]}({\cal M})$ is defined as the
space of multiforms whose components have the symmetry properties
of the diagram $D_{l_1,l_2,\ldots,l_S}:=\bigotimes\limits_{i=1}^S
Y^{(1)}_{(l_i)}$ which represents the above product of $S$ columns
with respective lengths $l_1$, $l_2$, ..., $l_S$.

The tensor field
$\a_{[\mu^1_1\ldots\mu^1_{l_1}]\ldots[\mu^S_1\ldots\mu^S_{l_S}]}(x)$
defines a multiform $\a\in\Omega^{l_1,\ldots,l_S}_{[S]}({\cal M})$
which explicitly reads \be
\a=\a_{[\mu^1_1\ldots\mu^1_{l_1}]\ldots[\mu^S_1\ldots\mu^S_{l_S}]}(x)\,\,d_1x^{\mu^1_1}\wedge\ldots\wedge
d_1x^{\mu^1_{l_1}}\otimes\ldots\otimes
d_Sx^{\mu^S_1}\wedge\ldots\wedge d_S x^{\mu^S_{l_S}}
\,.\label{multif}\ee In the sequel, when we refer to the multiform
$\a$ we will speak either of (\ref{multif}) or of its components.
More accurately, we will identify $\Omega_{[S]}({\cal M})$ with
the space of the smooth tensor field components.

We endow $\Omega_{[S]}({\cal M})$ with the structure of a
(multi)complex by defining $S$ anticommuting differentials \be
d_i:\Omega^{l_1,\ldots,l_i,\ldots,l_S}_{\,\,[S]}({\cal
M})\rightarrow\Omega^{l_1,\ldots,l_i+1,\ldots,l_S}_{\,\,[S]}({\cal
M})\,,\quad 1\leq i\leq S \,,\ee defined by adding a box
containing the partial derivative in the $i$-th column. For
instance, $d_2$ acting on the previous diagrammatic example is
\vspace{0.3cm}
\begin{center}\footnotesize
\begin{picture}(0,37)(0,-24)
\multiframe(-40,10)(10.5,0){1}(10,10){}
\multiframe(-40,-0.5)(10.5,0){1}(10,10){}
\put(-35.5,-12){$\vdots$}\put(-35.5,-23.5){$\vdots$}
\multiframe(-40,-35)(10.5,0){1}(10,10){} \put(-20,7){$\bigotimes$}
\multiframe(0,10)(10.5,0){1}(10,10){}
\multiframe(0,-0.5)(10.5,0){1}(10,10){} \put(4.5,-12){$\vdots$}
\multiframe(0,-25)(10.5,0){1}(10,10){}
\multiframe(0,-35.5)(10.5,0){1}(10,10){$\partial$}
\put(20,7){$\bigotimes$} \put(40,7){$\ldots$}
\put(60,7){$\bigotimes$}
\multiframe(80,10)(10.5,0){1}(10,10){}\put(85,-1.5){$\vdots$}
\multiframe(80,-15)(10.5,0){1}(10,10){}
\end{picture}\normalsize\end{center}
\vspace{0.3cm}

\noindent\,\,{\bf Summary of notations:} The complex
$\Omega_{[S]}({\cal M})$ is the subspace of $S$-uple multiforms.
The space $\Omega_{(S)}({\cal M})$ is the $(S+1)$-complex of
tensors represented by Young diagrams with at most $S$ columns. It
is the direct sum of subcomplexes
$\Omega^{(l_1,l_2,\ldots,l_S)}_{(S)}({\cal M})$. The space
$\Omega_S({\cal M})=\oplus_p\Omega^p_S({\cal M})$ is the space of
maximal tensors. Thus we have the chain of inclusions
$\Omega_{S}({\cal M})\subset \Omega_{(S)}({\cal M})
\subset\Omega_{[S]}({\cal M})$.

\subsection{Hodge and trace operators}

We introduce the following notation for the $S$ possible Hodge
dual definitions \be
*_i:\Omega^{l_1,\ldots,l_i,\ldots,l_S}_{\,\,[S]}({\cal
M})\rightarrow\Omega^{l_1,\ldots,D-l_i,\ldots,l_S}_{\,\,[S]}({\cal
M})\,,\quad 1\leq i\leq S \,.\ee The operator $*_i$ is defined as
the action of the Hodge operator on the indices of the $i$-th
column. To remain in the space of covariant tensors requires the
use of the flat metric to lower down indices.

Using the metric, another simple operation that can be defined is
the trace. The convention is that we always take the trace over
indices in two different columns, say the $i$-th and $j$-th. We
denote this operation by\be
\mbox{Tr}_{ij}:\Omega^{l_1,\ldots,l_i,\ldots,l_j,\ldots,l_S}_{\,\,[S]}({\cal
M})\rightarrow\Omega^{l_1,\ldots,l_i-1,\ldots,l_j-1,\ldots,l_S}_{\,\,[S]}({\cal
M})\,,\quad i\neq j \,.\ee The Schur module definition (see
appendix \ref{Young}) gives the necessary and sufficient set of
conditions for a (covariant) tensor $T_{\m_1\m_2\ldots\m_p}(x)$ of
rank $p$ to be in the irreducible representation of $GL(D,{\mathbb
R})$ associated to the Young diagram $Y$ (with $|Y|=p$). Every
index of $T_{\m_1\m_2\ldots\m_p}(x)$ is placed in one box of $Y$.
The set of conditions is the following :
\begin{quote}

$(i)$ $T_{\m_1\m_2\ldots\m_p}(x)$ is completely antisymmetric in
the entries of each column of $Y$,

$(ii)$ complete antisymmetrization of $T_{\m_1\m_2\ldots\m_p}(x)$
in the entries of a column of $Y$ and another entry of $Y$ which
is on the right-hand side of the column vanishes.

\end{quote}
Using the previous definitions of multiforms and of the Hodge dual
and trace operators, this set of conditions can translate the
proposition \ref{Schurprop} in the
\begin{proposition}(Schur module)

Let $\a$ be a multiform in $\Omega^{l_1,\ldots,l_S}_{[S]}({\cal
M})$. If \ba l_j\leq l_i<D\,,\quad\forall\,
i,j\in\{1,\ldots,S\}\,,\nn\ea then one has the equivalence\ba
\mbox{Tr}_{ij}\,\{\,*_i\,\a\,\}\,=\,0\quad\forall\, i,j:\,\, 1\leq
i<j\leq S \quad\Longleftrightarrow\quad \a\in
\Omega^{(l_1,\ldots,l_S)}_{(S)}({\cal M})\,.\nn\ea\label{Schurm}
\end{proposition}
The condition (i) is satisfied since $\a$ is a multiform. The
condition (ii) is written in terms of tracelessness conditions.
Let us mention that such rewriting is only possible for metric
spaces, not required by the study of representations of
$GL(G\mathbb R)$.

Another useful property, that generalizes the derivation followed
in the chain of equations (\ref{*R})-(\ref{p*R}), is for any
$i,j\in\{1,\ldots,S\}$ \be
\bullet\quad\left\{\begin{array}{lll}\mbox{Tr}_{ij}\,\a\,=\,0&& \\
d_i \a\,=\,0&&\end{array}\right.\quad\Longrightarrow\quad
d_j\,(*_j\,\a)\,=\,0\,.\label{makeuse}\ee

The following property on powers of the trace operator will also
be useful later on. We state it as the
\begin{proposition}\label{tracepower}
Let $\a\in\Omega^{l_1,\ldots,l_S}_{(S)}({\cal M})$ be a multiform
such that $l_j\leq l_i$. For any $m\in\mathbb N$ ( $0\leq m\leq
D-l_i$ ), one has the equivalence \ba
\big(\mbox{Tr}_{ij}\big)^m\{*_i*_j\a\}=0\quad\quad\Longleftrightarrow\quad
\big(\mbox{Tr}_{ij}\big)^{m+l_i+l_j-D}\,\{\,\a\,\}=0\,.\nn\ea
\end{proposition} \proof{The proposition \ref{tracepower} is a direct consequence of
(\ref{also}). \begin{description}
\item[$\underline{\,\bf\Rightarrow:}$] We proceed by
recurrence on the number of traces (acting on $\a$). More
precisely, the recurrence hypothesis is that, for any
$p\in{\mathbb N}_0$, if $\big(\mbox{Tr}_{ij}\big)^{l_j-n}\{\a\}=0$
for all $n\geq p-1$, then
\be\big(\mbox{Tr}_{ij}\big)^{D-l_i-p}\{*_i*_j\a\}=0\quad\Rightarrow\quad
\big(\mbox{Tr}_{ij}\big)^{l_j-p}\{\a\}=0\,.\label{bogol}\ee This
is true because $\big(\mbox{Tr}_{ij}\big)^{D-l_i-p}\{*_i*_j\a\}$
is equal to a sum of terms proportional to
$\big(\mbox{Tr}_{ij}\big)^{l_j-n}\{\a\}$ for all $n\geq p$ (due to
(\ref{also})), and these last terms vanish by hypothesis except
the one proportional to $\big(\mbox{Tr}_{ij}\big)^{l_j-p}\{\a\}$.
And the final step to proof the recurrence hypothesis is that the
vanishing of $\big(\mbox{Tr}_{ij}\big)^{D-l_i-p}\{*_i*_j\a\}$
implies $\big(\mbox{Tr}_{ij}\big)^{l_j-p}\{\a\}=0$.

The recurrence hypothesis will proof the proposition because
$\big(\mbox{Tr}_{ij}\big)^m\{*_i*_j\a\}=0$ implies that
$\big(\mbox{Tr}_{ij}\big)^{\overline{m}}\{*_i*_j\a\}=0$ for all
$\overline{m}\geq m$. Indeed, we start from
$\big(\mbox{Tr}_{ij}\big)^{D-l_i}\{*_i*_j\a\}=0$ to deduce that
$\big(\mbox{Tr}_{ij}\big)^{l_j}\{\a\}=0$, as follows immediately
from (\ref{also}). The recurrence (\ref{bogol}) can be applied
from $p=1$ to $p=D-l_i-m$ and one goes down from
$\big(\mbox{Tr}_{ij}\big)^{l_j}\{\a\}=0$ to the desired vanishing
of $\big(\mbox{Tr}_{ij}\big)^{m+l_i+l_j-D}\,\{\,\a\,\}$.
\item[$\underline{\,\bf\Leftarrow:}$] That the sufficiency
is also true is obvious from the implication (\ref{bogol}) since
$*_i*_j(*_i*_j\a)=\pm\a$.
\end{description}
}

We end up this section by a result on the irreducible
representations of $O(D-1,1)$ \cite{Hamermesh} which is a
corollary of the proposition \ref{tracepower}.
\begin{corollary}The (completely) traceless tensors corresponding to Young diagrams
in which the sum of the lengths of the first two columns exceeds
$D$ must be identically zero.

In other words, let $\a\in\Omega^{(l_1,\ldots,l_S)}_{(S)}({\cal
M})$ be an irreducible tensor such that $l_1+l_2\leq D$. Then, one
has the equivalence\ba
\big(\mbox{Tr}_{ij}\big)\{\a\}=0\quad\forall i,j
\quad\Longleftrightarrow\quad \a\equiv 0\,.\nn\ea\label{Trvanish}
\end{corollary}The hypotheses of the corollary \ref{Trvanish} are different with
the ones of the proposition \ref{Schurm} because $l_j\leq l_i$
implies that the dual $*_i\,\a$ has the sum of the lengths of the
$i$-th and $j$-th columns equal to $\ell_i+\ell_j=D-l_i+l_j\leq
D$.

Let us now apply all these new tools to the specific case of
linearized gravity.

%%%%%%%%%%%%%%%%%%%%%%%%%%%%%%%%%%%%%%%%%%%%%%%%%%%%%%%%%%%%%%%%%%%%

\section{Linearized gravity field equations}\label{fieldequs}

From now on, we restrict ourselves to the case of linearized
gravity, i.e. $S=2$ symmetric gauge fields. So there are two
possible Hodge operations, denoted by $*$, acting on the first
column if it is written on the left and on the second column if it
is written on the right. Since we are no longer restricted to
maximal Young diagrams the notation $d$ is ambiguous (we do not
know a priori on which Young symmetry type we should project in
the definition (\ref{defd})). Instead we use the above
differentials $d_i$. For linearized gravity there exist only two
of them : $d_1$ called the (left) differential $d^L$ and $d_2$,
the (right) differential $d^R$. With these differentials it is
possible to rewrite (\ref{d*R'}) in the compact form
$d^L*R=d^RR*=0$. The second Bianchi identity reads $d^L R=d^R
R=0$.

Our convention is that we take the trace over indices in the same
row, using the flat background metric $\eta_{\mu\nu}$. We denote
this operation by Tr (which is $\mbox{Tr}_{12}$ according to the
definition given in the previous section). In this notation the
Einstein equation (\ref{Einstein}) takes the form Tr$R=0$ while
the first Bianchi identity (\ref{BianchiI}) reads Tr$*R=0$. From
the proposition \ref{Schurm}, the following property holds : Let
$B$ be a ``biform" in $\Omega^{p,q}_{[2]}({\cal M})$ which means
$B$ is a tensor with symmetry
\begin{center}\footnotesize
\begin{picture}(20,37)(0,-24)
\multiframe(0,10)(10.5,0){1}(10,10){}
\multiframe(0,-0.5)(10.5,0){1}(10,10){}
\put(4.5,-12){$\vdots$}\put(4.5,-23.5){$\vdots$}
\multiframe(0,-35)(10.5,0){1}(10,10){} \put(20,7){$\bigotimes$}
\multiframe(40,10)(10.5,0){1}(10,10){} \put(45,-1.5){$\vdots$}
\multiframe(40,-15)(10.5,0){1}(10,10){}
\end{picture}\normalsize\end{center}
Then, $B$ obeys the (first) ``Bianchi identity"
\be\mbox{Tr}(*B)=0\ee if and only if $B\in
\Omega^{(p,q)}_{(2)}({\cal M})$. Pictorially it is described by
the diagram
\begin{center}\footnotesize
\begin{picture}(20,37)(0,-24)
\multiframe(0,10)(10.5,0){1}(10,10){}
\multiframe(0,-0.5)(10.5,0){1}(10,10){}
\put(4.5,-12){$\vdots$}\put(4.5,-23.5){$\vdots$}
\multiframe(0,-35)(10.5,0){1}(10,10){}
\multiframe(10.5,10)(10.5,0){1}(10,10){} \put(15.5,-1.5){$\vdots$}
\multiframe(10.5,-16)(10.5,0){1}(10,10){}
\end{picture}\normalsize\end{center}
that is, the two columns of the product are glued together.

With all this new artillery, it becomes easier to extend the
concept of EM duality for linearized gravity. First of all we
emphasize the analogy between the Bianchi identities and the field
equations by rewriting them respectively as
\be\left\{\begin{array}{lll}\mbox{Tr}*R=0&& \\
d^LR=d^RR=0&&\end{array}\right. ,\label{BianchiIII}\ee and
\be\left\{\begin{array}{lll}\mbox{Tr}R=0&& \\
d^L*R=d^RR*=0&&\end{array}\right. ,\label{fieldequ}\ee where
$R_{\mu\nu\,\rho\sigma}\equiv\mbox{ \footnotesize
\begin{picture}(50,15)(0,0)
\multiframe(-3,6.5)(10.5,0){1}(10,10){$\mu$}
\multiframe(-3,-4)(10.5,0){1}(10,10){$\nu$}\put(15,1){$\bigotimes$}
\multiframe(31,6.5)(10.5,0){1}(10,10){$\rho$}
\multiframe(31,-4)(10.5,0){1}(10,10){$\sigma$}
\end{picture}}$. We recall that $d^L*R=d^RR*=0$ was obtained in section \ref{Lingravi} by
using the second Bianchi identity.

As discussed at the end of the subsection \ref{Hodged}, the first
Bianchi identity implies that $R$ effectively has the symmetry
properties of the Riemann tensor, i.e.
$R_{\mu\nu\,\rho\sigma}\equiv\mbox{ \footnotesize
\begin{picture}(25,15)(0,0)
\multiframe(-3,6.5)(10.5,0){1}(10,10){$\mu$}
\multiframe(-3,-4)(10.5,0){1}(10,10){$\nu$}
\multiframe(7.5,6.5)(10.5,0){1}(10,10){$\rho$}
\multiframe(7.5,-4)(10.5,0){1}(10,10){$\sigma$}
\end{picture}}$. Using this symmetry
property, the two equations $d^LR=d^RR=0$ can now be rewritten as
the single equation $dR=0$. Therefore if the manifold ${\cal M}$
is of trivial topology then, for a given multiform $R\in
\Omega^{2,2}_{[2]}({\cal M})$, one has the equivalence
\be\left\{\begin{array}{lll}\mbox{Tr}*R=0&& \\
d^LR=d^RR=0&&\end{array}\right.\Leftrightarrow\quad
\left\{\begin{array}{lll}R=d^2h &&\\
h\in\Omega^2_2({\cal M})&&\end{array}\right. \,,\label{Bibi}\ee
due to proposition \ref{GenPoinrect}.

\subsection{Dual linearized Riemann tensor}

By proposition \ref{Schurm}, the (vacuum) Einstein equation
Tr$R=0$ can then be translated into the assertion that the dual of
the Riemann tensor has (on-shell) the symmetries of a diagram
$(D-2,2)$, in other words $*R\in\Omega^{(D-2,2)}_{(2)}({\cal M})$.
The second Bianchi identity $d^R R=0$ together with the Einstein
equations imply the equation $d^L*R=0$. Furthermore the second
Bianchi identity $d^R R=0$ is equivalent to $d^R*R=0$, therefore
we have the equivalence
\be\left\{\begin{array}{lll}\mbox{Tr}R=0&& \\
d^RR=0&&\end{array}\right.\Leftrightarrow\quad
\left\{\begin{array}{lll}*R\in\Omega^{(D-2,2)}_{(2)}({\cal M}) &&\\
d^L*R=d^R*R=0 &&\end{array}\right. \,.\ee In addition
$d^L*R=d^R*R=0$ now imply $*R=d^2\tilde{h}$ (where we denote the
non-ambiguous product $d^L\,d^R$ by $d^2$). The tensor field
$\tilde{h}\in \Omega^{(D-3,1)}_{(2)}({\cal M})$ is the dual gauge
field of $h$ obtained in (\ref{dualcurva}). This property
(\ref{dualcurva}) which holds for manifolds $\cal M$ with the
topology of ${\mathbb R}^D$ is a direct application of the
corollary \ref{coro} of the generalized Poincar\'e lemma given in
the next subsection; we anticipate this result here in order to
motivate the theorem \ref{GenPoin} by a specific example. We have
an analogue of (\ref{Bibi}) that exhibits a duality symmetry
similar to the EM duality of electrodynamics under the interchange
of Bianchi identities and field equations,
\be\left\{\begin{array}{lll}\mbox{Tr}R=0&& \\
d^L*R=d^R*R=0&&\end{array}\right.\Leftrightarrow\quad
\left\{\begin{array}{lll}*R=d^2\tilde{h} &&\\
\tilde{h}\in\Omega^{(D-3,1)}_{(2)}({\cal M})&&\end{array}\right.
\,.\ee Tensor gauge fields in $\Omega^{(D-3,1)}_{(2)}({\cal M})$
have mixed symmetry and are discussed above in section
\ref{mixed}. The right-hand-side of (\ref{dualcurva}) is
represented by \vspace{0.5cm}
\begin{center}\footnotesize
\begin{picture}(0,30)(40,-20)
\multiframe(0,10)(10.5,0){1}(10,10){$\partial$}
\multiframe(0,-0.5)(10.5,0){1}(10,10){}
\multiframe(0,-11)(10.5,0){1}(10,10){}
\put(4.5,-22.5){$\vdots$}\put(4.5,-34){$\vdots$}
\multiframe(0,-45.5)(10.5,0){1}(10,10){}
\multiframe(10.5,10)(10.5,0){1}(10,10){$\partial$}
\multiframe(10.5,-0.5)(10.5,0){1}(10,10){}
\end{picture}\normalsize\,.\end{center}\vspace{1cm}
The appropriate symmetries are automatically implemented by the
antisymmetrizations in (\ref{dualcurva}) since the dual gauge
field $\tilde{h}$ has already the appropriate symmetry
\vspace{0.5cm}
\begin{center}\footnotesize
\begin{picture}(0,25)(40,-20)
\multiframe(0,10)(10.5,0){1}(10,10){}
\multiframe(0,-0.5)(10.5,0){1}(10,10){}
\put(4.5,-12){$\vdots$}\put(4.5,-23.5){$\vdots$}
\multiframe(0,-35)(10.5,0){1}(10,10){}
\multiframe(10.5,10)(10.5,0){1}(10,10){}
\end{picture}\normalsize\,.\end{center}\vspace{0.5cm}
In other words, the two explicit antisymmetrizations in
(\ref{dualcurva}) are sufficient to ensure that $*R$ has the
symmetries associated with $Y_{(D-2,2)}^{(2)}$. A general
explanation of this fact will be given at the end of the next
subsection.

The dual linearized Riemann tensor is invariant under the
transformation \be\delta \tilde{h}=d(S+A)\,\quad\mbox{with}\quad
S\in\Omega^{(D-4,1)}_{(2)}({\cal M})\,,\quad
A\in\Omega^{(D-3,0)}_{(2)}({\cal M})\cong \Omega^{D-3}({\cal
M})\,.\ee The right-hand-side of this gauge transformation,
explicitly written in (\ref{gtransfoht}), is represented by
\begin{center}\footnotesize
\begin{picture}(0,37)(40,-20)
\multiframe(0,10)(10.5,0){1}(10,10){}
\multiframe(0,-0.5)(10.5,0){1}(10,10){}
\put(4.5,-12){$\vdots$}\put(4.5,-23.5){$\vdots$}
\multiframe(0,-35)(10.5,0){1}(10,10){$\partial$}
\multiframe(10.5,10)(10.5,0){1}(10,10){}\put(25,7){$\bigoplus$}
\multiframe(40,10)(10.5,0){1}(10,10){}
\multiframe(40,-0.5)(10.5,0){1}(10,10){}
\put(44.5,-12){$\vdots$}\put(44.5,-23.5){$\vdots$}
\multiframe(40,-35)(10.5,0){1}(10,10){}
\multiframe(50.5,10)(10.5,0){1}(10,10){$\partial$}
\end{picture}\normalsize\end{center}\vspace{0.5cm}
In this formalism, the reducibilities (\ref{red1}) and
(\ref{red2}) respectively read (up to coefficient redefinitions)
\ba&\stackrel{(i-1)}{S}=d(\stackrel{(i)}{S}+\stackrel{(i)}{A})\,,\quad\stackrel{(i-1)}{A}=-d\stackrel{(i)}{A} \,,\quad (i=2,\ldots, D-2)\,,&\nn\\
&\stackrel{(i)}{S}\in\Omega^{(D-3-i,1)}_{(2)}({\cal M})\,,\quad
\stackrel{(i)}{A}\in\Omega^{D-2-i}({\cal M})\,.&\label{redu}\ea
%The reducibility relations (\ref{redu}) can be written in the compact form
%$\stackrel{(i-1)}{S}+\stackrel{(i-1)}{A}=d(\stackrel{(i)}{S}+\stackrel{(i)}{A})$
%where the operator $d$ acts in two different ways on $\stackrel{(i)}{A}$.
These reducibilities are a direct consequence of the corollary
\ref{corol}.

\subsection{Comparison with electrodynamics}

Compared to electromagnetism, linearized gravity presents several
new features. First, there are now two kinds of Bianchi
identities, some are algebraic relations (Bianchi I) and the other
are differential equations (Bianchi II). In electromagnetism, only
the latter are present. Second (this is perhaps more important),
the equation of motion of linearized gravity theory is an
algebraic equation for the curvature ($\mbox{Tr}R=0$). This is
natural since the curvature tensor already contains two
derivatives of the gauge field. Moreover for higher spin gauge
fields $h\in\Omega^{(1,\ldots,1)}_S$ ($S\geq 3$) the natural gauge
invariant curvature $d^Sh\in\Omega^{(2,\ldots,2)}_S$ contains $S$
derivatives of the completely symmetric gauge field, hence natural
second order equations of motion cannot contain this
curvature\footnote{In some recent work \cite{Francia:2002},
Sagnotti et al. wrote the (second order) e.o.m. of
\cite{Fronsdal:1978,deWit:1980} in terms of the curvature, by
formally dividing by d'Alembertian. This new formulation
\cite{Francia:2002} avoids the tedious trace conditions of
\cite{Fronsdal:1978,deWit:1980} but the price paid is high :
non-locality of field equations.}. Third the current conservation
in electromagnetism is a direct consequence of the field equation
while for linearized gravity the Bianchi identities play a crucial
role.

In relation with the first remark, the introduction of sources for
linearized gravity seems rather cumbersome to deal with. A natural
proposal is to replace the Bianchi I identities by equations\be
\mbox{Tr}*R=\hat{T}\,,\quad\hat{T}\in\Omega^{D-3,1}_{[2]}({\cal
M})\,.\ee If one uses the terminology of electrodynamics, it is
natural to call $\hat{T}$ a ``magnetic" source. If a dual source
is effectively present, i.e. $\hat{T}\neq 0$, the tensor $R$ is no
longer irreducible under $GL(D,\mathbb R)$, that is to say $R$
becomes a sum of tensors of different symmetry types and only one
of them has the symmetries of the Riemann. This case is not
considered in this work. The linearized Einstein equations reads
\be \mbox{Tr}R=\overline{T}\,,\quad
\overline{T}\in\Omega^{1,1}_{[2]}({\cal M})\,.\ee The sources
$\overline{T}$ and $\hat{T}$ respectively couple to the gauge
fields $h$ and $\tilde{h}$. The ``electric" source $\overline{T}$
is a symmetric tensor (related to the energy-momentum tensor) if
the dual source $\hat{T}$ vanishes, since $R\in\Omega^4_2({\cal
M})$ in that case. An other intriguing feature is that a violation
of Bianchi II identities implies a non-conservation of the
linearized energy-momentum tensor because \be \partial^\m
T_{\m\n}=\frac32\,
\partial_{[\m}^{\,}R_{\n\r]}^{\quad\m\r}\,,\ee according to the
linearized Einstein equations (\ref{Einsteinwsource}).

Let us now stress some peculiar features of $D=4$ dimensional
spacetime. From our previous experience with electromagnetism and
our definition of Hodge duality, we naturally expect this
dimension to be privileged. In fact, the analogy between
linearized gravity and electromagnetism is closer in four
dimensions because less independent equations are involved: $*R$
has the same symmetries as the Riemann, thus the dual gauge field
$\tilde{h}$ is a symmetric tensor in $\Omega^2_2({\cal M})$. So
the Hodge duality is a symmetry of the theory only in four
dimensional spacetime. The dual tensor $*R$ is represented by a
Young diagram of rectangular shape and the proposition
\ref{GenPoinrect} can be used to derive the existence of the dual
potential as a consequence of the field equation $d*R=0$.

%%%%%%%%%%%%%%%%%%%%%%%%%%%%%%%%%%%%%%%%%%%%%%%%%%%%%%%%%%%%%%%%%%

\section{Generalized Poincar\'e lemma}\label{GenPl}

In the previous subsection, the Hodge duality operation enforced
the use of the space $\Omega_{(S)}({\cal M})$ of tensors with at
most $S$ columns. This unavoidable fact asks for an extension to
general irreducible tensors in $\Omega_{(S)}({\cal M})$ of the
proposition \ref{GenPoinrect}.

\subsection{Generalized nilpotency}

Let $Y_p$ be well-included\footnote{See appendix \ref{Young} for
the definition of this stronger notion of inclusion.} into
$Y_{p+q}$, that is $Y_p\Subset Y_{p+q}$. We ``generalize"
(\ref{defd}) by introducing the differential operators $d^I$ with
$I \subset \{1,2, \dots, S\}$ ($\# I = q$) by
\begin{equation}  d^I\equiv c_I\,{\bf Y}_{p+q}\circ \big(\prod_{i\in I} \partial_i\big):\Omega^p_{(Y)}({\cal M})\rightarrow
\Omega^{p+q}_{(Y)}({\cal M})
\end{equation}
where $\partial_i$ means that the index corresponding to this
partial derivative is placed at the bottom of the $i$-th
column\footnote{The other choice of Young symmetrizer is more
convenient because for any $\a\in\Omega_{(Y)}({\cal M})$, one has
\be\big(\overline{\bf Y}\circ \prod_{i\in I}
\partial_i\big)\a=\big(\overline{\bf Y}\circ \prod_{i\in I}
d_i\big)\a\ee since one starts by the antisymmetrization in the
columns. Other kind of gentle properties relating the action of
$d^I$ and the one of $d^{\#I}$ in the space of maximal tensors can
be found in \cite{Dubois-Violette:2001}.} and $c_I$ is a
normalization factor such that in the proposition \ref{commutprop}
we have strict equalities\footnote{The precise expression for the
constant $c_I$ was obtained in \cite{Olver:1983}. A summary of the
results of the unpublished \cite{Olver:1983} can be found in the
contribution ``Invariant theory and differential equations" of
Olver in \cite{Olver:1987}.}. When $I$ contains only one element
($q = 1$) we recover the definition of $d$. The operator $d^I$
will be represented by the Young diagram $Y_{p+q}$ where we put a
partial derivative symbol $\partial$ in the $q$ boxes which do not
belong to the subdiagram $Y_p\Subset Y_{p+q}$.

The product of operators $d^I$ is commutative: $d^I\circ
d^J=d^J\circ d^I$ for all $I,J\subset\{1,\ldots,S\}$ ($\#I=q$,
$\#J=r$) such that the product maps to a well-defined Young
diagram $Y_{p+q+r}$. The following proposition gathers all these
properties
\begin{proposition}\label{commutprop}
Let $I$ and $J$ be two subsets of $\subset\{1,\ldots,S\}$. Let
$\a$ be an irreducible belonging to $\Omega_{(2)}({\cal M})$. The
following properties are satisfied\footnote{According to the
terminology of \cite{Olver:1983}, these properties means that the
set ${\mathbb Y}^{(S)}$ is endowed with the structure of
hypercomplex by means of the maps $d^I$.}
\begin{itemize}
  \item $d^I\a\,=\,d^{\#I}\a$
  \item If $I\cap J=\phi$, then $(d^I\circ d^J)\a\,=\,d^{\,I\cup
  J}\a$.
  \item If $I\cap J\neq\phi$, then $(d^I\circ d^J)\a=0$.
\end{itemize}
\end{proposition}
This proposition \ref{commutprop} is proved in \cite{Olver:1983}.
In general, the operator $d^I$, is represented by the diagram
$\widetilde{Y}_{p+q}$ with a partial derivative in the cells at
the bottom of the $q$ columns. The last property states that the
product of $d^I$ and $d^J$ identically vanishes if it is
represented by a diagram $\widetilde{Y}_{p+q+r}$ with at least one
column containing two partial derivatives. The proposition
\ref{commutprop} proves that the operator $d^I$ provides the most
general non-trivial way of acting with partial derivatives in
$\Omega_{(S)}({\cal M})$.

The proposition \ref{commutprop} is also helpful because it makes
contact with the definition (\ref{defd}) by noticing that the
operator $d^I$ can be identified, up to a constant factor, with
the (non-trivial) $q$-th power of the operator $d$. Despite this
identification, we frequently make use of the notation $d^I$
because it contains more information than the notation $d^{\# I}$.

The space $\Omega_{(2)}({\cal M})$ can be pictured analogously to
the representation (\ref{Ylatt}) of the set of Young diagrams
${\mathbb Y}^{(2)}$
\ba\xymas{&&&&\ldots\\&&&\Omega^{(3,3)}_{(2)}({\cal
M})\ar[r]\ar[ur]&\cdots
\\&&\Omega^{(2,2)}_{(2)}({\cal M}) \ar[r]\ar[ur]&
\Omega^{(3,2)}_{(2)}({\cal M}) \ar[u]\ar[r]\ar[ur]&\cdots \\
& \Omega^{(1,1)}_{(2)}({\cal M}) \ar[r]\ar[ur]&
\Omega^{(2,1)}_{(2)}({\cal M}) \ar[r]\ar[u]\ar[ur]&
\Omega^{(3,1)}_{(2)}({\cal M})\ar[u]\ar[r]\ar[ur]&\cdots \\
\Omega^{(0,0)}_{(2)}({\cal M}) \ar[r]\ar[ur]&
\Omega^{(1,0)}_{(2)}({\cal M})\ar[u]\ar[r]\ar[ur]&
\Omega^{(2,0)}_{(2)}({\cal M})\ar[u]\ar[r]\ar[ur]
&\Omega^{(3,0)}_{(2)}({\cal M})\ar[u]\ar[r]\ar[ur] &\cdots
}\nn\ea\vspace{.5cm}

\noindent From the previous discussions, the definitions of the
arrows should be obvious:
\begin{itemize}
  \item[$\rightarrow$ :] Horizontal arrows are maps $d\propto d^{\{1\}}$.
  \item[$\uparrow$ :] Vertical arrows are maps $d\propto d^{\{2\}}$.
  \item[$\nearrow$ :] Diagonal arrows are maps $d^2\propto d^{\{1,2\}}$.
\end{itemize}
The proposition \ref{commutprop} translates in terms of this
diagram into the fact that
\begin{itemize}
  \item this diagram is completely commutative (modulo numerical factors), and
  \item the composition of any two arrows with at least one common
  direction maps to zero identically.
\end{itemize}
Of course, these diagrammatic properties hold for arbitrary $S$
(the corresponding picture is a mere higher-dimensional
generalization since ${\mathbb Y}^{(S)}\subset {\mathbb R}^S$).

\subsection{Generalized cohomology}

The {\bf generalized cohomology}\footnote{This definition of
generalized cohomology extends the definition of
``hypercohomology" introduced in \cite{Olver:1983}.} of the
generalized complex $\Omega_{(S)}({\cal M})$ is defined to be the
family of graded vector spaces $H_{(m)}(d)=\oplus_{{\mathbb
Y}^{(S)}} H^{(l_1,\ldots,l_S)}_{(m)}(d)$ with $1\leq m \leq S$
where $H^{(l_1,\ldots,l_S)}_{(m)}(d)$ is the set of $\a\in
\Omega^{(l_1,\ldots,l_S)}_{(S)}({\cal M})$ such that\be d^I\a = 0
\; \; \; \; \forall I \subset \{1,2, \dots, S\} \; \,\vert\, \, \#
I = m\,,\,\,d^I\a\in\Omega^*_{(S)}({\cal M})\ee with the
equivalence relation \be\a\,\,\,\sim\,\,\,\a\,\,\,+
\sum_{\begin{array}{c}
J \subset \{1,2, \dots, S\}\\
\#J = S - m +1
\end{array}}d^J\b_J\,,\quad\quad \b_J\in
\Omega^*_{(S)}({\cal M})\,.\label{equivrelatio}\ee Let us stress
that each $\b_J$ is a tensor in an irreducible representation of
$GL(D,\mathbb R)$ and such that $d^J\b_J\in
\Omega^{(l_1,\ldots,l_S)}_{(S)}({\cal M})$. In other words, each
irreducible tensor $d^J\b_J$ is represented by a specific diagram
$Y_{(l_1,\ldots,l_S)}^J$ constructed in the following way:
\begin{description}
  \item[1st] One starts from the Young diagram $Y_{(l_1,\ldots,l_S)}^{(S)}$ of the irreducible tensor field
  $\a$.
  \item[2nd] One removes one cell in $S-m+1$ columns of the
  diagram, making sure that the reminder is still a Young diagram.
  \item[3rd] One replaces all the removed cells by cells
  containing a partial derivative.
\end{description}
This procedure is a direct application of the lemma
\ref{commutprop}. The irreducible tensors $\b_J$ are represented
by a diagram obtained at the second step. They are well included
into the one of $\a$; more precisely all these diagrams belong to
Ker$\Subset\a$. A less explicit definition of the generalized
cohomology is by the following quotient\be
H_{(m)}(d)=\frac{\bigcap\,\mbox{Ker}\,d^m}{\sum\,\mbox{Im}\,d^{S-m+1}}\,.\ee

We can now state a generalized version of the Poincar\'e lemma,
the proof of which will be postponed to the next subsection
because it is rather lengthy and technical. We have the
\cadre{theorem}{GenPoin}{(Generalized Poincar\'e lemma)

Let $Y^{(l_1,\ldots,l_S)}_{(S)}$ be a Young diagram with $l_S\neq
0$ and columns of lengths strictly smaller than $D$ : $l_i<D$,
$\forall i\in\{1,2, \dots, S\}$. For all $m\in\mathbb N$ such that
$1\leq m\leq S$ one has that \ba
H^{(l_1,\ldots,l_S)}_{(m)}\left(\Omega_{(S)}({\mathbb
R}^D)\right)=0\,.\nn\ea }
\noindent The theorem \ref{GenPoin}
extends the proposition \ref{GenPoinrect}; the latter can be
recovered retrospectively by the fact that, for rectangular
tensors, there exist only one $d^I\a$ and one $\b_J$.

\subsection{Applications to gauge theories}

In linearized gravity, one considers the action of nilpotent
operators $d_i$ on the tensors instead of the distinct operators
$d^{\{i\}}$. But it is possible to show the
useful\begin{proposition}Let $\a$ be an irreducible tensor of
$\Omega_{(S)}({\cal M})$. We have the implication\ba
&&\big(\prod_{ i\in I}d_i\,\,\big)\,\,\a=0\,, \quad\forall
I\subset \{1,2, \dots, S\} \; \,\vert\, \, \# I = m
\nn\\&&\Longrightarrow\quad d^I\a=0\,, \quad\forall I\subset
\{1,2, \dots, S\} \; \,\vert\, \, \# I = m\,.\nn\ea
\label{weaker}\end{proposition}\noindent Therefore, the conditions
appearing in symmetric tensor gauge theories are stronger than the
cocycle condition of $H_{(m)}(d)$ and the coboundary property also
applies.

Now we enunciate the following corollary which is a specific
application of the theorem \ref{GenPoin} together with the
proposition \ref{weaker}. Its interest resides in its
applicability in linearized gravity field equations (we
anticipated the use of this corollary in the previous subsection).
\begin{corollary}\label{coro}
Let $\kappa\in\Omega_{(S)}({\cal M})$ be an irreducible tensor
field represented by a Young diagram with at least one row of $S$
cells and without any column of length $\geq D-1$. If the tensor
$\kappa$ obeys \ba d_i \kappa=0\,\quad \forall
i\in\{1,\ldots,S\}\,,\nn\ea then
\ba\kappa=\big(\prod\limits^S_{i=1} d_i\big)\lambda\nn\ea where
$\lambda$ belongs to $\Omega_{(S)}({\cal M})$ and the tensor
$\kappa$ is represented by a Young diagram where all the cells of
the first row are filled by partial derivatives.
\end{corollary}
The proof amounts to notice that the two tensors with
diagrams\vspace{0.3cm}
\begin{center}\footnotesize
\begin{picture}(0,37)(0,-24)
\multiframe(-100,10)(10.5,0){1}(10,10){}
\multiframe(-100,-0.5)(10.5,0){1}(10,10){}
\put(-95.5,-12){$\vdots$}\put(-95.5,-23.5){$\vdots$}
\multiframe(-100,-35)(10.5,0){1}(10,10){$\partial$}
\multiframe(-89.5,10)(10.5,0){1}(10,10){}
\multiframe(-89.5,-0.5)(10.5,0){1}(10,10){}
\put(-85,-12){$\vdots$}
\multiframe(-89.5,-24.5)(10.5,0){1}(10,10){$\partial$}
\put(-74.5,7){$\ldots$}
\multiframe(-58,10)(10.5,0){1}(10,10){}\put(-53.5,-1){$\vdots$}
\multiframe(-58,-14)(10.5,0){1}(10,10){$\partial$}

\multiframe(30,10)(10.5,0){1}(10,10){$\partial$}
\multiframe(30,-0.5)(10.5,0){1}(10,10){}
\put(34.5,-12){$\vdots$}\put(34.5,-23.5){$\vdots$}
\multiframe(30,-35)(10.5,0){1}(10,10){}
\multiframe(40.5,10)(10.5,0){1}(10,10){$\partial$}
\multiframe(40.5,-0.5)(10.5,0){1}(10,10){} \put(45,-12){$\vdots$}
\multiframe(40.5,-24.5)(10.5,0){1}(10,10){} \put(55.5,7){$\ldots$}
\multiframe(72,10)(10.5,0){1}(10,10){$\partial$}\put(76.5,-1){$\vdots$}
\multiframe(72,-14)(10.5,0){1}(10,10){}
\end{picture}\normalsize\end{center}\vspace{0.3cm}
are proportional since the initial symmetrization of the partial
derivatives $\partial$ in each row will be killed by the
antisymmetrization in the columns that immediately follows for two
glued columns of different length (if they have the same length
the partial derivatives are in the same row and the symmetrization
is automatic). By induction, starting from the left, one proves
that this is true for an arbitrary number of columns. This
argument remains true if we add smaller columns at the right of
the Young diagram.

The last subtlety in the corollary is that antisymmetrization in
each row ($\kappa=(\prod d_i)\lambda$) automatically provides the
appropriate Young symmetrization since $\lambda$ has the
appropriate symmetry. This can easily be checked by taking a
complete antisymmetrization of the tensor $\kappa$ in the entries
of a column and another entry which is on the right-hand side.
This automatically vanishes because the index in the column at the
right is either
\begin{itemize}
\item attached to a partial derivative, in which case the antisymmetrization contains two
partial derivatives, or
\item attached to the tensor $\lambda$.
Therefore the antisymmetrization over the indices of the column
except the one in the first row (containing a partial derivative
symbol $\partial$) takes an antisymmetrization of the tensor
$\lambda$ in the entries of a column and another entry which is on
the right-hand side. This vanishes since $\l$ has the symmetry
properties corresponding to the diagram obtained after eliminating
the first row of $\kappa$.
\end{itemize}
This last discussion can be summarized by the operator formula\be
\bf{Y}_{(l_1+1,\ldots,l_S+1)}^{(S)} \circ\partial^S
\circ\bf{Y}_{(l_1,\ldots,l_S)}^{(S)}\,\propto\,\prod\limits^S_{i=1}
d_i\circ\bf{Y}_{(l_1,\ldots,l_S)}^{(S)} \,,\ee where
$\partial^S$ are $S$ partial derivatives with indices
corresponding to the first row of a given Young diagram.

We present another corollary, which determines the reducibility
identities for the mixed symmetry type gauge
field.\begin{corollary}\label{corol} Let $\l\in\Omega_{(2)}({\cal
M})$ be a sum of two irreducible tensors
$\l_1\in\Omega^{(l_1-1,l_2)}_{(2)}({\cal M})$ and
$\l_2\in\Omega^{(l_1,l_2-1)}_{(2)}({\cal M})$ with $l_1\geq l_2$
($\l_1=0$ if $l_1=l_2$). One has the implication \ba
\sum\limits^2_{i=1}d^{\{i\}}\l_i=0\,\quad\Rightarrow\quad
\l_i\,=\,\sum\limits_{j=1}^2d^{\{j\}} \m_{ij} \quad
(i=1,2)\,,\nn\ea where
\begin{description}
  \item[-] $\m_{11}\in\Omega^{(l_1-2,l_2)}_{(2)}({\cal M})$ (which vanishes
if $l_1\leq l_2+1$),
  \item[-] $\m_{12},\m_{2,1}\in\Omega^{(l_1-1,l_2-1)}_{(2)}({\cal M})$
  ($\m_{1,2}=0$ if $l_1= l_2$), and
  \item[-] $\m_{22}\in\Omega^{(l_1,l_2-2)}_{(2)}({\cal M})$.
\end{description}
Furthermore, if $l_1> l_2$ we can assume, without loss of
generality, that \ba\m_{21}=-\m_{12}\,.\nn\ea
\end{corollary}
\proof{We apply $d^{\{1\}}$ and $d^{\{2\}}$ to the equation
$\sum\limits^2_{i=1}d^{\{i\}}\l_i=0$ and get $d_1 d_2 \l_i=0$ in
view of the remarks made below the corollary \ref{coro}. From the
theorem,
%\ref{GenPoin},
we deduce that $\l_i\,=\,\sum\limits_{j=1}^2d^{\{j\}} \m_{ij} $
with tensors $\m_{ij}$ in the appropriate spaces given in the
corollary \ref{corol}. Their vanishing agree with the above-given
rule.

To finish the proof we should consider the case $l_1>l_2$.
Assembling the results together,
$\sum\limits^2_{i=1}d^{\{i\}}\l_i=d^{\{1,2\}} (\m_{12}+\m_{21})=0$
because $d^{\{1\}}$ and $d^{\{2\}}$ commute in $\Omega_{(2)}({\cal
M})$. Thus, $d_1 d_2 (\m_{12}+\m_{21})=0$. Using again the
corollary \ref{corol}, one obtains
$\m_{12}+\m_{21}=\sum_{k=1}^2d^{\{k\}}\n_k$ with
$\n_1\in\Omega^{(l_1-2,l_2-1)}_{(2)}({\cal M})$ and
$\n_2\in\Omega^{(l_1-1,l_2-2)}_{(2)}({\cal M})$ ($\n_1=0$ if
$l_1=l_2$). Hence we can make the redefinitions
$\m_{12}\rightarrow\m_{12}-d^{\{2\}}\n_2$ and
$\m_{21}\rightarrow\m_{21}-d^{\{1\}}\n_1$ which do not affect
$\l$, in such a way that we have (without loss of generality)
$\m_{21}=-\m_{12}$.} This proposition can be generalized to give a
full proof of the gauge reducibility rules given in
\cite{Labastida:1986} and reviewed in next section.

%%%%%%%%%%%%%%%%%%%%%%%%%%%%%%%%%%%%%%%%%%%%%%%%%%%%%%%%%%%%%%%%%%%%%

\section{Arbitrary Young symmetry type gauge field
theories}\label{arbitraryYoung}

We now generalize the section \ref{fieldequs} for arbitrary
irreducible tensor representations of $GL(D,\mathbb R)$. The
discussion presented below fits in the approach followed by
\cite{Aulakh:1986} for two columns ($S=2$) and by
\cite{Labastida:1986,Hull:2001} for an arbitrary number of
columns. The interest of this section is its translation in the
present mathematical language and the use of the generalized
Poincar\'e lemma to give a more systematic mathematical foundation
to this approach.

\subsection{Bianchi identities}

First of all, we generalize our previous discussion on linearized
gravity by introducing a will-be curvature $K$, which is a priori
a multiform in $\Omega^{l_1,\ldots,l_S}_{[S]}({\cal M})$ ($l_S\neq
0$) with $1\leq l_j\leq l_i<D$ for $i\leq j$. Secondly, suppose
the (algebraic) {\bf Bianchi I} relations to be
\be\mbox{Tr}_{ij}\{\,*_i\,K\}=0\,,\quad \forall i,j:\,\, 1\leq
i<j\leq S\,,\label{BianchI}\ee in order to obtain, from the
proposition \ref{Schurm}, that $K$ is an irreducible tensor
belonging to $\Omega^{(l_1,\ldots,l_S)}_{(S)}({\cal M})$. Thirdly,
decide that the (differential) {\bf Bianchi II} relations are \be
d_iK=0\,,\quad\forall i:\,\, 1\leq i\leq S\,,\label{BianchII}\ee
in such a way that, from corollary \ref{coro}, one gets that \be
K=d_1d_2\ldots d_S \k\,.\ee In that case, the curvature is indeed
a natural object for describing a theory with gauge fields $\k\in
\Omega^{(l_1-1,\ldots,l_S-1)}_{(S)}({\cal M})$. The gauge
invariances are then \be \k\rightarrow
\k+d^{\{i\}}\b_i\,,\label{gtransfo}\ee where the gauge parameter
$\b_i$ is an irreducible tensors $\b_i$ in
$\Omega^{(l_1-1,\ldots,l_i-2,\ldots,l_S-1)}_{(S)}({\cal M})$ for
any $i$ such that $l_i\geq 2$, as follows from theorem
\ref{GenPoin} and proposition \ref{weaker}.

\subsection{Reducibilities}

In general the gauge transformations (\ref{gtransfo}) are
reducible, i.e. $d^{\{j\}}\b_j\equiv 0$ for non-vanishing
irreducible tensors $\b_j\neq 0$. The procedure followed in the
proof of the corollary \ref{corol} can be applied in the general
case. We then recover the rules of \cite{Labastida:1986} to form
the $(i+1)$-th generation reducibility parameters from the $i$-th
generation Young diagrams :
\begin{description}
  \item[(A)] Start with the curvature Young diagram
  $Y_{(l_1,\ldots,l_S)}^{(S)}$.
  \item[(B)] Remove a box in a row which has not had a box removed previously
(in forming the lower generations of reducibility parameters) such
as the result is a standard Young diagram.
  \item[(C)] There is one and only one
reducibility parameter for each Young diagram.
\end{description}
Of course the gauge parameters are taken to be the first
reducibility parameters. The Labastida rules provide the complete
BRST spectrum with the full tower of ghosts of ghosts.

More explicitly, the chain of reducibilities is \be
\stackrel{(i)}{\b}_{j_1j_2\ldots
j_i}=d^{j_{i+1}}\stackrel{(i+1)}{\b}_{j_1j_2\ldots
j_ij_{i+1}}=0\,,\quad (i=1,2,\ldots,r)\ee where $r=l_1-1$ is the
number of rows of $\k$. The chain is of length $r$ because at each
step one removes a box in a row which has not been chosen before.
We take $\b_j=\stackrel{(1)}{\b}_j$ since the gauge parameters are
the first reducibility parameters. We can see that the order of
reducibility of the gauge transformations (\ref{gtransfo}) is
equal to $l_1-2$.

The subscripts of the $i$-th reducibility parameter
$\stackrel{(i)}{\b}_{j_1j_2\ldots j_i}$ belong to the set
$\{1,\ldots,S\}$. They determine the Young diagram corresponding
to the irreducible tensor $\stackrel{(i)}{\b}_{j_1j_2\ldots j_i}$
: reading from the left to the right, the subscripts give the
successive columns in which one removed the bottom box following
the rules (A) and (B). A reducibility parameter
$\stackrel{(i)}{\b}_{j_1j_2\ldots j_i}$ vanishes if these two
rules are not fulfilled. Furthermore, they are antisymmetric in
any pair of different indices\be \stackrel{(i)}{\b}_{j_1\ldots
j_k\ldots j_l\ldots j_i}=-\stackrel{(i)}{\b}_{j_1\ldots j_l\ldots
j_k\ldots j_i}\,,\quad \forall j_l\neq j_k\,.\label{minusign}\ee
This property ensures the rule (C) and provides the correct signs
to fulfill the reducibilities. Indeed, \be
d^{j_i}\stackrel{(i)}{\b}_{j_1j_2\ldots
j_i}=d^{j_i}d^{j_{i+1}}\stackrel{(i+1)}{\b}_{j_1j_2\ldots
j_ij_{i+1}}=0\,,\ee due to proposition \ref{commutprop} and
(\ref{minusign}).
\subsection{Field equations}

We make the important assumption \be l_i+l_j\leq D\,,\quad
\forall\, i,j\label{hypoth}\ee and take the field equations to be
in that case \be \mbox{Tr}_{ij}\{K\}=0\,,\quad \forall
\,i,j\,.\label{FIeld}\ee The assumption (\ref{hypoth}) comes from
the corollary \ref{Trvanish}. Indeed, since $l_j\leq l_i$ for
$i\leq j$, the condition (\ref{hypoth}) reduces to $l_1+l_2\leq D$
which is necessary to allow a non-trivial curvature, $K\neq 0$.

Together with the Bianchi I identities, we deduce that, for any
non-empty subset $I\subset\{1,2, \dots, S\}$ ($\#I=m$), \be
\mbox{Tr}_{ij}\,\{\,\big(*_i \prod_{k \in I} *_k
\big)K\,\}\,=\,0\,,\quad \forall i,j:
\ell_j\leq\ell_i\label{from}\ee with $\ell_i$ the length of the
i-th column of the tensor $\big(\prod_{k \in I} *_k
\big)K$ : \be \ell_i\equiv\left\{\begin{array}{lll}l_i\,\quad&\mbox{if}\,\,i\not\in I\,,&\\
D-l_i\,\quad&\mbox{if}\,\,i\in I\,.&\end{array}\right. \ee
Indeed, let be $i$ and $j$ such that $\ell_j\leq\ell_i$. There are
essentially four possibilities:
\begin{itemize}
  \item $i\not\in I$ and
    \begin{itemize}
    \item $\underline{j\not\in I}$: Then $l_j\leq l_i$ and the Bianchi I identity (\ref{BianchI}) is
equivalent to (\ref{from}) since $\mbox{Tr}_{ij}$ and $*_k$
commute if $i\neq k$ and $j\neq k$.
    \item $\underline{j\in I}$: Then one should have $D-l_j\leq l_i$
    which means that $D\leq l_i+l_j$ which is in contradiction
    with the hypothesis (\ref{hypoth}) except if $l_i+l_j=D$. From the proposition \ref{tracepower},
    we deduce that in such case the field equation (\ref{FIeld}) is equivalent to (\ref{from}).
\end{itemize}
  \item $i\in I$ and
    \begin{itemize}
    \item $\underline{j\not\in I}$: We have $l_j\leq D-l_i$ which is
    equivalent to $l_i+l_j\leq D$. The field equation
    (\ref{FIeld}) is of course equivalent to (\ref{from}) since $*_i^2 K=\pm K$.
    \item $\underline{j\in I}$: We have $D-l_j\leq D-l_i$ which is
    equivalent to $l_i\leq l_j$. The Bianchi I identity $\mbox{Tr}_{j\,i}\{*_j
    K\}=0$ is therefore satisfied and equivalent to (\ref{from}) because $\mbox{Tr}_{ij}=\mbox{Tr}_{ji}$.
\end{itemize}
\end{itemize}

Due to the proposition \ref{Schurm}, it follows from (\ref{from})
that \be \big( \prod_{k \in I} *_k \big)K\,\in\,
\Omega^{(l_1,\ldots,D-l_{k_1},\ldots,D-l_{k_m},\ldots,l_S)}_{\quad(S)}({\cal
M})\,,\label{inom}\ee with \be k_i\in I\,, \quad\forall
i\in\{1,\ldots,m\}\,,\ee and $k_i<k_j$ for any
$i,j\in\{1,\ldots,m\}$.

Now we use the property (\ref{makeuse}) to deduce from the Bianchi
II identities (\ref{BianchII}) and the field equations
(\ref{FIeld}) that $d_i*_iK=0$ for any $i$. Therefore \be d_i\big(
\prod\limits_{k \in I} *_k \big)K=0\,,\quad \forall
i\in\{1,\ldots,S\}\,,\label{folows}\ee because $d_i$ and $*_j$
commute if $i\neq j$, and either
\begin{itemize}
  \item $i\not\in I$, and (\ref{folows}) follows from $d_iK=0$,
  or
  \item $i\in I$ and then (\ref{folows}) is a consequence of $d_i*_iK=0$.
\end{itemize}
In other words, any $\big( \prod_{k \in I} *_k \big)K$ satisfies
(on-shell) its own Bianchi II identity (\ref{folows}) which,
together with (\ref{inom}), implies the (local) existence of a
dual gauge field $\tilde{\k}_I$ such that \be\big( \prod_{k \in I}
*_k \big)K=d_1d_2\ldots d_S \tilde{\k}_I\,,\ee with
\be\tilde{\k}_I\in
\Omega^{(l_1-1,\ldots,D-l_{k_1}-1,\ldots,D-l_{k_m}-1,\ldots,l_S-1)}_{\quad(S)}({\cal
M}) \,.\ee The Hodge operators therefore relates different free
field theories of arbitrary tensor gauge fields, extending the EM
duality property of electrodynamics.

The field equations of the dual theory are \be
\mbox{Tr}^{m_{ij}}_{ij}\,\{\,\big(\prod_{k \in I} *_k
\big)K\,\}\,=\,0\,,\quad \forall i,j:\,i<j\,\,\,\mbox{if}\,\,\,
i\in I\,,j\not\in I \label{genfieldequ}\ee where \be m_{ij}\equiv
\left\{\begin{array}{lll}1+D-l_i-l_j\quad&\mbox{if}\,\,i\,\,\mbox{and}\,\,j\in
I\,,&\\
1\,\quad&\mbox{if}\,\,i\,\,\mbox{or}\,\,j\not\in
I\,.&\end{array}\right.\ee Since the trace is symmetric in $i$ and
$j$ we consider only three distinct cases:
\begin{itemize}
  \item $\underline{i\not\in I}$ and $\underline{j\not\in I}$: The starting field equation
    (\ref{FIeld}) is of course equivalent to the dual field
    equation (\ref{genfieldequ}).
  \item $\underline{i\in I}$ and
    \begin{itemize}
    \item $\underline{j\not\in I}$: If $i<j$ the Bianchi I
    relation (\ref{BianchI}) is satisfied and it implies (\ref{genfieldequ}).
    \item $\underline{j\in I}$: A direct use of proposition
    \ref{tracepower} leads from the field equation
    (\ref{FIeld}) to (\ref{genfieldequ}).
\end{itemize}
\end{itemize}
As one can see, a seemingly odd feature of some dual field
theories is that their field equations are not of the same type
that (\ref{FIeld}). In fact, the dual field equations are of the
type (\ref{FIeld}) for all $I$ only in the exceptional case where
$D$ is even and $l_i=l_j=D/2$. Notice that this is satisfied for
free gauge theories of completely symmetric tensors in $D=4$ flat
space. The point is that $\ell_i+\ell_j=2D-l_i-l_j\geq D$ for
$i,j\in I$, therefore the property (\ref{hypoth}) is in general
not satisfied by the dual tensor $\big(\prod_{k \in I} *_k
\big)K$.

A natural idea is that when $l_i+l_j> D$ for a tensor
$K\in\Omega^{l_1,\ldots,l_S}_{[S]}({\cal M})$ ($l_S\neq 0$) with
$i$ and $j$ two elements of $\in\{1,\ldots,S\}$, the corresponding
field equation is modified \cite{Hull:2001} to follow\be
\mbox{Tr}^{1+l_i+l_j-D}_{ij}\,\{\,K\,\}\,=\,0\,.\label{tracexp}\ee
This condition is gentle because for $m\geq l_i+l_j-D$ the
irreducible tensor $\mbox{Tr}^m_{ij}\,\{\,K\,\}$ is represented by
a diagram whose the sum of the lengths of $i$-th and $j$-th column
(equal to $l_i+l_j-2m$) is smaller than $D$. Therefore, the trace
power in (\ref{tracexp}) is precise the critical exponent such
that (\ref{tracexp}) does not necessarily imply $K\equiv 0$ (due
to the proposition \ref{Trvanish}).

%%%%%%%%%%%%%%%%%%%%%%%%%%%%%%%%%%%%%%%%%%%%%%%%%%%%%%%%%%%%%%%%%%%%%
%%%%%%%%%%%%%%%%%%%%%%%%%%%%%%%%%%%%%%%%%%%%%%%%%%%%%%%%%%%%%%%%%%%%%

\chapter{Duality-symmetric actions and chiral forms}\label{manifestd}

\par

To make EM-duality symmetry manifest in the action itself is an
old goal of theoretical physicists. Efforts in this direction have
been undertaken long time ago \cite{Zwanziger:1971,Deser:1976},
but more substantial results have been achieved during the last
decade, when the connections with supergravity and string theory
prompted further investigation.

As a result one counts different formulations. Firstly, quadratic
but non-covariant versions \cite{Deser:1976,Schwarz:1994}, and
also quadratic and covariant actions but with an infinite number
of auxiliary fields \cite{McClain:1990,Berkovits:1997}. In 1994,
Khoudeir and Pantoja proposed a manifestly Lorentz invariant
duality symmetric action \cite{Khoudeir:1996} by covariantizing
the action of \cite{Deser:1976,Schwarz:1994} but this attempt
failed because this action does not really describe a single gauge
vector field \cite{Pasti:1997}. Still, by introducing an auxiliary
field to their action, Pasti, Sorokin and Tonin were able to
construct non-polynomial Lagrangians with manifest space-time
symmetry \cite{Pasti:1995,Pasti:1995i}, the so-called PST model.
Obviously, it appears that (in any of the formulations) a high
price has to be paid in order to implement the duality symmetry at
the level of the action.

The field equations of electrodynamics and linearized gravity are
presented in a duality-symmetric form in section \ref{double}. A
link is also made with chiral fields, which is used in section
\ref{noncovact} to derive the duality-symmetric equations from an
action principle where the Lorentz symmetry is not manifest. In
section \ref{s:dual}, we review the PST construction. We end this
chapter by rederiving the boundary rules of $D$-brane
electrodynamics from compact formulae in section
\ref{D-branelectro}.

%%%%%%%%%%%%%%%%%%%%%%%%%%%%%%%%%%%%%%%%%%%%%%%%%%%%%%%%%%%%%%%%%%%

\section{Duality-symmetric equations of motion}\label{double}

\subsection{Electrodynamics (Double-potential
formulation)}\label{doublep}

One of the basic ingredients of the duality-symmetric formulation
resides in increasing the number of gauge fields (doubling in that
case) to make global symmetries manifest. At the same time, the
number of gauges symmetries increases in such a way that the
theory still possesses the same number of physical degrees of
freedom.

Let us start with $D$-dimensional electrodynamics equations
(\ref{Max+}) for electric $p$-branes, written as \be d{\cal
F}=*\cal{J}\label{boxJ}\ee where we defined \be {\cal
F}\equiv\left(
\begin{array}{c} F\\ $*$F
\end{array}
\right)\,,\quad \cal{J}\equiv\left( \begin{array}{c} J_m\\ J_e
\end{array}
\right)\,.\label{calF}\ee We define the graded space of
differential form doublets $\overline{\Omega}({\cal
M})\equiv\bigoplus\limits_{p,q=0}^D\overline{\Omega}^{(p,q)}({\cal
M})$ with $\overline{\Omega}^{(p,q)}({\cal M})\equiv
\Omega^p({\cal M})\otimes\Omega^{q}({\cal M})$.

Formally, one can solve the sourceless Maxwell equations by using
the Poincar\'e lemma in $\overline{\Omega}({\cal M})$ and obtain
\begin{equation}
{\cal F}=d{\cal A}+*{\cal G}
\end{equation}
with
\begin{equation}
{\cal A}\equiv\left( \begin{array}{c} A^1\\
A^2
\end{array}
\right)\,,\quad {\cal G}\equiv\left( \begin{array}{c} G^1\\
G^2
\end{array}
\right)\end{equation} where $G^1=G$ the usual Dirac brane current
and $G^2$ its electric analogue such that $d*G^2=*J_e$. The Hodge
dual $*$ is defined on $\overline{\Omega}({\cal M})$ by acting
simultaneously on the two components. The electric and magnetic
variables are now on the same footing. In this formulation, there
is a pair of gauge fields $A^\a$ $(\a=1,2)$.

The Hodge square relation (\ref{*2}) \be *^2{\cal F}=\sigma {\cal
F}\,,\quad \sigma=(-)^{p(D-p)+1}\,,\ee imposes for consistency
\begin{equation}
{\cal F}=(*J){\cal F}.\label{self}
\end{equation}
This equation, which is an identity for an ${\cal F}$ defined by
(\ref{calF}), is a self-duality condition with involution operator
$*J$ acting on the space\footnote{If we define the product in the
graded space $\underline{\Omega}({\cal
M})\equiv\oplus_p\overline{\Omega}^{(p,D-p)}({\cal M})$ by \be
\left(
\begin{array}{c} A_{[p]}\\ B_{[D-p]}
\end{array}
\right)\wedge\left(
\begin{array}{c} C_{[q]}\\ D_{[D-q]}
\end{array}
\right)\equiv \left(
\begin{array}{c} \sigma_p\,\sigma_q\,A_{[p]}\wedge C_{[q]}\\
*\left[(*D)_{[p]}\wedge (*B)_{[q]}\right]
\end{array}
\right)\,,\ee then the operator $*J$ is an involution of the
(supercommutative) algebra $\underline{\Omega}({\cal M})$. This
algebra is not a complex for the nilpotent operator $d$ since this
operator is not internal in $\underline{\Omega}({\cal M})$.}
$\overline{\Omega}^{(p,D-p)}({\cal M})$ ($(*J)^2=$Identity), where
$J$ is the matrix
\be J=\left( \begin{array}{c c} 0 & \sigma\\
1 & 0
\end{array}
\right) \,\label{Jdef}\ee which satisfies the useful identities
\be J^2=\sigma I\,,\quad J^T=\sigma J\,.\label{J2}\ee

As one can see, for a self-dual field, the field equation now
follows from the Bianchi identity; the dynamics is really produced
by the self-duality condition. Therefore, the idea is simply to
obtain this self-duality equation as an e.o.m. derived from an
action principle (this gives some hint that a relation might exist
between EM-duality symmetry and chiral forms, as will be explained
later on).

\subsubsection{Duality symmetry}

In dimensions $D=2(p+2)$ with $p$ even, the EM-duality rotations
are symmetries of the action. The matrix $J$ is an antisymmetric
matrix, invariant under orthonormal change of basis, namely
$J\rightarrow R^{-1} J R=J$, $R\in SO(2)$. Consequently the e.o.m.
(\ref{self}) is manifestly invariant under the EM-duality rotation
(\ref{transfo}) rewritten in terms of the new variables
\begin{equation}
{\cal A} \rightarrow R \,{\cal A}\,, \quad{\cal G}\rightarrow R
\,{\cal G}\,,\quad R\in SO(2).
\end{equation}
Besides the manifest symmetry, a nice feature of the
duality-symmetric formulation is that the duality rotation is an
\emph{off-shell local} transformation of the gauge fields $A^\a$.

With a theta-term in the action, we have seen that the duality
group is enhanced to $SL(2,{\mathbb R})$. The natural
two-component fieldstrength is now \begin{equation} {\cal
F}=\mbox{Re}(\psi\,\overline{F}) \equiv \,\left( \begin{array}{c} F\\
\frac{1}{g^2}*F-\frac{\theta}{2\pi} F
\end{array}
\right)\,.
\end{equation} due to (\ref{MaxIIII}). It fulfills a self-duality condition
\be {\cal F}\,=\,\frac{1}{\sqrt{\det\gamma}}\,\,\gamma\,
J\,*\,{\cal F}\label{selfd}\ee where $\gamma$ is the symmetric
matrix \be \gamma=  \left(
\begin{array}{cc}
1 & \tau_1 \\
\tau_1 & |\tau|^2
\end{array}
\right) = \left(
\begin{array}{cc}
1 & -\frac{\theta}{2\pi} \\
-\frac{\theta}{2\pi} & \frac{1}{g^{4}}+\frac{\theta}{2\pi}^2
\end{array}
\right)\,. \label{gammaij}\ee of determinant equal to
$(\tau_1)^2=g^{-4}$.

Using a trick given in \cite{Gibbons:1996} the matrix
$\frac{1}{\sqrt{\det\gamma}}\gamma$ may be written as \be \gamma =
{\psi\psi^\dagger + c.c.\over\sqrt{{\rm det}\left(\psi\psi^\dagger
+ c.c.\right)}}.\ee The choice of the first component of $\psi$
equal to $1$ fixes the representation of $\gamma$. The
$SL(2,{\mathbb R})$ duality transformation may then be constructed
so that it automatically leaves the self-duality equation
invariant: \ba \psi \rightarrow \psi^\prime \propto S\psi \quad &
\Rightarrow & \quad \gamma\rightarrow S\,\gamma \,S^T \nonumber
\ea where $S\in SL(2,{\mathbb R})$. This is the $SL(2,{\mathbb
R})$  duality symmetry of the classical theory. In the quantum
theory, the invariance of the charge lattice restricts the
$SL(2,{\mathbb R})$ group to its discrete $SL(2,{\mathbb Z})$
subgroup, which is by-now usually referred as $S$-\emph{duality}
(where $S$ stands for strong/weak coupling duality corresponding
to the transformation $\tau\rightarrow-1/\tau$ that relates a
strongly coupled electrodynamics to its weakly coupled
counterpart).

\subsection{Linearized gravity (Four-potential formulation)}

Following \cite{Hull:2001} we introduce a matrix $\cal R$ defined
by
\be {\cal R}=\left( \begin{array}{c c} R & R*\\
*R & *R*
\end{array}
\right) \,,\label{calR}\ee with $R\in \Omega^{2,2}_{[2]}({\cal
M})$. A tensor ${\cal R}$ of the form (\ref{calR}) fulfills the
identities \be{\cal R}=*J\,\,{\cal R}=\sigma{\cal
R}*J.\label{selfgrav}\ee One may start without assuming
(\ref{calR}) and see (\ref{selfgrav}) as self-duality conditions
with involution operator $*J$.

We can now rewrite the linearized Bianchi I and Einstein equations
in the empty Minkowski space as
\be\left\{\begin{array}{lll}{\cal R}\in \Omega_{(2)}({\cal M})&& \\
d^{L,R}\,{\cal R}=0&&\end{array}\right. \,.\label{fieldgrav}\ee As
explained in section \ref{fieldequs}, on a spacetime manifold with
the topology of ${\mathbb R}^D$ the system (\ref{fieldgrav}) for
an arbitrary $\cal R$ is equivalent to \be {\cal R}=d^2{\cal
H}\,,\quad
{\cal H}=\left( \begin{array}{c c} h & \bar{h}\\
\tilde{h} & \hat{h}
\end{array}
\right) \,.\ee with $h\in \Omega_{2}({\cal M})$,
$\tilde{h}\in\Omega^{(D-3,1)}_{(2)}({\cal M})$,
$\bar{h}\in\Omega^{(1,D-3)}_{(2)}({\cal M})$ and
$\hat{h}\in\Omega^{(D-3,D-3)}_{(2)}({\cal M})$. This leads to a
four-potential formulation of linearized gravity. With the two
self-duality equations (\ref{selfgrav}), we recover one
independent graviton among the four.

\subsubsection{Duality symmetry}

To deal with duality symmetry, we immediately focus to $D=4$
Minkowski background space ${\mathbb M}_4$. The matrix $J$ is now
an antisymmetric two-by-two matrix. This matrix $J$ is invariant
under an orthonormal change of basis. Therefore the self-duality
equations (\ref{selfgrav}) are manifestly invariant under the
transformations \be{\cal R}\rightarrow R_1 \,{\cal R}\,
R_2\,,\quad R_1,R_2\in SO(2)\,,\ee which are the \emph{duality
transformations for linearized gravity}.

As pointed earlier, a nice feature in four dimensions is that we
can solve (\ref{fieldequ}) in the same way as (\ref{BianchiIII}),
in the sense that we remain in the space of rectangular tensors.
For an arbitrary ${\cal R}$ satisfying (\ref{fieldgrav}) with
entries in $\Omega^{(2,2)}_2({\cal M})$ we have ${\cal R}=d^2
{\cal H}$ with ${\cal H}$ the two by two matrix with entries in
$\Omega^2_2({\cal M})$ \cite{Hull:2001}, hence the entries of
$\cal R$ belong to $\Omega^4_2({\cal M})$. Making use of ${\cal
H}$ a duality transformation becomes the local transformation of
the gauge field: ${\cal H}\rightarrow R_1 \,{\cal H} R_2$,
$R_1,R_2\in SO(2)$. In the (off-shell) four-potential formulation,
the duality group is therefore $SO(2)\times SO(2)$. The discrete
duality subgroup corresponds to rotations of angles that are
multiple of $\pi/2$ is ${\mathbb Z}_4\times {\mathbb Z}_2$.

The linearised Riemann tensor $R$ belongs to $R\in\Omega^2_2({\cal
M})$. In four dimensions, this implies $*R=R*$, which itself
implies $R=-*R*$. Therefore, there is one supplementary constraint
in four dimensions: the two by two matrix ${\cal R}$ must be
\emph{symmetric and traceless}. It can be checked that if ${\cal
R}$ is symmetric, then one self-duality condition in
(\ref{selfgrav}) implies the other. Furthermore, we can assume
$\tilde{h}=\bar{h}$ and $h=-\hat{h}$ without loss of generality
since the matrix ${\cal R}$ is symmetric and traceless, hence
${\cal H}$ can be assumed to have the same properties. This again
leads to a two-potential formulation. In this case, the effective
duality group is $SO(2)$ because if $R^1$ and $R^2$ rotates with
the same angle, a symmetric traceless matrix is unaffected by a
duality transformation.

\subsection{Link with chiral fields}\label{linkwchiral}

\subsubsection{Electrodynamics}

In a spacetime of dimension $D=2$ modulo $4$, it is possible to
define a strict Hodge self-duality for $D/2$-forms since the Hodge
dual operator $\star$ is an involution in the space
$\Omega^{\frac{D}{2}}({\cal M})$. In other words, an odd $p$-brane
in $2p+4$ dimensions can be charged under a \emph{chiral}
$(p+1)$-forms $A$, the fieldstrength of which is self-dual:
\be\mathrm{F}=\star\, \mathrm{F}\,.\label{seldu}\ee

Let us assume that the $D$-dimensional spacetime ${\cal M}_D$ is a
direct product involving a spacelike two-torus, ${\cal M}_D={\cal
M}_{D-2}\times T^2$. Our dimensional reduction ansatz are that (i)
the six-dimensional metric is the direct product the two-torus
metric with a ($D-2$)-dimensional metric and (ii) the self-dual
field decomposes as the sum \be\mathrm{F} = F^1 dx^4+F^2dx^5\,,\ee
where $F^1$ and $F^2$ belong to $\Omega^{D/2-1}({\cal M}_{D-2})$.
The 2-metric on the two-torus is taken to be \be
ds_{2-torus}^2=(dy^1)^2+(dy^2)^2\,.\label{metric}\ee The complex
modulus $\tau$ is the complex structure of the two-torus \be y\sim
y+1\,,\quad y\sim y+\tau\,,\quad y=y^1+iy^2\,.\ee Therefore if one
takes a coordinate system $(x^1,x^2)$ with identifications
\be(x^1,x^2)\sim (x^1+1,x^2)\,,\quad(x^1,x^2)\sim
(x^1,x^2+1)\,,\ee one can make the substitution $dy^1=dx^1+\tau_1
dx^2$ and $dy^2=\tau_2\, dx^2$ in (\ref{metric}) to get
(\ref{gammaij}). The $SL(2,{\mathbb Z})$ acts as large
diffeomorphisms on the two-torus, i.e. \emph{modular
transformations}. The $D$-dimensional self-duality equation
becomes (\ref{selfd}) and the quantum $S$-duality symmetry finds a
geometrical ``explanation" as the two-torus symmetry
\cite{Verlinde:1995}.

\subsubsection{Linearised gravity}

Having the previous example in mind, Hull argued that something
analogous happens for four-dimensional linearised gravity
\cite{Hull:2000,Hull:2000i}. Indeed, let us define the chiral
field\footnote{This object appears in the bosonic sector of the
(4,0) superconformal theory \cite{Hull:2000}.}
$\mathrm{h}\in\Omega^4_2({\cal M}_6)$ on the six-dimensional
spacetime manifold ${\cal M}_6$. It possesses the symmetries of
the Riemann tensor and its fieldstrength
$\mathrm{R}=d^2\mathrm{h}$ is a self-dual tensor of
$\Omega^6_2({\cal M}_6)$ with respect to the two Hodge dual
operations:\be
\mathrm{R}=\star\,\,\mathrm{R}=\mathrm{R}\star\,.\label{selfgrrr}\ee
In fact, one self-duality condition imply the other since
$\star\mathrm{R}=\mathrm{R}\star$ for
$\mathrm{R}\in\Omega^6_2({\cal M}_6)$.

Our dimensional reduction ansatz are the following: (i) the flat
six-dimensional spacetime manifold is the direct product ${\cal
M}_6={\mathbb M}_4\times T^2$ and (ii) the self-dual field
decomposes as the sum \be\mathrm{R} = R^1\,(d^Lx^4\otimes
d^Rx^4-d^Lx^5\otimes d^Rx^5)+R^2\,(d^Lx^4\otimes
d^Rx^5+d^Lx^5\otimes d^Rx^4)\,,\ee where $R^1$ and $R^2$ are
biforms whose components have the symmetries of the Riemann, i.e.
they belong to $\Omega^4_2({\mathbb M}_4)$. With these ansatz, the
self-duality equations (\ref{selfgrrr}) translate into the
equations (\ref{selfgrav}) if the torus is rectangular
and if one defines $\cal R$ as \be {\cal R}=\left( \begin{array}{c c} R^1 & R^2\\
R^2 & -R^1
\end{array}
\right) \,.\ee The general case for which the torus is not assumed
to be rectangular can be found in \cite{Hull:2000ii}. To
summarize, the duality-symmetric four dimensional massless
spin-two theory can be understood as the dimensional reduction of
the exotic chiral theory for the gauge field $\mathrm{h}$ with
symmetries of the Riemann tensor. As before, the duality symmetry
finds a geometric origin in this picture.

%%%%%%%%%%%%%%%%%%%%%%%%%%%%%%%%%%%%%%%%%%%%%%%%%%%%%%%%%%%%%%%%%%

\section{Non-covariant action}\label{noncovact}

The implementation of the self-duality condition in variational
principles has been achieved in several steps and approaches. By
the way we can refer to the following main stages
\cite{Deser:1976,Floreanini:1987,Henneaux:1988,Schwarz:1994,Pasti:1995,Pasti:1995i}
(given in chronological order). We begin with the non-covariant
approach
\cite{Deser:1976,Floreanini:1987,Henneaux:1988,Schwarz:1994}. The
natural framework for this non-covariant action is the Hamiltonian
formalism. Still, in this section we will directly present the
result.

\subsection{Self-duality condition}

A starting point to derive the self-duality condition (\ref{self})
from an action principle is the important remark that the set of
equations (\ref{self}) is redundant; it contains twice the same
information since the covariant set of equations \be {\cal
F}_{\m_1\ldots\m_{p_\a}}^\a=J^{\a\b}\,(*{\cal
F})_{\m_1\ldots\m_{p_\a}}^\b\label{equivalent}\ee is equivalent to
the non-covariant conditions \be {\cal E}_{i_1\ldots
i_{p_\a-1}}^\a=J^{\a\b}\,{\cal B}_{i_1\ldots i_{p_\a-1}}^\b\,.\ee
with \be{\cal E}_{i_1\ldots i_{p_\a-1}}^\a={\cal F}_{0i_1\ldots
i_{p_\a-1}}^\a\,,\quad {\cal B}_{i_1\ldots i_{p_\a-1}}^\b=(*{\cal
F})_{0i_1\ldots i_{p_\a-1}}^\b\,,\label{defEB}\ee and
\begin{equation}
p_\a\equiv\left( \begin{array}{c} p+2\\
D-p-2
\end{array}
\right)\,.\end{equation} In curved space, one foliates the
spacetime manifold $\cal M$ by spacelike surfaces $\Sigma$ with
normal $v=dx^0=dt$. The set of equations (\ref{equivalent}) is
equivalent to its contraction with $v^{\m_1}$ as is proved in
subsection \ref{selfcondit}.

\subsection{Self-dual gauge fields}

From now on, we focus on theeven dimensions $D=2(p+2)$ with
\emph{self-dual fields}, that is on duality-symmetric theories.
Since the time direction is privileged it is natural to work with
spatial forms of the vector space ${\Omega}(\Sigma)$. The spatial
differential $\tilde{d}\equiv dx^i\partial_i$ endows
${\Omega}(\Sigma)$ with a differential complex structure. For
simplicity, we assume the absence of charged sources.

In form notations (\ref{defEB}) reads\be {\cal F}=dt\wedge {\cal
E}-\sigma\,*(dt\wedge{\cal B})\,,\label{defcalF}\ee where $\cal E$
and $\cal B$ belong to $\overline{\Omega}(\Sigma)$. The gauge
field doublet can also be decomposed in a part containing $dt$ and
a purely spatial part \be {\cal A}=dt\wedge \widetilde{\cal
A}_0+\widetilde{\cal A}\,,\label{defcalA}\ee where
$\widetilde{\cal A}_0$ and $\widetilde{\cal A}$ belong to
$\overline{\Omega}(\Sigma)$. Combining (\ref{defcalF}) with
(\ref{defcalA}) and the operatorial identity in
$\Omega^p(\Sigma)$\be *\,dt=(-)^{D-p}\,\tilde{*}\ee where
$\tilde{*}$ is the Hodge dual operator on $\Sigma$, one finds in
the absence of charged sources \be{\cal
E}=\partial_0\widetilde{\cal A}-\tilde{d}\widetilde{\cal
A}_0\,,\quad {\cal B}=\tilde{*}(\tilde{d}\widetilde{\cal A})\,.\ee
In components, we have\ba {\cal E}_{i_1\ldots
i_{p+1}}^\a&=&\partial_0 A_{i_1\ldots
i_{p+1}}^\a\,+\,(p+1)\,(-)^{p+1}\,\partial_{[i_1}A_{i_2\ldots
i_{p+1}]\,0}^\a\,,\nn\\\quad{\cal B}^{i_1\ldots
i_{p+1}\,\a}&=&\frac{1}{p!}\varepsilon^{i_1\ldots
i_{D-1}}\partial_{i_{p+2}}A_{i_{p+3}\ldots i_{D-1}}^\a\,.\ea

\subsection{Action principle}

We introduce an obvious notation for the (matrix \emph{and} wedge)
product of two arbitrary elements of $\overline{\Omega}(\Sigma)$
\begin{equation}
A\equiv\left( \begin{array}{c} A_{[p]}\\
A_{[q]}
\end{array}
\right)\,,\quad
B\equiv\left( \begin{array}{c} B_{[r]}\\
B_{[s]}
\end{array}
\right)\,.\end{equation} The following notation is useful for
formulating duality-symmetric actions\be A^T\wedge B\equiv
A_{[p]}\wedge B_{[r]}+ A_{[q]}\wedge B_{[s]}\,.\ee

In the absence of sources, the non-covariant action for self-dual
gauge fields is taken to be \ba &S_{non-cov}&\equiv\,-\frac12\int
dt \int\limits_{\Sigma} \tilde{*}\,{\cal B}^T\wedge(J\,{\cal
E}+{\cal B })\nn\\&=&-\frac{1}{2(p+1)!}\int\limits_{\cal M} d^Dx
\,\,{\cal B}^{i_1\ldots i_{p+1}}_\a\,(J^{\a\b}\,{\cal
E}_{i_1\ldots i_{p+1}\,\b}+{\cal B }_{i_1\ldots
i_{p+1}}^\a)\label{e:SchwarzSen}\,.\ea This action possesses the
gauge invariances \be \widetilde{\cal A}\rightarrow\widetilde{\cal
A}+\tilde{d}\widetilde{\Lambda}\,,\quad \widetilde{\cal
A}_0\rightarrow\widetilde{\cal
A}_0+\tilde{d}\widetilde{\Lambda}_0\,.\ee

In the presence of charged sources, one should add a term
$\frac12\int{\cal A}^T\wedge *\,J\,\cal{J}$.

The e.o.m. is \be\tilde{d}\,(J\,{\cal E}+{\cal B })\,=\,0\,.\ee If
the de Rham cohomology group $H^{p+1}(\Sigma)$ is trivial, then
the e.o.m. is equivalent to ${\cal E}-J\,{\cal
B}=\tilde{d}\tilde{\Phi}$. In an appropriate gauge, we have the
looked-for self-duality equation ${\cal E}-J\,{\cal B}=0$.

In \cite{Leao:2001}, the authors have shown how to implement the
$SL(2,\mathbb Z)$ symmetry of the Maxwell and BI theories
off-shell, that is, with transformations implemented on the
potential, along the lines proposed be by Deser and Teitelboim
many years ago.

%%%%%%%%%%%%%%%%%%%%%%%%%%%%%%%%%%%%%%%%%%%%%%%%%%%%%%%%%%%%%%%%%%

\section{PST action}\label{s:dual}

The nice feature of the PST action is that besides the EM duality
it also exhibits manifest Lorentz covariance. The PST action
(\ref{e:pst}) has been obtained precisely by ``covariantizing" the
previous non-covariant action. The price paid for keeping the
Lorentz covariance as a symmetry for the action $S_0$ is given by
an extra gauge transformation compared to the non-covariant
formulation. It is precisely this symmetry that will be deformed
when looking at the interacting theory.

The new feature of the PST formulation is the presence of the
auxiliary field $a$. In the sequel, we will always assume that
$\partial a$ is timelike ($(\partial a)^2<0$) to make contact with
the previous section, in which the time direction is privileged.
Hence the normalized vector field ($v^2=-1$) is
\begin{equation}
v\equiv\frac{da}{\sqrt{-(\partial a)^2}}\,. \label{v}
\end{equation}

\subsection{Self-duality condition}\label{selfcondit}

The PST treatment applies to all cases, in particular to the even
dimensions $D=2(p+2)$ with self-dual fields, that is the
duality-symmetric ($p$ even) and chiral ($p$ odd) case. Here we
will focus on the duality-symmetric treatment in dimensions
$D=2(p+2)$, after we will consider the chiral forms. We introduce
the interior product of the unit timelike vector $v$ with the
anti-self-dual part of the $(p+2)$-form field strength
\begin{eqnarray} \mathrm{F}\equiv
i_v ({\cal F}-\tilde{*}{\cal F})\,, \label{calH}
\end{eqnarray}
where $\tilde{*}:=J*$ is the involution operator in
$\bar{\Omega}({\cal M})$. The interest of $\mathrm F$ is that the
self-duality of $\cal F$ is exactly equivalent to the vanishing of
$\mathrm F$, which generalizes the discussion made at the
beginning of section \ref{noncovact}. \be {\cal F}=\tilde{*}{\cal
F}\,\quad\Leftrightarrow \quad \mathrm{F}=0\,.\ee This is easily
proved for any $v$ by exhibiting the identity
\begin{equation}\label{F-*F} \bullet\quad {\cal F}-\tilde{*}{\cal
F}+v\wedge{\mathrm F}+\tilde{*}(v\wedge{\mathrm F})=0\,.
\end{equation} This identity itself is straightforwardly proven
\begin{eqnarray}
{\cal F}-\tilde{*}{\cal F}\stackrel{ \mbox{\tiny (\ref{id})} }{=}
-(vi_v-\tilde{*}vi_v\tilde{*})({\cal F}-\tilde{*}{\cal
F})\stackrel{ \mbox{\tiny (\ref{calH})} }{=}-v\wedge {\mathrm
F}-\tilde{*}(v\wedge{\mathrm F})\nonumber
\end{eqnarray}
In conclusion $v\wedge\mathrm{F}=0$ is an equation that one would
get derived from an action principle. This is the task of next
subsection.

\subsection{Action principle}\label{subone}

In order to obtain this equation from an action principle in the
sequel we need a sample of convenient identities and conventions.

\subsubsection{Useful identities}

First of all we need some identities for the (matrix \emph{and}
wedge) product of two arbitrary elements of
$\overline{\Omega}({\cal M})$. More specifically, one is
interested in \be A^T\wedge J\, B\equiv A_{[q]}\wedge
B_{[r]}+\sigma A_{[p]}\wedge B_{[s]}\,.\ee For biforms
$\mathcal{A}$ and $\mathcal{B}$ with the same parities as
$\mathcal{F}$ we have\be \bullet\quad{\cal A}^T\wedge J\, {\cal
B}\,=\,-\,{\cal B}^T\wedge J\, {\cal A}\,.\label{BC}\ee For
biforms $\mathrm{A}$ and $\mathrm{B}$ with the same parities as
$\mathrm{F}$ one has instead \be \bullet\quad{\mathrm A}^T\wedge
J\, {\mathrm B}\,=\,{\mathrm B}^T\wedge J\, {\mathrm
A}\,.\label{AwB}\ee Furthermore,\be \bullet\quad d({\mathrm
A}^T\wedge J\,{\mathrm B})=d{\mathrm A}^T\wedge J\,{\mathrm
B}+\,d{\mathrm B}^T\wedge J\,{\mathrm A}\,.\label{dajb}\ee

\subsubsection{PST action}

The PST action is defined as \be S_{PST}\equiv\frac12\int
\left(v\wedge{\mathrm F}^T\wedge J\,{\cal F}+{\cal A}^T\wedge
\tilde{*}\cal{J}\right)\,.\label{isdefinedas}\ee The PST action
can be rewritten in a form more convenient for computational
purposes, using the identity
\begin{equation}\label{F*F}
\bullet\quad 2\sigma  \,v\wedge{\mathrm F}^T\wedge J\,{\cal
F}={\cal F}^T\wedge *{\cal F}+{\mathrm F}^T\wedge *{\mathrm F}
\end{equation}
The proof of this identity takes several elementary steps
\begin{eqnarray}
{\mathrm F}^T\wedge*{\mathrm F} &\stackrel{
\mbox{\tiny(\ref{calH})} }{=}& {\mathrm F}^T\wedge * i_v({\cal
F}-\tilde{*}{\cal F}) \stackrel{ \mbox{\tiny (\ref{*iv})} }{=}
-v\wedge{\mathrm F}^T\wedge *({\cal F}-\tilde{*}{\cal F}) \nonumber\\
&\stackrel{ \mbox{\tiny (\ref{J2})} }{=}& \sigma(v\wedge{\mathrm F})^T\wedge J({\cal F}-\tilde{*}{\cal F})\nonumber\\
&\stackrel{ \mbox{\tiny (\ref{a*b})} }{=}& \sigma\,[\,
(v\wedge{\mathrm F})^T\wedge J\,{\cal F}\,-\,{\cal F}^T\wedge
J\tilde{*}(v\wedge{\mathrm F})\,]\nn\\
&\stackrel{ \mbox{\tiny (\ref{F-*F})}
}{=}&\sigma\,[\,v\wedge{\mathrm F}^T\wedge J\,{\cal F}+ {\cal
F}^T\wedge J({\cal F}-\tilde{*}{\cal
F}+v\wedge{\mathrm F})\,]\nonumber\\
&\stackrel{ \mbox{\tiny (\ref{J2})} }{=}&
\sigma(2\,v\wedge{\mathrm F}^T\wedge J\,{\cal F}+\underbrace{{\cal
F}^T\wedge J\,{\cal F}}_{\stackrel{ \mbox{\tiny (\ref{BC})}
}{=}0})-{\cal F}^T\wedge *{\cal F}\,.\nn
\end{eqnarray}
Therefore the PST action also reads in form notation\ba
S_{PST}&\stackrel{ \mbox{\tiny (\ref{F*F})}
}{=}&\frac14\,\sigma\int \left[{\cal F}^T\wedge*{\cal F}+{\mathrm
F}^T\wedge \tilde{*}{\mathrm F}\right]+ \frac12\int{\cal
A}^T\wedge \tilde{*}\cal{J}\,,\label{PSTact}\ea and in components,
\ba S_{PST} &\stackrel{ \mbox{\tiny (\ref{a*b})}
}{=}&-\frac{1}{4(p+2)!}\,\int {\cal
F}_{\m_1\ldots\m_{p+2}}^\a{\cal
F}^{\m_1\ldots\m_{p+2}}_\a\nn\\&&\quad+\frac{1}{4(p+1)!}\,\int
J_\a^{\,\,\b}\,{\mathrm F}_{\m_1\ldots\m_{p+1}}^\a {\mathrm
F}_\b^{\m_1\ldots\m_{p+1}}\nn\\&&\quad + \frac{1}{2(p+1)!}\,\int
J_\a^{\,\,\b}\,{\cal A}_{\m_1\ldots\m_{p+1}}^\a {\cal
J}^{\m_1\ldots\m_{p+1}}_\b\,.\nn\ea

\subsubsection{Variation of the action with respect to the Dirac brane $\cal G$}

The variation of the action under a variation of the Dirac brane
positions can be read from (\ref{PSTact}) by inserting $\d {\cal
F}=0$ coming from $\d*{\cal G}=d*{\cal V}$ and $\d{\cal A}=-*{\cal
V}$ to obtain
\begin{equation}\label{varG} \delta_{\cal G}
S_{PST}=-\frac12\int{\cal V}^T\wedge \tilde{*}\cal{J}\,.
\end{equation}

To obtain the general variation of (\ref{PSTact}) we first proceed
in a way similar to the derivation of (\ref{F*F})
\begin{eqnarray}
{\mathrm F}^T\wedge *\delta_{\cal F}{\mathrm F} &=& \sigma\,v\wedge{\mathrm F}^T\wedge J\,\delta({\cal F}-\tilde{*}{\cal F}) \nonumber\\
&\stackrel{ \mbox{\tiny (\ref{J2})} }{=}& \sigma\,v\wedge{\mathrm
F}^T\wedge J\,\delta{\cal
F}-v\wedge{\mathrm F}^T\wedge *\delta{\cal F}\nn\\
&\stackrel{ \mbox{\tiny (\ref{a*b})} }{=}& \sigma\,v\wedge{\mathrm
F}^T\wedge J\,\delta{\cal F}-\delta {\cal F}^T\wedge
*(v\wedge{\mathrm F})\nn\\ &\stackrel{ \mbox{\tiny (\ref{F-*F})}
}{=}&\sigma\,[ v\wedge{\mathrm F}^T\wedge J\,\d{\cal F}+ \d{\cal
F}^T\wedge J({\cal
F}-\tilde{*}{\cal F}+v\wedge{\mathrm F})]\nonumber\\
&\stackrel{ \mbox{\tiny (\ref{J2})} }{=}&
\sigma(2\,v\wedge{\mathrm F}^T\wedge J\,\d{\cal F}+\d{\cal
F}^T\wedge J\,{\cal F})-\d{\cal F}^T\wedge *{\cal F}\nn
\end{eqnarray}
and we find
\begin{equation}\label{deltachiral}
\delta_{\cal F} S_{PST}\stackrel{ \mbox{\tiny (\ref{BC})} }{=}\int
\left(\,v\wedge{\mathrm F}^T - \frac12\,{\cal F}^T\right)\wedge
J\,\d {\cal F}
\end{equation}
We can now treat the

\subsubsection{Variation of the action with respect to the gauge
field $\cal A$}

It is necessary to evaluate
\begin{eqnarray}
\int {\cal F}^T\wedge J\, \d_{\cal A}{\cal F}&\stackrel{
\mbox{\tiny (\ref{BC})} }{=}& -\int d\,\d{\cal A}^T\wedge J\,
{\cal F}\stackrel{ \mbox{\tiny (\ref{dajb})} }{=}(-)^D\int d{\cal
F}^T \wedge J\,\d{\cal A}\nn\\ &\stackrel{ \mbox{\tiny
(\ref{AwB})} }{=}&\int \d{\cal A}^T \wedge \tilde{*}\cal{J}\nn
\end{eqnarray}
With (\ref{deltachiral}) this implies
\begin{equation}\label{varA}
\d_{\cal A} S_{PST}=\int v\wedge{\mathrm F}^T \wedge J\,\d_{\cal
A} {\cal F}\stackrel{ \mbox{\tiny (\ref{AwB})} }{=}-\int \d{\cal
A}^T\wedge J\,d(v\wedge{\mathrm F})\,.
\end{equation}
We are also interested in the

\subsubsection{Variation of the action with respect to the auxiliary
field $a$}

The variation of $v$ is given by
\begin{eqnarray}
\delta v &\stackrel{ \mbox{\tiny (\ref{v})}}{=}&\frac{d(\delta
a)}{\left(-(\partial
a)^2\right)^\frac12}+\frac{da}{\left(-(\partial
a)^2\right)^{\frac32}}i_{da}d(\delta a)\nonumber\\
&=& \frac{1}{\sqrt{-(\partial a)^2}}\,(I+vi_v)d(\delta a)\nonumber\\
&\stackrel{ \mbox{\tiny (\ref{ivv})}}{=}&-
\frac{1}{\sqrt{-(\partial a)^2}}\,i_v\left(v\wedge d(\delta
a)\right)\label{deltav}
\end{eqnarray}
For the second term in (\ref{PSTact}), containing $v$, we evaluate
the variation \be \sigma{\mathrm F}^T\wedge*\delta_a{\mathrm
F}\stackrel{ \mbox{\tiny (\ref{*iv})} }{=} \delta v\wedge{\mathrm
F}^T\wedge *({\cal F}-\tilde{*}{\cal F}) \ee

\begin{eqnarray} \sigma{\mathrm
F}^T\wedge*\delta_a{\mathrm F}&\stackrel{
\mbox{\tiny(\ref{deltav})}}{=}&-\frac{1}{\sqrt{-(\partial
a)^2}}\,i_v\left(v\wedge d(\delta a)\right)\wedge{\mathrm
F}^T\wedge
*({\cal F}-\tilde{*}{\cal F}) \nn\\
&\stackrel{ \mbox{\tiny(\ref{a*b}), $i_v{\cal F}=0$
}}{=}&-\frac{1}{\sqrt{-(\partial a)^2}}({\cal F}-\tilde{*}{\cal
F})^T\wedge
*i_v\left(v\wedge d(\delta a)\wedge{\mathrm F}\right)\nonumber\\
&\stackrel{ \mbox{\tiny (\ref{*iv})}
}{=}&\frac{1}{\sqrt{-(\partial a)^2}}\,v\wedge({\cal
F}-\tilde{*}{\cal F})^T\wedge *\left(v\wedge d(\delta
a)\wedge{\mathrm F}\right)\nonumber\\
&=& -\frac{d(\delta a)}{\sqrt{-(\partial a)^2}} \wedge
v\wedge {\mathrm F}^T\wedge *\left(v\wedge({\cal F}-\tilde{*}{\cal F})\right)\nonumber\\
&\stackrel{ \mbox{\tiny (\ref{*iv})} }{=}&-\frac{d(\delta
a)}{\sqrt{-(\partial a)^2}}\wedge v\wedge{\mathrm F}^T\wedge
J\,{\mathrm F} \label{iceage}\end{eqnarray}

We need some intermediate steps before arriving at the variation
in a close form. Directly from (\ref{v}) we get $dv =
d(\frac{1}{\sqrt{-(\partial a)^2}})\wedge da$. Therefore,
\begin{equation}\label{dv}
\frac{dv}{\sqrt{-(\partial a)^2}}=d(\frac{1}{\sqrt{-(\partial
a)^2}})\wedge v
\end{equation}
This simple identity allows us to deduce
\begin{equation}\label{dvhh}
\bullet\quad d\left(\frac{1}{\sqrt{-(\partial a)^2}}\,
v\wedge{\mathrm F}^T \wedge J\,{\mathrm
F}\right)=\frac{2}{\sqrt{-(\partial a)^2}}\, d\left(v\wedge
{\mathrm F}^T\right)\wedge J\,{\mathrm F}
\end{equation}
A direct check is
\begin{eqnarray}
&&d\left(\frac{1}{\sqrt{-(\partial a)^2}} \,v\wedge{\mathrm F}^T
\wedge
J\,{\mathrm F}\right)\nonumber\\
&&\quad\stackrel{ \mbox{\tiny
(\ref{dajb})}}{=}d(\frac{1}{\sqrt{-(\partial a)^2}})\,v\wedge
{\mathrm F}^T \wedge J\,{\mathrm F}+\frac{1}{\sqrt{-(\partial
a)^2}}\,(dv\wedge{\mathrm F}^T
\wedge J\,{\mathrm F}-2v\wedge d{\mathrm F}^T \wedge J\,{\mathrm F})\nonumber\\
&&\quad\quad\stackrel{ \mbox{\tiny
(\ref{dv})}}{=}\frac{2}{\sqrt{-(\partial a)^2}}\,(dv\wedge{\mathrm
F}^T -v\wedge d{\mathrm F}^T)\wedge J\,{\mathrm F}
=\frac{2}{\sqrt{-(\partial a)^2}}\, d\left(v\wedge {\mathrm
F}^T\right)\wedge {\mathrm F}\,.\nn
\end{eqnarray}
Eventually, from (\ref{iceage}) and (\ref{dvhh}) we get
\begin{equation}\label{delta2} \delta_a S_{PST}=\int\frac{\delta a}{\sqrt{-(\partial a)^2}}\,
d\left(v\wedge {\mathrm F}^T\right)\wedge J\,{\mathrm
F}\end{equation}

\subsection{Gauge invariance}

Putting together (\ref{varA}) and (\ref{delta2}) the general
variation of the PST action gives \be \d S_{PST}\stackrel{
\mbox{\tiny (\ref{AwB})}}{=}\int(\frac{\delta a}{\sqrt{-(\partial
a)^2}}\,{\mathrm F}^T -\d{\cal A}^T)\wedge J\,d(v\wedge{\mathrm
F})\,.\ee We have already anticipated the gauge transformations
associated to the arbitrariness of Dirac branes. Besides of this
one, there are three different type of gauge transformations that
leave the action invariant up to a boundary:

\subsubsection{Gauge transformations I}

First, one has the standard transformations of Abelian
antisymmetric gauge fields \be \d_I a=0\,,\quad \d_I {\cal
A}=d\varphi\,. \ee

\subsubsection{Gauge transformations II}

The second type of gauge transformations are really specific to
the PST model \be \d_{II} a=\phi\,,\quad \d_{II} {\cal
A}=\frac{\phi}{\sqrt{-(\partial a)^2}}\,{\mathrm F}\,. \ee We see
explicitly that $a$ is pure gauge. A convenient gauge-choice is to
choose $v$ as a unit timelike vector, with which one recovers the
non-covariant action since one can take a frame such that
$v^\m=(1,0,\ldots,0)$. The residual gauge transformations II then
corresponds to the non-manifest invariance under Lorentz boosts.
This convenient gauge-choice proves that the {\emph PST and
non-covariant actions are equivalent at the classical level}.

\subsubsection{Gauge transformations III}

The trivial identity $da\wedge da\equiv 0$ imply the gauge
transformations \be \d_{III} a=0\,,\quad \d_{III} {\cal
A}=da\wedge {\cal E}\,. \label{varIII}\ee Once the previous gauge
choice is done, the transformations III correspond to a shift of
the temporal component of the two gauge fields.

\subsection{Equations of motion}\label{subthree}

From (\ref{varA}) and (\ref{delta2}), we obtain the e.o.m. for
$\cal A$ and $a$. The equation for $a$ is a consequence of the
equation for $\cal A$, which is consistent with the fact that the
former is pure gauge. Hence the e.o.m. reduce to \be d(v\wedge
{\mathrm F})=0\,.\label{eomA}\ee If $H^{p+2}({\cal
M})=H^{D-p-2}({\cal M})$ are trivial, then (\ref{eomA}) is
equivalent to $v\wedge {\mathrm F}=d\Sigma$ but the left-hand-side
is proportional to $da$ therefore $v\wedge {\mathrm F}=da\wedge
d{\cal E}'$. But the variation of $v\wedge \mathrm F$ under a
gauge transformation III is equal to \be\d_{III}(v\wedge \mathrm
F)\stackrel{ \mbox{\tiny (\ref{calH})}}{=}v i_v\, \d_{III}({\cal
F}-\tilde{*}{\cal F}) \stackrel{ \mbox{\tiny (\ref{ivv}),
(\ref{*iv})}}{=}-(I+i_v v +v*v\,J^T )\d_{III}{\cal
F}\,.\label{varIIIuse}\ee Inserting (\ref{varIII}) in
(\ref{varIIIuse}), one finds $\d_{III}(v\wedge \mathrm F)=da\wedge
d\cal E$ since $v\wedge da\equiv 0$. Therefore, with an
appropriate gauge choice we can always arrive at the desired
equation $v\wedge \mathrm F=0$, which is equivalent to the
self-duality condition.

\subsection{Chiral decomposition}

When $\sigma=1$ we can diagonalize the matrix $J$ by the
orthonormal change of basis defined by
\be V=\left( \begin{array}{c c} \frac{1}{\sqrt{2}} & \frac{1}{\sqrt{2}}\\
\frac{1}{\sqrt{2}} & -\frac{1}{\sqrt{2}}\end{array}
\right)=V^T\,,\quad\quad V^T\,V=I\,,\label{Vdef}\ee i.e.
\be V\,J\,V^T=\left( \begin{array}{c c} 1 & 0\\
0 & -1\end{array} \right)=\sigma_3 \,.\label{s3}\ee

In dimensions equal to $2$ mod $4$, the field strength is equal to
the sum of a self-dual
$\stackrel{(+)}{F}=\frac12(F+*F)=\frac12({\cal F}^1+{\cal F}^2)$
and an anti-self-dual part
$\stackrel{(-)}{F}=\frac12(F-*F)=\frac12({\cal F}^1-{\cal F}^2)$,
i.e. \be F=\stackrel{(+)}{F}+\stackrel{(-)}{F}\,,\quad
*\stackrel{(\pm)}{F}=\pm \stackrel{(\pm)}{F}\,.\ee In the diagonal
basis, the PST action (\ref{isdefinedas}) reads \be
S_{PST}=\frac12\int (v\wedge{\overline{\mathrm F}}^T\wedge
\sigma_3\,\overline{{\cal F}}+\overline{\cal A}^T\wedge
*\,\sigma_3\overline{\cal J})\,,\ee
where
\ba\overline{{\cal F}}&\equiv &V\,{\cal F}=\sqrt{2}\left(\begin{array}{c} \stackrel{(+)}{F}\\
\stackrel{(-)}{F}\end{array}\right)=d\overline{\cal A}\nn\\
\overline{\cal A}&\equiv &V{\cal
A}\equiv\sqrt{2}\left(\begin{array}{c}
\stackrel{(+)}{A}\\\stackrel{(-)}{A}\end{array}\right)\,,\nn\\\nn\\
\overline{{\mathrm F}}&=&i_v(\overline{{\cal
F}}-\sigma_3\overline{{\cal F}})\nn\\\nn\\
\overline{{\cal J}}&\equiv &V\,{\cal J}=d\overline{\cal G}\nn\\
\overline{\cal G}&\equiv &V{\cal
G}\equiv\sqrt{2}\left(\begin{array}{c}
\stackrel{(+)}{G}\\\stackrel{(-)}{G}\end{array}\right)\,,\nn\\
\ea Consequently the action is equal to the difference of two
actions: one describing a chiral $(p+1)$-form and one describing
an antichiral $(p+1)$-form\be
\frac12\,S_{PST}=\stackrel{(+)}{S}[\stackrel{(+)}{A},\stackrel{(+)}{G}]\,-\stackrel{(+)}{S}[\stackrel{(-)}{A},\stackrel{(-)}{G}]\,.\ee
The action for (anti)-chiral forms is \be
\stackrel{(\pm)}{S}[\stackrel{(\pm)}{B}_{\m_1\ldots\m_{p+1}},\stackrel{(\pm)}{G}_{\m_1\ldots\m_{p+2}}]=\frac12\int
v\,\wedge\stackrel{(\pm)}{\cal H}\wedge \stackrel{(\pm)}{H}
+\stackrel{(\pm)}{B}\wedge\,
*\stackrel{(\pm)}{J}\,,\label{chiract}\ee
with\be\stackrel{(\pm)}{\cal H}\equiv i_v(\stackrel{(\pm)}{H}\mp
*\stackrel{(\pm)}{H} )\,,\quad \stackrel{(\pm)}{H}
=d\stackrel{(\pm)}{B}\,.\ee Following the steps of subsections
\ref{subone}-\ref{subthree}, one easily shows that the e.o.m.
derived from $\stackrel{(\pm)}{S}$ by varying
$\stackrel{(\pm)}{B}$ is $\stackrel{(\pm)}{\cal H}=0$.

%%%%%%%%%%%%%%%%%%%%%%%%%%%%%%%%%%%%%%%%%%%%%%%%%%%%%%%%%%%%%%%%%%%%%%%%

\section{D-brane electrodynamics}\label{D-branelectro}

Of particular interest in the modern understanding of type I and
II string theories are the Dirichlet branes (D-branes) which are
(extended) solitons carrying Ramond-Ramond (R$\otimes$R) charges.
Their name arises from the fact that they support the endpoints of
open strings.

The low-energy dynamics of a pointlike soliton can be approximated
by quantum mechanics in the moduli space of zero modes. The
generalization for $p$-dimensional extended defect is that the
zero modes give rise to massless worldvolume fields ; the quantum
mechanics becomes a $(p+1)$-dimensional field theory. More
precisely, let ${\cal M}_{p+1}$ be the worldvolume of a single
D$p$-brane. The worldvolume massless modes of the open string
theory form a supersymmetric $U(1)$ gauge theory with a vector
$A_\mu$ ($\m=0,\ldots,p$), $9-p$ real scalars $\Phi^i$ and their
fermionic superpartners\footnote{For reviews on D-branes, the
reader may read \cite{Bachas:1998,Argurio:1998,Johnson:2000}.}.

\begin{figure}[ht!]
     \centerline{\includegraphics[width=0.3\linewidth]{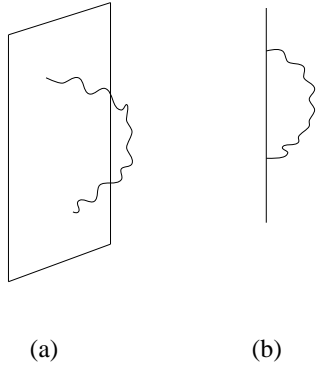}}
\caption{Fundamental open strings can end on D-brane.}
\end{figure}

The bosonic sector of this effective superstring theory will be
called the ``D-brane electrodynamics", the dynamics of which is
described by the action $S_D$, that is a sum of three distinct
terms: \be S_D=S_{SUGRA}+S_{DBI}+S_{WZ}\,,\ee where $S_{SUGRA}$
describes the bulk dynamics of the fields in the bosonic sector of
supergravity (SUGRA). The so-called \emph{Dirac-Born-Infeld} (DBI)
\emph{action} $S_{DBI}$ gives the coupling of the worldvolume
modes to the massless \emph{Neveu-Schwarz} (NS) \emph{fields} of
the bulk closed string theory, i.e. the metric $G_{\mu\nu}$, the
dilaton $\phi$ and the NS two-form $B_{\mu\nu}$. Finally, $S_{WZ}$
is the \emph{Wess-Zumino} (WZ) \emph{term} which is the D-brane
extension of the minimal coupling between the R$\otimes$R form
gauge fields and the D-brane worldvolume.

\subsection{Worldsheet action}

The bosonic sigma model corresponding to the introduction of
background fields for the massless NS bulk fields: the metric, the
antisymmetric tensor and the dilaton:
\begin{equation}
S_\sigma={1\over4\pi\alpha^\prime}\int\limits_{{\cal W}_s}
d^2\!\sigma\, g^{1/2} \,\left\{(g^{ab}G_{\mu\nu}(X)+\epsilon^{ab}
B_{\mu\nu}(X))\partial_a X^\mu\partial_b X^\nu+\alpha^\prime\phi R
\right\}\ , \label{curved}
\end{equation}
where $B_{\mu\nu}$ is the background antisymmetric tensor field
and $\Phi$ is the background value of the dilaton.

This worldsheet theory defined by the sigma model (\ref{curved})
is classically Weyl invariant. In the quantum theory, Weyl
invariance imposes condition on the massless background fields,
which can be seen as e.o.m. derived from the bulk action \be
S_{NS}=\int\limits_{\cal M} e^{-2\phi}\left(R*1+4 *(d\phi)\wedge
d\phi-\frac12 *H\wedge H\right)+O(\alpha^\prime)\,,\ee where $R$
is the scalar curvature of the metric $G$ and $H=dB$ is the field
strength of the NS two-form $B$ in the absence of the NS$5$-brane
(magnetically charged).

Turning to the open string sector, we may also write the effective
action which summarizes the leading order (in $\alpha^\prime$)
open string physics at tree level:
\begin{equation}
{\rm S}=-\frac12\int e^{-\Phi} *F\wedge F+O(\alpha^\prime)\ ,
\label{yangmills}
\end{equation}
for an Abelian gauge field $A$ living in the bulk, i.e.
$A\in\Omega^1({\cal M})$.

\subsubsection{Minimal coupling}

The sigma model action provides the minimal coupling
$1/2\pi\a'\int B$ to the NS two-form $B$. A fundamental string may
end on any D-brane on whose worldvolume ${\cal W}_p$ it couples as
an elementary electrically charged particle, $\partial{\cal
W}_s\subset {\cal W}_p$. A gauge field $A\in\Omega^1({\cal W}_p)$
is coupled in the sigma-model fashion to the boundary $\partial
{\cal W}_s$ of the fundamental string worldsheet by
$\int_{\partial{\cal W}_s}A$. Mathematically speaking this is
natural because in terms of the definition (\ref{naturalint}),
this coupling \be 1/2\pi\a'\int\limits_{{\cal
W}_s}B\,+\int\limits_{\partial{\cal
W}_s}A=\int\limits_{(\partial{\cal W}_s,{\cal
W}_s)}(1/2\pi\a'\,B,-A)\,,\label{mct}\ee is the integral of
$(1/2\pi\a'\,B,-A)\in\Omega^2({\cal M},{\cal W}_p)$ on the
relative $1$-cycle $(\partial{\cal W}_s,{\cal W}_s)\in H_1({\cal
W}_p,{\cal M})$.

The NS two-form $B$ possesses the following gauge transformation
$B\rightarrow B+d\a$, where $\a\in\Omega^1({\cal M})$. For closed
strings, one does not need any supplementary term to have gauge
invariance. But for open strings, gauge invariance requires that
the gauge field possesses the gauge transformations $A\rightarrow
A-(2\pi\a^{\prime})^{-1}\a+d\b$, where $\b\in\Omega^0({\cal
W}_p)$. In terms of the relative de Rham complex $\Omega^*({\cal
M},{\cal W}_p)$, we have \be
(B,-2\pi\a'A)\rightarrow(B,-2\pi\a'A)+d(\a,2\pi\a'\b)\,,\ee which
leaves the minimal coupling term (\ref{mct}) invariant due to a
generalization of Stokes' theorem.

All this suggests the following gauge invariant field strength:
\be F=dA+(2\pi\a^{\prime})^{-1}\,i^*B+\bar{*}g\,,\ee in the
presence of a magnetically charged brane living on the D-brane
worldvolume of current $\bar{*}j_m$ such that
$d\bar{*}g=\bar{*}j_m$. We define the form $A$ to be the formal
sum of all the spacetime extensions of the vector gauge field
living in D$p$-brane worldvolumes. We anticipate the knowledge of
the set $\upsilon$ of integers $p$ such that the corresponding
D$p$-branes are known to be stable. One
has fot type II superstring theories \be\upsilon\equiv \,\,\left\{\begin{array}{lll}&\{0,2,4,6\}& \,\mbox{for IIA}\\
&\{1,3,5,7\}&\,\mbox{for IIB}\end{array}\right. \,.\ee The
previosuly defined gauge field $A$ belongs to
$\Omega^1\big(\bigcup\limits_{p\in\upsilon}{\cal W}_p\big)$.

\subsection{Supergravity sector}

To avoid self-duality condition subtleties in case IIB, we
restrict to the IIA string theory. The bosonic part of the
low-energy action ($D=10$) of which is the sum \be
S_{SUGRA}=S_{NS}+S_{R}+S_{CS}\,.\ee In the absence of magnetically
charged brane, it reads (in the string frame) \ba S_R=-\frac12\int
*F_{[2]}\wedge F_{[2]}-\frac12\int *F_{[4]}\wedge
F_{[4]}\,,\\S_{CS}=-\frac12\int B \wedge dC_{[3]}\wedge
dC_{[3]}\,, \ea where $F_{[2]}=dC_{[1]}$ and
$F_{[4]}=dC_{[3]}+H\wedge C_{[1]}$ are the R$\otimes$R field
strengths. In the sequel we will focus on the ``D-brane"
electrodynamics by considering all the NS gauge fields that appear
in $S_{NS}$ to be non-dynamical.

The gauge transformations which leave the field strengths and the
action invariant are \ba B&\rightarrow& B+d\Sigma_{[1]}\,,\\
C_{[1]}&\rightarrow& C_{[1]}+d\Lambda_{[0]}\,,\\
C_{[3]}&\rightarrow& C_{[3]}+d\Lambda_{[2]}-B\wedge d\Lambda_{[0]}
\,.\ea

Now we assume the absence of any charged D$p$-brane, in which case
the WZ term vanishes. The ``modified" Bianchi identities are \ba
dF_{[2]}&=&0\,,\label{b1}\\dF_{[4]}+H\wedge F_{[2]}&=&0\ea The
e.o.m. obtained from varying $C_{[1]}$ and $C_{[3]}$ in the
absence of any charged
D-brane \ba d*F_{[2]}-H\wedge *F_{[4]}&=&0\,,\\
d*F_{[4]}+H\wedge F_{[4]}&=&0\,,\label{b4}\ea where we used
$H\wedge H\equiv 0$ since $H$ is of odd parity. Let us introduce
the non homogeneous form \cite{Cederwall:1998}\be R\equiv
F_{[2]}+F_{[4]}+*F_{[4]}-*F_{[2]}\,. \ee We notice that $e^B R$ is
closed since \be dR+H\wedge R=0\ee is a compact form of the set of
Bianchi and field equations (\ref{b1})-(\ref{b4}). From this, we
deduce the exactness of $e^B R$ and derive the potential $C$ of
inhomogeneous form degree:\be e^B \wedge R\equiv d(e^B\wedge
C)\,.\ee Since this means that\be R=dC+H\wedge C\,,\ee we identify
the first two terms in $C$ as $C_{[1]}$ and $C_{[3]}$. The other
two terms are defined to be the two dual potentials
$\tilde{C}_{[5]}$ and $\tilde{C}_{[7]}$ (up to a sign) :\be
C=C_{[1]}+C_{[3]}+\tilde{C}_{[5]}-\tilde{C}_{[7]}\,.\ee

\subsubsection{Double potential formulation of the field equations}

As previously seen from type IIA SUGRA, the R$\otimes$R tensor
field fulfills Bianchi identities which are ``modified". In fact,
for the two type II ten dimensional SUGRAs, it is convenient to
collect all the R$\otimes$R gauge fields in terms of a non
homogeneous element of the de Rham complex \be C\equiv
\sum_{p\in\upsilon} C_{[p+1]}\,,\ee which contains at the same
time the potential $C_{[p+1]}$ and its dual $\tilde{C}_{[7-p]}$
\cite{Cederwall:1998}. Still, it has a well-defined parity
opposite to the one of any element of $\upsilon$ (In
ten-dimensional string theories, only branes with rank of definite
parity $\upsilon$ are stable). We anticipate the knowledge of the
charge non-conservation rule and define the field strength as \be
e^B\wedge R\equiv d(e^{B}\wedge
C)+e^{-2\pi\a'\,dA}*G\,,\label{defR}\ee where $*G$ is a formal
Dirac brane, associated to all the D-branes, such that
$d*G=e^{-\bar{*}g}*J$ with \be *J\equiv\sum_{p\in\upsilon}*J_p\ee
the sum of all D$p$-brane currents and $g$ the sum of the Dirac
brane currents associated to the magnetic charge living on the
D-brane worldvolumes, that is, $\bar{*}g$ is a two-form such that
$d\bar{*}g=\bar{*}j_m$. The natural gauge transformations are
$C\rightarrow C+ e^{-B}d\Lambda$. Of course SUGRA e.o.m. are not
EM duality symmetric but the double potential formulation still
allows a more symmetric treatment of D-branes charged either
electrically or magnetically under the R$\otimes$R gauge fields.

Applying $d$ on the definition (\ref{defR}) we get \be dR+H\wedge
R=e^{-2\pi\a'\,F}\wedge *J\,.\ee which is the compact form of type
II SUGRA Bianchi and field equations in the presence of D-branes
charged under the R$\otimes$R gauge fields. If we apply $d$ once
more we get \be \mbox{\begin{tabular}{|c|}
  \hline\\
  $ d*J=(2\pi\a')\,\bar{*}j_m\wedge *J\,,$ \\\\
  \hline
\end{tabular}}\label{chargenon}
\end{equation} which provides the \emph{D-brane-boundary
rule} \cite{Argurio:1997}. The charge non-conservation is entirely
determined by other \emph{currents} and we can therefore apply the
prescribed surgical brane operation. By the way, decomposing
(\ref{chargenon}) in form degree and comparing with
(\ref{comparedwith}) we get the well-known rule that: \emph{The
D$p$-brane worldvolume boundary happens to lie on the worldvolume
of a D$(p+2)$-brane, on which it appears as the worldvolume of
magnetic $(p-1)$-brane}.

\subsection{Dirac-Born-Infeld action}

At leading order, the low-energy effective action of a single
D$p$-brane corresponds to the sum of the NG action $p$-brane
action (\ref{NGact}) with the dimensional reduction of
ten-dimensional $U(1)$ super-YM action (\ref{yangmills}). As usual
in string theory, there are higher order $\a'=\ell_s^2$
corrections, where $\ell_s$ is the string length scale. The action
incorporating these corrections to all orders in the slowly
varying field strength is
\cite{Fradkin:1985,Abouelsaood:1987,Leigh:1989}\be
S_{DBI}=-T_p\int\limits_{{\cal M}_{p+1}}d^{p+1}\xi\,
e^{-\phi}\,\sqrt{-\det(i^*G+2\pi \a' F)}\,.\label{DBI}\ee where
$T_p$ is the D$p$-brane tension and $F$ is the gauge field
strength. It can be trusted as long as derivatives of the gauge
field strength and second derivatives of the scalar are small on
the string scale $\ell_s$. If $\phi$ and $B$ vanish, (\ref{DBI})
is closely related to the BI action for $b=2\pi \a'$. The DBI
action reveals that the D-branes are dynamical objects, whose
transverse displacements are described by $\Phi^i$, i.e. $\D
X^i(\xi)=2\pi \a' \Phi^i$ in the transverse gauge
$(X^\mu)=(\xi^a,X^i)$. It is implicit in the pull-back of the bulk
spacetime tensors on the D-brane worldvolume; more precisely in
the static gauge \ba i^*(G+B)_{ab}&=&G_{ab}+B_{ab}+4\pi
\a^{\prime}(G_{i(a}\partial_{b)} \Phi^i+B_{i[a}\partial_{b]}
\Phi^i)\nn\\&&+4\pi^2 \a^{\prime
2}(G_{ij}+B_{ij})\partial_a\Phi^i\partial_b\Phi^j\,.\ea

\subsection{Wess-Zumino term}

Since a D-brane carries a R$\otimes$R charge
\cite{Polchinski:1995}, there must exist couplings to the massless
R$\otimes$R states of the closed string. These interactions are
incorporated in a CS term equal to \be S_{WZ}=\int\limits_{\cal M}
*J\wedge e^{-2\pi\a'F}\wedge C\,.\ee The integral on $\cal M$
picks up from the integrand the component of form degree equal to
ten. An other way to derive the D-brane-boundary rule would be to
consider the dictum of the gauge invariance under\be C\rightarrow
C+ e^B d\Lambda\,.\label{youpiiie}\ee

When the R$\otimes$R gauge fields are not magnetically charged,
there is no need to complete the supergravity sector by Dirac
brane currents and the WZ term is gauge invariant by
itself\footnote{The WZ term is gauge invariant because the
supergravity action with no Dirac currents is gauge invariant by
construction.} under (\ref{youpiiie}). In general, the WZ
variation is equal to \be \int\limits_{\cal M}
e^{-2\pi\a'(dA+\bar{*}g)}\wedge *J\wedge d\Lambda=0\,.\ee If we
ask this to be true for all $\Lambda$, the simple brane-boundary
rule (\ref{chargenon}) follows.

\subsection{Coinciding branes}

\begin{figure}[ht!]
     \centerline{\includegraphics[width=0.4\linewidth]{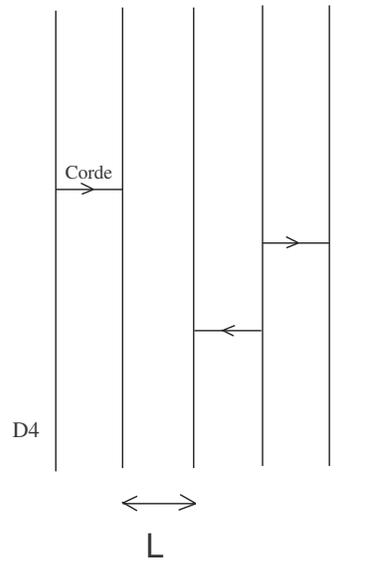}}
\caption{Parallel D4-branes separated by a distance of the order
of $L$ with strings stretching from one to
another.}\label{compare}
\end{figure}
As previously seen, much can be deduced from an analysis of
low-energy SUGRA solutions. However, when several, say $n$,
identical D-branes coincides (i.e. they are closely spaced,
$L\sim\ell_s$) a typically stringy effect appears: the worldvolume
theory gauge group is enhanced to $U(n)$. This results from the
appearance of extra massless degrees of freedom corresponding to
the open strings stretching between different D$p$-branes. The
theory becomes a non-Abelian gauge theory which is not yet fully
understood. We will briefly come back to this issue in section
\ref{Nogoss}.

%%%%%%%%%%%%%%%%%%%%%%%%%%%%%%%%%%%%%%%%%%%%%%%%%%%%%%%%%%%%%%%%%%%
%%%%%%%%%%%%%%%%%%%%%%%%%%%%%%%%%%%%%%%%%%%%%%%%%%%%%%%%%%%%%%%%%%%

\chapter{BRST quantization of duality-symmetric Maxwell's
theory}\label{BRSTquantization}

The quantization of theories containing chiral $p$-forms has been
already performed for several values of $p$ and different
formulations of the systems. The covariant Hamiltonian BRST
(Becchi-Rouet-Stora-Tyutin) quantization of one chiral boson was
realized in \cite{McClain:1990} and generalized to chiral
$p$-forms in \cite{Devecchi:1996}, applying the formulation of
infinitely many ghosts. On the other hand, chiral $2$--forms in
$6$ dimensions have been recently quantized
\cite{VanDenBroeck:1999,Kor} within the covariant
Batalin-Vilkovisky (BV) treatment making use of various
gauge-fixing conditions. The BV method has been also adopted
\cite{Girotti:1997} in proving the quantum equivalence of the
non-covariant \cite{Schwarz:1994} and the Maxwell
\cite{Pasti:1995,Pasti:1995i} theories. Nevertheless, the
generating functionals derived in \cite{Girotti:1997} do not
exhibit a manifest Lorentz covariance.

The present chapter presents the path-integral quantization for
the covariant duality-symmetric Maxwell theory, that is the PST
formulation. The presence of the auxiliary field non-polynomially
coupled to the two gauge potentials makes the gauge algebra
non-Abelian, with field-dependent \emph{structure functions}. As a
consequence, we must choose a suitable quantization procedure. The
antifield-BRST method \cite{Batalin:1983} is considered here
because it is a powerful quantization technique applicable also
for models with open and/or non-Abelian (field-dependent)
algebras. This approach resides in compensating all the
(commuting) gauge symmetries of the original system by some
(fermionic) variables, called ghosts. After extending the action
to a suitable chosen non-minimal sector, we have to fix the gauge.
We are able to perform two different gauge-fixings
\cite{Bekaert:2000}. The covariant one preserves Lorentz
invariance but it has the disadvantage of an intricate form in the
ghost sector which makes its integration difficult. On the other
hand, giving up explicit Lorentz symmetry we present also a
non-covariant gauge which has a simple structure in its fermionic
part leading us the (formal) equality of the generating functional
of PST action with the non-covariant action.

In section \ref{quantoscill}, we briefly illustrate the duality
symmetry concepts in its most simple realization: the good old
harmonic oscillator. The Hamiltonian BRST gauge-fixing of
Maxwell's action is performed in section \ref{a:Max}. This shows
that the generating functional of Maxwell's theory is equal to its
non-covariant (Hamiltonian) but duality-symmetric formulation,
which proves their quantum equivalence \cite{Girotti:1997}. In
section \ref{BVPST} we compute the minimal solution of the master
equation corresponding to the four-dimensional PST action. We also
infer the BRST symmetry. By a well chosen non-minimal sector and
an adequate gauge-fixing fermion the remaining gauge invariances
are fixed. The non-covariant gauge is the starting point in
proving that the generating functional of the PST and
non-covariant formulations of duality-symmetric Maxwell's theory
are formally equal \cite{Bekaert:2001}. The latter is quantum
mechanically equivalent to the ordinary Maxwell theory, so that
all these models are physically related (on-shell) at the
classical and quantum level, even if their off-shell descriptions
are different. Still, we perform a ``formal" path integral
quantization in the sense that the possible UV divergences due to
the non-Gaussian character of the integral were not considered.
Incidentally, the absence of anomalies and non-trivial
counterterms has been shown recently in \cite{DelCima:2000} by
power counting argument. On this basis, it seems that the PST
model can be trusted at the quantum level also.

%%%%%%%%%%%%%%%%%%%%%%%%%%%%%%%%%%%%%%%%%%%%%%%%%%%%%%%%%%%%%%%%%%

\section{Quantum oscillator}\label{quantoscill}

\subsection{Harmonic oscillator}

The good old harmonic oscillator is a gentle toy model which
shares analogous properties to duality symmetric electrodynamics.
For instance, its manipulation suggests a convenient track to
follow for Maxwell's theory.

The harmonic oscillator Schr\"odinger equation kernel is equal to
\be <q_2,t_2|q_1,t_1>= \int{\cal D}q\,{\cal D}p\,\exp
\left(i\int_{t_1}^{t_2}
dt\,[p(t)\dot{q}(t)-H\left(q(t),p(t)\right)]\right)\label{beu} \ee
where the Hamiltonian is\be H(q,p)=\frac12\,p^2+\frac12\,
q^2\,.\label{Ham}\ee The boundary conditions are in the
configuration space: $q(t_1)=q_1$ and $q(t_2)=q_2$. In the same
way, in momentum space the kernel is equal to the path integral
\be <p_2,t_2|p_1,t_1>= \int{\cal D}q\,{\cal D}p\,\exp
\left(i\int_{t_1}^{t_2}
dt\,[-q(t)\dot{p}(t)-H\left(q(t),p(t)\right)]\right)
\label{beu2}\ee with boundary conditions: $p(t_1)=p_1$ and
$p(t_2)=p_2$. In other words, the quantum harmonic oscillator
Hamiltonian is invariant under the ``duality" transformation
\begin{eqnarray}
q&\rightarrow& - p,\nonumber\\
p&\rightarrow&\,\, q,\label{Sqp}
\end{eqnarray}
and kernel (\ref{beu}) goes into kernel (\ref{beu2}). This
configuration-momentum space duality of the harmonic oscillator
could seem pretty trivial, nevertheless, the electric-magnetic
duality is (in some sense) nothing else than its generalization
for Abelian gauge field theory. Indeed, in the Hamiltonian
formalism the electric field ${\bf E}$ is ``conjugate" to the
magnetic field ${\bf B}$, and the Hamiltonian density
$\frac12\,({\bf E}^2+{\bf B}^2)$ has the quadratic form
(\ref{Ham}).

The (classical) Hamiltonian equations are \be
\dot{q}-p=0\,,\quad\dot{p}+q=0\,.\quad \ee They are left invariant
by the ``duality rotation"\begin{eqnarray}
q&\rightarrow&\,\, \cos\alpha\, q -\, \sin\alpha\,p,\nonumber\\
p&\rightarrow& - \,\sin\alpha \,q
+\cos\alpha\,p.\label{rotqp}\end{eqnarray} Moreover, the
Hamiltonian itself is manifestly invariant under this rotation.
This is natural since the duality symmetry is essentially
Hamiltonian in nature. We already met this fact when we looked for
manifestly duality-symmetric actions. The duality rotation is a
canonical transformation generated by $qp$.

To make a connection with Maxwell's theory, we take a look at the
Lagrangian formalism. The Green functions generating functional
is\be Z[J]= \int{\cal D}q\,\exp \left(i\int
dt[L\left(q(t)\right)+J(t)q(t)]\right) \ee where \be
L(q)=\frac12\,\dot{q}^2-\frac12\, q^2\,.\label{Lag}\ee The
boundary conditions are $q(t_1 \rightarrow -\infty)=0$ and $q(t_2
\rightarrow +\infty)=0$. The Euler-Lagrange equation derived from
(\ref{Lag}) is\be \ddot{q}+q=0\,.\ee We have \be
Z[J]=Z[0]\,\exp\left(\frac{-i}{2}\int dt\,dt'\,J(t)\Delta(t-t')
J(t')\right)\ee where \be
\Delta(t-t')=\frac{1}{2\pi}\int\limits_{-\infty}^{+\infty}d\omega\,\frac{e^{-i\omega(t-t')}}{\omega^2-1+i\epsilon}\ee

To go back to a first order formalism, we can insert\footnote{This
is a constant in $q(t)$ by the change of variable
$\overline{p}(t)=p(t)-\dot{q}(t)$ and a constant in $\bar{p}(t)$
by the integration.} \be\int{\cal D}p\,\exp \left(-\frac{i}{2}\int
dt\,(p-\dot{q})^2\right)=const,\ee in the path integral, and get
\be Z[0]= \int{\cal D}q\,{\cal D}p\,\exp \left(i\int
dt\,[p(t)\dot{q}(t)-H\left(q(t),p(t)\right)]\right) \ee with
supplementary boundary conditions $p(t_1 \rightarrow -\infty)=0$
and $p(t_2 \rightarrow +\infty)=0$. If we integrate by part the
kinetic term and use\be\int{\cal D}q\,\exp \left(-\frac{i}{2}\int
dt\,(\dot{p}- q)^2\right)=const.\ee we arrive at the dual
formulation\be Z[0]= \int{\cal D}p\,\exp \left(i\int dt
L\left(p(t)\right)\right)\,, \ee which is exactly identical to the
original one as previously stated.

As we will see in the sequel, the Maxwell theory possesses two
significant new features: it is a \begin{itemize}\item Linear
field theory; hence its dynamics is described by an infinite set
of decoupled harmonic oscillators.
\item Gauge theory; i.e. it is a system with constraints
and we need to gauge-fix the action in the path integral.
\end{itemize}

In this picture non-linear electrodynamics or Yang-Mills theories
appear as sophisticate systems of coupled anharmonic oscillators,
which make them difficult to study outside the scope of
perturbative treatment. Unfortunately, many of the beautiful
mathematical results on EM duality concerns the analogue of
harmonic oscillators because we mainly deal with free (=
quadratic) theories. A seductive conjecture is that some (unknown)
duality symmetry exists for self-interacting systems with
non-Abelian gauge symmetries, like Yang-Mills theories. We will
address this question in chapter \ref{deformation}.

\subsection{Anharmonic oscillator}

By the way, as we have seen previously with electrodynamics, some
results can be generalized to some extent for non-linear theories.
Presently, one can consider a general Lagrangian $L(q,\dot{q})$
and look for a necessary condition leading to a duality rotation
invariant Hamiltonian: $H(q,p)=h(q^2+p^2)$. The variation of the
Legendre transform of the Lagrangian under an infinitesimal
duality rotation is asked to vanish \be\d H= \d p\, \dot{q}-\d
q\frac{\partial L}{\partial q}=0\quad\Leftrightarrow\quad
q\dot{q}+\frac{\partial L}{\partial q}\frac{\partial L}{\partial
\dot{q}}=0\,.\label{CH0+1}\ee We recognize the Courant Hilbert
equation (\ref{duality}) if we define $u_+\equiv\frac12
\dot{q}^2$, $u_-\equiv-\frac12q^2$, $f(u_+,u_-)\equiv
L\left(q(u_+,u_-),\dot{q}(u_+,u_-)\right)$. The
configuration-momentum duality symmetry will also hold for the
quantum dynamics of models satisfying the equation (\ref{CH0+1}).
A specific example is provided by the BI-like oscillator of
Lagrangian \be L =\frac{1}{b}\left( 1-\sqrt{\left( 1-b\,
\dot{q}^2\right) \left( 1+b\,q^2\right) }\right) \label{LBIosc}\ee
arising from the solution (\ref{BIuseful}) of the Courant Hilbert
equation, where we defined $b=\frac{\alpha}{2}$. Its corresponding
Hamiltonian is given by \be H =\frac{1}{b}\left(
\sqrt{1+b\,(p^2+q^2)}-1\right)\,.\label{HamBI}\ee

The following proposition also proves that there exists an
infinite class of physically interesting duality-symmetric
anharmonic oscillators.
\begin{proposition}Let $f(u_+,u_-)$ be a solution of:
$f_+f_-=1$. The function $L(q,\dot{q})\equiv
f\left(u_+(q,\dot{q}),u_-(q,\dot{q})\right)$
\begin{itemize}
\item is analytic near $(q,\dot{q})=(0,0)$ and
\item satisfies $L(q,\dot{q})=\frac12(\dot{q}^2-q^2)+O(q^4,\dot{q}^4)$,
\end{itemize}
{\bf if and only if} the Legendre transform $h(q^2+p^2)\equiv p
dot{q}-L(x(q,\dot{q}),y(q,\dot{q}))$, where $p=\frac{\partial
L}{\partial \dot{q}}$ defines implicitly $\dot{q}(q,p)$, is such
that the function $h(z)$
\begin{itemize}
\item[(i)] is analytic near the origin $z=0$ and
\item[(ii)] satisfies $h(z)=\frac12 z+O(z^2)$.
\end{itemize}
Furthermore, if $h(z)$ is strictly increasing (i.e. $h'(z)\neq 0$)
then the boundary condition defines the Hamiltonian:
$L(q,0)\equiv-h(q^2)$. \label{ploplop}
\end{proposition}

\proof{First of all, the Courant Hilbert equation (\ref{duality})
is a necessary and sufficient requirement to have a
duality-invariant Hamiltonian $H(q,p)\equiv
p\,\dot{q}-L(q,\dot{q})$.

The dynamical system is has no constraint by hypothesis. Indeed,
the two functions $p(q,\dot{q})$ and $\dot{q}(q,p)$ are invertible
and are inverse of each other \be p=\frac{\partial L}{\partial
\dot{q}}\quad\Leftrightarrow\quad \dot{q}=\frac{\partial
H}{\partial p}\,.\ee This comes from the fact that, at the origin
$\frac{\partial^2 L}{\partial \dot{q}^2}=1 \neq 0$ and
$\frac{\partial^2 H}{\partial p^2}=1 \neq 0$, due to the
hypotheses of the proposition \ref{ploplop}. From standard
theorems on implicit functions \cite{Dieu} we find also that the
two functions $p(q,\dot{q})=\dot{q}+O(q^2,\dot{q}^2)$ and
$\dot{q}(q,p)=p+O(q^2,p^2)$ are analytic near the origin.

\begin{description}
\item[$\underline{\,\bf\Rightarrow:}$] We first prove the necessity.
The composition of two analytic functions in a neighborhood is
also an analytic function (in the corresponding neighborhood).
Since $\dot{q}(q,p)=p+O(q^2,p^2)$ and
$L(q,\dot{q})=\frac12(\dot{q}^2-q^2)+O(x,y)$ are analytic near the
origin, the Hamiltonian $H(q,p)=\frac12(q^2+p^2)+O(q^4,p^4)$ is
analytic near the origin $(q,p)=(0,0)$.

Furthermore, since $f$ is a solution of $f_+f_-=1$ we know that
the Hamiltonian is duality-invariant: $H(q,p)=h(p^2+q^2)$.
Therefore the function $h(z)=z+O(z^2)$ is analytic near $z=0$.

When $q\neq 0$, the variables $y$ vanishes if and only if
$\dot{q}=0$. If $p$ vanishes, then also does $\dot{q}$ since
$\dot{q}=2p\, h'(p^2+q^2)$. Conversely, if $h'(z)\neq 0$ then
$\dot{q}=0$ implies $p=0$ for consistency, which proves
$L(q,0)\equiv-h(q^2)$.

\item[$\underline{\,\bf\Leftarrow:}$] By the same argument than before one proves that
the function
$L(q,\dot{q})=\frac12(\dot{q}^2-q^2)+O(q^4,\dot{q}^4)$ is analytic
near the origin (with the appropriate behaviour).
\end{description}
}

In general, it is untractable to write a Lagrangian path integral
corresponding to a duality rotation invariant Hamiltonian:
$H=h(q^2+p^2)$ because the explicit integration over the momentum
$p$ is rather awkward for an arbitrary function $h$. One has to
resort to approximate methods because the classical Lagrangian
receives quantum corrections, as follows from the stationary phase
method (e.g. see \cite{Henneauxbook}, pp.344-346). Fortunately, it
is easy compute the energy spectrum in the Fock basis for a
duality-invariant anharmonic oscillator. First, one introduces the
destruction and creation operators\be a=\frac12(x+ip)\,,\quad
a^+=\frac12(x-ip)\,.\ee Second, one defines the operator\be
N=a^+a=\frac12(x^2+p^2-1)\,,\ee the spectrum of which is known to
be $\mathbb N$ if there exists a lowest state $|0>$, the
``vacuum". Thus the energy spectrum of the duality-symmetric
anharmonic oscillator is given by the formula\be
E_n=h(2n+1)\,,\quad n\in{\mathbb N}\,.\ee As an example, for the
BI-like oscillator defined by the Lagrangian (\ref{LBIosc}) its
energy spectrum is read directly from (\ref{HamBI}): \be
E_n=\frac{1}{b}\left( \sqrt{2b\,n+(b+1)}-1\right)\,.\ee For large
$n$, the energy levels go as $\sqrt{n}$ while they go as $n$ for
the harmonic oscillator.

%%%%%%%%%%%%%%%%%%%%%%%%%%%%%%%%%%%%%%%%%%%%%%%%%%%%%%%%%%%%%%%%%%

\section{Hamiltonian BRST gauge-fixing of Maxwell's action}\label{a:Max}

As a warm-up, let us start with Maxwell's theory in its standard
form, i.e. non-manifestly covariant. This rather standard exercise
will reveal useful by itself (i) in providing a better
understanding of duality symmetry in its most natural (=
Hamiltonian) formulation and (ii) in deriving the correct path
integral measure from the Hamiltonian formalism (the most
foolproof treatment of gauge systems).

We review the standard Hamiltonian BRST procedure by a direct
glance at this specific example. Most of the material presented in
the two next subsections is taken from \cite{Henneauxbook},
chapter 19.

\subsection{Hamiltonian analysis}

The sourceless Maxwell action (\ref{Maxy}) is equal to \be
S^M=-\frac14\int d^4x\,F^{\m\n}F_{\m\n}\,.\label{Maxi}\ee The
action is left invariant by the gauge transformations\be\d
A_\m=\partial_\m\epsilon\label{e:gge}\ee The canonical momenta are
\ba\pi^0=0\,\quad\pi^i=F^{i0}=-E^i\,,\\\{A_\m(\bf{x}),\pi^\n(\bf{x}')\}=\d^\n_\m\,\d(\bf{x}-\bf{x}')\ea
where $i$, $j$, $\dots$ stand for spatial indices in the 3
dimensional hyperplane $x^0$ constant and $\{\,,\,\}$ is the
canonical Poisson bracket.

The canonical Hamiltonian $H_C$ (only well-defined on the
submanifold defined by the primary constraint $\pi^0\approx 0$) is
given by \be H_C=H_0-\int d^3x\,A_0\,\partial_i\pi^i\ee where
$H_0=\int d^3x\,{\cal H}_0$. The Hamiltonian density ${\cal H}_0$
is equal to
\begin{equation}
{\cal H}_0 = \frac{1}{2} (\pi^i\pi_i + B^i B_i)\,.\label{canH}
\end{equation}
The consistency condition $\dot{\pi}^0\approx 0$ leads to the
secondary constraint
\be\{\pi^0,H_C\}=\partial_i\pi^i\approx0\,,\ee which one
recognizes as the Gauss law $\partial_i E^i=0$. There is no
further constraint. The two constraints are first class, they
generate the gauge transformations (\ref{e:gge}).

The extended Hamiltonian $H_E$ reads \ba H_E&=&H_C+\int
d^3x(\lambda\, \pi^0+\m \,\partial_i\pi^i)\nn\\&=&H_0+\int
d^3x(\lambda\, \pi^0+\m' \,\partial_i\pi^i)\,,\label{extf}\ea
where $\lambda$, $\m$ and $\m'$ play the role of Lagrange
multiplier for the constraints. The extended Hamiltonian generates
the time evolution and exhibits all the degrees of freedom in the
equations of motion. In the present case, the equations contain
indeed two arbitrary functions of time corresponding to the two
first class constraints.

\subsection{BRST charge and cohomology}

The central idea of the BRST scheme is to replace the original
gauge symmetries by a fermionic rigid symmetry acting on an
appropriately extended phase space, enlarged by the ``ghosts".

For each first class constraint, we introduce a ghost: $\bar{c}$
corresponds to $\pi^0\approx0$ and $c$ corresponds to the Gauss
law. Their respective momenta are $\bar{\cal P}$ and $\cal P$. The
Poisson bracket is extended to a (graded) bracket by the
relations\be\{\bar{c}(\bf{x}),\bar{\cal
P}(\bf{x}')\}=\d(\bf{x}-\bf{x}')\,,\quad\{c(\bf{x}),{\cal
P}(\bf{x}')\}=\d(\bf{x}-\bf{x}')\,.\ee

A conventional distinction can be drawn between what is called the
minimal and non-minimal sectors. The minimal sector is usually
taken to be the conjugate pairs $(A_i,\pi^i)$ and $(c,{\cal P})$.
The addition of the conjugate pairs $(A_0,\pi^0)$
$(\bar{c},\bar{\cal P})$ provides the non-minimal sector. This
distinction arises naturally if one considers that $A_0$ plays the
role of a Lagrange multiplier for the Gauss law in the canonical
Hamiltonian (\ref{canH}).

The different gradings of the non-minimal sector of fields, ghosts
and their momenta dictated by the set of first class constraints
is listed in Table~\ref{t:mingh'}.
\begin{table}[ht]
\begin{center}
\begin{tabular}{|c|c|c|c|c|c|c|}\hline
field \& ghosts&$A_\mu$&$\bar{c}$&$c$&$\pi^\m$&$\bar{\cal P
}$&${\cal P}$\\ \hline
pure ghost&$0$&$1$&$1$&$0$&$0$&$0$\\
\hline antighost&$0$&$0$&$0$&$0$&$1$&$1$\\
\hline statistic &$+$&$-$&$-$&$+$&$-$&$-$\\
\hline
\end{tabular}
\caption{Pure ghost number, antighost number and statistics of the
(non-minimal) extended phase space. \label{t:mingh'}}
\end{center}
\end{table}
The ghost number is defined to be the difference between the pure
ghost number and the antighost number.

The BRST charge is found to be\be \Omega=\int d^3x\,(\partial_i
\pi^i c- i\,{\bar c}\,\pi_0 ).\label{BRSTch}\ee Its ghost number
is equal to one. It fulfills
\be\{\Omega,\Omega\}=0\label{mastr}\ee

The BRST transformation is the canonical transformation
$s\,\cdot=\{\cdot,\Omega\}$. From the Jacobi identity, the
equation (\ref{mastr}) is equivalent to the nilpotency of the BRST
operator, $s^2=0$.

The (classical) cohomology of the BRST differential is easily
worked out if one introduces the transverse and longitudinal
components of $A_i$ and $\pi^i$ according to the general
decomposition of a vector $Z_i=\partial_iZ^L+Z_i^T$ \be
Z_i^T=Z_i-\partial_i\bigtriangleup^{-1}(\partial_jA^j)\,,\quad
Z^L= \bigtriangleup^{-1}(\partial_jA^j)\,,\ee where
$\bigtriangleup^{-1}$ is the inverse Laplace operator. The BRST
symmetry takes then a form where all the contractible pairs are
manifest\ba sA^L=c\,,\quad
sc=0\,,&\quad& s{\cal P}=-\pi^L\,,\quad s\pi^L=0\,,\nn\\
sA_0=-i\bar{c}\,,\quad s\bar{c}=0\,,&\quad&
s\bar{\cal P}=-i\pi^0\,,\quad s\pi^0=0\,,\nn\\
sA_i^T=0\,,&\quad& s\pi_i^T=0\,. \ea Therefore the BRST cohomology
$H(s)$ is only non-trivial in vanishing ghost number, the set of
observables, and is the space of functions on the reduced phase
space, that is the quotient of the constraint surface by the gauge
orbits, which is generated by the (two) conjugate pairs
$(A_i^T,\pi_i^T)$.

The previous non-minimal set of ghosts and momenta is necessary to
implement the covariant Lorentz gauge. However, for later use we
will consider the Coulomb gauge. To go to the temporal gauge, we
will make use of a singular change of variable.

\subsection{Path integral quantization of Maxwell's theory}

For the parallel with the next section, we explicitly treat our
two favorite gauge-choices: the Lorentz and Coulomb gauges.

The generating functional for the Maxwell theory in the
Hamiltonian approach can be obtained from
\begin{equation}
Z[0]= \int {\cal D} A_\mu \,{\cal D}\pi^\mu \,{\cal D}c \, {\cal
D}{\cal P} \,{\cal D}\bar c\,{\cal D} \bar {\cal P}\, \exp iS^M_\P
\end{equation}
The Hamiltonian gauge-fixed action is given by
\begin{equation}
S^M_\Psi= \int d^4 x\,(\pi^\mu\dot A_\mu + {\cal P} \dot{c} +
\bar{\cal P} \dot{\bar c}  - {\cal H}_0
-\{\Psi,\Omega\})\end{equation} where the function $\Psi$ is the
gauge-fixing fermion ($gh(\Psi)=-1$). The gauge fixed Hamiltonian
density is the sum of the gauge-fixing term $\{\Psi,\Omega\}$ and
the density of the BRST-invariant extension of the extended
Hamiltonian in the multiplier gauge $\l=\m'=0$ (In that case, it
is simply equal to ${\cal H}_0$ since $\{H_0,\Omega\}=0$). It
generates a nonambiguous time evolution in the extended phase
space with no arbitrariness once the gauge-fixing fermion $\Psi$
is chosen.

The Fradkin-Vilkovisky theorem ensures that the path integral is
independent on the choice of gauge-fixing fermion if the boundary
conditions select definite physical states at the endpoints (in
such a way that the path integral stands for an expectation value
between physical states). Moreover, the path integral associates
well-defined (gauge choice independent) quantum average to all the
BRST invariant functions of the extended phase space that belong
to the same BRST cohomological class (\cite{Henneauxbook}, chapter
16). In other words, a change of $\Psi$ does not modify the
dynamics of the BRST-invariant functions at the cohomological
level.

\subsubsection{Lorentz gauge-fixing}

The gauge-fixing fermion corresponding to the Lorentz gauge is
given by \be\Psi=i\,\bar{\cal P}\,\partial_i A^i+{\cal P}A_0
\label{gff}\ee With such a choice, the gauge-fixed Hamiltonian
density reads
\begin{equation} {\cal H}={\cal H}_0- A_0
\partial_i \pi^i - \pi_0\partial^i A_i - i {\cal P}\bar {c} +
i\bar{\cal P} \bigtriangleup c\,.\label{gfHam}
\end{equation}

One can easily integrate the fields $\pi^0$, the conjugate momenta
${\cal P}$, $\bar {\cal P}$ as well as their corresponding ghosts
$c$, $\bar c$. Then, we obtain that
\begin{equation}
Z[0]= \int {\cal D} A_\m \,{\cal D}\pi^i \,{\rm
det}(\Box)\,\delta(\partial^\m A_\m) \, \exp i{\tilde S}^M_\P
\end{equation}
with
\begin{equation}
{\tilde S}^M_\Psi= \int d^4 x\,(\pi^i\dot A_i - {\cal
H}_0-A_0\partial_i \pi^i).
\end{equation}
The integration on $\pi_0$ gives the delta-function enforcing the
Lorentz gauge while the determinant of the d'Alembert operator
$\Box$ comes from the integration on the fermionic ghosts and
momenta.

A last (Gaussian) integration over $\pi^i$ leads to a Lagrangian
generating functional that could have been derived \'a la
Fadeev-Popov (see Faddeev formula in \cite{Henneauxbook}, chapter
16)
\begin{equation}
Z[0]= \int {\cal D} A_\m \,{\rm det}(\Box)\,\delta(\partial^\m
A_\m) \, \exp i S^M\,.
\end{equation}
The Maxwell action $S^M$ is given in (\ref{Maxi}). One can
completely eliminate the momenta in the gauge-fixed action, the
terminology is that it is a ``propagating gauge".

\subsubsection{Coulomb gauge-fixing}

We will now follow the road chosen in \cite{Girotti:1997} to make
a connection with the non-covariant duality-symmetric formulation
of Maxwell's theory. We will first implement the Coulomb gauge in
the path integral and then we will solve the Gauss law in terms of
a new potential vector.

One starts by redefining the gauge-fixing function in the
following way \be\partial_iA^i\rightarrow \frac{1}{\epsilon}\,
\partial_iA^i\,.\ee At the same time one performs the following
field redefinition \be\pi^0\rightarrow \epsilon
\pi^0\,,\quad\bar{\cal P}\rightarrow \epsilon \bar{\cal P}\,,\ee
which is of trivial(=1) super-Jacobian\footnote{Fermionic volume
elements have an inverse transformation law under scale
transformation.}. The gauge-fixing fermion (\ref{gff}) is not
modified by this change of variable. One can now get rid of the
kinetic term containing $\pi^0$ by taking the limit
$\epsilon\rightarrow 0$.

One easily integrates the field $A_0$ and then, as before, one
integrates $\pi^0$, the conjugate momenta ${\cal P}$, $\bar {\cal
P}$ as well as their corresponding ghosts $c$, $\bar c$. After all
this one finds
\begin{equation}
Z[0]= \int {\cal D} A_i \,{\cal D}\pi^i \,{\rm
det}(\Box)\,\delta(\partial_i \pi^i)\,\delta(\partial^i A_i) \,
\exp iS_\P
\end{equation}
with
\begin{equation}
S_\Psi= \int d^4 x\,(\pi^i\dot A_i - {\cal H}_0).
\end{equation}
The integration on $A_0$ and $\pi^0$ gave the delta-functions
enforcing, respectively, the Gauss law and the Coulomb gauge.

The previous field redefinition and limit process is a trick which
works for any internal gauge symmetries (= constraints linear and
homogeneous in the momenta) to get the reduced phase space path
integral (see \cite{Henneauxbook}, p.398) from the non-minimally
extended BRST path integral. More accurately, in our case we get
\begin{equation}
Z[0]= \int {\cal D} A^T_i \,{\cal D}\pi^{T\,i}\,{\rm
det}^2(\bigtriangledown) \,{\rm det}(\Box) \,{\rm
det}^{-2}(\bigtriangleup) \, \exp i\tilde{S}_\P
\end{equation}
with
\begin{equation}
\tilde{S}_\Psi= \int d^4 x\,\left(\pi^{T\,i}\dot A^T_i - {\cal
H}_0(A^T_i,\pi^{T\,i})\right)\,
\end{equation}
since the Jacobian of
$Z_i=\partial_iZ^L+Z_i^T\rightarrow(Z^L,Z_i^T)$ is equal to
$\det(\nabla)$ and \be\delta(\partial_i Z^i)=\d(\triangle
Z^L)={\rm det}^{-1}(\bigtriangleup) \,\d(Z^L)\,.\ee

\subsubsection{Duality-invariant gauge-fixed path integral}

In order to make the connection with the gauge-fixed
duality-symmetric action, we have to move to a two-potential
formulation. For this purpose, we solve the Gauss constraint
$\partial_i \pi^i=0$ by introducing a potential $Z^i$ such that
\begin{equation}
\pi^i=\epsilon^{ijk}\partial_jZ_k.\label{e:pi}
\end{equation}
The potential $Z_i$ can be decomposed into a sum of a longitudinal
and a transverse part: $Z_i=Z_i^L+Z_i^T$. When $Z_i$ is transverse
($Z_i=Z^T_i$), the equation (\ref{e:pi}) is invertible (with
appropriate boundary conditions). More precisely, in that case one
expresses $Z_i$ as
\begin{equation}
Z_i=-\bigtriangleup^{-1}(\epsilon_{ijk}\partial^j
\pi^k)\,.\label{e:zed}
\end{equation}

We can introduce the field $Z^i$ in the path integral in the
following way
\begin{equation}
Z[0]= \int {\cal D} A_i \,{\cal D}\pi^i \,{\cal D}Z_i\, {\rm
det}(\Box)\,\delta(\partial_i \pi^i)\,\delta(\partial^i
A_i)\,\delta(Z^i+\bigtriangleup^{-1}\epsilon^{ijk}\partial_j
\pi_k) \, \exp i{\tilde S}^M_\P\,.
\end{equation}
In order to make the comparison with the non-covariant approach we
will use the relation
\begin{equation}
\delta(Z^i+\bigtriangleup^{-1}\epsilon^{ijk}\partial_j \pi_k) =
\delta(Z^{L\,i})\delta(Z^{T\,i}+\bigtriangleup^{-1}\epsilon^{ijk}\partial_j
\pi_k)\,.
\end{equation}
We also notice that
\begin{eqnarray}
&&\delta(Z^{T\,i}+\bigtriangleup^{-1}\epsilon^{ijk}\partial_j
\pi_k)\,=\,\underbrace{{\rm det}^{-1}\,(\bigtriangleup^{-1}{\rm
curl})}_{={\rm det}({\rm
curl})}\,\delta(\pi^{T\,i}-\epsilon^{ijk}\partial_jZ^T_k)
\nonumber\end{eqnarray} where ``${\rm curl}$" stands for the
operator $\epsilon^{ijk}\partial_j\,,$ and
$\partial_i\pi^{T\,i}=0$. Indeed, ${\rm det}(\bigtriangleup)={\rm
det}^2({\rm curl})$.

In addition we have the relation \be
\delta(\partial_i\pi^i)\delta(Z^{L\,i})=\delta(\pi^{L\,i})\delta(\partial_iZ^i).\ee
We finally identify the two potentials as follows
\begin{equation}
A^{1}_{i} =  A_i \,, \quad A^{2}_{i} =  Z_i.
\end{equation}
Putting all these remarks together we can integrate out the
$\pi_i$ to obtain the non-covariant duality-symmetric generating
functional found in \cite{Girotti:1997}
\begin{equation}\label{e:genfunc}
Z[0]= \int {\cal D} A^\a_i \, {\rm det}(\Box)\,{\rm det}({\rm
curl})\,\delta(\partial^i A^\a_i) \, \exp iS^{non-cov}\,,
\end{equation}
where $S^{non-cov}$ is given in the equation (\ref{e:SchwarzSen}).

%%%%%%%%%%%%%%%%%%%%%%%%%%%%%%%%%%%%%%%%%%%%%%%%%%%%%%%%%%%%%%%%%%%%%%

\section{Batalin-Vilkovisky quantization of PST action}\label{BVPST}

\subsection{Gauge symmetries of the classical
action}\label{s:action}

We start our discussion by considering the PST action proposed to
manifestly implement two symmetries in the description of free
Maxwell's theory: Lorentz invariance and electric-magnetic
duality. Since we are interested in studying the gauge invariances
for a system with this two symmetries we will neglect here a
possible supersymmetrization of the model which can be attained
through  a kinetic fermionic contribution of type
$i{\bar\p}\dslash\p$ in the Lagrangian. After fixing the notation,
we emphasize the physical content of this model. Next, we briefly
present its gauge algebra.

The sourceless PST action \cite{Pasti:1995i} constructed for the
description of self-dual vector field is
\begin{eqnarray}
S_0=\int d^4x \left(-{1\over 8}F^\alpha_{\m\n}F^{\alpha \,\m\n}
+{1\over{4(-u_\l u^\l)}}u^\mu{\cal F}^\alpha_{\m\n}{\cal
F}^{\alpha\, \n\rho}u_\rho \right)\,. \label{e:pst}
\end{eqnarray}
As explained before the Lagrangian contains two gauge potentials
$(A_\mu^\a)_{\a=1,2}$ and one auxiliary field $a$, appearing here
only as the gradient $u_\mu=\partial_\mu a$. The notation used
throughout this paper is
\begin{eqnarray}
u^2=u^\mu u_\mu \,,\quad\quad v_\mu=\frac{u_\mu}{\su}\,,\\
F^\a_{\m\n}=2\partial_{[\m}A^\a_{\n]}\,,\quad\quad *F^{\a}_{\m\n}=\half\e_{\m\n\rho\sigma}F^{\a \,\rho\sigma} \,,\\
\cF^\a_{\m\n}=\cL^{\a\b}F^\b_{\m\n}-*F^{\a}_{\m\n}\,,\quad\quad %=\half\e_{mnpq}\cL^{\a\b}\cF^{\b pq}\,,\\
\dH{\a}{\m}=\cF^\a_{\m\n}v^\nu  %=H_\mu^\a -H_\mu^{*\a}\,.
\label{e:not}
\end{eqnarray}
with $\cL^{\a\b}$ being the antisymmetric matrix of $so(2)$ with
$\cL^{12}=+1$. The equations of motion associated to \ref{e:pst}
read
\begin{eqnarray}
\d A_\mu^\a &:\quad\quad & \e^{\m\n\r\s}\partial_\nu(v_\r
H^{(-)\a}_\s)=0\,,
\label{e:eomAA}\\
\d u_\mu & :\quad\quad & \frac{1}{2\sqrt{-u^2}}\left(H^{(-)\a}_\nu
\cF^{\a \,\m\n}-H^{(-)\a}_\nu H^{(-)\a
\,\n}v^\mu\right)=0\,.\label{e:eomu}
\end{eqnarray}

It is straightforward to check the following gauge invariances of
the action (\ref{e:pst})
\begin{eqnarray}
\d_IA_\mu^\a=\partial_\mu\varphi^\a\,, &\,\d_I a=0 \,,\label{e:ginv1}\\
\d_{II}A_\mu^\a=-\cL^{\a\b}\dH{\b}{\m}\frac{\f}{\sqrt{-u^2}}\,, &\,\d_{II}a=\f \,,\label{e:ginv2}\\
\d_{III}A_\mu^\a=u_\mu\ve^\a\,, &\, \d_{III}a=0 \,.\label{e:ginv3}
\end{eqnarray}
They are irreducible. Pasti-Sorokin-Tonin have shown
\cite{Pasti:1995,Pasti:1995i} that this model is in fact
classically equivalent to the non-covariant action
\cite{Schwarz:1994} describing the dynamics of a single Maxwell
field. Indeed, using the equations of motion (\ref{e:eomAA}) we
can fix the gauge degrees of freedom of (\ref{e:ginv3}) in such a
way that the self-duality condition
\begin{equation}
\cF^{\a}_{\m\n}=0\label{e:duality}
\end{equation}
is satisfied. Such a consequence of the equations of motion allows
us to express one of the gauge fields $A_\mu^\a$ as function of
the other one yielding the usual Maxwell Lagrangian (with
remaining symmetry (\ref{e:ginv1})) plus a contribution of $u_\mu$
field. Further, we remark from the second invariance
(\ref{e:ginv2}) that $a$ is pure gauge. Another way to see that is
by expressing the field equation for $a$ as a consequence of the
equation of motion for $A_\mu^\a$. That is why $a$ can be easily
fixed away using a clever gauge condition (avoiding the
singularity $u^2=0$). So, the field $u_\mu$ as well as one of the
two $A_\mu^\a$ are auxiliary in the sense that one needs them only
to lift self-duality and Lorentz invariance at the rank of
manifest symmetries of the action. But, they can be removed on the
mass-shell taking into account the gauge invariances of the new
system. The manner of fixing the unphysical degrees of freedom in
the BRST formalism will be clarified in subsection
\ref{s:g-fixing}.

For the gauge algebra we get
\begin{eqnarray}
[\d_{II}(\f_1),\d_{II}(\f_2)]&=&\d_{III}\left(\frac{\cL^{\a\b}\dH{\b}{\m}}{(-u^2)^{3/2}}(\f_1\partial^\m\f_2-\f_2\partial^\m\f_1)\right)\,,\label{e:galg1}\\
\left[\d_{II}(\f),\d_{III}(\ve^\a)\right]&=&\d_I(\f\ve^\a)+\d_{III}\left(\frac{u^\m\f}{(-u^2)}
\partial_\m\ve^\a\right)\,.\label{e:galg2}
\end{eqnarray}
Thus, our system describes a non-Abelian gauge theory with
structure constants replaced by non-polynomial structure \emph{
functions}.

\subsection{Basic ingredients of antifield-BRST
formalism}\label{a:BRST}

Here we give only some of the main ideas underlying the Lagrangian
BRST method. For more details we refer the reader to
\cite{Batalin:1983,Henneauxbook,Henneaux:1989,Gomis:1995} and
references therein).

Let $S_0[\phi^i]$ be an action with the following bosonic gauge
transformations\footnote{We use the DeWitt notation.}
\begin{equation}\label{e:GT}
\delta_\varepsilon \phi^i = R^i_\alpha\epsilon^\alpha
\end{equation}
which are irreducible. Then, one has to enlarge the ``field''
content to
\begin{equation}
\{\Phi^A\}=\{\phi^i,C^{\alpha}\}.
\end{equation}
The fermionic ghosts $C^{\alpha}$ correspond to the parameters
$\varepsilon^\a$ of the gauge transformations (\ref{e:GT}). To
each field $\Phi^A$ we associate an antifield $\Phi_A^*$ of
opposite parity. The set of associated antifields is then
\begin{equation}
\{\Phi_A^*\}=\{\phi^{*}_i,C^*_\alpha\}.
\end{equation}
The fields possess a vanishing antighost number (\emph{ antigh})
and a nonvanishing pureghost number (\emph{ pgh})
\begin{equation}
pgh(\phi^i)=0,\quad pgh(C^\alpha)=1.\end{equation} The \emph{ pgh}
number of the antifields vanish but their respective \emph{antigh}
number is equal to
\begin{eqnarray}
antigh(\Phi_A^*)=1+pgh(\Phi^A).
\end{eqnarray}
The total ghost number (\emph{ gh}) equals the difference between
the \emph{ pgh} number and the \emph{ antigh} number. The
\emph{antibracket} of two functionals $X[\Phi^A,\Phi_A^*]$ and
$Y[\Phi^A,\Phi_A^*]$ is defined as
\begin{equation}
(X,Y)=\int d^nx\left( \frac{\delta^RX}{\delta\Phi^A(x)}
\frac{\delta^LY}{\delta\Phi_A^*(x)}
    -\frac{\delta^RX}{\delta\Phi_A^*(x)}\frac{\delta^LY}{\delta\Phi^A(x)}\right)\,,
\end{equation}
where $\delta^R/\delta Z(x)$ and $\delta^L/\delta Z(x)$ denote
functional right- and left-derivatives.

The \emph{extended action} $S$ is defined by adding to the
classical action $S_0$ terms containing the antifields coupled to
the BRST variations of the fields in such a way that the classical
\emph{master equation},
\begin{equation}(S,S)=0\,,\label{e:master}\end{equation}
is satisfied, with the following boundary condition:
\begin{equation}
S=S_0+\phi^*_iR^i_\alpha C^{\alpha}+...
\end{equation}
This imposes the value of terms quadratic in ghosts and
antifields. The extended action has also to be of vanishing \emph{
gh} number. If the algebra is non-Abelian, we know that we have to
add other pieces of \emph{ antigh} number two in the extended
action with the general form (due to structure functions)
\begin{equation}
S_2^{2a}=\frac{1}{2}C_\alpha^* f_{\beta \gamma}^\alpha C^\beta
C^\gamma\,.
\end{equation}
If the algebra is open, other terms in \emph{ antigh} number must
be added, quadratic in $\phi^*_i$'s. Furthermore, other terms in
higher \emph{ antigh} number could be necessary, e.g. when the
structure functions depend on the fields $\phi^i$.

The extended action captures all the information about the gauge
structure of the theory: the Noether identities, the (on-shell)
closure of the gauge transformations and the higher order gauge
identities are contained in the master equation.

The BRST transformation $s$ in the antifield formalism is a
canonical transformation, i.e. $sA=(A,S)$. It is a differential:
$s^2=0$, its nilpotency being equivalent to the master equation
(\ref{e:master}). The BRST differential decomposes according to
the \emph{ antigh} number as
$$s=\d + \gamma + "more"$$
and provides the gauge invariant functions on the stationary
surface, through its cohomology group at \emph{ gh} number zero
$H_0(s)$. The Koszul-Tate differential $\delta$ defined by
\begin{equation}\label{e:KT}
\d \F^*_i=(\F^*_i,S)\vert_{\F_A^*=0}
\end{equation}
implements the restriction on the stationary surface, and the
exterior derivative along the gauge orbits $\gamma$
$$\gamma\F^i=(\F^i,S)\vert_{\F^*_A=0}$$
picks out the gauge-invariant functions.

The solution $S$ of the master equation possesses gauge
invariance, and thus, cannot be used directly in a path integral.
There is one gauge symmetry for each field-antifield pair. The
standard procedure to get rid of these gauge degrees of freedom is
to use the \emph{gauge-fixed action} $S_\Psi$ defined by
\begin{equation}
S_\Psi=S_{non-min}\left[
\F^A,\F^*_A=\frac{\d\Psi[\F^A]}{\d\F^A}\right].
\label{e:GAUGEFIXED}
\end{equation}
The functional $\Psi[\F^A]$ is known as the \emph{gauge-fixing
fermion} and must be such that $S_\Psi[\Phi]$ is non-degenerate,
i.e. the equations of motion derived from the gauge-fixed action
$\d S_\Psi[\F^A]/\d\F^A=0$ have unique solution for arbitrary
initial conditions, which means that all gauge degrees of freedom
have been eliminated. It also has to be local in order that the
antifields be given by local functions of the fields.

The generating functional of the theory is then
\begin{equation}\label{e:PATHINT}
Z=\int[\cD\F^A] \exp iS_\Psi
\end{equation}
The value of the path integral is independent of the choice of the
gauge-fixing fermion $\Psi$. The notation $[\cD\Phi]$ stands for
$\cD\Phi\,\mu [\Phi]$, where $\mu[\F]$ is the measure of the path
integral. It is important to notice that the expression of the
measure $\mu[\Phi]$ in this path integral is not completely
determined by the Lagrangian approach. A correct way to determine
it, would be to start from the Hamiltonian approach for which the
choice of measure is trivial, indeed it is known to be
$\cD\F\cD\Pi$, that is the product over time of the Liouville
measure $d\Phi^Ad\Pi_A$. The Hamiltonian formalism meets no
problem because the Hamiltonian gauge-fixed action has no gauge
invariance. It can be proved that, if correctly handled, the two
approaches are equivalent (see \cite{Henneauxbook} for instance).
This justifies a posteriori the choice of the measure $\mu [\Phi]$
in (\ref{e:PATHINT}).

%%%%%%%%%%%%%%%%%%%%%%%%%%%%%%%%%%%%%%%%%%%%%%%%%%%%%%%%%%%%%%%%%%%%%

\subsection{Minimal solution of the master equation}\label{s:BRST}

Having made the classical analysis of the model, we can start now
the standard BRST procedure. The first step is to construct the
minimal solution of the master equation with the help of the gauge
algebra. In order to reach that end we will introduce some new
fields called ghosts and their antibracket conjugates known as
antifields.

The minimal sector of fields and antifields dictated by the gauge
invariances (\ref{e:ginv1})-(\ref{e:ginv3}) as well as their ghost
numbers and statistics are listed in Table~\ref{t:mingh}.
\begin{table}[ht]
\begin{center}
\begin{tabular}{|c|c|c|c|c|c|c|c|c|c|c|}\hline
$\Phi$&$A_\mu^\a$&$a$&$A_\mu^{\a
*}$&$a^*$&$c^\a$&$c$&$c'^\a$&$c^{\a *}$&$c^*$&$c'^{\a *}$\\ \hline
$gh(\Phi)$&$0$&$0$&$-1$&$-1$&$1$&$1$&$1$&$-2$&$-2$&$-2$\\ \hline
$antigh(\Phi)$&$0$&$0$&$1$&$1$&$0$&$0$&$0$&$2$&$2$&$2$\\ \hline
\mbox{stat($\Phi$)}&$+$&$+$&$-$&$-$&$-$&$-$&$-$&$+$&$+$&$+$\\
\hline
\end{tabular}
\caption{Ghost number, antighost number and statistics of the
minimal fields and their antifields. \label{t:mingh}}
\end{center}
\end{table}

The transformations (\ref{e:ginv1})-(\ref{e:ginv3}) determine
directly the $antigh$ number one piece of the extended action,
i.e.
\begin{eqnarray}
S_1&=&\int d^4\,x\left[A^{\a *\m}\left(\partial_\mu c^\a-
\cL^{\a\b}\dH{\b}{\m}\frac{c}{\sqrt{-u^2}}+u_\mu c'^\a\right)+ a^*
c\right]\,. \label{e:ext1}
\end{eqnarray}
In order to take into account the structure functions one has to
insert in the solution of the master equation a contribution with
$antigh$ number two of the form
\begin{eqnarray}\label{e:ext2}
S_2 = \int d^4\,x \left[c'^{\a *}\left(
\frac{\cL^{\a\b}\dH{\b}{\m}}{(-u^2)^{3/2}}\,c\,\partial^\m c
+\frac{v_\m}{\su}\,c\,\partial^\m c'^{\a}\right) +c^{\a *}c\,c'^\a
\right]\,.
\end{eqnarray}

Due to the field dependence of the structure functions one should
expect that $S_1$ and $S_2$ are not enough to completely determine
the extended action and one will need an extra piece of $antigh$
number three to do the job. Indeed, that was already the case for
chiral 2--forms in 6 dimensions discussed in
\cite{VanDenBroeck:1999}. Nevertheless, one can readily check that
in the present situation $S_{min}=S_0+S_1+S_2$ is the minimal
solution of the classical master equation $(S_{min},S_{min})=0$,
i.e.
\begin{eqnarray}
(S_1,S_1)_1 + 2(S_1,S_1)_1 =0\,,\nn \\
(S_2,S_2)_2 + 2(S_1,S_2)_2 =0\,.
\end{eqnarray}
This follows also as a consequence of the irreducibility of our
model.

Once $S_{min}$ has been derived, we can infer the BRST operator
$s$, which is the sum of three operator of different \emph{
antigh} number
\begin{equation}
s=\delta+\gamma+\rho\,.
\end{equation}
For instance, the non-trivial action of the Koszul-Tate
differential, of \emph{ antigh} number $-1$, is in our case given
by
\begin{eqnarray}
\d A^{\a *\m} &=& \e^{\m\n\r\s}\partial_\nu(v_\r H^{(-)\a}_\s)\,,\\
\d a^* &=& \partial_\mu\left(\frac{1}{2\sqrt{-u^2}}\left(H^{(-)\a}_\nu\cF^{\a\, \m\n}-H^{(-)\a}_\nu H^{(-)\a \,\n}v^\mu\right)\right)\,,\\
\d c^{\a *}&=&-\partial^\mu A^{\a *}_\mu\,,\label{e:KT1} \\
\d c^* &=&-\frac{\cL^{\a\b}\uH{\b}{\m}}{\su} A^{\a *}_\m +a^*\,,\label{e:KT2} \\
\d c'^{\a *}&=&u^\mu A^{\a *}_\mu\,.\label{e:KT3}
\end{eqnarray}
The third piece, $\rho$, of \emph{antigh} number $+1$ is present
also because the structure functions determined by
(\ref{e:ginv1})-(\ref{e:ginv3}) depend explicitly on the fields.

In this way the goal of this section, i.e. the construction of the
minimal solution for the master equation, has been achieved.

\subsection{The gauge-fixed action}\label{s:g-fixing}

The minimal solution $S_{min}$ will not suffice to fix all the
gauge invariances of the system and, before fixing the gauge, we
need a non-minimal solution for $(S,S)=0$ in order to take into
account the trivial gauge transformations. In this section we
first construct such a non-minimal solution and, afterwards, we
propose two possible gauge-fixing conditions which will yield two
versions for the gauged-fixed action: a covariant and a
non-covariant one.

\subsubsection{Non-minimal sector}

Inspired by the gauge transformations
(\ref{e:ginv1})-(\ref{e:ginv3}) and their irreducibility we
propose a non-minimal sector described in Table
\ref{t:non-minsect}.
\begin{table}[ht]
\begin{center}
\begin{tabular}{|c|c|c|c|c|c|c|c|c|c|c|c|c|}\hline
$\Phi$&$B^\a$&$B$&$B'^\a$&${\bar C}^\a$&${\bar C}$&${\bar C}'^\a$&
$B^{\a *}$&$B^*$&$B'^{\a *}$&${\bar C}^{\a *}$&${\bar C}^*$&${\bar
C}'^{\a *}$\\ \hline
$gh(\Phi)$&$0$&$0$&$0$&$-1$&$-1$&$-1$&$-1$&$-1$&$-1$&$0$&$0$&$0$\\
\hline
\mbox{stat($\Phi$)}&$+$&$+$&$+$&$-$&$-$&$-$&$-$&$-$&$-$&$+$&$+$&$+$\\
\hline
\end{tabular}
\caption{Ghost number and statistics of the non-minimal fields and
their antifields. \label{t:non-minsect}}
\end{center}
\end{table}

They satisfy the following equations
\begin{eqnarray}
s{\bar C}^{\cdots} = B^{\cdots}\,,\quad\quad sB^{\cdots}=0\,,\nn \\
sB^{\cdots *}={\bar C}^{\cdots *}\,,\quad\quad s{\bar C}^{\cdots
*} =0\,.
\end{eqnarray}
The dots are there to express that these relations are valid for
the correspondingly three kinds of non-minimal fields. We
immediately see that $\bar{C}^{\cdots}$'s and $B^{\cdots}$'s
constitute trivial pairs, as well as their respective antifields,
in such a way that they do not enter in the cohomology of $s$.
Hence, they are called non-minimal. Their contribution to a
solution of the master equation is
\begin{equation}\label{e:non-min}
S_{non-min}=S_{min} +\int d^4\,x \left({\bar C}^{\a *}B^\a +{\bar
C}^* B +{\bar C}'^{\a *}B'^\a \right)\,.
\end{equation}

\subsubsection{Covariant gauge fixing}

We will first try a covariant gauge fixing leading in principle to
a covariant gauge-fixed action: we will see what is the main
problem that occurs. We can consider the following \emph{
covariant} gauge choices
\begin{eqnarray}
\d_{I} &\ra & \partial^\mu A^\a _\mu =0\,,\label{e:gc1} \\
\d_{II} &\ra & u^2 +1=0\,,\label{e:gc2} \\
\d_{III} &\ra & u^\mu A^\a _\mu =0\,.\label{e:gc3}
\end{eqnarray}
The gauge choice (\ref{e:gc1}) is analogous to the Lorentz gauge.
In its turn (\ref{e:gc2}) allows to take a particular Lorentz
frame in which $u^\mu(x)$ is the unit time vector at the point
$x$. In such a case, at the point $x$, (\ref{e:gc3}) is the
temporal gauge condition for the two potentials.

A gauge-fixing fermion corresponding to the gauge choices
(\ref{e:gc1})-(\ref{e:gc3}) is
\begin{equation}\label{e:gaugeferm}
\P[\F^A]=-\int d^4\,x \left[{\bar C}^\a \partial^\mu A^\a _\mu
+{\bar C} (u^2 +1) +{\bar C}'^\a u^\mu A^\a _\mu \right]\,.
\end{equation}
We express now all the antifields with the help of $\P[\F]$, i.e.
\begin{equation}\label{e:antifields}
\F^*_A =\frac{\d\P[\F^A]}{\d\Phi^A}
\end{equation}
getting \ba
A_\mu^{\a *} &=& \partial_\mu {\bar C}^\a -u_\mu {\bar C}'^\a\,,\\
a^*&=& 2\partial_\mu (u^\mu {\bar C}) +\partial^\mu(A_\mu^\a {\bar C}'^\a) \,, \label{e:antifield1}\\
c^{\cdots *}&=& 0\,, \\ B^{\cdots *}&=& 0\,, \label{antifield2}\\
{\bar C}^{\a *} &=& -\partial^\mu A^\a _\mu \,, \\
{\bar C}^* &=& -(u^2 +1) \quad\quad {\bar C}'^{\a *} = -u^\mu A^\a
_\mu \,.\label{e:antifield3} \ea Using the last relations we can
find the gauge fixed action as in (\ref{e:GAUGEFIXED}) which in
our case reads
\begin{eqnarray}\label{e:fixedaction}
S_\P &=& S_0 + \int d^4\,x \left[ -{\bar C}^\a \Box c^\a -\frac{\cL^{\a\b}\uH{\b}{\m}}{\sqrt{-u^2}}\partial_\mu {\bar C}^\a \cdot c +u^\mu \partial_\mu {\bar C}^\a \cdot c'^\a\right. \nn\\
&&  -u_\mu {\bar C}'^\a \partial^\mu c^\a - u^2{\bar C}'^\a c'^\a -(2 u_\mu {\bar C} + A_\mu^\a {\bar C}'^\a)\partial^\mu c\nn\\
&&\left. -(\partial^\mu A^\a _\mu )B^\a -(u^2+1)B -(u^\mu A^\a
_\mu )B'^\a \right]\,.
\end{eqnarray}
Writing down the path integral (\ref{e:PATHINT}), we integrate
directly the fields $B^{\cdots}$ producing the gauge conditions
(\ref{e:gc1})-(\ref{e:gc3}). A further integration of ${\bar C}$,
${\bar C}'^\a$ and $c'^\a$ (in this order) leads to
\begin{equation}\label{e:fixed1}
Z[0] = \int\cD A^\a_\m\cD a\cD c^\a \cD c\cD {\bar C}^\a  \,\d
(\partial^\mu A_\mu^\a) \,\d (u^2+1) \,\d (u^\mu A_\mu^\a) \,\d
(u^\mu
\partial_\mu c) \exp iS'_\Psi
 \end{equation}
where
\begin{equation} \label{e:covgfa}
S'_\Psi =\int d^4x\left[ -{\bar C}^\a \Box c^\a +\left(
\frac{\cL^{\a\b}\uH{\b}{\m}}{\sqrt{-u^2}}\, c +\frac{u^\mu}{u^2}
(u^p\partial_p c^\a +A_\n^\a\partial^\n c )\right)\partial_\mu
{\bar C}^\a\right].
\end{equation}
Of course, the next step in getting a covariant generating
functional from which we would read out the \emph{ covariant}
propagator for the fields $A_\mu^\a$ would be the elimination of
$c$ and $a$ in (\ref{e:fixed1}). Due to the ``gauge condition" for
the ghost $c$ (i.e. $u^\mu \partial_\mu c =0$) and the way it
enters the gauge-fixed action $S'_\Psi$, this integration is
technically difficult. What one could try is to integrate both $c$
and $a$ at the same time. This is also not straightforwardly
possible as a consequence of the gauge condition (\ref{e:gc2}).
This requirement was necessary to \emph{ covariantly} fix the
symmetry (\ref{e:ginv2}). Nevertheless, one can attempt to find
the general solution to this equation (\ref{e:gc2}), which reduces
to the integration of $\partial^\mu a=\L^\mu{}_p(x)n^p $ (with
$\L^\mu{}_p(x)$ a point-dependent Lorentz boost and $n_p$ a
constant time-like vector, i.e. $n_p n^p =-1$). Such a solution is
still unconvenient due to $x$-dependence of the Lorentz
transformation matrix $\L^\mu{}_p(x)$.

A way to overcome this sort of complication is to choose a
particular form for this matrix, breaking Lorentz symmetry. It is
precisely this price that we have to pay in order to explicitly
derive the propagator of $A_\mu^\a$ fields. As it will be
explained in the next subsection, by taking a particular solution
for (\ref{e:gc2}), i.e. by giving up Lorentz invariance, we will
be able to express the gauged-fixed action in a more convenient
form for our purposes.

\subsubsection{Non-covariant gauge fixing}

As it was remarked in the previous subsection in order to
explicitly derive the Feynman rules for the PST model one has to
break up its Lorentz symmetry by taking a specific solution of the
equation (\ref{e:gc2}). In this subsection we present a
non-covariant gauge of the theory and the advantages for such a
choice will become clear in the next sections. A possible \emph{
non-covariant} gauge fixing is
\begin{eqnarray}
\d_{I} &\ra & \partial^\mu A^\a _\mu =0\,,\label{e:gnc1} \\
\d_{II} &\ra & a-n_\mu x^\mu=0\,,\quad\quad n_\mu n^\mu =-1 \,,\label{e:gnc2} \\
\d_{III} &\ra & n^\mu A^\a _\mu =0\,.\label{e:gnc3}
\end{eqnarray}
By (\ref{e:gnc2}), the gradient $\partial^\mu a$ becomes equal to
the vector $n^\mu$ introduced above. In a Lorentz frame where
$n^\mu=(1,0,0,0)$ the requirement (\ref{e:gnc3}) is the temporal
gauge condition and (\ref{e:gnc1}) the Coulomb gauge condition for
the two potentials $A^\a_\mu$.

Then, the gauge-fixing fermion will be
\begin{equation}\label{e:gaugeferm-non}
\P[\F^A]=-\int d^4\,x \left[{\bar C}^\a \partial^\mu A^\a _\mu
+{\bar C} (a-n_\mu x^\mu) +{\bar C}'^\a n^\mu A^\a _\mu \right]\,.
\end{equation}
%This choice yields
%\begin{displaymath}\begin{array}{cclccl}
%A_\mu^{\a *} &=& \partial_\mu {\bar C}^\a -u_\mu {\bar C}'^\a &\quad\quad {\bar C}^{\a *} &=& -\partial^\mu A^\a _\mu\,,\label{e:non-antifield1}\\
%a^*&=&  -{\bar C} +\partial^\mu(A_\mu^\a {\bar C}'^\a) &\quad\quad {\bar C}^* &=& -(a-n_\mu x^\mu) \,,\label{e:non-antifield2}\\
%&&&\quad\quad {\bar C}'^{\a *} &=& -u^\mu A^\a _\mu\,.\label{e:non-antifield3}
%\end{array}\end{displaymath}

Using the same non-minimal contribution $S_{non-min}$ as before,
the non-covariant gauge-fixed action is
\begin{eqnarray}\label{e:gauge-fixedaction}
S_\P = S_0 &+& \int d^4\,x \left[\left(\partial^\mu {\bar C}^\a -u^\mu {\bar C}'^\a \right)\left(\partial_\mu c^\a- \cL^{\a\b}\dH{\b}{\m}\frac{c}{\sqrt{-u^2}}+u_\mu c'^\a\right) \right.\nn\\
&& +\left(-{\bar C}+\partial^\mu(A_\mu^\a {\bar C}'^\a)\right)\cdot c-\partial^\mu A^\a _\mu B^\a \nn\\
&&\left.- (a-n_\mu x^\mu)B -u^\mu A^\a _\mu B'^\a \right]\,.
\end{eqnarray}

This action is by far more convenient in deriving the propagator
of the gauge fields than its covariant expression (\ref{e:covgfa})
because we can completely integrate the ghost sector. Also, the
bosonic part takes a more familiar form. The quantum equivalence
of the PST model with ordinary Maxwell's theory will be based also
on this non-covariant action.

\subsection{Path integral}\label{s:qmequiv}

The gauge-fixed action corresponding to the non-covariant gauge
choice can be used to recover the non-covariant theory, which is
itself equivalent to the Maxwell theory. The generating functional
is taken to be
\begin{equation}
Z= \int {\cal D} A^\a_\mu \,\cD a \,\cD c^{\cdots} \,\cD
B^{\cdots} \,\cD \bar{C}^{\cdots} \, {\rm det}(\Box)\,{\rm det}
^{-1}({\rm curl})\, \exp iS_\P
\end{equation}
where $S_\P$ is given by (\ref{e:gauge-fixedaction}).

After integrating out some fields, in the following order
($B^{\cdots},$ $\bar{C},$ $c,$ $\bar{C}'^\a,$ $c'^\a ,$ $a$), we
obtain the path integral
\begin{equation}
Z= \int {\cal D} A^\a_\mu \,\cD {\bar C}^\a\,\cD c^\a \,{\rm
det}(\Box)\, \,{\rm det} ^{-1}({\rm curl})\,\delta(\partial^\mu
A^\a_\mu)\,\delta(n^\mu A^\a_\mu)\, \exp iS'_\P
\end{equation}
where the gauge-fixed action reduces now to
\begin{equation}\label{e:non-fixedaction1}
S_\P ' =\int d^4\,x \left[ -\frac12 n^\mu *F^\a_{\m\n}\cF^{\a\,
\n\r}n_\r -{\bar C}^\a \Box c^\a -{\bar C}^\a n^\m
n^\n\partial_\m\partial_\n c^\a \right]
\end{equation}
If we place ourselves in a Lorentz frame where $n^\mu=(1,0,0,0)$,
the functional $S'_\P$ assumes the form of the sum of
non-covariant gauge-fixed action (\ref{e:SchwarzSen}) with a ghost
term
\begin{equation}\label{e:ghosts}
-\int d^4 x\, {\bar C}^\a \bigtriangleup c^\a \,.
\end{equation}
At this point we can integrate out the ghosts $\bar{C}^\a$ and
$c^\a$, and the two fields $A^\a_0$, obtaining exactly the
generating functional (\ref{e:genfunc}) of the Maxwell theory in
the non-covariant formulation (see section \ref{a:Max}).

This proves a formal (quantum) equivalence of the PST action
(\ref{e:pst}) with the Maxwell theory \cite{Bekaert:2001} which
was already known at the classical level. The equality of the two
generating functionals was not obvious \emph{a priori} because the
PST action of the free Maxwell's theory is not quadratic (and so
the path integral is not Gaussian) and the pure gauge field $a$ is
not, strictly speaking, an auxiliary field (its equation of motion
is not an \emph{algebraic} relation which allows its elimination
from the action). The absence of anomalies and non-trivial
counterterms has been proved in \cite{DelCima:2000} which strongly
supports the true equivalence of the PST model with the Maxwell
theory.

%%%%%%%%%%%%%%%%%%%%%%%%%%%%%%%%%%%%%%%%%%%%%%%%%%%%%%%%%%%%%%%%%
%%%%%%%%%%%%%%%%%%%%%%%%%%%%%%%%%%%%%%%%%%%%%%%%%%%%%%%%%%%%%%%%%

\chapter{Quantization conditions}\label{chargeq}

\par

Yang-Mills gauge theory in $2+1$ dimensions with a CS term (also
called topological mass term) is known to have its topological
mass quantized for gauge groups with a non-trivial third homotopy
group \cite{Deser:1982}. In general, the CS coefficient is
quantized since $\pi_3(G)\cong \mathbb Z$ for any semi-simple Lie
group $G$. However, in the Abelian case with compact group
manifold $G=U(1)$ all the homotopy groups higher than $\pi_1$ are
trivial and, a priori, the CS coefficient can take arbitrary
values. The authors of \cite{Henneaux:1986i} have pointed out that
when electric and magnetic charges are present the CS coefficient
must also be quantized, just as in the non-Abelian case. The
quantization arises from two key features. The first one is that,
for non-vanishing CS coefficient, electric worldlines can end on
magnetic sources, thereby relating the electric charge to the
magnetic charge via the CS coefficent. The second key property is
the usual Dirac charge quantization condition that was shown to
remain valid in the presence of a CS term by the authors. As they
noticed, their construction can be straightforwardly generalized
to $p$-brane electrodynamics in $(2p+3)$-dimensional spacetime
where the electric $p$-branes have the same dimensionality than
the Dirac branes attached to the magnetic $(p-1)$-brane, and for
which equations of motion are linear.

The next step would be to consider theories with a cubic CS term
in the action. This is the subject of section \ref{CCS}. For
$p$-brane electrodynamics, a cubic term exists only in
$(3p+5)$-dimensional spacetime. The extended dual objects carrying
magnetic charge have $2p+1$ spacelike dimensions. With a
non-vanishing CS coefficient, electric branes can end on magnetic
branes. A new feature is that the boundary of the electric brane
creates a dyonic $(p-1)$-brane living on the worldvolume of the
magnetic brane. In the cubic case, it is the electric brane and
the Dirac brane attached to the dyonic brane, who shares the same
dimensionality but a modified version of the previous argument
\cite{Henneaux:1986i} can provide a new quantization of the
corresponding CS coefficient.

In supergravity theories, the CS coefficent is usually fixed by
supersymmetry. Therefore the fact that the CS coefficient is
quantized (without any help of supersymmetry) can be understood as
another supporting evidence of the M-theory conjecture, since the
bosonic sector of the effective quantum theory somehow ``feels"
supersymmetry.

%%%%%%%%%%%%%%%%%%%%%%%%%%%%%%%%%%%%%%%%%%%%%%%%%%%%%%%%%%%%%%%%%%

\section{Charge quantization condition}

There exists such a huge number of different derivations of the
charge quantization condition that we just give few examples. The
most standard approach might be the Dirac's 1948 derivation
\cite{Dirac:1948} based on the unobservability of the Dirac
string. But there are other formulations in which one does not
even need to introduce a Dirac brane: for instance, if one
proceeds \`a la Wu-Yang \cite{Wu:1976}, or if one uses the
electromagnetic momentum quantization \cite{Jackson}. In 1968,
Zwanziger derived the charge quantization condition for dyons by
using the Dirac-Schwinger criterion
\cite{Dirac:1962,Schwinger:1963} to enforce Lorentz invariance
\cite{Zwanziger:1968}. Consistency under charge shifts produced by
(generalized) $\theta$-terms or compactification effects were also
used in \cite{Deser:1998ii} to derive the quantization condition
for dyonic branes. Recently, it was reobtained geometrically from
the classification of $p$-gerbes \cite{Caicedo:2002}. Here we will
follow the Lagrangian path-integral approach chosen by the authors
of \cite{Lechner:1999,Lechner:2000}. With the aim of simplicity we
will work without manifest duality symmetry. For a
duality-symmetric treatment we refer to
\cite{Lechner:1999,Lechner:2000}.

\subsection{Dual branes}

For simplicity, we will consider the electrodynamics for a system
with a single brane only electrically charged and a single
(distinct) brane only magnetically charged. We take an action
inspired from section \ref{non-linear} which reads
\begin{eqnarray}
&&S[A_{\mu_1\ldots\mu_{p+1}},G_{\mu_1\ldots\mu_{D-p-2}},J_e^{\mu_1\ldots\mu_{p+1}}]=\nn\\
&&\quad=\int\limits_{\cal M}{\cal L}
(F_{\mu_1\ldots\mu_{p+2}})+(-)^{D-p-1}\int\limits_{\cal M}
*J_e\wedge A-T_e\int\limits_{{\cal M}_e}
(*1)-T_m\int\limits_{{\cal M}_m} (*1)\,.\nn
\end{eqnarray}
with the same definitions as before. Furthermore, $T_e$ and $T_m$
are the brane tensions. The Dirac brane is assumed to obey the
Dirac veto. As seen in subsection \ref{Diracproc}, one has the
following gauge freedom \be*G\rightarrow*G+d*V\,,\quad
A\rightarrow A+d\Lambda-*V\,,\label{largeg}\ee corresponding to a
change $\Delta{\cal M}_D=\partial{\cal V}$ in the Dirac brane,
$*V=gP({\cal V})$. The variation of the action under such a large
gauge transformation is equal to \be \Delta S=(-)^{D-p}
\,eg\,L({\cal M}_e,\Delta {\cal M}_D)\,,\ee where $L({\cal
M}_e,\Delta {\cal M}_D)$ is the linking number between the
electric brane worldvolume ${\cal M}_e$ and the Dirac brane
variation $\Delta {\cal M}_D$ in the spacetime $\cal M$. As
pointed earlier, since this is a topological quantity, the physics
is independent of the choice of Dirac brane at the classical
level. But at the quantum level, the path integral would not be
invariant in general and there would be some kind of anomaly. If
there exist a (purely) electric brane and a (purely) magnetic
brane, quantum consistency enforces the product of the electric
charge density $e$ (of unit $L^{D/2-p-2}$) with the magnetic
charge density $g$ (of unit $L^{p-D/2+2}$) to be equal to a
multiple of $2\pi$, i.e. \be eg=2\pi n,\quad \quad n\in\mathbb
Z\,.\label{Diracqc}\ee This requirement is the so-called
\emph{Dirac quantization condition}.\footnote{The first appearance
of this path integral derivation of the Dirac quantization
condition for closed branes was in \cite{Lechner:1999}. To our
knowledge, the path integral derivation in the case of open branes
given below in sections \ref{YMmassive} and \ref{CCS} is new.}

{\bf Remark:} There is no quantization condition if the electric
and magnetic branes are in a configuration such that the variation
of the Dirac brane worldvolume $\Delta {\cal M}_D$ cannot link
around the electric charge worldvolume ${\cal M}_e$. An example of
such a configuration is when infinite magnetic and electric branes
have some common asymptotic spatial direction. Such configurations
were called \emph{Dirac insensitive} by the authors of
\cite{Bremer:1998}. As they have shown, this property is essential
for consistency with Kaluza-Klein reductions.

\subsubsection{Application in M-theory}

As an example, it is known that the $F$-string and the
NS$5$-brane, the dual D$p$-branes, the M$2$-brane and M$5$-brane
obeys the Dirac quantization condition (remarkably, for the
minimal value $|n|=1$). One of the non-trivial consistency check
of the $M$-theory conjecture has been to check if they fulfill the
Dirac charge quantization condition\footnote{For instance, Schwarz
investigated what could be learned from identifying either
M-theory on ${\mathbb R}^9\times T^2$ with IIB string theory on
${\mathbb R}^{10}\times S^1$ or M-theory on ${\mathbb
R}^{10}\times S^1$ with IIA string theory on ${\mathbb R}^{10}$
\cite{Schwarz:1996}.}. Consistency with the web of dualities
determined the BPS brane tensions in terms of the fundamental
length scale (this result is reviewed in \cite{deAlwis:1997}).

Further, by considering the implications of consistency with
$T$-duality together with some special ``scale setting" (e.g. for
the dyonic D$3$-brane in type IIB superstring theory), charge
scales in $D\leq 10$ supergravity theories were determined without
recourse to M-theory \cite{Bremer:1998}. In fact the derived
relations can be interpreted as supporting evidence for the
M-theory conjecture.

\subsection{Dyonic branes}

We now restrict ourselves to the $p$-brane electrodynamics in
$D=2p+4$ (the case $p$ even was considered in section \ref{sl2r})
where dyonic branes may exist. For simplicity, let us assume that
there are only two dyons of charge density $(e_1,g_1)$ and
$(e_2,g_2)$, of worldvolume ${\cal M}^2=\partial{\cal N}^1$ and
${\cal M}^1=\partial{\cal N}^2$ with Dirac brane ${\cal N}^1$ and
${\cal N}^2$. One takes the following action\footnote{The
conventions are different from the one chosen in section
\ref{sl2r}.}
\begin{eqnarray}
&&
S[A_{\mu_1\ldots\mu_{p+1}},G^{\mu_1\ldots\mu_{p+2}},J_e^{\mu_1\ldots\mu_{p+1}}]=\nn\\&&=\int\limits_{\cal
M } L(F_{\mu_1\ldots\mu_{2(p+1)}})+\frac12\,\theta
\int\limits_{\cal M } F\wedge F+(-)^{p+1}\int\limits_{\cal M }
*J_e\wedge A+I_K
\end{eqnarray}
For $p$ odd, the $\theta$-term is not really present since
$F\wedge F\equiv 0$. The variation of the action under a large
gauge transformation ({largeg}) corresponding to a change of Dirac
brane\be {\cal N}^1\rightarrow{\cal N}^{'1}={\cal
N}^1+\partial{\cal V}_1\,,\quad{\cal N}^2\rightarrow{\cal
N}'_2={\cal N}_2+\partial{\cal V}_2\,,\ee is equal to \be\Delta
S=(-)^p\int\limits_{\cal M } *J_e\wedge *V\ee Now, we use the
identity $*J_e\wedge *V=*J^1_e\wedge *V^2+*J^2_e\wedge *V^1$ to
get \be S'-S=(-)^p\left[e_1g_2\,L({\cal M}^1,{\cal N}^{'2}-{\cal
N}_2)+e_2g_1\,L({\cal M}^2,{\cal N}^{'1}-{\cal N}^1)\right]\ee We
have the topological relations \ba L({\cal M}^1,{\cal
N}^{'2}-{\cal N}_2)=I({\cal N}^1,{\cal N}^{'2})-I({\cal N}^1,{\cal
N}_2)\,,\\L({\cal M}^2,{\cal N}^{'1}-{\cal N}^1)=I({\cal
N}'_2,{\cal N}^{'1})-I({\cal N}^{'2},{\cal N}^1)\,,\ea where
$I({\cal B},{\cal C})$ is the intersection number between the
submanifolds ${\cal B}$ and ${\cal C}$ such that $dim({\cal
B})+dim({\cal C})=D$ (see the section \ref{linking} for precise
definitions).

Let us now assume that the initial and final Dirac branes do not
touch each other, it follows that\be I({\cal N}^1,{\cal
N}_2)=I({\cal N}^{'2},{\cal N}^{'1})=0\,.\ee Eventually, we use
\be I({\cal N}^1,{\cal N}^{'2})=(-)^{p^2}I({\cal N}^{'2},{\cal
N}^1)\ee to find that\be S'-
S=(-)^p\,\left(e_1g_2+(-)^{p+1}e_2g_1\right)\,I({\cal N}^1,{\cal
N}'_2)\ee Applying the same argument than in previous section, we
find that if there exist two dyonic branes of (dimensionless)
charge density $(e_1,g_1)$ and $(e_2,g_2)$, quantum consistency
imposes \be e_1g_2+(-)^{p+1}e_2g_1=2\pi n,\quad \quad n\in\mathbb
Z\,.\label{DZcond}\ee This requirement is the so-called
\emph{Dirac-Schwinger-Zwanziger quantization
condition}\footnote{Schwinger and Zwanziger derived this condition
in four dimensions in 1968 \cite{Schwinger:1968,Zwanziger:1968}.
Some years ago, it was extended to $p$-brane electrodynamics in
\cite{Deser:1997,Deser:1998iii}.}, which is hopefully left
invariant by the corresponding duality symmetries. If one of the
two dyonic brane has a (electric or magnetic) vanishing charge
density, we recover the previous quantization condition
(\ref{Diracqc}). Conversely, the condition (\ref{DZcond}) is
weaker than (\ref{Diracqc}) since the Dirac condition imply the
Dirac-Schwinger-Zwanziger (DSZ) condition. The Dirac condition
(\ref{Diracqc}) becomes necessary if one admit the Dirac branes to
touch.

\subsection{Charge lattice}

Let be a set of dyonic branes numbered by $i$ with electric and
magnetic charges pictured on a plane $(e_i,g_i)$. We are looking
for the set of points on the plane $(e,g)$ determined by the DSZ
condition. Can be applied to any couple of charged $p$-brane to
find the general relation\be e_ig_j+(-)^{p+1}e_jg_i=2\pi
n_{ij},\quad \quad n_{ij}\in\mathbb Z\,,\quad
n_{ij}=(-)^{p+1}n_{ji}\,.\label{DiZwcond}\ee

\subsubsection{Even dyonic branes}

The most interesting cases arise in dimensions $0$ modulo $4$, the
DSZ quantization condition is \be e_ig_j-e_jg_i=2\pi n_{ij},\quad
\quad n_{ij}\in\mathbb Z\,,\quad
n_{ij}=-n_{ji}\,.\label{sl2Dcond}\ee The minus sign in
(\ref{sl2Dcond}) plays an important role for several reasons. For
instance the dyonic branes are ``Dirac insensitive" to themselves.
Also the DSZ condition (\ref{sl2Dcond}) allows the existence of a
Witten effect since such the respective shifts in the electric
charges cancel out. The minus sign is also responsible for the
invariance of the quantization condition under the action of the
duality group $SL(2,{\mathbb R})$.

We now look for the general solution to the DSZ condition. We
represent the complex plane to represent the charges: $q=e+ig$.
Let be the dyonic brane with minimal strictly positive magnetic
charge in the universe. We normalize its magnetic charge to $2\pi$
in order to make link with conventions of section \ref{sl2r}. Let
be a purely electrically charged particle of charge $e$. Then the
condition (\ref{sl2Dcond}) gives $e=m\in\mathbb Z$. Therefore $1$
is the minimal non-vanishing electric charge magnitude of a purely
electrically charged particle and all of them have an electric
charge which is an integer. Now let us consider an arbitrary
dyonic brane represented by the complex number $q$ in the charge
plan. If there exists a purely electrically charged particle with
unit minimal charge. The DSZ condition imply that its magnetic
charge $g$ is a multiple of $2\pi$, i.e. $q=e+i\,2\pi n$ where
$n\in\mathbb Z$. Let now be the dyonic brane with minimal magnetic
charge $2\pi$ \emph{and} minimal non-vanishing positive electric
charge $\theta/2\pi$. The DSZ quantization condition imposes that
$e=\theta n+2\pi m$ with $m\in\mathbb Z$. To sum up, one finds the
following charge lattice in the complex plane\be q=(m+\theta
n)+i\,2\pi n\,,\quad (m,n)\in {\mathbb Z}^2\,.\label{latt}\ee It
can be easily checked that this lattice obeys the DSZ quantization
condition. The fundamental cell of this lattice is a two torus
defined by the following identifications in $\mathbb C$ \be q\sim
q+1\,\quad q\sim q+2\pi\tau\,.\ee

The duality group $SL(2,\mathbb R)$ is broken at the quantum level
to the discrete subgroup $SL(2,\mathbb Z)$ if one wants to
preserve the charge lattice. If $\theta$ is constrained to vanish,
the charge lattice is rectangular and the $SO(2)$ duality rotation
group is broken to the discrete subgroup ${\mathbb Z}_4$.

\subsubsection{Odd dyonic branes}

In dimensions $2$ modulo $4$, there is a plus sign in the DSZ
quantization condition (\ref{DiZwcond}) and the integers $n_{ij}$
are symmetric under the exchange of two dyonic branes, \be
e_ig_j+e_jg_i=2\pi n_{ij},\quad \quad n_{ij}\in{\mathbb
Z}\,.\label{Mexic}\ee Thus one can take $i=j$ to find \be
e_ig_i=\pi n_{i},\quad \quad n_i\in\mathbb Z\,.\label{Mexico}\ee
In this case, dyonic branes are ``sensitive" to themselves.

Reproducing the previous discussion, one obtains the following
charge lattice\be (e,\frac{g}{2\pi})\in {\mathbb Z}^2\,,\ee
because the DSZ quantization condition (\ref{Mexic}) preserves the
lattice (\ref{latt}) if and only if $\theta=0$. Indeed, the
condition (\ref{Mexic}) for $i\neq j$ imposes that $\theta$
vanishes or is equal to one half. The condition (\ref{Mexico})
finally constrain $\theta$ to vanish. As it should, the charge
lattice is left invariant by the action of the duality group
${\mathbb Z}_2\times{\mathbb Z}_2$. The fundamental cell of this
lattice is a rectangular two-torus defined by the following
identifications in $\mathbb C$ \be q\sim q+1\,\quad q\sim
q+i\,.\ee

\subsubsection{Duality symmetry}

We recall that the DSZ condition (\ref{DiZwcond}) is
duality-invariant. This can be seen easily in a duality-symmetric
fashion by noticing that, for a dyonic brane of charge $(e_i,g_i)$
and worldsheet ${\cal M}_i$ the bicurrent definition (\ref{calF})
gives \be *{\cal J}\equiv\left(
\begin{array}{c} *J_m\\ *J_e
\end{array}
\right)=\sum_i {\cal Q}_i P({\cal M}_i)\,,\ee where \be {\cal
Q}_i\equiv\left(
\begin{array}{c} g_i\\ e_i
\end{array}
\right)\,.\ee The DSZ quantization (\ref{DiZwcond}) can be
rewritten as \be {\cal Q}^T_i\,J\,{\cal Q}_j=2\pi n_{ij}\,.\ee The
left-hand-side is obviously invariant under a duality rotation
transformation \be {\cal Q}_i\rightarrow R{\cal Q}_i\,,\quad
J\rightarrow R^{-1} J R=J\,, R\in SO(2) \,.\ee (This works also
for $SL(2,\mathbb R)$.) In fact, the DSZ condition could have been
deduced in this form directly from the variation (\ref{varG}) of
the duality-symmetric action.

While the DSZ condition (\ref{DiZwcond}) is invariant under the
duality group, the lattice is only invariant under one of its
discrete subgroup. More precisely, in dimensions $0$ modulo $4$
the $SL(2,\mathbb R)$ duality group is broken to $SL(2,\mathbb
Z)$. If the vacuum angle $\theta$ is fixed to zero, $SO(2)$ is
broken at the quantum level to ${\mathbb Z}_4$.

It is remarkable that this symmetrical viewpoint \emph{does not}
contradict the empirical assymmetry between electric and symmetric
charges, as pointed by Schwinger in 1975. More accurately, for
Schwinger ``{\it what is special about the world thus far
disclosed by experiment is simply that no large charges have yet
been produced}" \cite{Schwinger:1975}. To understand properly
Schwinger's statement, one should understand ``small charge" as
duality-invariant statement. The most natural choice available is
the norm $||(e_i,g_i)||$. Schwinger considered the work
hypothesis\be ||(e_i,g_i)||^2=e_i^2+g_i^2\ll 1\,,\quad \forall
i\,.\label{Schwsm}\ee The Cauchy-Schwarz inequality implies
that\be |e_ig_j- e_jg_i|^2\leq (e_i^2+g_i^2)(e_j^2+g_j^2)\,.\ee
The DSZ condition (\ref{DiZwcond}) combined with (\ref{Schwsm})
leads to $2\pi |n_{ij}|\ll 1$, but $n_{ij}$ is an integer thus it
must strictly vanish. Therefore, any two particles that obeys
(\ref{Schwsm}) have the same ratio of electric to magnetic charge,
i.e. $e_i/g_i\,=\,e_j/g_j$ and, as explained in subsection
\ref{Magnmonop}, the magnetic charges $g_i$ can be made to vanish
by a duality rotation.

\subsection{Chiral forms}\label{chiralfdq}

Chiral gauge fields live in dimensions $2$ modulo $4$, for which
there is a plus sign in the DSZ quantization condition
(\ref{DiZwcond}). Using the action (\ref{chiract}), it can be
proved that one gets \be \n_i\,\n_j=2\pi n_{ij},\quad \quad
n_{ij}\in{\mathbb Z}\,,\label{chirqc}\ee with $n_{ij}=n_{ji}$.
Surprisingly this is twice the quantization (\ref{Mexic}) of the
product $\n_i\,\n_j$ with $\n_i:=e_i=g_i$ (the fieldstrength is
self-dual). One can take $i=j$ to find \be \n_i^2=2\pi n_{i},\quad
\quad n_i\in\mathbb N\,.\label{selfqc}\ee As one can see, this
self-sensitiveness is different from what one would have obtained
naively from the Dirac condition (\ref{Diracqc}). This remark will
be used below.

We claim that if there exists a self-dual brane with minimal
charge $\n=\sqrt{2\pi}$, then (\ref{chirqc}) enforces the
self-dual charges $\n_i$ to satisfy\be \n_i=m_i\,\n\,\quad
m_i\in\mathbb Z\,.\ee Indeed, (\ref{selfqc}) says that
$\n^2_i=n_i\,\n^2$. Furthermore, (\ref{chirqc}) implies
$\n_i\n=2\pi m_i$ ($m_i$ is an integer), the square of which gives
$n_i=m^2_i$.

%%%%%%%%%%%%%%%%%%%%%%%%%%%%%%%%%%%%%%%%%%%%%%%%%%%%%%%%%%%%%%%%%%

\section{Topological mass quantization condition}\label{YMmassive}

\subsection{Topologically massive Yang-Mills theory}

This theory is described by the usual YM gauge action with an
added topological term which preserves local gauge invariance.

\subsubsection{$2+1$ dimensions}

Yang-Mills gauge theory in $2+1$ dimensions with a CS term have
their topological mass quantized for gauge groups with a
non-trivial third homotopy group \cite{Deser:1982}. The argument
goes as follows. The CS term added to the Yang-Mills action
preserves local gauge invariances for spacetime manifold without
boundary, but is not invariant under large gauge transformations.
If we assume a compactification of space time to a three-sphere
$S^3$, large gauge transformations provide a map between $S^3$ to
the group manifold $G$. They fall into topological classes of the
third homotopy group $\pi_3(G)$. Under topologically non-trivial
large gauge transformations, the total action varies by a term
proportional to the instanton number representing $\pi_3(G)$. If
we ask for the invariance of the path integral under topologically
non-trivial gauge transformations when $\pi_3(G)$ is non-trivial,
we get a quantization condition of the topological mass, called
here CS coefficient. Stronger quantization conditions holds in
specific cases \cite{Pisarski:1986}.

\subsection{Topologically massive electrodynamics}

The topologically massive $2k$-brane electrodynamics action in
spacetimes of dimension $D=4k+3$ is (see section \ref{qCS}) \ba
&&S[A_{\mu_1\ldots\mu_{2k+1}},G^{\mu_1\ldots\mu_{2k+1}},J_e^{\mu_1\ldots\mu_{2k+1}}]=\nn\\&&=-\frac12\int
*F\wedge F- \frac{\a}{2}\int A\wedge dA+\int (*J_e-\a *G)\wedge A
+I_K\,.\nn\ea The gauge field e.o.m. is \be d*F+\a F=*J_e \,,\ee
which implies \be d*J_e=\a *J_m\,,\quad d*J_m=0\,,\ee in a way
such that the relation \be e+\a g=0\,,\label{bzzzzzzz}\ee is
imposed. A large gauge transformation (\ref{gaugetr}) leads to
\begin{equation}
\Delta S=e\int\limits_{{\cal M}_e -{\cal M}_D}(d\Lambda-
*V)\,=\,-egL({\cal M}_e -{\cal M}_D,\Delta{\cal M}_D) \,.
\end{equation}
Therefore the Dirac quantization condition (\ref{Diracqc}) also
applies here. As pointed in \cite{Henneaux:1986i} we can now
combine (\ref{bzzzzzzz}) with (\ref{Diracqc}) to get a
quantization of the CS coefficient \be
\a=\frac{2\pi}{g^2}\,m\,,\quad m\in{\mathbb Z}\,,\ee which is
reminiscent of the non-Abelian case.

The quantization of the Abelian CS coefficient was rederived in
\cite{Polychronakos:1987} with spacetime compactified to
$S^1\times M^2$ in the presence of a non-vanishing total magnetic
flux on $M^2$. The Abelian group manifold is $U(1)$ so we expect
some quantization from $\pi_1\left(U(1)\right)\cong {\mathbb Z}$.
There exist another physical situation for which we have the same
quantization: at finite temperature\footnote{An heuristic argument
is that when the theory is formulated at finite temperature, the
time direction is effectively compactified into a circle.}
\cite{Bralic:1996} and on the non-commutative plane
\cite{Nair:2001}.

\section{Chern-Simons coefficient quantization condition}\label{CCS}

\subsection{Non-Abelian gauge group}

A non-Abelian cubic CS term can be found in gauged
five-dimensional ${\cal N}=8$ supergravity
\cite{Gunaydin:1986,Andrianopoli:2000}. The CS term can exist only
with the gauge vectors in the adjoint of the gauge group, which is
$SO(6)$ in the compact version.

It is possible to get the quantization of the CS coefficent for
specific non-Abelian theories by using an argument similar to the
original one of Deser, Jackiw and Templeton \cite{Deser:1982}. The
variation of the action in $D$ gives the Cartan-Maurer integral
invariant, the integral of the trace of the Cartan $D$-form
corresponding to the gauge transformation, which might provide a
fidel representation of the $D$-th homotopy group of $G$ (see for
instance section 23.4 of \cite{Weinberg}). In the compact version
of gauged five-dimensional maximal supergravity, we expect the CS
coefficient to be quantized since $\pi_5\left(SO(6)\right)$ is
isomorphic to $\mathbb Z$.

%\AC{look at WZ in Petervan's book}

\subsection{M-theoretic electrodynamics}

Now we comeback to the section \ref{cCS} where we consider a
$p$-brane electrodynamics with a non-linear Chern-Simons coupling
in $(3p+5)$-dimensional spacetimes with $p$ even, for which
electric $p$-branes can end on magnetic $(2p+1)$-branes, creating
thereby a self-dual $(p-1)$-brane. For definiteness, we consider
again the five-dimensional toy model. With the experience gained
from the previous examples, the derivation of the CS quantization
condition is straightforward.

\subsubsection{Quantization condition in the bulk}

For a finite displacement $\delta {\cal M}_3=\partial{{\cal V}_4}$
of the Dirac membrane, we have $*G\rightarrow*G+d*V$ where
$*V=-gP({\cal V}_4)$ with ${\cal V}_4$ the manifold described by
the displacement of the Dirac brane. The action changes by
\begin{equation}
\delta_V I = \int\limits_{{\cal M}_5} *V\wedge *J_e
\end{equation}
It is proportional to the linking number in the space-time between
the variation $\delta {\cal N}_3$ of the Dirac membrane
worldvolume and the electric charge worldline ${\cal M}_1$.
\begin{equation}
\delta_V I = eg\,L(\delta {\cal N}_3,{\cal M}_1)
\end{equation}
The important factor $eg$ is present because the Poincar\'e duals
are normalized to unity. Obviously, for an infinitesimal variation
the linking number vanishes, hence this anomaly appears only for
large gauge transformations. If we require that the change of
phase in the path integral vanishes for $L(\delta {\cal N}_3,{\cal
M}_1)=1$, we obtain the usual Dirac quantization condition
\begin{eqnarray}
eg &=& 2\pi n\,,\quad n\in\mathbb Z\,.\label{aa}
\end{eqnarray}

\subsubsection{Quantization condition on the worldsheet}

In the case of the instanton, what one could call the Dirac
anomaly is equal to
\begin{eqnarray}
\delta_{v} I &=& \frac{\alpha g}{2}\int\limits_{{\cal M}_2}
\bar{*}v\,\wedge \bar{*}j\,.
\end{eqnarray}
The integral is equal to the square of the instanton charge $\nu$
times the ``linking number" on the magnetic string worldsheet
between the variation of the Dirac point worldline and the
instanton. Inserting this in the path integral, consistency at the
quantum level imposes at first sight $\frac{\alpha g
\nu^2}{2}=2\pi m$ with $m$ integer. The quantization of an
instanton in two dimensions was previously considered in
\cite{Henneaux:1985}. An point to take into account is that the
instanton is in fact self-dual. Since we impose this by hand in
the equations of motion, this cannot appear by a Lagrangian
analysis. Fortunately, the analysis of subsection \ref{chiralfdq}
shows that the quantization condition for chiral gauge fields is
equal to the one obtained by a blind application of Dirac formula.
Therefore, one has
\begin{equation} \alpha g \nu^2=4\pi m\,,\quad m\in\mathbb
Z\,.\label{aaa}
\end{equation}

\subsubsection{Quantization of the CS coefficient}

From (\ref{charges}), (\ref{aa}) and (\ref{aaa}) we obtain
\cite{Bekaert:2002i}
\begin{equation}
\mbox{\begin{tabular}{|c|}
  \hline\\
  $\alpha = e^3/8\pi^2p\,,$ \\\\
  \hline
\end{tabular}}  \quad\quad p\in{\mathbb
Z}\,. \label{quant}
\end{equation}
Up to a different choice of normalization, this is the
quantization condition obtained by Bachas \cite{Bachas:1998},
except that the integer is in the denominator in our relation
(\ref{quant}) while it is in the numerator for
Bachas\footnote{Compare the relation (\ref{quant}) with equation
(4.22) of \cite{Bachas:1998}. The different normalizations can be
translated into $\alpha=sqrt{2}\kappa_{(5)}\,k$ and
$e=\sqrt{2}\kappa_{(5)}\,q$.}. This is not paradoxal because
Bachas' derivation was based on a different argumentation, using
compactification to four dimensions together with the Witten
effect. A possible viewpoint on this issue is that the two
quantization conditions together determines uniquely the value
$\alpha = e^3/8\pi^2$ of the CS coefficient, which is precisely
the only value consistent with supersymmetry \cite{Cremmer:1978i}.

%%%%%%%%%%%%%%%%%%%%%%%%%%%%%%%%%%%%%%%%%%%%%%%%%%%
%%%%%%%%%%%%%%%%%%%%%%%%%%%%%%%%%%%%%%%%%%%%%%%%%%%

\chapter{Consistent deformations}\label{deformation}

\par

Glancing at the huge list of possible extensions of EM-duality
symmetry (some of which are presented in chapter \ref{journey}),
one might be tempted to try to generalize EM-duality to
non-Abelian gauge theories. The M-theory conjecture even brings in
some argument that such a generalization could exist in a way or
another, as is reviewed in section \ref{Nogoss}. Still, we proved
the no-go theorem \ref{Noway} which shows that the standard
Noether procedure will not provide such a non-Abelian theory as a
local deformation of the Abelian self-dual gauge field theory.
Since the two main assumptions of this no-go theorem are locality
and continuity of the deformations, in order to escape its
conclusion, one should perhaps leave the standard formalism of
perturbative local field theory.

The theoretical problem of determining consistent interactions is
an hard task in general. The equations for the consistent
interactions are rather intricate because they may be non linear
and involve simultaneously not only the deformed action, but also
the deformed structure functions of the deformed gauge algebra (as
well as the deformed reducibility coefficients if the gauge
transformations are reducible). Furthermore, ``trivial"
interactions that are simply induced by a change of variables
should be factored out. As we review in section \ref{defos}, one
can reformulate the problem as a cohomological problem
\cite{Barnich:1993}. This approach systematizes the perturbative
construction of the consistent interactions (the Noether method)
and, furthermore, enables one to use the powerful tools of
homological algebra. We present the BRST machinery used in the
proof of our theorem \ref{Noway} in the section \ref{LBRSTco}. In
section \ref{s:self}, we determine all consistent, continuous,
local and Lorentz invariant deformations of a system of Abelian
self-dual vector gauge fields, with the help of the previously
obtained results.

A better understanding of non-Abelian duality remains a seducing
goal. Indeed, duality lead to an indirect approach of
non-perturbative phenomena. For instance, a satisfactory
definition of EM duality for Non-Abelian gauge theories should
provide an explicit definition of 'tHooft topological operators
and a full proof of their commutation relation, thereby
approaching closer to a rigorous proof of quark confinement
\cite{'tHooft:1978}. Full understanding of non-Abelian duality is
probably also an unavoidable step to achieve the following
ultimate dream: write all equations of ``M-theory" with all
duality symmetries manifest.

The authors of \cite{Chan:1995,Chan:1996} proposed an interesting
tentative of definition of EM duality for non-Abelian Yang-Mills
gauge fields in terms of loop space variables (which are
intrinsically non-local), but their lack of completely concrete
mathematical expressions makes their full success unclear
yet\footnote{For reviews of their work and its possible
phenomenological applications one might see
\cite{Chan:1998,Chan:1999,Chan:1999i,Tsou:2001}.}.

%%%%%%%%%%%%%%%%%%%%%%%%%%%%%%%%%%%%%%%%%%%%%%%%%%%%%%%%%%%%%%%%%%

\section{Toroidal compactifications of M-theory}\label{Nogoss}

\subsection{M5-branes}

In the low energy limit where bulk gravity decouples, a single
M5-brane is described by a six-dimensional ${\cal N}=(2,0)$
superconformal field theory. Its field content consists of five
scalar fields and a single chiral two-form. A Lorentz
non-covariant action was constructed in
\cite{Perry:1997,Schwarz:1997,Aganagic:1997}. A covariant action
was obtained in \cite{Pasti:1997i,Bandos:1997}. The covariant
action contains appropriate extra auxiliary fields and gauge
symmetries. Partial gauge fixing of the covariant action yields
the non-covariant action. Once $n$ M$5$-branes coincide, the
situation changes and little is known about the underlying
physics.

\subsection{Compactification over a circle}

\begin{figure}[ht!]
     \centerline{\includegraphics[width=0.6\linewidth]{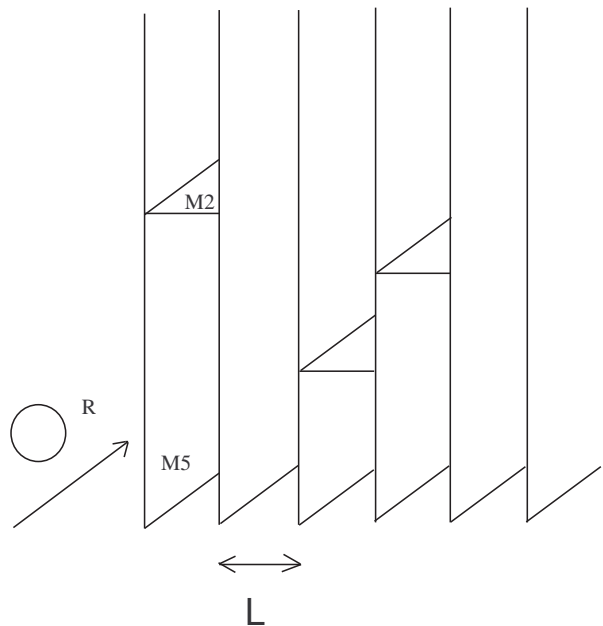}}
\caption{Parallel M$5$ separated by a distance of the order of $L$
with M$2$ stretching from one to the other.}\label{comparedto}
\end{figure}

Let us compactify one direction of the eleven dimensional space on
a circle of radius $R$. For small radius, the resulting theory is
the weakly coupled type IIA string theory. When the M5-branes are
transversal to the circle, they appear in the type IIA theory as
$n$ coinciding NS5-branes. In fact, not much is explicitly known
about this system. However, when the M$5$-branes are longitudinal
to the circle, they emerge as $n$ coinciding D$4$-branes. The
figure \ref{comparedto} should be compared to the figure
\ref{compare}. The effective action for such a system is a $U(n)$
non-Abelian Born-Infeld action. Ignoring higher derivative terms
and focusing on the leading term, one gets that the dynamics of
the D$4$ system is governed by $5$ scalar fields in the adjoint
representation of $U(n)$ coupled to a 5-dimensional $U(n)$ gauge
theory.

\begin{figure}[ht!]
     \centerline{\includegraphics[width=0.7\linewidth]{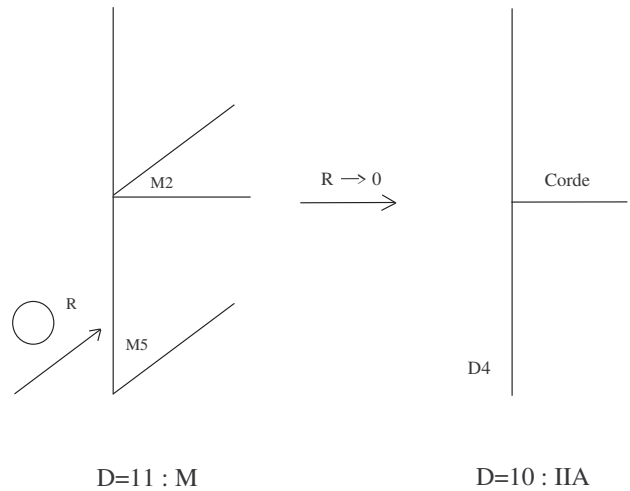}}
\caption{A M$5$ wrapped along a shrinked circle becomes a D4
brane.}
\end{figure}

Going back to the eleven dimensional picture, this observation
suggests the existence of a non-Abelian extension of chiral
two-forms living on the worldvolume of $n$ coinciding M$5$-branes
worldvolumes, the theory of which is believed to be a non-trivial
superconformally invariant quantum theory. The $5$-dimensional
coupling arises as the radius of compactification $g^2_{YM}=R$.

M-theoretical considerations indicate that $n$ coinciding
M5-branes constitute a highly unusual physical system. Indeed, the
supergravity description of $n$ M5-branes predicts that both the
entropy \cite{Klebanov:1996} and the two-point function for the
stress-energy tensor \cite{Gubser:1997} scale as $n^3$ in the
large $n$ limit. Anomaly considerations lead to a similar
behaviour \cite{Henningson:1998}. Some people expect the
interacting $(2,0)$ theory to fall outside the scope of
perturbative covariant local field theory
\cite{Hull:2000,Hull:2000i,Hull:2000ii}. The same could be
expected to occur for $SL(2,{\mathbb Z})$ duality symmetry of four
dimensional non-Abelian vector theory, as is reviewed in
subsection \ref{nogo}.

\subsubsection{No-go theorem for self-dual fields}

By free \emph{self-dual theories} we understand a
$2(p+1)$-dimensional linear $p$-brane electrodynamics described
for
\begin{itemize}
  \item \underline{$p$ even} by two gauge fields dual to each other, as considered in the duality-symmetric formulation of subsection
\ref{doublep}.
  \item \underline{$p$ odd} by a single chiral gauge field, as discussed in subsection \ref{linkwchiral}.
\end{itemize}
In some sense, self-dual gauge fields satisfy the more restrictive
self-duality condition possible in their spacetime. We refer to
the section \ref{double} for a complete discussion of their
self-duality equations.

Due to the following theorem, no local field theory seems to be
able to describe a system of coinciding M5-branes.
\cadre{theorem}{Noway}{(No-go theorem for self-dual gauge fields)

No consistent, local interactions of a set of free self-dual
fields can deform the Abelian gauge transformations if the local
deformed action (free action + interaction terms) continuously
reduces to a sum of free, duality-symmetric, non-covariant actions
in the zero limit for the coupling constant.} The proof of this
theorem is essentially algebraic
\cite{Bekaert:1999,Bekaert:2000i,Bekaert:2000} and is based on a
cohomological reformulation of the Noether procedure for
constructing consistent deformations \cite{Barnich:1994}. The next
section elaborates on the techniques required for the above proof.

Local deformations of the action cannot modify the Abelian nature
of the algebra of the two-form gauge symmetries. The no-go theorem
\ref{Noway} holds under the assumption that the deformed action is
continuous in the coupling constant (i.e., possible
non-perturbative ``miracles" are not investigated) and reduces, in
the limit of vanishing coupling constant, to the action describing
free chiral $2$-forms. In particular, no assumption is made on the
polynomial order (cubic, quartic, ...) of the interaction terms.

\subsection{Compactification over a torus}\label{nogo}

New insights on EM-duality symmetry have been provided by
M-theory. The type IIB string theory reduced on a circle is known
to be T-dual to M-theory on a two-torus. Accordingly, the
S-duality symmetry of the IIB string theory is a consequence of
the invariance of the torus under large diffeomorphisms, the
$SL(2,{\mathbb Z})$ symmetry of the IIB theory being associated
with the modular group of the torus. Thus, in the M-theory context
the S-duality of IIB string theory arises elegantly from simple
geometric arguments. Likewise, the system of a single M5-brane
system provides an appealing geometric understanding of EM-duality
symmetry.

The worldvolume of the M$5$-brane supports a self-interacting
chiral two-form potential which couples minimally to dyonic
strings located at the intersection of the M$5$-brane and some
M$2$-branes ending on it. If the M$5$-brane is wrapped around the
torus, the T-dual picture in IIB theory is a D$3$-brane with
fundamental strings ending on it. As a consequence the D$3$-brane
itself is inert under the modular group $SL(2,{\mathbb Z})$. In
terms of the D$3$-brane worldvolume theory, this symmetry
translates into the EM-duality symmetry of Abelian BI theory
\cite{Berman:1997i,Nurmagambetov:1998}. This sheds some light on
the relation between duality-symmetric theories and chiral forms.
Indeed, the e.o.m. (\ref{self}) in four dimensions finds its
origin in the self-duality equation of the three-form field
strength living on the wrapped M5-brane. A ${\mathbb Z}_4$-duality
transformation then corresponds to the ``exchange" of the two
circles in the compactification from M to IIB theory.

The next step of interest is to consider a system where $n$
wrapped M5-branes coincide. Unfortunately little is known about
this interacting (2,0) superconformal theory. In the T-dual
picture, the M5-branes appear as a set of coinciding D3-branes.
Their dynamics is governed by a four-dimensional $U(n)$
supersymmetric DBI theory which, in the weak field limit, is an
ordinary $U(n)$ non-Abelian gauge theory with ${\cal N}=4$
supersymmetry. The determination of all higher order terms in
$\alpha'$ is still an open question but some progress has been
made recently in this direction (see
\cite{Bergshoeff:2000,Bergshoeff:2000i,Koerber:2001,Refolli:2002}
and references therein). In any case, from the same arguments as
before, the non-Abelian gauge theory on the coinciding D3-brane
worldvolumes should possess an $SL(2,\mathbb Z)$
symmetry\footnote{This conjectured duality symmetry could be a new
constraint to be imposed in order to derive the full non-Abelian
BI action.}.

The BRST techniques can be applied in a straightforward fashion to
prove the results in \cite{Bekaert:1999i} as well. There,
deformations of chiral four-forms in ten dimensions were analyzed
leading to the conclusion that the only consistent deformation in
the type IIB coupling of the chiral four-form to the NS$\otimes$NS
and the R$\otimes$R two-forms familiar from IIB supergravity.

\subsubsection{No-go theorems for Yang-Mills fields}

To conclude, let us consider the sourceless Yang-Mills equations
and look after any duality symmetry property. They read
\begin{equation}
D_A \left( \begin{array}{c} F\\ $*$F
\end{array}
\right) = 0,\label{YM}
\end{equation}
where $A$ is Lie-algebra valued one-form, its curvature is equal
to $F= D_A A=dA+A^2$ and the covariant derivative acts as $D_A
=d+[A,\,]$ on other forms. Naively, one could think that the
Yang-Mills equations (\ref{YM}) are left invariant by the duality
rotation (\ref{transfo}). However, that is not the case since the
covariant derivative $D_A$ depends on $A$, which is not inert
under the duality rotation.

In the seventies, two no-go theorems \cite{Deser:1976,Gu:1975}
were found to prevent such a trivial generalization of EM-duality
for Abelian gauge fields:
\begin{proposition}Generically, there is
no infinitesimal transformation $\delta A$ which is able to
implement the infinitesimal duality rotation
\begin{eqnarray}
\delta F=*F\,\delta\alpha\,,\quad\delta
*F=-F\,\delta\alpha.\nonumber
\end{eqnarray}
\end{proposition}
\begin{proposition}There exists an example of gauge field $A$
with curvature $F=D_AA$ solution of the Yang-Mills equation
$D_A*F=0$ but without ``dual" gauge field $\tilde A$. In other
words, there exists no vector field $\tilde A$ such that
$*F=D_{\tilde{A}}\tilde{A}$.
\end{proposition}
The second no-go theorem is essentially a counterexample obtained
by Gu and Yang \cite{Gu:1975}. It teaches us that the Poincar\'e
lemma does not generalize straightforwardly to covariant
derivative $D_A$ (Anyway, $D_A$ is not nilpotent since
$D^2_A=[F,\,\,]$). This prevents a direct application of the
scheme of the section \ref{double} to obtain a duality-symmetric
formulation of Yang-Mills theory.

Of course, no-go theorems have the weakness of their hypotheses.
In consequence, a priori nothing prevents less ``trivial"
generalizations of EM-duality symmetry for non-Abelian gauge
theories. The no-go theorem of the previous subsection further
restricts the generalization possibilities \cite{Bekaert:2001}.
Below, the proposition \ref{t:classification} gives a complete
classification of the possible local interactions of a sum of
duality-symmetric non-covariant actions.

\subsection{$(4,0)$ superconformal theory}

Hull considered the $5$-dimensional ${\cal N}=8$ (ungauged)
supergravity, arising as a massless sector of M-theory
compactified on a six-torus $T^6$. Using a line of arguments
similar to those used previously, Hull conjectured that its strong
coupling dual field theory should be a $6$-dimensional $(4,0)$
superconformal theory \cite{Hull:2000,Hull:2000i,Hull:2000ii}
containing an exotic fourth-rank tensor described in subsection
\ref{linkwchiral}. In \cite{Hull:2000}, it was shown that the
dimensional reduction of the free $6$-dimensional $(4,0)$ theory
on a circle indeed gives the linearized $5$-dimensional ${\cal
N}=8$ supergravity theory, with Planck length given by the circle
radius $l_P=R$.

In the dimensional reduction, the $6$-dimensional self-dual gauge
field in the irrep. of $GL(6,\mathbb R)$ associated to the Young
diagram $\mbox{\footnotesize
\begin{picture}(25,15)(0,0)
\multiframe(0,6.5)(10.5,0){1}(10,10){}
\multiframe(0,-4)(10.5,0){1}(10,10){}
\multiframe(10.5,6.5)(10.5,0){1}(10,10){}
\multiframe(10.5,-4)(10.5,0){1}(10,10){}
\end{picture}}$ reduces, in the $5$-dimensional picture,
to gauge fields in irrep. of $GL(5,{\mathbb R})$ corresponding to
the diagrams $\mbox{\footnotesize
\begin{picture}(25,15)(0,0)
\multiframe(0,0)(10.5,0){1}(10,10){}
\multiframe(10.5,0)(10.5,0){1}(10,10){}
\end{picture}}$, $\mbox{\footnotesize
\begin{picture}(25,15)(0,0)
\multiframe(0,6.5)(10.5,0){1}(10,10){}
\multiframe(0,-4)(10.5,0){1}(10,10){}
\multiframe(10.5,6.5)(10.5,0){1}(10,10){}
\end{picture}}$, and $\mbox{\footnotesize
\begin{picture}(25,15)(0,0)
\multiframe(0,6.5)(10.5,0){1}(10,10){}
\multiframe(0,-4)(10.5,0){1}(10,10){}
\multiframe(10.5,6.5)(10.5,0){1}(10,10){}
\multiframe(10.5,-4)(10.5,0){1}(10,10){}
\end{picture}}$. To show that, on-shell, these three gauge fields describe a single $5$-dimensional graviton it
was necessary to dualize the $5$-dimensional gauge field
$\mbox{\footnotesize
\begin{picture}(25,15)(0,0)
\multiframe(0,6.5)(10.5,0){1}(10,10){}
\multiframe(0,-4)(10.5,0){1}(10,10){}
\multiframe(10.5,6.5)(10.5,0){1}(10,10){}
\end{picture}}$ to a symmetric tensor $\mbox{\footnotesize
\begin{picture}(25,15)(0,0)
\multiframe(0,0)(10.5,0){1}(10,10){}
\multiframe(10.5,0)(10.5,0){1}(10,10){}
\end{picture}}$ (see section \ref{MixedY} for more explanations).

\subsubsection{No-go theorem for mixed symmetry type fields}

Here also, we proved a no-go theorem which shows once again that
dualization of non-Abelian gauge fields falls outside the
conventional pattern. Indeed, the $5$-dimensional duality between
a symmetric gauge field $h_{\m\n}$ and a mixed Young symmetry type
gauge field $\tilde{h}_{\m\n\,|\,\r}$ (see section \ref{MixedY})
holds at the linearized level, but the following no-go theorem
says that it seems difficult to reconstruct a self-interacting
theory for $\tilde{h}_{\m\n\,|\,\r}$ corresponding to the Einstein
theory for the metric $g_{\m\n}=\eta_{\m\n}+h_{\m\n}$.

\cadre{theorem}{nogothree}{In $D\geq 5$ dimensions and under the
hypothesis of \begin{itemize}
\item locality,
\item Lorentz invariance,
\item smooth deformation,
\end{itemize}
there is no deformation of the free theory of the gauge field
$M_{\m\n\,|\,\r}\equiv \mbox{\footnotesize
\begin{picture}(25,15)(0,0)
\multiframe(0,6.5)(10.5,0){1}(10,10){$\m$}
\multiframe(0,-4)(10.5,0){1}(10,10){$\n$}
\multiframe(10.5,6.5)(10.5,0){1}(10,10){$\r$}
\end{picture}}$ which would modify the gauge algebra. If one adds the restriction
that there should be \begin{itemize}
\item no deformation involving four
derivatives or more in the Lagrangian,
\end{itemize} then there are no
deformation of the free theory which would modify the gauge
transformations without modifying the gauge algebra.} Of course,
one could add higher derivative terms involving product of the
field strength and its derivatives. Such deformations modify
neither the gauge algebra nor the gauge transformations.

%%%%%%%%%%%%%%%%%%%%%%%%%%%%%%%%%%%%%%%%%%%%%%%%%%%%%%%%%%%%%%%%%%%%%%%%%

\section{Constructing deformations as a cohomological
problem}\label{defos}

This section follows closely \cite{Barnich:1993,Henneaux:1998i}
and the section 2.2 of \cite{Knaepen:1999}.

\subsection{Noether method}

Let $\stackrel{(0)}{S_0} [\varphi^i]$ be a ``free" action with
``free" gauge symmetries\footnote{We use de Witt's notation.}
\begin{equation}
\delta_\varepsilon \varphi^i = \stackrel{\smash{(0)}}{R}^i_\alpha
\varepsilon^\alpha\,.
\end{equation}
They lead to the so-called \emph{Noether identities}
\cite{Henneauxbook} of the ``free" theory
\begin{equation}
{\delta\stackrel{(0)}{S}\over\delta\varphi^i}
\stackrel{\smash{(0)}}{R}^i_\alpha=0\,.
\end{equation}
We now start to perturb the ``free" theory by introducing new
terms $\stackrel{(i)}{S_0}$ ($i=1,2,\ldots$) to the action. The
perturbation is controlled by a coupling constant $g$ which
appears in the expansion of the ``deformation" $S_0$,
\begin{equation}
S_0 = \stackrel{(0)}{S_0} + g \stackrel{(1)}{S_0} + g^2
\stackrel{(2)}{S_0} + ...\label{fullaction}
\end{equation}
By \emph{consistent deformations}, one means that the deformed
action should be gauge invariant under the (possibly deformed)
gauge transformation rules,
\begin{equation}
R^i_\alpha = \stackrel{\smash{(0)}}{R}^i_\alpha + g
\stackrel{\smash{(1)}}{R}^i_\alpha + g^2
\stackrel{\smash{(2)}}{R}^i_\alpha + ...\label{fullsymmetries} .
\end{equation}
As in the ``free" case, it may be translated into the requirement
that the Noether identities should hold to all orders
\begin{equation}
{\delta{S}\over\delta\varphi^i}{R}^i_\alpha=0,\label{as1}
\end{equation}
where,
\begin{equation}
\delta_\varepsilon \varphi^i = R^i_\alpha \varepsilon^\alpha.
\end{equation}
As one can see by expanding (\ref{as1}) order by order in the
coupling constant, there are a priori an infinite number of
consistency conditions (of increasing complexity) to take into
account.

For reducible theories, which is the case relevant to chiral
$p$-forms, there are even additional constraints.  The gauge
transformations of the free theory are not independent,
\begin{equation}
\stackrel{\smash{(0)}}{R}^i_\alpha
\stackrel{\smash{(0)}}{Z}^\alpha_A = 0
\end{equation}
(possibly on-shell). There could be higher reducibility conditions
(e.g. reducibilities of reducibilities). One must then also impose
that the gauge transformations remain reducible, possibly in a
deformed way. This yields further constraints. The conclusion is
that to be consistent, the deformation should preserve the gauge
structure of the theory.

\subsection{BRST formulation}\label{BRSTdef}

A convenient toolkit for translating the problem of consistent
interactions into a cohomological problem is the antifield
formalism \cite{Batalin:1983}.\footnote{For reviews, we refer to
\cite{Henneauxbook,Gomis:1995}. The section \ref{a:BRST} also
presents a short review of the BV formalism.} If the undeformed
theory is actually free, it is a relatively easy task to solve the
master equation for the ``free" theory. Let $\stackrel{(0)}{S}$ be
a solution of the undeformed master equation
$(\stackrel{(0)}{S},\stackrel{(0)}{S})=0$. Since the master
equation captures all the information about the gauge structure of
the theory, the existence of a consistent deformation $S_0$ of the
original gauge invariant action $\stackrel{(0)}{S}_0$ is
equivalent to the existence of a deformation $S$ of
$\stackrel{(0)}{S}$,
\begin{equation}
S = \stackrel{(0)}{S} + g \stackrel{(1)}{S} + g^2
\stackrel{(2)}{S} + ...\label{fullmasterequation}
\end{equation}
which is a solution to the master equation $(S,S)=0$. Expanding
this equation order by order in the coupling constant yields
various consistency relations,
\begin{eqnarray}
(\stackrel{(0)}{S},\stackrel{(0)}{S})&= 0 \label{deformation1}\\
(\stackrel{(0)}{S},\stackrel{(1)}{S})&= 0 \label{deformation2}\\
2(\stackrel{(0)}{S},\stackrel{(2)}{S}) +
(\stackrel{(1)}{S},\stackrel{(1)}{S})&= 0 \label{deformation3}\\
&\vdots\ \ \ .\nonumber
\end{eqnarray}
The first equation is satisfied by assumption. Now comes the
cohomology. Let $\stackrel{(0)}{s}$ be the ``free" BRST
differential defined by \be\stackrel{(0)}{s}\equiv
(\stackrel{(0)}{S},\cdot)\ee The crucial point for us is now that
the second equation states that $\stackrel{(1)}{S}$ is a cocycle
for $\stackrel{(0)}{s}$. Furthermore, if $\stackrel{(1)}{S}$ were
a coboundary, $\stackrel{(1)}{S}=
(\stackrel{(1)}{T},\stackrel{(0)}{S})$, we can show that this
would correspond to a trivial deformation of $\stackrel{(0)}{S}$
(i.e. a deformation which amounts to a simple redefinition of the
fields). The solution of the master equation $S_0$ is of definite
vanishing ghost number, therefore the cohomology group of interest
in the study of first order deformations is
$H^{0}(\stackrel{(0)}{s})$.

For practical purpose, we more precisely consider deformations
which are local in spacetime, i.e., we impose that
$\stackrel{(1)}{S}, \stackrel{(2)}{S}, ...$ be \emph{local}
functionals. Reformulating the equations in terms of the Lagrange
densities takes care of this problem. Then, in the local context,
the proper cohomology group to evaluate is
$H^{0,D}(\stackrel{(0)}{s}\mid d)$ where the first and second
superscripts denote the ghost number and form degree,
respectively.

Let us consider briefly the case when all the representatives of
$H^{D,0}(\stackrel{(0)}{s}\mid d)$ can be taken not to depend on
the antifields, which is the situation met for the systems
considered in this chapter. We may take the first-order
deformations $\stackrel{(1)}{\cal{S}}$ to be
antifield-independent. In this case equation (\ref{deformation2})
reduces to $(\stackrel{(0)}{S},\stackrel{(2)}{S})=0$ and implies
that the deformation at order $g^2$ defines also an element of
$H^{D,0}(\stackrel{(0)}{s}\mid d)$.  We can thus take
$\stackrel{(2)}{S}$ not to depend on the antifields either.
Proceeding in this manner order by order in the coupling constant,
we conclude that the additional terms in $S$ are all independent
of the antifields. Since the antifield-dependent terms in the
deformation of the master equation are related to the deformations
of the gauge transformations, this means that there is no
deformation of the gauge transformations. Summarizing, if there is
no non-trivial dependence on the antifields in
$H^{D,0}(\stackrel{(0)}{s}\mid d)=0$, the only possible consistent
interactions are of the first class and do not modify the gauge
symmetry.

%%%%%%%%%%%%%%%%%%%%%%%%%%%%%%%%%%%%%%%%%%%%%%%%%%%%%%%%%%%%%%%%%%%%

\section{Local BRST cohomology}\label{LBRSTco}

We will illustrate the computation of the local BRST cohomology in
vanishing ghost number on a particular example: a system of N free
chiral 2-forms $A^A_{ij}$ ($A=1,\ldots,N$),\footnote{The integer
$N$ could be a function of the number $n$ of coincident M5-branes
(e.g., $N \sim n^3$) in an M-theoretic picture.} the dynamics of
which is described by the non-covariant action
\cite{Henneaux:1987,Henneaux:1988},
\begin{equation}
S_0[A^A_{ij}]=\sum_A\int dt\,d^5x\,
B^{A\,ij}(\dot{A}^A_{ij}-B^A_{ij}),\quad (A=1,\ldots,N) ,
\label{flipflap}
\end{equation}
where
\begin{equation}
B^{A\,ij}=\frac{1}{6}\epsilon^{ijklm}F^A_{klm}=\frac{1}{2}\epsilon^{ijklm}
\partial_k A^A_{lm}.
\end{equation}
The action (\ref{flipflap}) differs from the one in
\cite{Perry:1997,Aganagic:1997} where a space-like dimension was
singled out. Here we take time as the distinguished direction;
from the point of view of the PST formulation
\cite{Pasti:1997i,Bandos:1997}, the two approaches simply differ
in the gauge fixation.

We work in Minkowski spacetime. This implies, in particular, that
the topology of the spatial sections are the same as ${\mathbb
R}^5$. Most of our considerations would go unchanged in a curved
background of the product form ${\mathbb R} \times \Sigma$
provided the De Rham cohomology groups $H^2(\Sigma)$ and
$H^1(\Sigma)$ of the spatial sections $\Sigma$ vanish. [If
$H^2(\Sigma)$ is non-trivial, there are additional gauge
symmetries besides (\ref{cling}) below, given by time-dependent
spatially closed $2$-forms; similarly, if $H^1(\Sigma)$ is
non-trivial, there are additional reducibility identities besides
(\ref{clong}) below. We would thus need additional ghosts and
ghosts of ghosts. These, however, would not change the discussion
of \emph{local} Lagrangians because they would be global in space
(and local in $t$).]

The action $S_0$ is invariant under the following gauge
transformations
\begin{equation}
\delta_{\Lambda}A^A_{ij}=\partial_i \Lambda^A_j -
\partial_j \Lambda^A_i, \label{cling}\end{equation} because
$B^{A\,ij}$ is gauge-invariant and identically transverse
($\partial _i B^{A\,ij} \equiv 0$)
\footnote{%
Since $A^A_{0i}$ does not occur in the action -- even if one replaces $%
\partial_0 A^A_{ij}$ by $\partial_0 A^A_{ij} - \partial_i A^A_{0j} - \partial_j
A^A_{i0}$ (it drops out because $B^{A\,ij}$ is transverse) --, the
action is of course invariant under arbitrary shifts of
$A^A_{0i}$.}. As $\delta A_{ij}^A=0$ for
\begin{equation}
\Lambda_i^A=\partial_i\varepsilon^A, \label{clong}
\end{equation}
this set of gauge transformations is reducible. This exhausts
completely the redundancy in $\Lambda_i^A$ since $H^1({\mathbb
R}^5) = 0$.

The equations of motion obtained from $S_0[A^A_{ij}]$ by varying
$A^A_{ij}$ are
\begin{equation}
\epsilon^{ijklm}\partial_k \dot{A}^A_{lm}-2\partial_k F^{A\,ijk}=0
\Leftrightarrow \epsilon^{ijklm}\partial_k
(\dot{A}^A_{lm}-B^A_{lm})=0. \label{bam}
\end{equation}
Using $H^2({\mathbb R}^5)=0$, one finds that the general solution
of (\ref{bam}) is
\begin{equation}
\dot{A^A_{ij}}-B^A_{ij}=\partial_i \Lambda^A_j -
\partial_j \Lambda^A_i. \label{solution}
\end{equation}
The ambiguity in the solutions of the equations of motion is thus
completely accounted for by the gauge freedom (\ref{cling}). Hence
the set of gauge transformations is complete.

We can view $\Lambda^A_i$ as $A^A_{0i}$, so the equation
(\ref{solution}) can be read as the self-duality equation
\begin{equation}
F^A_{0ij}-*F^A_{0ij}=0,
\end{equation}
where $F^A_{0ij}=\dot{A^A_{ij}}+\partial_i A^A_{j0}+\partial_j
A^A_{0i}$. Alternatively, one may use the gauge freedom to set
$\Lambda^A_i=0$, which yields the self-duality condition in the
temporal gauge.

\subsection{Fields - Antifields - Solution of the master equation}

The solution of the master equation is easy to construct in this
case because the gauge transformations are Abelian.

The fields in presence here are
\begin{equation}
\{\Phi^M\}=\{A^A_{ij},C^A_i,\eta^A\}.
\end{equation}
The ghosts $C^A_i$ correspond to the gauge parameters
$\Lambda^A_i$, and the ghosts of ghosts $\eta^A$ correspond to
$\epsilon^A$.

Now, to each field $\Phi^M$ we associate an antifield $\Phi^*_M$.
The set of antifields is then
\begin{equation}
\{\Phi^*_M\}=\{A^{*Aij},C^{*Ai},\eta^{*A}\}.
\end{equation}
The fields and antifields have the respective parities
\begin{eqnarray}
&&\epsilon(A^A_{ij})=\epsilon(\eta^A)=\epsilon(C^{*Ai})=0\\
&&\epsilon(C^A_i)=\epsilon(A^{*Aij})=\epsilon(\eta^{*A})=1.
\end{eqnarray}
The antibracket is defined as
\begin{equation}
(X,Y)=\int d^nx\left( \frac{\delta^RX}{\delta\Phi^M(x)}
\frac{\delta^LY}{\delta\Phi^*_M(x)}
-\frac{\delta^RX}{\delta\Phi^*_M(x)}\frac{\delta^LY}{\delta\Phi^M(x)}\right)
\end{equation}
where $\delta^R/\delta Z(x)$ and $\delta^L/\delta Z(x)$ denote
functional right- and left-derivatives.

Because the set of gauge transformations is complete and defines a
closed algebra, the (minimal, proper) solution of the master
equation $(S,S)=0$ takes the general form
\begin{equation}
S=S_0+\sum_M\int (-)^{{\epsilon}(M)}\Phi^*_M s\Phi^M,
\end{equation}
where ${{\epsilon}(M)}$ is the Grassmann parity of $\Phi^M$. More
explicitly, we have
\begin{equation}
S=S_0 +\sum_A \int dtd^5x(A^{*Aij}\partial_i C^A_j -
C^{*Ai}\partial_i \eta^A)
\end{equation}

The solution $S$ of the master equation captures all the
information about the gauge structure of the theory : the Noether
identities, the closure of the gauge transformations algebra and
the higher order gauge identities are contained in the master
equation. The existence of $S$ reflects the consistency of the
gauge transformations.

\subsection{BRST operator} \setcounter{equation}{0}

The BRST operator $s$ is obtained by taking the antibracket with
the proper solution $S$ of the classical master equation,
\begin{equation}
s\, X   = (S,X).
\end{equation}

The BRST operator can be decomposed as
\begin{equation}
s=\delta+\gamma
\end{equation}
where $\delta$ is the Koszul--Tate differential
\cite{Henneauxbook}. What distinguishes $\delta$ and $\gamma$ is
the antighost number ($antigh$) defined through
\begin{eqnarray}
antigh(A^A_{ij})=antigh(C^A_i)=antigh(\eta^A)=0,\\
antigh(A^{*A\,ij})=1,\quad antigh(C^{*A\,i})=2,\quad
antigh(\eta^{*A})=3.
\end{eqnarray}

The ghost number ($gh$) is related to the antighost number by
\begin{equation}
gh=puregh-antigh
\end{equation}
where $puregh$ is defined through
\begin{eqnarray}
puregh(A^A_{ij})=0, \quad puregh(C^A_i)=1,\quad puregh(\eta^A)=2,\\
puregh(A^{*A\,ij})=puregh(C^{*A\,i})=puregh(\eta^{*A})=0.
\end{eqnarray}

The differential $\delta$ is characterized by $antigh(\delta)=-1$,
i.e. it lowers the antighost number by one unit and acts on the
fields and antifields according to
\begin{eqnarray}
\delta A^A_{ij}&=&\delta C^A_{i}=\delta \eta^A=0,\\
\delta A^{*A\,ij}&=&2\partial_k
F^{A\,kij}-\epsilon^{ijklm}\partial_k
\dot{A}^A_{lm},\\
\delta C^{*A\,i}&=&\partial_j A^{*A\,ij},\\
\delta \eta^{*A} &=& \partial_i C^{*A\,i}.
\end{eqnarray}

The differential $\gamma$ is characterized by $antigh(\gamma)= 0$
and acts as
\begin{eqnarray}
\gamma A^A_{ij}&=&\partial_i C^A_j - \partial_j C^A_i,\\
\gamma C^A_i&=&\partial_i \eta^A,\label{yep}\\
\gamma \eta^A&=&0,\\
\gamma A^{*A\,ij}&=&\gamma C^{*A\,i}=\gamma \eta^{*A}=0.
\end{eqnarray}

Furthermore we have,
\begin{equation}
sx^\mu = 0, \; s(dx^\mu)=0.
\end{equation}

\subsection{Local forms - Algebraic Poincar\'e lemma}

A {\bf local function} is a function of the fields, the ghosts,
the antifields,  and their derivatives up to some finite order $k$
(which depends on the function),
\begin{equation}
f=f(\Phi,\partial_{\mu}\Phi,\ldots,\partial_{\mu_1}\ldots\partial_{\mu_k}\Phi).
\end{equation}
A local function is thus a function over a finite dimensional
vector space $J^k$ called  {\bf jet space}. A {\bf local form} is
an exterior polynomial in the $dx^\mu$'s with local functions as
coefficients. The algebra of local forms will be denoted by
${\cal{A}}$. In practice, the local forms are polynomial in the
ghosts and the antifields, as well as in the differentiated
fields, so we shall from now on assume that the local forms under
consideration are of this type. We can actually show that
polynomiality in the ghosts, the antifields and their derivatives
follows from polynomiality in the derivatives of the $A_{ij}$ by
an argument similar to the one used in \cite{Barnich:1995} for
$1$-forms; and polynomiality in the derivatives is automatic in
our perturbative approach where we work order by order in the
coupling constant(s).

Note that we also exclude an explicit $x$-dependence of the local
forms. We could allow for one without change in the conclusions.
In fact, as we shall indicate below, allowing for an explicit
$x$-dependence simplifies some of the proofs.  We choose not to do
so here since the interaction terms in the Lagrangian should not
depend explicitly on the coordinates in the Poincar\'e-invariant
context.

The following proposition describes the cohomology of $d$ in the
algebra $\cal A$ of local forms, in degree $q<D$.
\begin{proposition}
The cohomology of $d$ in the algebra of local forms of degree
$q<D$ is given by
\begin{eqnarray}
H^0(d,{\cal A}) &\cong& {\mathbb R}, \nonumber\\
H^q(d,{\cal A})&=& \{ \hbox{Constant Forms} \} , \; 0<q<D
.\nonumber
\end{eqnarray}
\end{proposition}
Constant forms are by definition polynomials in the $dx^\mu$'s
with constant coefficients. This proposition is called the
algebraic Poincar\'e lemma (for $q<D$). There exist many proofs of
this lemma in the literature.

Constant $q$-forms are trivial in degree $0<q<D$ in the algebra of
local forms with an explicit $x$-dependence; e.g., $dx^0 = df$,
where $f$ is the $x^0$-dependent function $f= x^0$.  Thus, in this
enlarged algebra, the cohomology of $d$ is simpler and vanishes in
degrees $0<q<D$. This is the reason why the calculations are
somewhat simpler when we allow for an explicit $x$-dependence.

We work in a formalism where the time direction is privileged. For
this reason, it is useful to introduce the following notation :
the $l$-th time derivative of a field $\Phi$ (including the ghosts
and antifields) is denoted by $\Phi^{(l)}$ ($=\partial_0^l\Phi$),
and the spatial differential is denoted by
$\tilde{d}=dx^i\partial_i$.

A {\bf local spatial form} is an exterior polynomial in the
spatial $dx^k$'s with coefficients that are local functions. The
algebra of local spatial forms will be denoted by
$\widetilde{\cal{A}}$. If we write the set of the generators of
the jet space $J^k$ as
\begin{equation}
\{\Phi^{(l_0)},\partial_{i_1}\Phi^{(l_1)},\ldots,
\partial_{i_1}\dots\partial_{i_k}\Phi^{(0)};\,\,l_j=0,\ldots,k-j\},
\end{equation}
it is clear that
\begin{proposition}
The cohomology of $\tilde{d}$ in the algebra $\widetilde{\cal{A}}$
of local spatial forms of degree $q<D-1$ is given by
\begin{eqnarray}
H^0(\tilde{d},\widetilde{\cal{A}}) &\cong& {\mathbb R}, \nonumber\\
H^q(\tilde{d},\widetilde{\cal{A}}) &=& \{ \hbox{Constant spatial
forms} \}, \; 0<q<D-1 .\nonumber
\end{eqnarray}
\end{proposition}

A similar decomposition of space and time derivatives occurs of
course in the Hamiltonian formalism. A discussion of the problem
of consistent deformations of a gauge invariant action has been
carried out in the Hamiltonian context in
\cite{Bizdadea:2000,Bizdadea:2000i}.

\subsection{Wess-Zumino consistency condition}

The local BRST cohomology is defined as the cohomology $H(s\mid
d,{\cal A})$ of $s$ modulo $d$ in the graded algebra $\cal A$ of
local forms. The algebra $\cal A$ is graded according to many
degrees ; for instance, the form degree, the pure ghost number,
the antighost number, the degree of polynomiality, etc. Two
degrees are of particular interest: the form degree $p$ and the
ghost number $g$. This grading of ${\cal A}=\oplus_{p,g} {\cal
A}^{p,g}$ induces a grading of the local BRST cohomology \be
H(s\mid d,{\cal A})\equiv\oplus_{p,g} \,H^{p,g}(s\mid d)\ee where
$H^{p,g}(s\mid d)$ is defined by (i) the so-called
\emph{Wess-Zumino consistency condition}, that is $\a^{p,g}\in
H^{p,g}(s\mid d)$ if and only if \ba
s\a^{p,g}+d\b^{p-1,g+1}=0\,,\quad \a^{p,g}\in {\cal
A}^{p,g}\,,\,\,\a^{p-1,g+1}\in {\cal A}^{p-1,g+1}\,, \ea and (ii)
the equivalence relation \ba \a^{p,g}\sim\a^{p,g} +
s\r^{p,g-1}+d\s^{p-1,g}\,,\quad \r^{p,g-1}\in {\cal
A}^{p,g-1}\,,,\,\,\s^{p-1,g}\in {\cal A}^{p-1,g}\,. \ea

Now, all the necessary tools have been presented to proof the
theorem \ref{Noway}. As explained in subsection \ref{BRSTdef}, it
follows from the \begin{proposition}There is no non trivial
dependence on the antifields for the elements of $H^{D,0}(s \mid
d)$.\label{noantif}\end{proposition} In conclusion the proof of
theorem \ref{Noway} reduces to the computation of the cohomology
$H^{D,0}(s\mid d)$ for the free theory. The detailed computation
for the chiral two-forms system, using homological perturbation
theory techniques, has been placed in the appendix \ref{pmain}
since the proof is rather lengthy and technical. It follows the
path cleared by the general theorems of
\cite{Barnich:1995i,Barnich:1995}.

%%%%%%%%%%%%%%%%%%%%%%%%%%%%%%%%%%%%%%%%%%%%%%%%%%%%%%%%%%%%%%%%%%%%

\section{Self-interactions of a single gauge vector}\label{s:self}

In this section we want to generalize the PST action describing a
single \emph{free} Maxwell field to an interacting theory with
only one on-shell gauge vector (the e.o.m. allows to express on of
the two gauge field in term of the other). The interacting model
should of course maintain the Lorentz covariance and should lead
to a deformed self-duality condition. Firstly, we review in
subsec. \ref{ss:C-Heq} how this has been achieved in the
non-covariant approach, then we extend it in subsec.
\ref{ss:defPST} for the covariant case, and finally we look for
the solutions of the Courant-Hilbert equation.

\subsection{The return of the Courant-Hilbert equation} \label{ss:C-Heq}

We want to introduce self-couplings for the non-covariant EM
duality-symmetric action. In \cite{Deser:1998} it was proposed to
attack this problem using the Hamiltonian formulation, which is
appropriate for first-order actions. Let us review their approach,
trying to justify completely the ansatz in the light of the
results obtained in the appendix \ref{pmain}, where a
classification of local consistent deformations is given for a set
of chiral two-forms. The analogue of proposition \ref{inter} for
the four-dimensional duality-symmetric theory under consideration
is the \begin{proposition}\label{t:classification}All consistent,
continuous, local deformations of a system of free Abelian vector
fields (labelled by indices $A,B=1,\cdots,N$) described by a sum
of $N$ free, duality-symmetric, non-covariant actions as the
coupling constant goes to zero, are only of two types:
\begin{itemize}
\item[I.]{Those which are strictly invariant under the original gauge transformations $A_i^{\a\, A}\rightarrow A_i^{\a\, A}+\partial_i\Lambda^{\a\, A}$; they are polynomials in
the gauge-invariant fieldstrength $F^{\a A}_{ij}=\partial_i
A_j^{\a\, A}-\partial_j A_i^{a\, A}$ and their partial
derivatives, i.e. they are of the form
\begin{equation}\int d^4x\,f(\partial_{\m_1\dots \m_r}F^{\a\, A}_{ij})\,,\quad (r=0,1,\ldots)\,.\end{equation}}
\item[II.]{Those which are invariant only up to a boundary term;
they are linear combinations of Chern-Simons like terms, i.e.
\begin{equation}\int d^4x\,\lambda_{rs\,\a\b\,AB}\,(\partial_0)^rA_i^{\a\, A}(\partial_0)^sB^{\b\, B\, i} \,,\end{equation}
where $\lambda_{rs\,\a\b\,AB}$ ($\a,\b=1,2$) are constants such
that $\lambda_{sr\,\b\a\,BA}=\lambda_{rs\,\a\b\,AB}\,$.}
\end{itemize}\end{proposition} For instance, the Hamiltonian of the free theory is a term of
type I, as well as the Hamiltonian for Born-Infeld theory. The
kinetic term is of type II.

From the very beginning we state three basic requirements made on
the model in this approach: (i) The deformation remains a
first-order action, (ii) manifest rotation invariance, (iii)
manifest duality-symmetry. The first assumption comes from the
fact that we work in Hamiltonian formalism. The second and third
requirements simply extend the properties of the free model.

The requirement (i) combined with proposition
\ref{t:classification} and with the fact that we deal with a
single on-shell vector gauge field ($N=1$) eliminates deformations
of the type $II$ besides the one without any derivative. Thus, the
non-linear action is
\begin{equation}
S=-{1\over 2}\int d^4x\,B^{i\alpha}J_{\alpha\beta}E^\beta_i - \int
dx^0 H,
\end{equation}
where
\begin{equation}
H = \int d^3x \left({\mathcal H}(\partial_{i_1\dots
i_{k-1}}B^{\alpha}_{i_k})+\chi A_i^\alpha B^{i\alpha}\right)
\end{equation}
stands for the Hamiltonian of the model. The constant factor of
the Chern-Simons like term is denoted by $\chi$. If we now
restrict ourself to (iv) Hamiltonian densities that do not
explicitly depend on the derivatives of the magnetic
field\footnote{In other words, we assume a slowly varying
fieldstrength.}: ${\mathcal H}={\mathcal H}(B^{\a A}_i)$, we can
deduce from (ii) and (iii) that the Hamiltonian density ${\mathcal
H}$ only depends on two independent space scalars that are
manifestly duality invariant, namely
\begin{equation}
y_1 = \frac12 B^{\alpha}_iB^{\alpha i}\,, \; \; y_2 = \frac{1}{4}
B^{\a}_iB^{\b i}B^{\a}_jB^{\b j}\,.
\end{equation}
We set ${\mathcal H}= f(y_1,y_2)$.

Now, all the symmetries are manifest except Lorentz symmetry. With
the help of tensor calculus, it is rather easy to construct
interactions that preserve Lorentz invariance. But there is an
alternative way to control Lorentz invariance. It is through the
commutation relations of the energy-momentum tensor components:
Dirac and Schwinger have established a sufficient condition for a
manifestly rotation and translation invariant theory (in space) to
be also Lorentz-invariant \cite{Dirac:1962,Schwinger:1963}. The
condition becomes also necessary when one turns to couplings with
gravitation \cite{Schwinger:1963} or if one asks for
diffeomorphism-invariance (``path independence of the dynamical
evolution") \cite{Teitelboim:1973}. The \emph{Dirac-Schwinger
criterion} yields in our case $\chi=0$ and a non-linear
first-order differential equation for $f$ \cite{Bekaert:1998}
\begin{equation}\label{e:C-H2}
f_1^2 +2 y_1 f_1 f_2 + 2 (y_1^2 -y_2 )f_2^2 = 1\,,
\end{equation}
where $f_i = \frac{\partial f}{\partial y_i}$ for $i=1,2.$ The
equation (\ref{e:C-H2}) can be given a simpler look by the change
of variables
\begin{eqnarray}\label{e:vartrans}\left\{\begin{array}{lll}
y_1 &=& u_+ + u_- \\ y_2 &=& u^2_+ + u^2_-
\end{array}\right.\,.
\end{eqnarray}
Denoting the function derivatives by $f_\pm \equiv {\partial
f\over \partial u_\pm}$, one gets again the Courant-Hilbert
equation (\ref{duality}). It is not surprising to find the same
equation for the Hamiltonian density and the Lagrangian density
because they are related by a Legendre transformation, and
Legendre transformation relates a model and its dual
\cite{Gaillard:1981}.

This rederivation of (\ref{duality}) proves that the class of
duality-symmetric non-linear electrodynamics considered in
subsection \ref{CHsols} covered the most general set of physically
natural possibilities, since the Dirac-Schwinger condition becomes
necessary when one turns to general covariance. Combining the
proposition \ref{t:classification} with the Dirac-Schwinger
criterion, we prove the stronger
\begin{corollary}\label{physrev}Any
\begin{itemize}
  \item non-trivial
  \item local
  \item consistent
  \item diffeomorphism-invariant
  \item analytic in the weakly-coupled regime
\end{itemize}
deformation of the four-dimensional free
non-(manifestly)-covariant self-dual gauge field theory is
entirely characterized by a function $\Psi(t)$ satisfying the
conditions $(i)-(iv)$ of theorem
\ref{t:analyticity}.\end{corollary} Therefore, the class
satisfying the reasonable physical requirements of the corollary
\ref{physrev} is precisely the class infinite class of
duality-symmetric theories of the corollary \ref{infiniteclass}.
Once again, we observe that manifest duality symmetry does not
easily combine with Lorentz symmetry and locality.

\subsection{Self-couplings of the PST model} \label{ss:defPST}

In this section, we ``covariantize" the results of the previous
section, generalizing the procedure of the free PST model. In this
way, we construct a deformed theory describing self-couplings of
one Maxwell field (on-shell) which is manifestly duality-symmetric
and Lorentz covariant.

Just as in the case of the free system the covariantization
involves an auxiliary field and an extra gauge invariance. As we
showed in \cite{Bekaert:2001}, it is the deformation of this
symmetry that leads to the Courant-Hilbert equation.

To recall the assumptions we stress that it seems natural to
require the interacting action to satisfy the same kind of
symmetries as the free one. In other words we should expect
besides the Lorentz invariance also a manifest duality-symmetry.
From the analysis on the consistent deformations of the
non-covariant action, we know that the interaction should depend
only on the field strengths and its derivatives. If we are in the
weakly varying field limit, this reduces the number of independent
invariants to only two (since we neglect derivative of the field
strength), namely
\begin{equation}
z_1 = \frac{1}{2}\tilde\cH^{\a}_\m \tilde\cH^{\a \,\m}\,, \; \;
z_2 = \frac{1}{4}\tilde\cH^{\a}_\m \tilde\cH^{\b
\,\m}\tilde\cH^{\a}_\n \tilde\cH^{\b \,\n}\,.
\end{equation}
Similarly to the non-covariant situation, we keep the ``kinetic"
term and add an interaction term $f(z_1 ,z_2)$ depending only on
the invariants of the theory. As an assumption, we take the
following action for a self-interacting gauge vector
\begin{eqnarray}
S^{I}&=& \int d^4x \, \left( \frac12 \cH^\a _\mu \tilde \cH^{\a
\,\m} - f(z_1 ,z_2) \right) \, \label{e:interpst}
\end{eqnarray}
A priori $f$ is a general function but the connection with the
free theory (\ref{e:pst}) imposes its analyticity at the origin
and its reduction to $f\rightarrow z_1$ in the weak field limit.
We are going to restrict the class of possible interactions by
demanding the field equations as well as the action to remain
invariant (up to a boudary term) under some modified
transformations of type (\ref{e:ginv1})-(\ref{e:ginv3}). In fact
we deform only the gauge symmetry\footnote{The other two gauge
symmetries remain the same and play the same role as in the free
model. The deformed action (\ref{e:interpst}) is trivially
invariant under these two transformations.} (\ref{e:ginv2}) to
\begin{eqnarray}
\delta A_\mu^\a = \cL^{\b\a} (\cH_\mu^\b - {\cal K}_\mu^\b
)\frac{\f}{\sqrt{-u^2}}\,, &\,\delta a=\f \,,\label{e:defginv2}
\end{eqnarray}
where we denoted the deforming contribution as ${\cal K}_\mu^\b =
\frac{\d f}{\d \tilde \cH^{\b \,\m}}\,.$ In the free limit case
the last transformation is nothing but (\ref{e:ginv2}), as it
should be.

Using an approach similar to the free case (i.e. gauge-fixing
(\ref{e:ginv3}) the solution to the equations of motion determined
by $S^I$ read
\begin{eqnarray}
\cH^\a_\m = {\cal K}^\a_\m = f_1 \tilde \cH^\a_\m + f_2 (\tilde
\cH^3)_\m^\a \,,\label{e:defduality}
\end{eqnarray}
where $f_i = \frac{\partial f}{\partial z_i}$ for $i=1,2$ and
$(\tilde \cH^3)_\m^\a = \tilde \cH^\b_\m \tilde \cH^{\b \,\n}
\tilde \cH^\a_\n$. The general variation of this equation is very
intricate but we can make use of the other symmetries of the
system (Lorentz and SO(2) rotation invariance) to choose a basis
in which the vector $u_\m = \d_\m^0$ (i.e. it is time-like) and
the only non-vanishing components of the tensor $\tilde \cH^\a_\m$
are
\begin{equation}
\tilde \cH^1_{1} = \l_+ \,, \quad\quad  \tilde \cH^2_{2} = \l_-\,.
\end{equation}
In this special choice the equation of motion reduces to only
\begin{eqnarray}
\g_{\pm} = \l_{\pm} (f_1 + f_2 \l_\pm^2)\label{e:e.o.m.}
\end{eqnarray}
where $\g_+ =\cH^1_{1}$, $\g_- =\cH^2_{2}$ are the non-zero
components of $\cH^\a_\m$ in this basis. After using the field
equations (\ref{e:defduality}), the variation (\ref{e:defduality})
takes the form
\begin{equation}
\d \cH^1_{3} = \l_-,\quad\d \tilde\cH^1_{3} = \g_-.
\end{equation}
The variation of the e.o.m. (\ref{e:e.o.m.}) is then
\begin{displaymath}
\l_- = f_1 \g_-  + f_2 \g_- \l_+^2 \,,
\end{displaymath}
which upon using once more the equations of motion gives
\begin{equation}\label{e:C-H1}
(f_1 + f_2 \l^2_+)(f_1 + f_2 \l^2_-) = 1\,.
\end{equation}
We can reformulate it in terms of the two invariants $z_1 =
\frac{1}{2} (\l^2_+ + \l^2_- )$ and $z_2 = \frac{1}{4}(\l^4_+ +
\l^4_-)$. Eventually performing the transformation
(\ref{e:vartrans}) we end up with the Courant-Hilbert equation
(\ref{duality}). Thus, we derived this equation by imposing the
invariance of the equations of motion under the gauge
transformation (\ref{e:defginv2}).

We now briefly show that the action is also invariant (up to total
derivatives) under the same gauge transformations
(\ref{e:defginv2}). Inserting (\ref{e:defginv2}) in the general
variation of the action
\begin{eqnarray}
\d S^I &=&  \int d^4x \, \left[ ( \cH^\a_\m - {\cal K}^\a_\m)
\d_{A,a} \tilde \cH^{\a \,\m} +\frac{1}{2} \d_a \cH^\a_\m \tilde
\cH^{\a \,\m} - \frac{1}{2} \cH^\a_\m \d_a \tilde \cH^{\a \,\m}
\right]\nn
\end{eqnarray}
and, after canceling the contributions of  first-order in the
derivatives of $f$ coming from the variation with respect to
$A_\mu^\a$, respectively $a$, we deduce
\begin{eqnarray}
\d S^I &=& \frac{1}{2} \int d^4x \, \e^{\m\n\l\rho} \cL^{\b\a} \d
v_\mu v_\nu \left( \tilde\cH^\a_{\l} \tilde\cH^\b_{\rho} - {\cal
K}^\a_{\l} {\cal K}^\b_{\rho} \right)\,.\nonumber
\end{eqnarray}
We evaluate this variation in the special basis mentioned above
and get
\begin{equation}
\d S^I = - 2 \int d^4x\, \d v_3 \,\l_+ \l_- \,[(f_1 + f_2
\l^2_+)(f_1 + f_2 \l^2_-)-1]\,.
\end{equation}
The restriction (\ref{e:C-H1}) is sufficient to guarantee the
invariance of both, field equations and action, since upon
applying the Courant-Hilbert equation (\ref{duality}) we get $\d
S^I =0$.

To conclude, we have constructed a modified Lorentz covariant
theory that also possesses a manifest EM duality. The allowed
self-interactions are restricted to a class of functions of two
variables that must satisfy the Courant-Hilbert equation. Our
analysis straightforwardly generalizes the PST formulation of BI
electrodynamics \cite{Bandos:1997} to any non-linear theory of a
single gauge vector in four dimensions.

%%%%%%%%%%%%%%Appendices%%%%%%%%%%%%%%%%%%%%%%%%%%%

%%%epsilon tensor defined as \varepsilon => check everywhere in the file%%%%

%%%definitions in boldfont => replace \it by \bf everywhere%%%

\appendix

%%%%%%%%%%%%%%%%%%%%%%%%%%%%%%%%%%%%%%%%%%%%%%%%%%%%%%%%%%%%%%%%%%
%%%%%%%%%%%%%%%%%%%%%%%%%%%%%%%%%%%%%%%%%%%%%%%%%%%%%%%%%%%%%%%%%%

\chapter{Algebraic preliminaries}\label{preliminaries}

\section{Fundamental structures}

A {\bf ring} $\Lambda$ is a set endowed with two internal laws $+$
and $\cdot$ for which\footnote{The following basic definitions are
taken from \cite{Lang}.}
\begin{description}
  \item[-] $(\Lambda,+)$ is an Abelian group
  \item[-] $\cdot$ is associative and admits a unit element $1$
  \item[-] $\cdot$ is distributive with respect to $+$
\end{description}
One speaks of a {\bf commutative} ring if $\cdot$ is commutative.
A ring $\mathbb K$ is a {\bf field} if (i) the unity $0$ for the
addition is distinct from the unity $1$ for the multiplication,
and (ii) $(\mathbb K_0,\cdot)$ is a commutative
group.\footnote{For instance let $n$ be in ${\mathbb N}_0$, then
$n\mathbb Z$ is a (commutative) ring only for $n=1$ (such that
$1\in n\mathbb Z$) but ${\mathbb Z}_n\equiv {\mathbb Z}/n{\mathbb
Z}$ is always a commutative ring under the congruence-class
addition and multiplication. Furthermore, when $n$ is a prime
integer ${\mathbb Z}_n$ is a field. Other familiar examples of
fields are ${\mathbb Q}$, ${\mathbb R}$ and ${\mathbb C}$ (under
the usual addition and multiplication laws).}

A (left) {\bf module} over $\Lambda$ is an (additive) Abelian
group $(V,+)$ together with an external law $\Lambda\times
V\rightarrow V:(\lambda, x)\rightarrow \lambda x$ which
\begin{description}
  \item[-] is distributive with respect to the addition in $V$
  \item[-] is distributive with respect to the addition in $\Lambda$
  \item[-] is associative: $(\lambda \mu) x=\lambda(\mu x)$
  \item[-] satisfies $1 x=x$.
\end{description}
A module $V$ over a field $\mathbb K$ is called a {\bf vector
space} (or, simply, a space). An {\bf algebra} $A$ over $\Lambda$
is a module over $\Lambda$ endowed with an internal law $*:A\times
A\rightarrow A$ which is a bilinear map.

Let $V$ be a vector space over $\mathbb K$. The set $End(V)$ of
linear transformations $V\rightarrow V$ is a vector space over
$\mathbb K$, whose elements are called {\bf endomorphisms}. The
invertible endomorphisms are called {\bf automorphisms}.

If $V$ is a vector space over $\mathbb R$, we say that an
automorphism $^*:V\rightarrow V: x\rightarrow x^*$ is said to be
an {\bf involution} if it is its own inverse, i.e. $(x^*)^*=x$ for
any $x$ in $V$. We call the equation $x=x^*$ a {\bf self-duality
condition}. Further, if $V$ is an algebra, the endomorphism $^*$
is an involution if it obeys the additional property:
$(x\,*\,y)^*=y^*\,*\,x^*$, $\forall x,y\in V$. The self-duality
condition defines a well-defined subspace (subalgebra) of
$V$.\footnote{The most simple example of involution is the
conjugation operation in the commutative algebra $\mathbb C$ over
$\mathbb R$. The self-duality condition is the reality condition
which defines the subalgebra ${\mathbb R}\subset{\mathbb C}$.}

\section{Grading}

Let $\{V_i\}_{i\in G}$ be a family of modules indexed by the
Abelian group $(G,+)$. The direct sum $V=\oplus_i V_i$ is called
the $G$-{\bf graded module} associated with the family
$\{V_i\}_{i\in G}$. An element $x$ which belongs to one of the
$V_i$ is said to be {\bf homogeneous}.

The algebra $A$ is $G$-{\bf graded}\footnote{The subsequent set of
definitions is taken from \cite{Henneauxbook}, chapter 8.} if (i)
it splits as $A=\oplus_{n}A_n$, (ii) the unit belongs to $A_0$,
(iii) the multiplication is such that $A_n *A_m\subset A_{n+m}$.
If $x\in A_n$, one says that the {\bf degree} of $x$ is equal to
$n\in G$.

The set $End(A)$ of endomorphisms $A\rightarrow A$ is an
associative algebra with unit. The grading of $A$ endows $End(A)$
with a natural grading. An endomorphism $N$ has degree $deg(M)$
iff for any homogeneous $x\in A$ \ba
deg(Mx)=deg(M)+deg(x).\nonumber\ea

A ${\mathbb Z}_2$-graded associative algebra $A=A_0\oplus A_1$ is
said to be {\bf supercommutative} if \ba
x*y=(-)^{\epsilon_x\epsilon_y}y*x\,,\nonumber\ea where the degree
is denoted by $\epsilon$ and is usually called the (Grassmann)
{\bf parity}.

For a ${\mathbb Z}_2$-graded vector space $V$ the ordinary
multiplication of endomorphisms preserves the grading of $End(V)$,
$\epsilon_{N_1 \,N_2}=\epsilon_{N_1}+\epsilon_{N_2}$. Given two
endomorphisms $N_1$ and $N_2$, their {\bf graded commutator}
$[N_1,N_2]$ is defined by \ba
[N_1,N_2]=N_1N_2-(-)^{\epsilon_{N_1}\epsilon_{N_2}}N_2 N_1
\label{commutator}\nn\ea The commutator (\ref{commutator}) endows
$End(V)$ with a graded Lie algebra structure (for which a graded
version of the Jacobi identity is used).

\section{Complexes}\label{complexes}

A {\bf differential complex} is defined to be a $\mathbb Z$-graded
module $V=\oplus_i V_i$ with a nilpotent endomorphism $d$ of
degree $\pm 1$, that is there is a chain of linear transformations
$d_i$
\ba\ldots\stackrel{d_{i-2\deg(d)}}{\longrightarrow}V_{i-\deg(d)}\stackrel{d_{i-\deg(d)}}{\longrightarrow}V_i\stackrel{d_i}{\longrightarrow}V_{i+\deg(d)}\stackrel{d_{i+\deg(d)}}{\longrightarrow}\ldots\nn\ea
such that $d_i\circ d_{i-\deg(d)}=0$, $\forall i$. The sequence is
said to be {\bf exact} if for all $i$ the kernel of $d_i$ is equal
to the image of its predecessor $d_{i-\deg{d}}$.

Let $(V,d)$ and $(W,\d)$ be two differential complexes with
grading $V=\oplus_i V_i$ and $W=\oplus_j W_j$. A {\bf morphism of
complexes} is a sequence \ba f_i:V_i \rightarrow W_j\nn\ea of
homomorphisms such that for all $i$ the following square commutes:
$$
\xyma{ V_i      \ar[r]^{f_i} \ar[d]^{d_i} &W_j
                             \ar[d]^{\d_j}\\
V_{i+\deg(d)}        \ar[r]^{f_{i+\deg(d)}}       &W_{j+\deg(\d)}
}
$$

Let $A$ be a supercommutative algebra, of particular interest are
the elements $D\in End(A)$ that obey the Leibnitz rule, \ba
D(x*y)=(Dx)*y+(-)^{\epsilon_D\epsilon_x}\,x*(Dy)\nn\ea These
endomorphisms are called {\bf derivations}. We adopt the left
action for derivations. A {\bf differential} $d$ is an odd
derivation ($\epsilon_d=1$) that is nilpotent of order two,
($d^2=0$). A {\bf graded differential algebra} is a graded
supercommutative algebra with a differential $d$ such that
$|\deg(d)|=1$.

An {\bf $N$-complex} is defined\footnote{One refers to
\cite{Dubois-Violette:2000} for a detailed exposure on generalized
differential complexes.} as a graded complex $V=\oplus_i V_i$
equipped with an endomorphism $d$ of degree $\pm 1$, that is
nilpotent of order $N$: $d^N=0$. For a generalized differential
algebra, the operator $d$ is called its {\bf $N$- differential}.

Let $(V,d)$ and $(W,\d)$ be two differential complexes with
$\deg(d)=\deg(\d)=\pm 1$ related by a complex morphism
$f:V\rightarrow W$. The graded module $U= \oplus_i U_i$ with
$U_i\equiv V_i\times W_{j\mp 1}$ equipped with \ba \Delta :
U_i\rightarrow U_{i\pm 1}:(v,w)\rightarrow \left(dv,f(v)-\d
w\right)\,,\nn\ea is a differential complex $Cf\equiv(U,\Delta)$,
that is called the {\bf mapping cone} of the complex morphism $f$.
The mapping cone of the identity $Id:V\rightarrow V$ is called
{\bf cone} of $V$ and is denoted by $CV$.\footnote{These two
definitions and the proposition \ref{coniso} are taken from
\cite{Dold}, chapter II.}

The following straightforward proposition is a mere application of
the definitions. This proposition is used in section
\ref{toutestrelatif}.
\begin{proposition}\label{relalgebra}Let $(A,d)$ be a differential algebra. Let $B=CA$ be the cone of $A$.
We equipped  this graded space $B= \oplus_m B_m$ (with $B_m\equiv
A_m\otimes A_{m-\deg(d)}$) by the external map $\star:B_m\otimes
B_n\rightarrow A_{m+n-\deg(d)}$ defined by \ba
(x,y)\star(z,t)=x*t+(-)^{(\epsilon_x+1)(\epsilon_z+1)}z*y\,.\nn\ea
This map obeys a kind of Leibnitz rule since \ba
&(x,y)\star(z,t)=(-)^{(\epsilon_x+1)(\epsilon_z+1)} (z,t)\star
(x,y)\,,&\nn\\ & d\left[
(x,y)\star(z,t)\right]=\left[\D(x,y)\right]\star (z,t)
+(-)^{(\epsilon_x+1)}\, (x,y)\star \left[\D (z,t)
\right]\,.&\nn\ea\end{proposition}

\section{(Co)homology}

Let $(V,d)$ be differential complex. One can define the quotient
$H(d)\equiv\frac{\mbox{Ker}_d}{\mbox{Im}_d}$ called the {\bf
cohomology} if $\deg(d)=+1$, and {\bf homology} if $\deg(d)=-1$.
This complex inherits the grading of $V$. The elements of $H(d)$
are called ({\bf co}-){\bf cycles}. An element of $\mbox{Ker}d$
that is not in $\mbox{Im}d$ is called a {\bf non-trivial}
(co)-cycle. Elements of $\mbox{Im}d$ are said to be exact trivial
or {\bf exact} (co)-cycles.

Let $A$ be a graded differential algebra with respect to two
differentials $d$ and $\delta$. If these two differentials
anticommute (i.e. $[d,\delta]=0$) we can define the {\bf
cohomology} of $d$ {\bf modulo} $\delta$. This cohomology is
denoted by $H(d|\delta)$ and defined as
\[
H(d|\delta)=\left\{ \a\in A\,\,|\,\, d\a+\delta \b=0,\,\,
\a\sim\a+d\rho+\d\sigma \right\}.
\]

The {\bf generalized cohomology} of the $N$-complex $V$ is the
family of graded modules $H_{(k)}(d))$ with $1\leq k \leq N-1$
defined by $H_{(k)}(d)=\mbox{Ker}(d^k)/\mbox{Im}(d^{N-k})$, i.e.
$H_{(k)}(d)=\oplus_iH^i_{(k)}(d)$ where
\[
H^i_{(k)}(d)=\left\{ \a\in V^i\,\,|\,\, d^k\a=0,\,\,
\a\sim\a+d^{N-k}\b,\,\,\b\in V^{p+k-N} \right\}.
\]

A straightforward proposition ends this section.
\begin{proposition}\label{coniso}Let $(V,d)$ and $(W,\d)$ be two differential
complexes. Let $Cf\equiv(U,\Delta)$ be the cone of the complex
morphism $f:V\rightarrow W$. Then \ba H(d)\cong
H(\d)\quad\Leftrightarrow\quad H(\Delta)=0\,.\nn\ea In particular,
$H(d)$ and $H(\d)$ are trivial if and only if $H(\Delta)$ is
trivial.
\end{proposition}

%%%%%%%%%%%%%%%%%%%%%%%%%%%%%%%%%%%%%%%%%%%%%%%%%%%%%%%%%%%%%%%%%%
%%%%%%%%%%%%%%%%%%%%%%%%%%%%%%%%%%%%%%%%%%%%%%%%%%%%%%%%%%%%%%%%%%

\chapter{Differential form toolkit: definitions, conventions and all that}\label{difform}

%%%%%%%%%%%%%%%%%%%%%%%%%%%%%%%%%%%%%%%%%%%%%%%%%%%%%%%%%%%%%%%%%%%%%%%%

\section{Wedge product}

Let ${\cal M}$ be a manifold of dimension $D$ with coordinates
$x^\mu$. The space of {\bf differential forms} on ${\cal M}$ is
denoted by $\Omega({\cal M})$ and is the graded vector space over
$\mathbb R$ of (covariant) antisymmetric tensor fields on ${\cal
M}$ with graduation induced by the tensor rank. A {\bf non
homogeneous} differential form $\a$ is an element of $\Omega({\cal
M})$ which is the sum of forms with different form degrees.
Obviously, one has $\Omega^p({\cal M})=0$ for $p>D$.

Let $\{1,dx^\m, dx^\m\wedge dx^\n,\ldots\}$ be a basis of the
exterior algebra $\Lambda{\mathbb R}^D$. The {\bf components} of a
$p$-form in $\Omega^p({\cal M})$ are defined through
\begin{equation}
\alpha=\frac{1}{p!}\,\alpha_{\mu_1\ldots\mu_p}(x)\,dx^{\mu_1}\wedge\ldots\wedge
dx^{\mu_p},\quad \alpha\in\Omega^p({\cal M}).
\end{equation} The definition of the {\bf wedge product} $\wedge$ is self-explanatory from the notation used
for the basis of $\Lambda {\mathbb R}^D$. With this product,
$\Omega({\cal M})$ becomes a supercommutative algebra graded since
\begin{equation} \a\wedge\b = (-)^{pq}\b\wedge\a\quad
\alpha\in\Omega^p({\cal M}),\,\b\in\Omega^q({\cal M})\,.
\end{equation}

\section{Pull-back}

Let $f:{\cal N}\rightarrow{\cal M}$ be a smooth map between the
two manifolds ${\cal N}$ and ${\cal M}$ with respective
coordinates $y^m$ and $x^\m$. If, in local coordinates, $f$ is
defined by $x^\m(y^m)$, the {\bf pull-back} $(i^*\a)$ of a
$p$-form $\a \in\Omega^p({\cal M})$ writes in
components\footnote{In this thesis, we frequently make an abuse of
notation by not doing an explicit distinction between a form and
its pull-back when the context makes it clear.}
\begin{equation}
(i^*\alpha)_{m_1\ldots m_p}=\alpha_{\m_1\ldots
\m_p}\left(x(y)\right)\,\frac{\partial x^{\m_1}}{\partial
y^{m_1}}\ldots\frac{\partial x^{\m_p}}{\partial y^{m_p}}\,.
\end{equation} It fulfills $i^*(\a\wedge\b)=i^*\a\wedge i^*\b$ for arbitrary $\a,\b\in\Omega({\cal M})$

\section{Exterior derivative}

We introduce the {\bf exterior derivative} $d$ defined as the
composition of the derivative operator $\partial$ with
antisymmetrization: \begin{equation} d\equiv { \bf
A}_{p+1}\circ\partial:\Omega^p({\cal
M})\rightarrow\Omega^{p+1}({\cal M})\label{dmap}
\end{equation} Due to the previous definitions, we can write $d=
dx^\mu\frac{\partial}{\partial x^\mu}$. Since $d$ is a
differential of degree $1$, the graded algebra $\Omega^*({\cal
M})$ is endowed with a differential algebra structure and is
called the {\bf de Rham complex}. The cohomology $H^*({\cal M})$
of the differential $d$ in the space $\Omega^*({\cal M})$ is
called the de Rham cohomology.

From the definitions, it follows that the components of $d\a\in
\Omega^{p+1}$ are
\begin{equation}
(d\a)_{\mu_1\ldots\mu_{p+1}}=(p+1)\,\partial_{[\mu_1}\a_{\mu_2\ldots\mu_{p+1}]}\,,
\end{equation}
where the {\bf complete antisymmetrization} over the indices is
denoted by the brackets; the weight is taken to be one, that is
\begin{equation}
\alpha_{\mu_1\ldots\mu_p}\equiv
\frac{1}{p!}\left(\alpha_{\mu_1\ldots\mu_p}+\rm{even}\,\rm{
perms}-\rm{odd}\,\rm{perms}\right),
\end{equation}
in such a way that:
$\alpha_{\mu_1\ldots\mu_p}=\alpha_{[\mu_1\ldots\mu_p]}$ for an
antisymmetric tensor. It can be explicitly checked that the
pull-back map \be i^*:\Omega^p({\cal M})\rightarrow \Omega^p({\cal
N})\ee is a morphism of complexes, i.e. $d\circ i^*=i^*\circ d$.

%%%%%%%%%%%%%%%%%%%%%%%%%%%%%%%%%%%%%%%%%%%%%%%%%%%%%%%%%%%%%%%%%%

\section{Epsilon densities and integration}

If ${\cal M}$ is an orientable manifold of dimension $D$, one can
define two numerically invariant tensor densities by
\be \epsilon_{\mu_1\ldots\mu_D}=\epsilon^{\mu_1\ldots\mu_D}=\left\{\begin{array}{c} 1 \quad\mbox{for any even permutation of the set }1,\ldots,D\\
0 \quad\mbox{if two indices are equal}\\
-1 \quad\mbox{for any odd permutation of the set }1,\ldots,D
\end{array}\right.\label{defeps}\ee
In form notation, the tensor density $\epsilon_{\mu_1\ldots\mu_D}$
gives the components of the ``volume element" $\epsilon=d^Dx$
(which is \emph{not} a differential form because its components
are densities). Since ${\cal M}$ is orientable, this $D$-form is
globally well-defined.

The two tensors densities naturally define the determinant of any
matrix $A$. For instance, \be
\epsilon_{\b_1\ldots\b_D}\det(A^\a_{\,\,\b})=\epsilon_{\a_1\ldots\a_D}A^{\a_1}_{\,\,\b_1}\ldots
A^{\a_D}_{\,\,\b_D}\label{det}\ee If we insert the Jacobian matrix
$J$ into (\ref{det}) one can immediately see that
$\epsilon_{\mu_1\ldots\mu_D}$ is a tensor density of weight one.
Analogously, inserting $A=J^{-1}$ one obtains that
$\epsilon^{\mu_1\ldots\mu_D}$ is a tensor density of weight minus
one.

All this provides the integration measure necessary to define the
{\bf integration} of $D$-forms on orientable manifolds ${\cal M}$.
Let $\a\in \Omega^D({\cal M})$ be equal to
\be\a=\a(x)\,\epsilon\,,\ee where $\a(x)$ is a scalar density of
weight minus one defined on $U\subset {\mathbb R}^D$, the
definition domain of the coordinates system on ${\cal M}$. The
integral of the differential form $\a$ on ${\cal M}$ is defined as
\be \int\limits_{\cal M}\a\equiv \int\limits_U\a(x) \,d^Dx\ee The
integral of a non homogeneous form picks up the component of
degree $D$ and integrates it according to the previous definition.

%%%%%%%%%%%%%%%%%%%%%%%%%%%%%%%%%%%%%%%%%%%%%%%%%%%%%%%%%%%%%%%%%%

\section{Epsilon tensors and volume form}

From now on ${\cal M}$ is taken to be a smooth \emph{orientable}
manifold of dimension $D$. Let a metric $g_{\mu\nu}$ be defined on
${\cal M}$ with $t$ timelike directions. The metric enables us to
define a tensor from the epsilon densities since
$|g|=(-)^t\det(g_{\mu\nu})$ is a (strictly positive) scalar
density of weight minus two. The {\bf Levi-Civita tensor} is
defined by
\begin{equation}\label{LeviCivita}
\varepsilon_{\mu_1\ldots\mu_D}=\sqrt{|g|}\,\epsilon_{\mu_1\ldots\mu_D}\,.
\end{equation}
The Levi-Civita tensor defines the (coordinate invariant) {\bf
volume form} \be \varepsilon=\sqrt{|g|}\,d^Dx\,.\ee Combining
(\ref{defeps}) with (\ref{det}) one get the contravariant form of
the Levi-Civita tensor
\begin{equation}
\varepsilon^{\mu_1\ldots\mu_D}=\frac{(-)^t}{\sqrt{|g|}}\,\epsilon^{\mu_1\ldots\mu_D}\,.
\end{equation}

The {\bf Kronecker symbols} are defined by \be
\d_{\mu_1\ldots\mu_p}^{\nu_1\ldots\nu_p}\equiv
\d^{[\nu_1}_{\mu_1}\ldots\d^{\nu_p]}_{\mu_p}=\d^{\nu_1}_{[\mu_1}\ldots\d^{\nu_p}_{\mu_p]}\ee
They are numerically invariant tensors which define projectors
onto the completely antisymmetric representation. For instance,
\be
\d_{\mu_1\ldots\mu_p}^{\nu_1\ldots\nu_p}A_{\nu_1\ldots\nu_p}=A_{[\mu_1\ldots\mu_p]}\,.\ee
It follows directly from the definition of the $\epsilon$
densities the extremely useful identity
\be\varepsilon_{\mu_1\ldots\mu_D}\,\varepsilon^{\nu_1\ldots\nu_D}=(-)^t\,
D!\,\,\d_{\mu_1\ldots\mu_D}^{\nu_1\ldots\nu_D} \,,\ee which is
most usually used with some contraction of indices
\be\varepsilon_{\mu_1\ldots\mu_p\,\rho_1\ldots\rho_{D-p}}\,\varepsilon^{\nu_1\ldots\nu_p\,\rho_1\ldots\rho_{D-p}}=(-)^t\,
p\,!\,(D-p)!\,\d_{\mu_1\ldots\mu_p}^{\nu_1\ldots\nu_p}\,.\label{also}\ee

%%%%%%%%%%%%%%%%%%%%%%%%%%%%%%%%%%%%%%%%%%%%%%%%%%%%%%%%%%%%%%%%%%

\section{Hodge $*$ operator}

The {\bf Hodge $*$} or {\bf dual} is a map from $p$-forms to
$(D-p)$-forms \be *:\Omega^p({\cal {\cal M}})\rightarrow
\Omega^{D-p}({\cal {\cal M}})\label{defHodge}\ee acting as
\begin{equation}
*(dx^{\mu_1}\wedge\ldots\wedge
dx^{\mu_p})\equiv\frac{1}{(D-p)!}\,\varepsilon_{\nu_1\ldots\nu_{D-p}}
^{\quad\quad\quad\mu_1\ldots\mu_p}\,\,dx^{\nu_1}\wedge\ldots
\wedge dx^{\nu_{D-p}}
\end{equation}
Note in particular that taking $p=0$ in (\ref{defHodge}) gives
$*1=\epsilon$. Therefore, with the definition (\ref{LeviCivita})
one derives \be\frac{\partial( *1)}{\partial g^{\mu \nu
}}=-\frac12\, g_{\m\n}\,(*1)\,,\label{pardetg}\ee since $\partial
g/\partial g^{\m\n}=-\,g\,g_{\m\n}$.

In terms of components the definition (\ref{defHodge}) translates
into \be
(*\a)_{\mu_1\ldots\mu_p}=\frac{1}{(D-p)!}\,\varepsilon_{\mu_1\ldots\mu_D}\a^{\mu_{p+1}\ldots\mu_{D-p}}\label{*a}
\ee Apply $*$ twice on a $p$-form, we get \be
*^2=(-)^{p(D-p)+t}\,.\label{*2}\ee As a consequence we find
\begin{equation}
dx^{\mu_1}\wedge\ldots\wedge
dx^{\mu_D}=(-)^t\,\varepsilon^{\mu_1\ldots\mu_D}\,*1\,.
\end{equation}

The Hodge dual definition implies the following identity for two
arbitrary $p$-forms $\alpha$ and $\beta$
\begin{equation}\label{a*b}
*\alpha \wedge \beta=*\beta \wedge
\alpha=\frac{1}{p!}\,\alpha_{\mu_1 \cdots \mu_p}\beta^{\mu_1
\cdots \mu_p}\,*1\,.
\end{equation}Therefore \be *\a\wedge\b=(-)^{p(D-p)}\a\wedge
*\b\,,\ee which, combined with (\ref{*2}), gives\be
(D-p)!\,\alpha_{\mu_1 \cdots \mu_p}\beta^{\mu_1 \cdots
\mu_p}=(-)^t\,p!\,(*\alpha)_{\mu_1 \cdots
\mu_{D-p}}(*\beta)^{\mu_1 \cdots \mu_{D-p}}\,.\label{niceid}\ee
This also follows from (\ref{also}) and the definition (\ref{*a}).

If the components of $\a$ and $\b$ are assumed not to depend
explicitly on the metric, then \be\frac{\partial(\alpha_{\r_1
\cdots \r_p}\beta^{\r_1 \cdots \r_p})}{\partial g^{\mu \nu
}}=p\,\,\alpha_{(\m|\r_2 \cdots \r_p|}\,\beta_{\n)}^{\,\,\r_2
\cdots \r_p}\,.\label{parab}\ee We get from (\ref{pardetg}) and
(\ref{a*b}) the useful identity
\be\frac{\partial(*\alpha\wedge\beta)}{\partial g^{\mu \nu
}}=*\alpha_{(\m}\wedge\beta_{\n)} - \frac12\,
g_{\m\n}\,(*\a\wedge\b)\,,\label{para*b}\ee where we introduced
the forms $\a_\m$ and $\b_\n$ defined in the same way by\be
\a_\m=\frac{1}{(p-1)!}\,\alpha_{\mu\r_1\ldots\r_{p-1}}(x)\,dx^{\r_1}\wedge\ldots\wedge
dx^{\r_{p-1}},\quad \alpha_\m\in\Omega^{p-1}({\cal
M}).\label{am}\ee Using carefully (\ref{niceid}) and
(\ref{para*b}) we obtain another useful identity \be
*(*\alpha)_{(\m}\wedge(*\beta)_{\n)} = (-)^t\, \frac12\,
g_{\m\n}\,*\a\wedge\b\,,\label{par*a*b}\ee

%%%%%%%%%%%%%%%%%%%%%%%%%%%%%%%%%%%%%%%%%%%%%%%%%%%%%%%%%%%%%%%%%

\section{Interior product}

Let $v=v^\mu\partial_\mu$ be a vector field. The interior product
of $v$ with a $p$-form is a map from $p$-forms to $p-1$-forms \be
i_v:\Omega^p({\cal {\cal M}})\rightarrow \Omega^{p-1}({\cal {\cal
M}})\ee defined by
\begin{equation}
i_v(dx^{\mu_1}\wedge\ldots dx^{\mu_p})\equiv
p\,\,v^{[\mu_1}dx^{\mu_2}\wedge\ldots \wedge dx^{\mu_p]}
\end{equation}
In terms of components the interior product of the vector field
$v$ with a $p$-form $\a$ writes
$$
(i_v\a)_{\mu_1\cdots\mu_{p-1}} = v^\mu\a_{\mu
\mu_1\cdots\mu_{p-1}}\,.
$$
As $d$, the interior product $i_v$ is a differential acting from
the left.

In this sequel of this subsection, the product between forms will
always be the wedge product and the symbol $\wedge$ will be
omitted for convenience since we are dealing with operatorial
identities. The interior product can be alternatively defined in
an operatorial way
\begin{equation}
i_v=(-)^{p(D-p+1)+t+1}\,*v*\,
\end{equation}
which holds on any $p$-form and for any vector field $v$. From
this we can derive the following operatorial
identities\footnote{All the following identities have been
inspired from a convenient set used in \cite{Lechner:1998} to
treat the PST model in form language.}:
\begin{equation}\label{*iv}
*i_v=(-)^{D-p}\,v*\,, \quad\quad i_v*=(-)^{D-p+1}\,*v\,.
\end{equation}

Obviously we have $i_v v=g_{\mu\nu}v^\mu v^\nu\equiv v^2$. It is a
particular case of the operatorial identity
\begin{eqnarray}\label{ivv}
i_v v + v i_v= v^2\,.
\end{eqnarray}
We can make the small remark that if we take for $v$ the
differential $d$, then $i_d$ is usually denoted by $\delta$ and
corresponds to the divergence. Furthermore, in that case the
equation (\ref{ivv}) defines the Laplacian $\Delta$ ($t=0$) or the
d'Alembertian $\Box$ ($t=1$).

From (\ref{ivv}) we derive $*i_v v* + *v i_v*= (-)^{p(D-p)+t}v^2$.
The identity (\ref{*iv}) gives $*(i_v v)*=(-)^{p(D-p)+t+1} v i_v$.
Particularly useful will be the following decomposition of the
identity $I$ that follows from the previous relations
\begin{equation}\label{id}
I ={1\over v^2}[vi_v+(-)^{p(D-p)+t+1}*vi_v*]
\end{equation}

%%%%%%%%%%%%%%%%%%%%%%%%%%%%%%%%%%%%%%%%%%%%%%%%%%%%%%%%%%%%%%%%%%
%%%%%%%%%%%%%%%%%%%%%%%%%%%%%%%%%%%%%%%%%%%%%%%%%%%%%%%%%%%%%%%%%%

\chapter{Some taste of algebraic topology}\label{topology}

%%%%%%%%%%%%%%%%%%%%%%%%%%%%%%%%%%%%%%%%%%%%%%%%%%%%%%%%%%%%%%%%%%

\section{Simplicial homology}\label{shomology}

To begin with one considers the notions of singular $q$-simplex
$\Delta_q$ and boundary map to be defined already\footnote{The
material presented in this section arises from the chapter 13 of
\cite{Frankel}.}. A formal finite sum of $q$-simplexes with
coefficients in an (additive) Abelian group $G$ is a {\bf
singular} $q$-{\bf chain}. If the coefficients are $\pm 1$, the
chain is called an {\bf integer} $q$-{\bf chain}. The singular
$q$-chain group is the Abelian group of $q$-chains and is denoted
by $\Omega_q({\cal M};G)$. The boundary operator for singular
$q$-simplexes induces the boundary homomorphism \be
\partial:\Omega_q({\cal {\cal M}};G)\rightarrow \Omega_{q-1}({\cal
{\cal M}};G)\ee The boundary operator $\partial$ is nilpotent.
Therefore one can define the quotient group $H_q({\cal M
};G)\equiv\frac{\mbox{Ker}_\partial}{\mbox{Im}_\partial}$ called
the $q$-th {\bf homology group}. For instance, the manifold $\cal
M$ is path-connected if $H_0({\cal M};G)$ is generated by a single
point.

If the $q$-chain coefficents are in a subring $\L$ of $\mathbb R$,
then $\Omega_q({\cal M};\L)$ becomes a module over $\L$. Therefore
$\Omega_*({\cal M};\L)$ is a $\mathbb Z$-graded module which,
endowed with the boundary homomorphism $\partial$, becomes a
complex. For simplicity, the complex of singular chains with real
coefficients will be denoted by $\Omega_*({\cal M})\equiv
\Omega_*({\cal M};\mathbb R)$ in the sequel. A singular $q$-chain
complex allows a convenient definition of the differential
$q$-form integration such that\be \int: \Omega_q({\cal
M};\L)\otimes \Omega^q({\cal M})\rightarrow {\mathbb
R}:(\r,\a)\rightarrow \int_\r \a \ee is a bilinear map, and such
that the \emph{Stokes theorem} \be\int\limits_\s
d\a=\int\limits_{\partial\s}\a\label{Stokes}\ee is true for all
$\a\in\Omega^q({\cal M};\L)$, $\s\in \Omega_{q+1}({\cal M})$.

Let us denote $\Lambda_{q}({\cal M})$ the set of all
$q$-dimensional smooth orientable submanifolds $\cal N$ in the
manifold ${\cal M}$. The subset of closed submanifold is denoted
by $\overline{\Lambda}_{q}({\cal M})$. Of course $\Lambda_0({\cal
M})=\overline{\Lambda}_0({\cal M})={\cal M}$ and $0\leq q\leq D$.
Each submanifold $\cal N$ is defined by an embedding $i:{\cal
N}\hookrightarrow {\cal M}$.

There exist an important correspondence between closed orientable
submanifolds and singular simplexes which can be observed in the
following two algebraic topology theorems. First, every closed
oriented submanifold $\cal N$ of $\overline{\Lambda}_q({\cal M})$
defines a $q$-cycle in $H_q({\cal M};G)$ by associating the same
coefficient $g$ to each oriented $q$-triangle in a suitable
triangulation of $\cal N$. In this thesis, we consider the
specific case where $g$ is only determined by the orientation. We
use the natural map \be tri:\overline{\Lambda}_q({\cal
M})\rightarrow \Omega_q({\cal M};{\mathbb Z}_2)\ee from the subset
$\overline{\Lambda}_q({\cal M})\subset{\Lambda}_q({\cal M})$ of
closed $q$-dimensional submanifolds to the module $\Omega_q({\cal
M};{\mathbb Z}_2)$ of integer $q$-chains $\Omega_q({\cal
M};{\mathbb Z}_2)$, which is such that \be tri(\cup_i {\cal
N}_i)=\sum_i (-)^{o(i)}tri({\cal N}_i)\ee where $o(i)$ is equal to
$0$ or $1$ according to the orientation of ${\cal N}_i$. Moreover,
R. Thom has proved a converse of the previous theorem in the case
of real coefficients \cite{Frankel}. In the sequel, with the
previous theorems in mind, one will seldom make the distinction
between an orientable submanifold and its associated simplex with
integer (or real) coefficients.

%%%%%%%%%%%%%%%%%%%%%%%%%%%%%%%%%%%%%%%%%%%%%%%%%%%%%%%%%%%%%%%%%%%%%%

\section{de Rham currents}\label{current}

From now on, ${\cal M}$ is a boundaryless orientable manifold of
dimension $D$. We point out that, up to now, we did not specified
the regularity properties of the differential $p$-forms. Usually
it is assumed implicitly that their components are smooth
functions (sometimes compact support is also required). However,
for physical applications it may be convenient to work with {\bf
de Rham currents} \cite{Lechner:1999,Lechner:2000}, i.e. linear
functionals on the space of smooth differential forms with compact
support, which are continuous in the sense of
distributions\footnote{For a rigorous treatment of currents, one
refers to the chapter III of \cite{deRham}.}. Also, the space of
currents with compact support is the topological dual of
$\Omega({\cal M})$. For any practical purpose, $p$-currents are
$p$-forms with distribution-valued components. Throughout the
text, we still call them ``forms" and we use the notation
$\Omega({\cal M})$ for the space of currents. This absence of
distinction is partly justified by the fact that any closed
current is cohomologous to a smooth differential form (theorem 14
of \cite{deRham}, p.94).

In the space of de Rham current, we define the {\bf extension} map
which is a complex morphism, \be e:\Omega^p({\cal N})\rightarrow
\Omega^{p}({\cal M})\,,\ee going in the direction inverse to the
pull-back. The current $e(\a)$ is defined as the current with
support on $\cal N$ and such that its pull-back on $\cal N$ is
equal to $\a$. In other words, one has $i^*\circ
e=Id\big(\Omega({\cal N})\big)$. From now on, the action of the
map $e$ will not be explicitly stated but understood from the
context.

\subsection{Poincar\'e duality}\label{PD}

With the space of de Rham currents we can define a map \be
P:\Omega_p( {\cal M})\rightarrow \Omega^{D-p}({\cal M}):{\cal
N}\rightarrow *\a \label{Pdualmap}\ee which associates to every
$p$-dimensional manifold ${\cal N}\in\Lambda_p({\cal M})$ a
$(D-p)$-current $*\a=P({\cal N})$. If $y^m$ are coordinates on
$\cal N$, the embedding $i:{\cal N}\hookrightarrow {\cal M}$ is
locally defined by $x^\m(y^m)$. The explicit expression for the
$p$-form $\a$ is
\begin{equation}
\a^{\mu_1\ldots\mu_p}(x) = \int_{\cal N}
%\frac{1}{\sqrt{-h}}\,
\,\delta^{(D)}\left(x-x(y)\right) dx^{\mu_1} \wedge \ldots\wedge
dx^{\mu_p} \,.
\end{equation}
We call the form $P({\cal N})=*\a$ the {\bf Poincar\'e dual} of
${\cal N}$. This nomenclature is justified below.

The map $P$ satisfies an interesting number of properties:
\begin{itemize}
\item By construction, the Poincar\'e dual $*\alpha$ has its support on the corresponding manifold ${\cal
N}$ and obeys the nice identity \be *\a\wedge *\a\equiv
0.\label{nice}\ee
\item The Poincar\'e dual
$P({\cal N})=*\a$ of ${\cal N}$ is such that for all
$\b\in\Omega^p({\cal M})$.
\begin{equation}
\int\limits_{\cal M}*\a\wedge \b=\int\limits_{\cal N}
\b\,.\label{PDdef}
\end{equation}
This is a direct consequence of the property (\ref{a*b}) and the
definition of the Dirac delta\footnote{In \cite{deRham} the map
$P$ is a mere definition of the current $*\a$ associated to the
chain ${\cal N}$ (see \cite{deRham}, p.40).}.
\item If $\cal M$ is closed, the Poincar\'e dual map is a morphism of complexes, that
is \be P\circ \partial_p+(-)^{D-p}\, d_{D-p}\circ
P=0\label{Pmorphism}\ee where $\partial_p:\Omega_p({\cal
M})\rightarrow\Omega_{p-1}({\cal M})$ is the boundary map acting
on $p$-dimensional manifolds and $d_p:\Omega^{D-p}( {\cal
M})\rightarrow\Omega^{D-p+1}( {\cal M})$ is the de Rham
differential acting on ($D-p$)-forms. Consequently, there is a
well-defined Poincar\'e dual map $P^*$ induced in (co)homology.
Loosely speaking, the Poincar\'e duality is our translator between
\emph{algebraic topology} and \emph{differential geometry}
concepts.
\item Furthermore, the induced Poincar\'e
dual map provides an isomorphism \be P^*:H_p(\partial,{\cal
M})\rightarrow H^{D-p}(d,{\cal M})\ee between a certain homology
group for the boundary operator $\partial$ and a corresponding De
Rham cohomology group. This is the origin of our Poincar\'e
duality definition. This isomorphism is also valid for smooth
differential forms if $P$ is instead defined by means of
(\ref{PDdef}) (see p.44 and pp.50-52 in \cite{Bott&Tu}).
\item Let $P_{\cal M}:\Omega_p( {\cal M})\rightarrow \Omega^{D-p}({\cal M})$ be the Poincar\'e dual map on ${\cal M}$
and let $P_{\cal N}:\Omega_p( {\cal N})\rightarrow
\Omega^{q-p}({\cal N})$ be the Poincar\'e dual map on ${\cal N}
\in\Lambda_q({\cal M})$. Then \be P_{\cal M}({\cal N})\wedge
P_{\cal N }=P_{\cal M}\,.\label{wedgeP}\ee
\end{itemize}

To prove the point (\ref{nice}), let us introduce the following
definition: A form $\beta\in\Omega^p({\cal M})$ is {\bf
decomposable} if $\beta=\Phi_1\wedge\cdots\wedge\Phi_p$ for some
$\Phi_i\in\Omega^1(M)$. In other words, in an appropriate system
of coordinates $\beta$ has only one non-vanishing independent
component. The property $\beta\wedge\beta= 0$ is valid for any
decomposable form $\beta$ since we find two times the same
Grassmannian $1$-form $\Phi_i$ in the exterior product. To end the
proof of (\ref{nice}) one remarks that volume forms are
decomposable. This is of interest because the Poincar\'e dual
$*\a$ evaluated at $x\in{\cal N}$ is proportional to the
($D$-dimensional) Hodge dual of the volume form of ${\cal N}$.
Furthermore the support of $*\a$ is on the corresponding manifold
${\cal N}$.

\subsection{Intersection and linking number}\label{linking}

Let ${\cal B}$ and ${\cal C}$ be two orientable submanifolds of
${\cal M}$ with respective dimensions $p$ and $D-p$ transversally
intersecting on a finite number of points only. Let $P({\cal B})$
and $P({\cal C})$ be their respective images by the Poincar\'e
dual map. A basic consequence of Poincar\'e duality is that the
integral of $P({\cal C})$ over ${\cal B}$ is an integer, counting
the number of intersections with sign between ${\cal B}$ and
${\cal C}$. We call it the {\bf intersection number}\footnote{It
is also called Kronecker index (for more details, see
\cite{deRham}, p.101).}, it is equal to
\begin{eqnarray}
I({\cal B},{\cal C})&=&\int\limits_{\cal M}P({\cal B})\wedge P({\cal C})\label{intersection}\\
&=&(-)^{p(D-p)}I({\cal C},{\cal B})\,.
\end{eqnarray}

Linear combinations of such Poincar\'e dual $p$-forms with integer
coefficients are called {\bf integer forms}
\cite{Lechner:1999,Lechner:2000} (they are the Poincar\'e images
of integer $(D-p)$-chains). The integral of a product of two
integer forms is an integer forms, whenever the integral is well
defined.

A nice geometric application of the intersection number is the
linking number. Let $\cal A$ and $\cal B$ be two trivial cycles of
$H({\cal M},\partial)$ such that (i) ${\cal A}\cap {\cal
B}=\emptyset$, (ii) $dim({\cal A})+dim({\cal B})+1=dim({\cal M})$
and (iii) $\partial {\cal M}=0$. Let $\eta_{\cal A}=d\omega_{\cal
C}$ and $\eta_{\cal B}=d\omega_{\cal D}$ be the respective
Poincar\'e duals of ${\cal A}=\partial {\cal C}$ and ${\cal
B}=\partial {\cal D}$. The {\bf linking number} $L({\cal B},{\cal
A})$ of ${\cal B}$ and ${\cal A}$ is defined to be the
intersection number between $\cal B$ and $\cal C$. It is given by
(pp. 229-234 of \cite{Bott&Tu})
\begin{eqnarray}
L({\cal B},{\cal A})&=&\int\limits_{\cal M}P({\cal B})\wedge P({\cal C})\label{linkingn}\\
&=&(-)^{\left[dim({\cal A})+1\right]\,dim({\cal B})}L({\cal
A},{\cal B})\,.
\end{eqnarray}
A nice property is that the linking number only depends on the
(co)homology classes chosen to compute it. The linking number is
therefore a well-defined \emph{topological} quantity since it only
depends on what we will call the {\bf linking homology class} of
$\cal A$ (the same is true for $\cal B$ of course), which is
defined by the equivalence relation \be {\cal A}\sim {\cal A}
+\partial {\cal K}\,,\quad I({\cal K},{\cal
B})=0\,.\label{equivr}\ee These properties can be checked easily
using the Stokes theorem together with the previous assumptions.

\section{``Everything is relative"}\label{toutestrelatif}

\subsection{Relative Homology}

The topological cycles that we have been involved with so far are
called {\bf absolute cycles}.\footnote{The section title quotes a
by-now famous sentence made by an amateur violin player who worked
as a third class technical expert at the Swiss Patent Office in
Bern. Lastly, he became famous for his early anticipation of
post-modern relativism.} Given ${\cal N}\in\Lambda_n({\cal M})$ a
submanifold of $\cal M$ without boundary we can define a {\bf
relative} $q$-{\bf cycle} to be a $q$-chain on $\cal M$ whose
boundary, if there is one, lies on $\cal N$. Every absolute cycle
is of course a relative cycle. We shall say that two relative
$q$-cycles are {\bf homologous} provided they differ by an
absolute boundary plus, perhaps, a $q$-chain that lies wholly on
$\cal N$,\be \a \sim\a+\partial\sigma-\rho\ee where
$\a\in\Omega_q({\cal M})$, $\rho\in\Omega_q({\cal N})$ and
$\sigma\in\Omega_{q+1}({\cal N})$. In other words a {\bf relative}
$q$-{\bf boundary} is an absolute $q$-boundary plus any $q$-chain
on $\cal N$ (The previous definitions are taken from
\cite{Frankel}).\footnote{A simple example arises in string
theory. Let $\cal N$ be a D$q$-brane and $\cal M$ the space. A
string with both ends sitting on the D$q$-brane is a relative
$1$-cycle homologous to a closed string living in the bulk. }

A more formal definition\footnote{The related set of definitions
and proposition is inspired from an illuminating explanation of J.
Kalkkinen on the use of relative (co)homology in $M$-theory,
followed by the reading of \cite{Figueroa-O'Farrill:2000}.} uses
the mapping cone definition. The inclusion map $i:{\cal
N}\hookrightarrow{\cal M}$ induces an obvious complex morphism
$i_*:\Omega({\cal N})\rightarrow \Omega({\cal M})$. The space of
relative $q$-chains is defined as $\Omega_q({\cal N},{\cal
M})\equiv \Omega_{q-1}({\cal N})\times\Omega_q({\cal M})$. The
graded complex $\Omega_*({\cal N},{\cal M})=\oplus_q\Omega_q({\cal
N},{\cal M})$ is equipped with the ``relative boundary operator"
$\partial$ defined by \be
\partial(\rho,\sigma)=(\partial\rho,i_*\rho-\partial\sigma)\,,\quad
\forall\rho\in \Omega_{q-1}({\cal N})\,,\,\forall\sigma\in
\Omega_q({\cal M})\,.\ee The interest of the relative chains is
that they keep track of the information on $\cal N$ also. It can
be checked that the {\bf relative homology group} $H_*({\cal
N},{\cal M})\equiv\frac{\mbox{Ker}_\partial}{\mbox{Im}_\partial}$
fits in the previous definitions.

There is an analogue construction for differential forms: the
relative de Rham cohomology\footnote{It is a particular case of
relative cohomology in the sense of \cite{Bott&Tu}, see pp.
78-79.}.

\subsection{Relative de Rham complex}

Following the mapping cone construction with the pull-back complex
morphism $i^*:\Omega^p({\cal M})\rightarrow \Omega^p({\cal N})$,
one defines the {\bf relative de Rham} complex $\Omega^*({\cal
M},{\cal N})=\oplus_p \Omega^p({\cal M},{\cal N})$ with
$\Omega^p({\cal M},{\cal N})\equiv\Omega^p({\cal
M})\otimes\Omega^{p-1}({\cal N})$. The nilpotent operator $d$ is
defined by \be d(\b,\a)=(d\b,i^*\b-d\a)\,,\quad \forall\b\in
\Omega^p({\cal M}),\,\forall\a\in \Omega^{p-1}({\cal N})\,.\ee The
relative de Rham cohomology $H^*({\cal M},{\cal N})$ is the module
of {\bf relative cocycles}, i.e. closed forms $\b$ on $\cal M$
which becomes exact when pulled-back on $\cal N$, \be
d\b=0\,,\quad i^*\b=d\a\,,\ee with the equivalence relation \ba
&&\b\sim\b+d\mu\,,\nn\\&&\a\sim\a+i^*\m-d\rho\,.\ea

The inclusion map $i:{\cal N}\hookrightarrow{\cal M}$ induces the
useful {\bf relative pull-back} from the cone $C\Omega^*({\cal
M})$ of $\Omega^*({\cal M})$ to the relative de Rham complex
$\Omega^*({\cal M},{\cal N})$ \be i^*:C\Omega^p({\cal
M})\rightarrow {\Omega}^p({\cal M},{\cal
N}):(\a,\b)\rightarrow(\a,i^*\b)\,.\ee The relative pull-back
provides a useful link with the relative differential forms. More
precisely, it is a morphism of complexes.

\subsection{Integration on relative chains}

A natural definition of the {\bf integration} of a relative
$q$-form $(\a,\b)$ on a relative $q$-chains $(\rho,\sigma)$ is
\be\int\limits_{(\rho,\sigma)}(\a,\b)=\int\limits_\sigma\a-\int\limits_\rho\b\,.\label{naturalint}\ee
In fact, this definition is natural because it obeys the analogue
of Stokes theorem\footnote{The minus sign is irrelevant and
originates from choice of orientation.} (\ref{Stokes})
\be\int\limits_{(\rho,\sigma)}d(\a,\b)=-\int\limits_{\partial(\rho,\sigma)}(\a,\b)\,.\ee
The integration is a bilinear map \be \Omega_*({\cal N},{\cal
M})\times\Omega^*({\cal M},{\cal N})\rightarrow\mathbb R\,.\ee

We define the complex morphism \be ext:\Omega^p({\cal
M})\rightarrow \Omega^{D+p-n}({\cal M}):\a\rightarrow \a\wedge
P({\cal N})\,.\label{ext}\ee The mapping cone of $ext\circ e$ is
the space $\Omega^*({\cal N},{\cal M})=\oplus^n_{p=0}\,
\Omega^p({\cal N},{\cal M})$ with $\Omega^p({\cal N},{\cal
M})\equiv\Omega^p({\cal N})\times\Omega^{D+p-n-1}({\cal N})$
endowed with nilpotent operator $d$ defined by \be
d(\a,\b)=(d\a,\a\wedge P({\cal N})-d\b)\,,\quad \forall\a\in
\Omega^p({\cal M}),\,\forall\b\in \Omega^{D+p-n-1}({\cal N})\,.\ee

\subsection{Relative Poincar\'e duality}

Nicely, the Poincar\'e duality also finds a place in this
framework. For this purpose, one will have to work with
differential forms on $\cal M$ only and make use of the
proposition \ref{relalgebra}. The product $ext\circ e$ induces the
natural complex morphism from the mapping cone $\Omega^*({\cal
N},{\cal M})$ to the cone $C\Omega({\cal M})$ \be
ext:\Omega^p({\cal N},{\cal M})\rightarrow C\Omega^{D+p-n-1}({\cal
M}):(\a,\b)\rightarrow(\,ext(\a)\,,\,\b)\,. \ee

We introduce a bilinear map \be \wedge: C\Omega^p({\cal M})\otimes
C\Omega^q({\cal M})\rightarrow \Omega^{p+q-1}({\cal M})\,,\ee that
we call the {\bf relative wedge product} of $C\Omega({\cal M})$.
It is defined by \be (\a,\b)\wedge(\a',\b')=
\a\wedge\b'+(-)^{(p+1)(q+1)}\b\wedge\a'\,.\ee From proposition
\ref{relalgebra}, one knows that \be
d\left[(\a,\b)\wedge(\a',\b')\right]=d(\a,\b)\wedge(\a',\b')+(-)^{p+1}(\a,\b)\wedge
d(\a',\b')\,.\ee

We define the natural {\bf Poincar\'e duality} map as the complex
morphism \be P_{{\cal N},{\cal M}}:\Omega_p({\cal N},{\cal
M})\rightarrow \Omega^{n-p}({\cal N},{\cal M})\label{definPoin}\ee
defined by \be P_{{\cal N},{\cal M}}\,(\rho,\sigma)=
\big(\,(-)^{(D-n)(n-p)+1}P_{\cal
N}(\rho)\,,\,(-)^{D-p}P(\sigma)\,\big)\,.\ee The {\bf relative
Poincar\'e duality} over the relative de Rham complex is the map
\be P\equiv ext\circ P_{{\cal N},{\cal M}}:\Omega_p({\cal N},{\cal
M})\rightarrow C\Omega^{D-p+1}({\cal M})\ee which, due to
(\ref{wedgeP}), acts as \be P(\rho,\sigma)=
\left(\,-P(\rho)\,,\,(-)^{D-p}P(\sigma)\,\right)\,.\label{relativeP}\ee
This definition seems satisfactory because it is a morphism of
complexes\footnote{More precisely, it obeys the same formula
(\ref{Pmorphism}).} and it obeys a generalization of
(\ref{PDdef}), \ba \int\limits_{\cal
M}P(\rho,\sigma)\wedge(\a,\b)=\int\limits_{(\rho,\sigma)}i^*(\a,\b)\,,\ea
for any $(\rho,\sigma)\in\Omega_p({\cal N},{\cal M})$, $(\a,\b)\in
C\Omega^p({\cal M})$.

Let $(\rho,\sigma)\in \Omega_{p+1}({\cal N},{\cal M})$ and
$(\lambda,\mu)\in \Omega_{D-p}({\cal N},{\cal M})$  be two chains
such that the intersections $\rho\cap\mu$ and $\sigma\cap \lambda$
contains only a finite number of points. We define the {\bf
relative intersection number} as
\begin{eqnarray}
I\left[(\rho,\sigma),(\lambda,\mu)\right]&=&\int\limits_{\cal M}P(\rho,\sigma)\wedge P(\lambda,\mu)\\
&=&(-)^{(p+1)(D-p)}I\left[(\rho,\sigma),(\lambda,\mu)\right]\,.\label{relint}
\end{eqnarray}
It can be written in terms of absolute intersection as \be
I\left[(\rho,\sigma),(\lambda,\mu)\right]=(-)^{p+1}I(\rho,\mu)+(-)^{(D-p)(p+1)}I(\sigma,\lambda)\,.\ee

%If $(\rho,\sigma)\in H_{p+1}({\cal M},{\cal N})$ and $(\lambda,\mu)\in H_{D-p}({\cal M},{\cal N})$  be two cycles. Their relative intersection number is proportional to the absolute linking number of $\partial\sigma$ and $\partial\m$. More precisely, it is equal to \be I\left[(\rho,\sigma),(\lambda,\mu)\right]=(-)^{p+1}\,2\,L(\partial\sigma,\partial\mu)\,.\label{rlinking}\ee Still a subtlety arises here since for $L(\partial\sigma,\partial\mu)$ to be well-defined one needs

%%%%%%%%%%%%%%%%%%%%%%%%%%%%%%%%%%%%%%%%%%%%%%%%%%%%%%%%%%%%%%%%%
%%%%%%%%%%%%%%%%%%%%%%%%%%%%%%%%%%%%%%%%%%%%%%%%%%%%%%%%%%%%%%%%%%

\chapter{Young diagrams}\label{Young}

\par

A {\bf Young diagram} $Y$ is a diagram which consists of a finite
number $S>0$ of columns of identical squares (referred to as the
{\bf cells}) of finite decreasing lengths $l_1\geq l_2\geq
\ldots\geq l_S\geq 0$.\footnote{The present set of definitions
comes essentially from \cite{Fulton,Dubois-Violette:2001}. The
fourth chapter of \cite{Bacry} gives a nice introduction to Young
diagrams and possible use in physics.} For instance,\ba Y\equiv
\mbox{ \footnotesize
\begin{picture}(38,15)(0,0)
\multiframe(10,14)(10.5,0){3}(10,10){}{}{}
\multiframe(10,3.5)(10.5,0){2}(10,10){}{}
\multiframe(10,-7)(10.5,0){2}(10,10){}{}
\multiframe(10,-17.5)(10.5,0){1}(10,10){}
\end{picture}\normalsize }
\nn\ea The total number of cells of the Young diagram $Y$ is
denoted by \be|Y|=\sum_{i=1}^S l_i\,.\ee

\section{Diagrams with at most $S$ columns}

For later use, one defines the subset ${\mathbb Y}^{(S)}$ of
${\mathbb N}^S$ by \be{\mathbb Y}^{(S)} \equiv
\{(n_1,\ldots,n_S)\in {\mathbb N}^S|\,n_1\geq n_2 \geq ...\geq
n_S\geq 0\}\,.\ee For two columns, the set ${\mathbb Y}^{(2)}$ is
in the plane ${\mathbb R}^2$ and can be pictured as the following
set of points in the plane ${\mathbb R}^2$\be
\begin{array}{ccccc}
   &  &  &  & $\ldots$ \\
   &  &  & $$\stackrel{(3,3)}{\bullet}$$ & $\ldots$ \\
   &  & $$\stackrel{(2,2)}{\bullet}$$ & $$\stackrel{(3,2)}{\bullet}$$ & $\ldots$ \\
   & $$\stackrel{(1,1)}{\bullet}$$ & $$\stackrel{(2,1)}{\bullet}$$ & $$\stackrel{(3,1)}{\bullet}$$ & $\ldots$ \\
   $$\stackrel{(0,0)}{\bullet}$$ & $$\stackrel{(1,0)}{\bullet}$$ & $$\stackrel{(2,0)}{\bullet}$$ & $$\stackrel{(3,0)}{\bullet}$$ & $\ldots$
\end{array}\ee
Let $Y_p$ be a diagram with $p$ boxes and $S$ columns of
respective lengths $l_1$, $l_2$, ..., $l_S$ ($\sum_{i=1}^S
l_i=p$). If $Y_p$ is a well defined Young diagram, then
$(l_1,l_2,\ldots,l_S)\in {\mathbb Y}^{(S)}$. Conversely, a Young
diagram $Y$ with at most $S$ columns is uniquely determined by the
gift of an element of ${\mathbb Y}^{(S)}$, and can therefore be
labeled unambiguously as $Y_{(l_1,l_2,\ldots,l_S)}^{(S)}$. The set
of all diagrams with at most $S$ columns is identified with
${\mathbb Y}^{(S)}$.

\section{Order relations}

There is natural definition of inclusion of Young diagrams \be
Y_{(m_1,\ldots,m_S)}^{(S)} \subset
Y_{(n_1,\ldots,n_S)}^{(S)}\,\Leftrightarrow\, m_1\leq
n_1\,,m_2\leq n_2\,,\ldots\,, m_S\leq n_S\,.\ee We can define a
stronger notion of inclusion. Let $Y^{(S)}_{(m_1,\ldots,m_S)}$ and
$Y^{(S)}_{(n_1,\ldots,n_S)}$ be two Young diagrams of ${\mathbb
Y}^{(S)}$. We say that $Y^{(S)}_{(m_1,\ldots,m_S)}$ is {\bf
well-included} into $Y^{(S)}_{(n_1,\ldots,n_S)}$ if
$Y^{(S)}_{(m_1,\ldots,m_S)}\subset Y^{(S)}_{(n_1,\ldots,n_S)}$ and
$n_i-m_i\leq 1$ for all $i\in\{1,\ldots,S\}$. In other words, the
difference of the two Young diagrams does not contain any column
(greater than a single box). We denote this particular inclusion
by $\Subset$, i.e. \be Y_{(m_1,\ldots,m_S)}^{(S)} \Subset
Y_{(n_1,\ldots,n_S)}^{(S)}\,\Leftrightarrow\, m_i\leq n_i\leq
m_i+1\quad\forall i\in\{1,\ldots,S\}\,.\ee This new inclusion
suggests the following diagram of ${\mathbb Y}^{(S)}$
\ba\xymas{&&&&\ldots\\&&&\stackrel{(3,3)}{\bullet}\ar[r]\ar[ur]&\cdots
\\&&\stackrel{(2,2)}{\bullet} \ar[r]\ar[ur]&
\stackrel{(3,2)}{\bullet} \ar[u]\ar[r]\ar[ur]&\cdots \\
& \stackrel{(1,1)}{\bullet}\ar[r]\ar[ur]&
\stackrel{(2,1)}{\bullet} \ar[r]\ar[u]\ar[ur]&
\stackrel{(3,1)}{\bullet}\ar[u]\ar[r]\ar[ur]&\cdots \\
\stackrel{(0,0)}{\bullet} \ar[r]\ar[ur]&
\stackrel{(1,0)}{\bullet}\ar[u]\ar[r]\ar[ur]&
\stackrel{(2,0)}{\bullet}\ar[u]\ar[r]\ar[ur]
&\stackrel{(3,0)}{\bullet}\ar[u]\ar[r]\ar[ur] &\cdots
}\label{Ylatt}\ea where all the arrows represent maps $\Subset$.
This diagram is of course is completely commutative.

The previous inclusions $\subset$ and $\Subset$ provide partial
order relations for ${\mathbb Y}^{(S)}$ because all Young diagrams
are not comparable. We now introduce a total order relation for
${\mathbb Y}^{(S)}$: the {\bf lexicographic order} $\ll$. If
$(m_1,\ldots,m_S)$ and $(n_1,\ldots,n_S)$ belong to ${\mathbb
Y}^{(S)}$, then \be Y_{(m_1,\ldots,m_S)}^{(S)} \ll
Y_{(n_1,\ldots,n_S)}^{(S)} \,\,\Leftrightarrow \,\,\exists
K\in\{1,\ldots,S\}:\,\,\left\{\begin{array}{lll}m_i=n_i\,,
~~\forall
i\in\{1,\ldots, K\}\,, \\
m_{K+1}\leq n_{K+1}\,.\end{array}\right. \ee We will call this
order the {\bf{$Y\!-$grading}}.

\section{Maximal diagrams}

A sequence of ${\mathbb Y}^{(S)}$ which is of physical interest is
the {\bf maximal sequence} denoted by $Y^S\equiv
(Y_p^S)_{p\in\mathbb N}$, with $|Y_p^S|=p$. It is defined as the
naturally ordered sequence of {\bf maximal diagrams}\footnote{The
subsequent notations for maximal sequences are different from the
one of \cite{Dubois-Violette:1999,Dubois-Violette:2001}. We have
shifted the index by one unit.} (The ordering is induced by the
inclusion of Young diagrams). Maximal diagrams are diagrams with
maximally filled rows, that is the Young diagram with $p$ cells
$Y^S_p$ defined in the following manner : we put cells in a row
until it contains $S$ cells and then we proceed in the same way
with the row below, and so on until all the $p$ cells have been
used. Consequently all rows but the last one are of length $S$
and, if $r_p$ is the rest of the division of $p$ by $S$
($r_p\equiv p\,\,\mbox{mod}\,S <S$) then the last row of the Young
diagram $Y^S_p$ contains $r_p\leq S$ cells (if $r_p\not= 0$). For
two columns ($S=2$) the maximal sequence is represented as the
following path in the plane ${\mathbb R}^2$
\ba\xymas{&&&\\&&\stackrel{(2,2)}{\bullet} \ar[r]^{\subset}&
\stackrel{(3,2)}{\bullet}\ar[u]^{\cup}\\
&\stackrel{(1,1)}{\bullet} \ar[r]^{\subset}&
\stackrel{(2,1)}{\bullet}\ar[u]^{\cup}\\
\stackrel{(0,0)}{\bullet} \ar[r]^{\subset}&
\stackrel{(1,0)}{\bullet}\ar[u]^{\cup} }\nn\ea Diagrams for which
all the rows have exactly $S$ cells are called {\bf rectangular
diagrams}. They sit at the diagonal of the diagram ${\mathbb
Y}^{(S)}$.

\section{Schur module}

Let $V$ be a finite-dimensional vector space of dimension $D$ and
$V^\ast$ its dual. The $n$-th tensor power $V^n$ of $V$ is
canonically identified with the space of multilinear forms
$(V^n)^\ast\cong (V^\ast)^n$, i.e. multilinear applications from
$V^n$ to $\mathbb R$. To a Young diagram, one associates
multilinear applications with a definite symmetry.

Let $Y$ be a Young diagram and let us consider that the $\vert
Y\vert$ copies of $V^\ast$ in $(V^\ast)^{\vert Y\vert}$ are
labeled by the cells of $Y$ so that an  element of
$(V^\ast)^{\vert Y\vert}$ is given by specifying an element of
$V^\ast$ for each cell of $Y$. The {\bf Schur module} $V^Y$ is
defined to be the vector space of all multilinear forms $T$ in
$(V^\ast)^{\vert Y\vert}$ such that:
\begin{quote}

$(i)$ $T$ is completely antisymmetric in the entries of each
column of $Y$,

$(ii)$ complete antisymmetrization of $T$ in the entries of a
column of $Y$ and another entry of $Y$ which is on the right-hand
side of the column vanishes.
\end{quote}
If $Y$ is single column, then the Schur module $V^Y$ is the
$|Y|$-th exterior product $\Lambda^{|Y|} V^\ast$, a subspace of
the exterior agebra $\Lambda V^\ast$. If $Y$ is a single row, then
$V^Y$ is the $|Y|$-th symmetric product $Sym^{|Y|}V^\ast$.

The main motivation of Schur modules comes from representation
theory because one has the
\begin{proposition}\label{Schurprop}
\begin{itemize}
  \item The Schur module $V^Y$ is an irreducible subspace invariant for the action of
$GL(D,\mathbb R)$ on $V^{\vert Y\vert}$.
  \item All finite dimensional representations of $GL(D,\mathbb R)$ can be described in terms
of representations $V^Y$.
\end{itemize}
\end{proposition}

Let $Y$ be a Young diagram and $T$ an arbitrary multilinear form
on $(V^\ast)^{\vert Y\vert}$, ($T\in V^{\vert Y\vert}$). Define
the multilinear form ${\cal Y}(T)$ on $(V^\ast)^{\vert Y \vert}$
by
\[
{\cal Y}(T)=T\circ{\cal A}\circ{\cal S}
\]
with
\[
{\cal A}=\sum_{c\in C}(-)^{\varepsilon(c)}c\,,\quad {\cal
S}=\sum_{r\in R} r
\]
where $C$ is the {\bf column group}, i.e. the group of the
permutations which permute the entries of each column and $R$ is
the {\bf row group} made of the permutations which permute the
entries of each row of $Y$. One has ${\cal Y} (T)\in V^Y$ and the
application ${\cal Y}$ of $V^{\vert Y\vert}$ satisfies ${\cal
Y}^2=\lambda{\cal Y}$ for some number $\lambda\not= 0$. Thus ${\bf
Y} = \lambda^{-1}{\cal Y}$ is a projection of $V^{\vert Y\vert}$
into itself, i.e. ${\bf Y}^2={\bf Y}$, with image Im$({\bf
Y})=V^Y$. The projection ${\bf Y}$ will be referred to as the {\bf
Young symmetrizer} of the Young diagram $Y$.

A less common choice of Young symmetrizer is also possible. One
defines the multilinear form $\overline{\cal Y}(T)$ on
$(V^\ast)^{\vert Y \vert}$ by
\[
\overline{\cal Y}(T)=T\circ{\cal S}\circ{\cal A}\,.
\]One has
$\overline{\cal Y} (T)\in V^Y$ and the application $\overline{\cal
Y}$ of $V^{\vert Y\vert}$ satisfies $(\overline{\cal
Y})^2=\m\,\overline{\cal Y}$ for some number $\m\not= 0$. Thus
$\overline{\bf Y} = \m^{-1}\overline{\cal Y}$ is a projection of
$V^{\vert Y\vert}$ into itself, i.e. $(\overline{\bf
Y})^2=\overline{\bf Y}$, with image Im$(\overline{\bf Y})=V^Y$.

%%%%%%%%%%%%%%%%%%%%%%%%%%%%%%%%%%%%%%%%%%%%%%%%%%%%%%%%%%%%%%
%%%%%%%%%%%%%%%%%%%%%%%%%%%%%%%%%%%%%%%%%%%%%%%%%%%%%%%%%%%%%%

\chapter{Lemmas for the analyticity
conditions}\label{l:analyticity}

We present here a number of lemmas necessary in the proof of
theorem \ref{t:analyticity}. The first lemma of this appendix
allows us to reformulate the analyticity of $L(x,y)$ as the
analyticity of the function $f(u_+,u_-)$ together with its
symmetry property. The function $f(u_+,u_-)$ is only an
intermediate tool necessary to achieve the proof of theorem
\ref{t:analyticity}. The second lemma provides a necessary and
sufficient requirement on the function $z(t)$ to generate
symmetric functions while the third one relates the analyticity of
$f(u_+,u_-)$ at the origin with the analyticity of $z(t)$ at the
origin, giving also the behaviors of $f$ and $z$. All together,
these lemmas lead us to the theorem \ref{t:analyticity}.

The property (\ref{roots}) suggests the use of Newton's theorem on
symmetric polynomials \cite{Newton}, which states (in particular)
that any symmetric polynomial $P(u_+,u_-)=P(u_-,u_+)$ in the roots
$u_+$ and $u_-$ can be re-expressed as a polynomial
$Q(x,y):=P\left(u_+(x,y),u_-(x,y)\right)$ in the coefficients $x$
and $y$. The following lemma provides a generalization of this
last property for functions of two variables, analytic at the
origin.
\begin{lemma}\label{l:analytic}
A function $L(x,y)$ is analytic at the origin $(x,y)=(0,0)$ if and
only if the symmetric function
\begin{equation}
f(u_+,u_-):=L\left(x(u_+,u_-),y(u_+,u_-)\right)
\end{equation} is analytic at the origin $(u_+,u_-)=(0,0)$.
\end{lemma}
\proof{Due to Newton's theorem, this lemma is obvious for formal
power series but convergence matters are rather intricate to
handle from that point of view, so we choose an other path.

\noindent$\bf\underline{\,\bf\Rightarrow:}$ The analyticity of
$f(u_+,u_-)$ at the origin is of course necessary since the
composition of two analytic functions in a neighborhood is also an
analytic function (in the corresponding neighborhood) and the
functions $x(u_+,u_-)$ and $y(u_+,u_-)$ are analytic at the
origin.

\noindent$\bf\underline{\,\bf\Leftarrow:}$ To prove that the
analyticity of $f(u_+,u_-)$ at the origin is sufficient to ensure
that $L(x,y)$ is also analytic at that point, we make the
following change of variables: $x=u_++u_-$, $z=u_+-u_-$. This is a
diffeomorphism everywhere, thus the function
\begin{equation}
h(x,z):=f\left(u_+(x,z),u_-(x,z)\right)
\end{equation}
is analytic at the origin $(x,z)=(0,0)$. But the symmetry of
$f(u_+,u_-)$ implies that $h(x,z)$ is even in $z$, i.e.
$h(x,z)=h(x,-z)$. Therefore, $h$ is only a function of $z^2$:
$h(x,z)=h(x,z^2)$. But $z^2=x^2 - 4 y$ is analytic function of $x$
and $y$, hence
\begin{equation}
L(x,y):=h\left(x(x,y),z^2(x,y)\right)
\end{equation}
is analytic at the origin $(x,y)=(0,0)$. }
\begin{lemma}
A necessary and sufficient condition for the symmetry of the
function $f(u_+,u_-)$ defined implicitly by (\ref{e:general}) is
that the function \begin{equation}\Psi(t)\equiv
-t\dot{z}^2(t)\end{equation} is equal to its inverse:
$\Psi\left(\Psi(t)\right)=t$.
\end{lemma}
\proof{It has been proved in \cite{Perry:1997} that it is
necessary. One can argue as follows that the requirement on $z$ to
satisfy $\Psi\left(\Psi(t)\right)=t$ suffices to have $f(u_+,u_-)$
symmetric. We begin by noticing that this last requirement implies
\begin{equation}\label{e:z}
\dot{z}\left(-t\dot{z}^2(t)\right) \dot{z}(t)=\pm 1
\end{equation}
Taking (\ref{e:z}) at $t=0$ will select the positive sign.
Consider $f(u_+,u_-)$ the function defined by (\ref{e:general}).
%\begin{equation}
%f(u_{+},u_{-})\equiv \left\{
%\begin{array}{c}
%f=\frac{2u_{+}}{\dot{z}\left( t\right) }+z\left( t\right) \\
%u_{-}=\frac{u_{+}}{\left( \dot{z}\left( t\right) \right) ^{2}}+t
%\end{array}
%\right.\,.
%\end{equation}
We have to show that this function is symmetric. From
\begin{equation}
u_{+}=\dot{z}^{2}(t)\left( u_{-}-t\right)  \label{S17}
\end{equation}
together with (\ref{e:z}), we deduce that
\begin{equation}
u_{+}
%=\frac{u_{-}}{\dot{z}^{2}\left( -t\dot{z}^{2}(t)\right) }-t\dot{z}^{2}(t)
=\frac{u_{-}}{\dot{z}^{2}\left( s\right) }+s  \,,\label{S14}
\end{equation}
with $s:=\Psi(t)$. With the help of (\ref{e:z}) and (\ref{S17}) we
also get
\begin{equation}
f
%=2\dot{z}(t)\left( u_{-}-t\right) +z(t)
=\frac{2u_{-}}{\dot{z}\left( -t\dot{z}^{2}(t)\right)
}-2t\dot{z}(t)+z(t) \,. \label{S13}
\end{equation}
%Since
%\begin{equation}
%\frac{d}{dt}\left( z(t)-2t\dot{z}(t)\right) =-\left( \dot{z}(t)+2t\ddot{z}(t)\right)  \label{S9}
%\end{equation}
%and
%\begin{equation}
%\frac{d}{dt}\left( z\left( -t\dot{z}^{2}(t)\right) \right) =\dot{z}\left( -t%
%\dot{z}^{2}(t)\right) \left( -\dot{z}^{2}(t)-2t\dot{z}(t)\ddot{z}(t)\right)
%=-\left( \dot{z}(t)+2t\ddot{z}(t)\right)  \label{S10}
%\end{equation}
% using (\ref{e:z}). As (\ref{S9}) and (\ref{S10}) are equal,
%we find
By taking the derivative, it can be checked that
\begin{equation}
z(t)-2t\dot{z}(t)=z\left( -t\dot{z}^{2}(t)\right) +K  \label{S11}
\end{equation}
where $K$ is a constant. By taking (\ref{S11}) at $t=0$, we find
that $K$ vanishes. Combining (\ref{S14}), (\ref{S13}) and
(\ref{S11}) together we infer
%that we can interchange $u_+$ and $u_-$ in the definition of %$f(u_+,u_-)$, that is, $f$ is symmetric:
the symmetry of $f$, i.e.
\begin{equation}
f\left( u_{+},u_{-}\right) =f\left( u_{-},u_{+}\right) \equiv
\left\{
\begin{array}{c}
f=\frac{2u_{-}}{\dot{z}\left( s\right) }+z\left( s\right) \\
u_{+}=\frac{u_{-}}{\left( \dot{z}\left( s\right) \right) ^{2}}+s
\end{array}
\right. \,.
\end{equation}}
\begin{lemma}\label{l:z}
Let $f(u_+,u_-)$ be a function defined implicitly by
(\ref{e:general}). It satisfies the following conditions
\begin{itemize}
\item[(i)] analyticity near $(u_+,u_-)=(0,0)$,
\item[(ii)] $f(u_+,u_-)=u_+ + u_- +O(|u_{\pm}|^2)$,
\end{itemize}
if and only if the generating function $z(t)$ is analytic near $0$
and $z(t)=t+O(t^2)$.
\end{lemma}
\proof{That the condition is necessary is trivial so we
immediately focus on the proof that it is sufficient. Let us
define the function
\begin{equation}
F(u_+,u_-,t) \equiv u_{-}-\frac{u_{+}}{\left( \dot{z}(t) \right)
^{2}}-t \,.
\end{equation}
The function $\frac{1}{\left( \dot{z}(t) \right) ^{2}}$ is
analytic near $t=0$ as $z(t)$ is also (near $t=0$) and
$\dot{z}(0)=1$. Hence, $F(u_+,u_-,t)$ is analytic near $(0,0,0)$.
Furthermore,
\begin{equation}
\frac{\partial F}{\partial t}(u_+,u_-,t)=\frac{2u_{+}
\ddot{z}(t)}{\left( \dot{z}(t) \right) ^{3}}-1 \rightarrow
\frac{\partial F}{\partial t}(0,0,0)=-1\neq 0\,.
\end{equation}
From standard theorems on implicit functions \cite{Dieu}, we find
that the function $g(u_+,u_-)$ defined by
$F\left(u_+,u_-,g(u_+,u_-)\right) =0$ is analytic near
$(u_+,u_-)=(0,0)$. The behavior of $g$ near the origin is
$g(u_+,u_-)= u_- - u_+ +O(|u_{\pm}|^2)$. Putting all things
together one derives the analyticity of
\begin{equation}
f(u_+,u_-)\equiv\frac{2u_{+}}{\dot{z}\left( g(u_+,u_-)\right)
}+z\left( g(u_+,u_-)\right)
\end{equation}
near $(u_+,u_-)=(0,0)$ and $f(u_+,u_-)=u_+ + u_- +O(|u_{\pm}|^2)$.
}

To finish the proof of theorem \ref{t:analyticity}, we show that
the four conditions: $\Psi\left(\Psi(t)\right)=t$, $\Psi(t)\neq
t$, $\Psi(0)=0$ and $z(0)=0$ imply the good behavior of $z(t)$,
i.e.
$$z(t)=t+O(t^2),$$ as in lemma \ref{l:z}. Indeed, if we take the derivative of the first
condition, we get $\dot{\Psi}\left(\Psi(t)\right)\dot{\Psi}(t)=1$.
This implies $\left(\dot{\Psi}(0)\right)^2=1$ (using the third
condition), hence $\dot{\Psi}(0)=\pm 1$. The positive sign
corresponds to the identity\footnote{The uniqueness of this
solution can be seen from a simple symmetry argument with respect
to the diagonal in the $(t,\Psi)$-plane.} $\Psi(t)=t$ that we
discard due to the second condition. The negative sign together
with $\Psi(0)=0$ gives: $\Psi(t)=-t\Phi(t)$ with $\Phi(t)=1+O(t)$.
Finally, from $\dot{z}^2(t)=\Phi(t)=1+O(t)$ and $z(0)=0$, we
arrive at the expected conclusion: $z(t)=t+O(t^2)$.

%%%%%%%%%%%%%%%%%%%%%%%%%%%%%%%%%%%%%%%%%%%%%%%%%%%%%%%%%%%%%%%%%%%
%%%%%%%%%%%%%%%%%%%%%%%%%%%%%%%%%%%%%%%%%%%%%%%%%%%%%%%%%%%%%%%%%%%

\chapter{Proof of the generalized Poincar\'e lemma}

We first provide an algorithm that recursively makes use of the
standard Poincar\'e lemma for multiforms. In practice, this
algorithm leads to a generalized Poincar\'e lemma. It is presented
for the specific case of the mixed symmetry type tensors of
section \ref{mixed}. In the second section, we give a more
abstract proof which is a mere adaptation of the one given in
\cite{Dubois-Violette:2001} for a more general sequence of Young
diagrams.

\section{Inductive proof}

The algorithm presented below is made of successive applications
of the standard Poincar\'e for multiforms. It provides a general
recipe for computing any generalized cohomology. In practice, the
algorithm amounts to play with Young diagram with some boxes
filled by partial derivatives.

This algorithm can provide a proof of the generalized Poincar\'e
lemma, which is recursive in several directions. First, there is
an induction on the number $S$ of columns (For instance we start
from the standard Poincar\'e lemma). Second, in order to compute
completely a cohomology group, we use an induction on the number
$l$ of cells in the new column. This induction in the case of
$H_{(*)}^{(n,l)}\left(\Omega_{(2)}({\mathbb R}^D)\right)$
corresponds to the first three subsections. The plan of the proof
for an arbitrary number of column $S$ is given in subsection
\ref{section3}.

\subsection{Generalized cohomology in $\Omega^{(*,1)}_{(2)}({\mathbb R}^D)$}

We begin by providing a proof that the two cohomologies
$H_{(1)}^{(n,1)}\left(\Omega_{(2)}({\mathbb R}^D)\right)$ and
$H_{(2)}^{(n,1)}\left(\Omega_{(2)}({\mathbb R}^D)\right)$ are
trivial for $0<n< D$, {\it{i.e.}} (1) that \be d^{\{i\}}~
\footnotesize
\begin{picture}(30,30)(0,0)
\multiframe(0,10)(10.5,0){1}(10,10){$1$}
\multiframe(10.5,10)(10.5,0){1}(17,10){$n\!\!+\!\!1$}
\multiframe(0,-0.5)(10.5,0){1}(10,10){$2$}
\multiframe(0,-18)(10.5,0){1}(10,17){$ $} \put(4,-13.5){$\vdots$}
\multiframe(0,-28.5)(10.5,0){1}(10,10){$n$}
\end{picture}\normalsize\vspace{1cm}
=0\, , ~~~~~~~~i=1,2 \label{cond1} \ee implies \be\vspace{1cm}
\footnotesize \begin{picture}(30,30)(0,0)
\multiframe(0,10)(10.5,0){1}(10,10){$1$}
\multiframe(10.5,10)(10.5,0){1}(17,10){$n\!\!+\!\!1$}
\multiframe(0,-0.5)(10.5,0){1}(10,10){$2$}
\multiframe(0,-18)(10.5,0){1}(10,17){$ $} \put(4,-13.5){$\vdots$}
\multiframe(0,-28.5)(10.5,0){1}(10,10){$n$}
\end{picture}\normalsize
=~~\footnotesize \begin{picture}(30,30)(0,0)
\multiframe(0,10)(10.5,0){1}(10,10){$1$}
\multiframe(10.5,10)(10.5,0){1}(10,10){$\partial$}
\multiframe(0,-0.5)(10.5,0){1}(10,10){$2$}
\multiframe(0,-18)(10.5,0){1}(10,17){$ $} \put(4,-13.5){$\vdots$}
\multiframe(0,-28.5)(10.5,0){1}(10,10){$\partial$}
\end{picture}\normalsize
\label{resu1} \ee and (2) that \be d^{\{1,2\}}~ \footnotesize
\begin{picture}(25,30)(0,0)
\multiframe(0,10)(10.5,0){1}(10,10){$1$}
\multiframe(10.5,10)(10.5,0){1}(17,10){$n\!\!+\!\!1$}
\multiframe(0,-0.5)(10.5,0){1}(10,10){$2$}
\multiframe(0,-18)(10.5,0){1}(10,17){$ $} \put(4,-13.5){$\vdots$}
\multiframe(0,-28.5)(10.5,0){1}(10,10){$n$}
\end{picture}\normalsize\vspace{1cm}
=~0\, \label{cond2} \ee implies \be \footnotesize
\begin{picture}(25,30)(0,0)
\multiframe(0,10)(10.5,0){1}(10,10){$1$}
\multiframe(10.5,10)(10.5,0){1}(17,10){$n\!\!+\!\!1$}
\multiframe(0,-0.5)(10.5,0){1}(10,10){$2$}
\multiframe(0,-18)(10.5,0){1}(10,17){$ $} \put(4,-13.5){$\vdots$}
\multiframe(0,-28.5)(10.5,0){1}(10,10){$n$}
\end{picture}\normalsize\vspace{1cm}
~=~~ \footnotesize \begin{picture}(25,30)(0,0)
\multiframe(0,10)(10.5,0){1}(10,10){$1$}
\multiframe(10.5,10)(10.5,0){1}(10,10){$\partial$}
\multiframe(0,-0.5)(10.5,0){1}(10,10){$2$}
\multiframe(0,-18)(10.5,0){1}(10,17){$ $} \put(4,-13.5){$\vdots$}
\multiframe(0,-28.5)(10.5,0){1}(10,10){$n$}
\end{picture}\normalsize
+~~ \footnotesize \begin{picture}(25,30)(0,0)
\multiframe(0,10)(10.5,0){1}(10,10){$1$}
\multiframe(10.5,10)(10.5,0){1}(17,10){$n\!\!+\!\!1$}
\multiframe(0,-0.5)(10.5,0){1}(10,10){$2$}
\multiframe(0,-18)(10.5,0){1}(10,17){$ $} \put(4,-13.5){$\vdots$}
\multiframe(0,-28.5)(10.5,0){1}(10,10){$\partial$}
\end{picture}\normalsize~~. \vspace{.5cm}
\label{resu2} \ee So far the notation introduced is
self-explanatory. The numbers in the cells are irrelevant, they
just signal the length of columns. For later convenience we review
our following convention : whenever a Young diagram $Y$ appears
with certain boxes filled in with partial derivatives $\partial$,
one takes a field with the representation of the Young diagram
obtained by removing from $Y$ all the $\partial$-boxes, one
differentiates this new field as many times as there are
derivatives in $Y$ and then project the result on the Young
symmetry of $Y$.

\subsubsection{First cohomology group}

For the two different possible values of $i$ in (\ref{cond1}) we
have the two conditions on the field $(n,1)$ :
\begin{itemize}
\item
\hspace*{2mm} \footnotesize \begin{picture}(25,30)(0,0)
\multiframe(0,10)(10.5,0){1}(10,10){$1$}
\multiframe(10.5,10)(10.5,0){1}(17,10){$n\!\!+\!\!1$}
\multiframe(0,-7.5)(10.5,0){1}(10,17){$ $} \put(3.5,-4){$\vdots$}
\multiframe(0,-18)(10.5,0){1}(10,10){$n$}
\multiframe(0,-28.5)(10.5,0){1}(10,10){$\partial$}
\end{picture}\normalsize\vspace{1cm}
$=0$~~~ for i=1
\\
and
\item
\hspace*{2mm} \footnotesize \begin{picture}(25,30)(0,0)
\multiframe(0,10)(10.5,0){1}(10,10){$1$}
\multiframe(10.5,10)(10.5,0){1}(17,10){$n\!\!+\!\!1$}
\multiframe(10.5,-0.5)(10.5,0){1}(17,10){$\partial$}
\multiframe(0,-7.5)(10.5,0){1}(10,17){$ $} \put(3.5,-4){$\vdots$}
\multiframe(0,-18)(10.5,0){1}(10,10){$n$}
\end{picture}\normalsize\vspace{1cm}
=~~0~~~ for i=2.
\end{itemize}

The first condition is treated now : one considers the index of
the second column as a spectator, which yields \ba \footnotesize
\begin{picture}(25,30)(0,0)
\multiframe(0,10)(10.5,0){1}(17,10){$1$}
\multiframe(0,-0.5)(10.5,0){1}(17,10){$2$}
\multiframe(17.5,10)(10.5,0){1}(17,10){$n\!\!+\!\!1$}
\multiframe(0,-17)(10.5,0){1}(17,16){$ $} \put(7,-13){$\vdots$}
\multiframe(0,-27.5)(10.5,0){1}(17,10){$n$}
\multiframe(0,-38)(10.5,0){1}(17,10){$\partial$}
\end{picture}\normalsize ~~~~\simeq~~~~ \footnotesize
\begin{picture}(25,30)(0,0)
\multiframe(0,10)(10.5,0){1}(17,10){$1$}
\multiframe(0,-0.5)(10.5,0){1}(17,10){$2$}
\multiframe(0,-17)(10.5,0){1}(17,16){$ $} \put(7,-13){$\vdots$}
\multiframe(0,-27.5)(10.5,0){1}(17,10){$n$}
\multiframe(0,-38)(10.5,0){1}(17,10){$\partial$}
\end{picture}\normalsize
\otimes ~~ \footnotesize \begin{picture}(15,15)(0,0)
\multiframe(0,-0.5)(10.5,0){1}(17,10){$n\!\!+\!\!1$}
\end{picture}\quad=\,0\nn\ea
\\
\\
\\
where the symbol $\simeq$ means that there is an implicit
projection using $\bf Y$ on the right-hand-side in order to agree
with the left-hand-side (in other words the symbol $\simeq$
replaces the expression $=\bf Y$). One uses the Poincar\'e
Lemma on the first column and write (The decomposition of the
tensor product of irrep. of $GL(D,\mathbb R)$ into a direct sum of
irrep. $GL(D,\mathbb R)$ is obtained by following the general
procedure given in \cite{Bacry}, pp. 91-92.)~~ \ba \footnotesize
\begin{picture}(25,30)(0,0)
\multiframe(0,10)(10.5,0){1}(10,10){$1$}
\multiframe(10.5,10)(10.5,0){1}(17,10){$n\!\!+\!\!1$}
\multiframe(0,-7.5)(10.5,0){1}(10,17){$ $} \put(3.5,-4){$\vdots$}
\multiframe(0,-18)(10.5,0){1}(10,10){$n$}
\end{picture}\normalsize
\vspace{2cm} ~~~~&=&~~~~ \footnotesize
\begin{picture}(25,30)(0,0)
\multiframe(0,10)(10.5,0){1}(17,10){$1$}
\multiframe(0,-0.5)(10.5,0){1}(17,10){$2$}
\multiframe(0,-17)(10.5,0){1}(17,16){$ $} \put(7,-13){$\vdots$}
\multiframe(0,-27.5)(10.5,0){1}(17,10){$n\!\!-\!\!1$}
\multiframe(0,-38)(10.5,0){1}(17,10){$\partial$}
\end{picture}\normalsize
\otimes ~~ \footnotesize \begin{picture}(15,15)(0,0)
\multiframe(0,-0.5)(10.5,0){1}(10,10){$n$}
\end{picture}
\nonumber \\\nonumber \\ \nonumber \\
&\simeq&~ \footnotesize \begin{picture}(15,15)(0,0)
\multiframe(0,-0.5)(10.5,0){1}(10,10){$\partial$}
\end{picture}\normalsize
\otimes \Big(~~ \footnotesize \begin{picture}(25,30)(0,0)
\multiframe(0,10)(10.5,0){1}(17,10){$1$}
\multiframe(0,-0.5)(10.5,0){1}(17,10){$2$}
\multiframe(0,-17)(10.5,0){1}(17,16){$ $} \put(7,-13){$\vdots$}
\multiframe(0,-27.5)(10.5,0){1}(17,10){$n\!\!-\!\!1$}
\end{picture}\normalsize
\otimes ~~ \footnotesize \begin{picture}(15,15)(0,0)
\multiframe(0,-0.5)(10.5,0){1}(10,10){$n$}
\end{picture}\normalsize
\Big) \ea
\\
\\
In the second line, we have undone the manifest antisymmetrization
with the index carrying the partial derivative; we are more
interested in the symmetries of the tensor under the derivative.

Now we first perform the product in the bracket to obtain a sum of
different types of irreducible tensors. Then, we perform the
product with the partial derivative to get \ba \footnotesize
\begin{picture}(25,30)(0,0)
\multiframe(0,10)(10.5,0){1}(10,10){$1$}
\multiframe(10.5,10)(10.5,0){1}(17,10){$n\!\!+\!\!1$}
\multiframe(0,-7.5)(10.5,0){1}(10,17){$ $} \put(3.5,-4){$\vdots$}
\multiframe(0,-18)(10.5,0){1}(10,10){$n$}
\end{picture}\normalsize
\vspace{2cm} ~~~~\simeq~~~~ \footnotesize
\begin{picture}(30,30)(0,0)
\multiframe(0,10)(10.5,0){1}(17,10){$1$}
\multiframe(17.25,10)(10.5,0){1}(17,10){$\partial$}
\multiframe(0,-0.5)(10.5,0){1}(17,10){$2$}
\multiframe(0,-17)(10.5,0){1}(17,16){$ $} \put(7,-13){$\vdots$}
\multiframe(0,-27.5)(10.5,0){1}(17,10){$n\!\!-\!\!1$}
\multiframe(0,-38)(10.5,0){1}(17,10){$n$}
\end{picture}\normalsize
\hspace*{2mm}\oplus\hspace*{2mm} \footnotesize
\begin{picture}(30,30)(0,0)
\multiframe(0,10)(10.5,0){1}(17,10){$1$}
\multiframe(17.25,10)(10.5,0){1}(17,10){$n$}
\multiframe(0,-0.5)(10.5,0){1}(17,10){$2$}
\multiframe(0,-17)(10.5,0){1}(17,16){$ $} \put(7,-13){$\vdots$}
\multiframe(0,-27.5)(10.5,0){1}(17,10){$n\!\!-\!\!1$}
\multiframe(0,-38)(10.5,0){1}(17,10){$\partial$}
\end{picture}\normalsize
\hspace*{2mm}\oplus\hspace*{2mm} \footnotesize
\begin{picture}(30,30)(0,0)
\multiframe(0,10)(10.5,0){1}(17,10){$1$}
\multiframe(0,-0.5)(10.5,0){1}(17,10){$2$}
\multiframe(0,-17)(10.5,0){1}(17,26){$ $} \put(7,-13){$\vdots$}
\multiframe(0,-27.5)(10.5,0){1}(17,10){$n\!\!-\!\!1$}
\multiframe(0,-38)(10.5,0){1}(17,10){$n$}
\multiframe(0,-48.5)(10.5,0){1}(17,10){$\partial$}
\end{picture}\normalsize.
\label{decom1} \ea
\\
\\

The last term on the above equation (\ref{decom1}) does not match
the symmetry of the left-hand-side, so it must vanish. Using the
Poincar\'e lemma, which is applicable since one is not in top form
degree ($n<D$), one gets \be \footnotesize
\begin{picture}(25,30)(0,0)
\multiframe(0,20)(10.5,0){1}(10,10){$1$}
\multiframe(0,9.5)(10.5,0){1}(10,10){$2$}
\multiframe(0,-17)(10.5,0){1}(10,26){$ $} \put(4,-10){$\vdots$}
\multiframe(0,-27.5)(10.5,0){1}(10,10){$n$}
\end{picture}\normalsize
=~~~~ \footnotesize \begin{picture}(30,30)(0,0)
\multiframe(0,20)(10.5,0){1}(10,10){$1$}
\multiframe(0,9.5)(10.5,0){1}(10,10){$2$}
\multiframe(0,-17)(10.5,0){1}(10,26){$ $} \put(4,-10){$\vdots$}
\multiframe(0,-27.5)(10.5,0){1}(10,10){$\partial$}
\end{picture}\normalsize.\vspace{1cm}
\ee Substituting this result in the decomposition (\ref{decom1})
yields \be\label{momo} \footnotesize \begin{picture}(30,30)(0,0)
\multiframe(0,10)(10.5,0){1}(17,10){$1$}
\multiframe(17.25,10)(10.5,0){1}(17,10){$n\!\!+\!\!1$}
\multiframe(0,-0.5)(10.5,0){1}(17,10){$2$}
\multiframe(0,-17)(10.5,0){1}(17,16){$ $} \put(7,-13.5){$\vdots$}
\multiframe(0,-27.5)(10.5,0){1}(17,10){$n\!\!-\!\!1$}
\multiframe(0,-38)(10.5,0){1}(17,10){$n$}
\end{picture}\normalsize\vspace{1.5cm}
\hspace*{2mm}\simeq~ \footnotesize \begin{picture}(30,30)(0,0)
\multiframe(0,10)(10.5,0){1}(17,10){$1$}
\multiframe(17.25,10)(10.5,0){1}(17,10){$\partial$}
\multiframe(0,-0.5)(10.5,0){1}(17,10){$2$}
\multiframe(0,-17)(10.5,0){1}(17,16){$ $} \put(7,-13.5){$\vdots$}
\multiframe(0,-27.5)(10.5,0){1}(17,10){$n\!\!-\!\!1$}
\multiframe(0,-38)(10.5,0){1}(17,10){$\partial$}
\end{picture}\normalsize
\hspace*{2mm}\oplus\hspace*{2mm} \footnotesize
\begin{picture}(30,30)(0,0)
\multiframe(0,10)(10.5,0){1}(17,10){$1$}
\multiframe(17.25,10)(10.5,0){1}(17,10){$n$}
\multiframe(0,-0.5)(10.5,0){1}(17,10){$2$}
\multiframe(0,-17)(10.5,0){1}(17,16){$ $} \put(7,-13.5){$\vdots$}
\multiframe(0,-27.5)(10.5,0){1}(17,10){$n\!\!-\!\!1$}
\multiframe(0,-38)(10.5,0){1}(17,10){$\partial$}
\end{picture}\normalsize\hspace*{3mm}\longrightarrow\hspace*{2mm} \footnotesize
\begin{picture}(30,30)(0,0)
\multiframe(0,10)(10.5,0){1}(17,10){$1$}
\multiframe(17.25,10)(10.5,0){1}(17,10){$n$}
\multiframe(0,-0.5)(10.5,0){1}(17,10){$2$}
\multiframe(0,-17)(10.5,0){1}(17,16){$ $} \put(7,-13.5){$\vdots$}
\multiframe(0,-27.5)(10.5,0){1}(17,10){$n\!\!-\!\!1$}
\multiframe(0,-38)(10.5,0){1}(17,10){$\partial$}
\end{picture}\normalsize\hspace*{3mm}
\ee where the arrow means that we performed a field redefinition.
Thus, without loss of generality, the right-hand-side can be
assumed to contain a partial derivative in the first column. With
this preliminary result, the second condition expressed in
(\ref{cond1}), \be d^{\{2\}} \hspace*{2mm}\vspace{.8cm}
\footnotesize
\begin{picture}(30,30)(0,0)
\multiframe(0,10)(10.5,0){1}(10,10){$1$}
\multiframe(10.5,10)(10.5,0){1}(17,10){$n\!\!+\!\!1$}
\multiframe(0,-0.5)(10.5,0){1}(10,10){$2$}
\multiframe(0,-18)(10.5,0){1}(10,17){$ $} \put(4,-13.5){$\vdots$}
\multiframe(0,-28.5)(10.5,0){1}(10,10){$n$}
\end{picture}\normalsize\equiv\hspace*{2mm}
\footnotesize \begin{picture}(25,30)(0,0)
\multiframe(0,10)(10.5,0){1}(10,10){$1$}
\multiframe(10.5,10)(10.5,0){1}(17,10){$n\!\!+\!\!1$}
\multiframe(10.5,-0.5)(10.5,0){1}(17,10){$\partial$}
\multiframe(0,-0.5)(10.5,0){1}(10,10){$2$}
\multiframe(0,-18)(10.5,0){1}(10,17){$ $} \put(4,-13.5){$\vdots$}
\multiframe(0,-28.5)(10.5,0){1}(10,10){$n$}
\end{picture}\normalsize~~=0
\ee gives
%\be
%\footnotesize \begin{picture}(30,30)(0,0)
%\multiframe(0,17)(10.5,0){1}(17,17){$1$}
%\multiframe(17.5,17)(10.5,0){1}(17,17){$n\!\!+\!\!1$}
%\multiframe(0,-0.5)(10.5,0){1}(17,17){$2$}
%\multiframe(17.5,-0.5)(10.5,0){1}(17,17){$\partial$}
%\put(7,-14){$\vdots$}
%\multiframe(0,-18)(10.5,0){1}(17,17){$ $}
%\multiframe(0,-35.5)(10.5,0){1}(17,17){$n\!\!-\!\!1$}
%\multiframe(0,-53)(10.5,0){1}(17,17){$\partial$}
%\end{picture}\normalsize\hspace*{4mm}=~~0
%\ee
\be\vspace{1.5cm} \footnotesize \begin{picture}(30,30)(0,0)
\multiframe(0,10)(10.5,0){1}(17,10){$1$}
\multiframe(17.5,10)(10.5,0){1}(17,10){$n$}
\multiframe(0,-0.5)(10.5,0){1}(17,10){$2$}
\multiframe(17.5,-0.5)(10.5,0){1}(17,10){$\partial$}
\put(7,-14){$\vdots$} \multiframe(0,-20)(10.5,0){1}(17,19){$ $}
\multiframe(0,-30.5)(10.5,0){1}(17,10){$n\!\!-\!\!1$}
\multiframe(0,-41)(10.5,0){1}(17,10){$\partial$}
\end{picture}\normalsize\hspace*{4mm}=~0\,.
\label{mouche} \ee The Poincar\'e lemma on the second column leads
to \be\vspace{1.5cm} \footnotesize \begin{picture}(30,30)(0,0)
\multiframe(0,10)(10.5,0){1}(17,10){$1$}
\multiframe(17.5,10)(10.5,0){1}(17,10){$n$}
\multiframe(0,-0.5)(10.5,0){1}(17,10){$2$} \put(7,-14){$\vdots$}
\multiframe(0,-20)(10.5,0){1}(17,19){$ $}
\multiframe(0,-30.5)(10.5,0){1}(17,10){$n\!\!-\!\!1$}
\multiframe(0,-41)(10.5,0){1}(17,10){$\partial$}
\end{picture}\normalsize\hspace*{4mm}\simeq~~
\footnotesize \begin{picture}(20,30)(0,0)
\multiframe(0,-0.5)(10.5,0){1}(17,10){$\partial$}
\end{picture}\normalsize
\otimes\hspace*{1mm} \footnotesize \begin{picture}(30,30)(0,0)
\multiframe(0,10)(10.5,0){1}(17,10){$1$}
\multiframe(0,-0.5)(10.5,0){1}(17,10){$2$} \put(7,-14){$\vdots$}
\multiframe(0,-20)(10.5,0){1}(17,19){$ $}
\multiframe(0,-30.5)(10.5,0){1}(17,10){$n\!\!-\!\!1$}
\multiframe(0,-40.5)(10.5,0){1}(17,10){$n$}
\end{picture}\normalsize
\simeq~~ \footnotesize \begin{picture}(25,25)(0,0)
\multiframe(0,10)(10.5,0){1}(17,10){$1$}
\multiframe(0,-0.5)(10.5,0){1}(17,10){$2$} \put(7,-14){$\vdots$}
\multiframe(0,-20)(10.5,0){1}(17,19){$ $}
\multiframe(0,-30.5)(10.5,0){1}(17,10){$n\!\!-\!\!1$}
\multiframe(0,-40.5)(10.5,0){1}(17,10){$n$}
\multiframe(0,-51)(10.5,0){1}(17,10){$\partial$}
\end{picture}\normalsize
\oplus\hspace*{2mm} \footnotesize \begin{picture}(30,30)(0,0)
\multiframe(0,10)(10.5,0){1}(17,10){$1$}
\multiframe(17.5,10)(10.5,0){1}(17,10){$\partial$}
\multiframe(0,-0.5)(10.5,0){1}(17,10){$2$} \put(7,-14){$\vdots$}
\multiframe(0,-20)(10.5,0){1}(17,19){$ $}
\multiframe(0,-30.5)(10.5,0){1}(17,10){$n\!\!-\!\!1$}
\multiframe(0,-41)(10.5,0){1}(17,10){$n$}
\end{picture}\normalsize~~~.
\label{moucheron} \ee The first totally antisymmetric component
vanishes since there is no component with the same symmetry on the
left-hand-side, implying that \be\vspace{1.4cm} \footnotesize
\begin{picture}(30,30)(0,0)
\multiframe(0,10)(10.5,0){1}(17,10){$1$}
\multiframe(0,-0.5)(10.5,0){1}(17,10){$2$} \put(7,-14){$\vdots$}
\multiframe(0,-20)(10.5,0){1}(17,19){$ $}
\multiframe(0,-30.5)(10.5,0){1}(17,10){$n\!\!-\!\!1$}
\multiframe(0,-41)(10.5,0){1}(17,10){$n$}
\end{picture}\normalsize
= ~~\footnotesize \begin{picture}(30,30)(0,0)
\multiframe(0,10)(10.5,0){1}(17,10){$1$}
\multiframe(0,-0.5)(10.5,0){1}(17,10){$2$} \put(7,-14){$\vdots$}
\multiframe(0,-20)(10.5,0){1}(17,19){$ $}
\multiframe(0,-30.5)(10.5,0){1}(17,10){$n\!\!-\!\!1$}
\multiframe(0,-41)(10.5,0){1}(17,10){$\partial$}
\end{picture}\normalsize
\ee which in turn, inserted in (\ref{moucheron}), gives
\be\label{momotu}\vspace{1.5cm} \footnotesize
\begin{picture}(30,30)(0,0)
\multiframe(0,10)(10.5,0){1}(17,10){$1$}
\multiframe(17.5,10)(10.5,0){1}(17,10){$n$}
\multiframe(0,-0.5)(10.5,0){1}(17,10){$2$} \put(7,-14){$\vdots$}
\multiframe(0,-20)(10.5,0){1}(17,19){$ $}
\multiframe(0,-30.5)(10.5,0){1}(17,10){$n\!\!-\!\!1$}
\multiframe(0,-41)(10.5,0){1}(17,10){$\partial$}
\end{picture}\normalsize
\hspace*{2mm}=\hspace*{2mm} \footnotesize
\begin{picture}(30,30)(0,0)
\multiframe(0,10)(10.5,0){1}(17,10){$1$}
\multiframe(17.5,10)(10.5,0){1}(17,10){$\partial$}
\multiframe(0,-0.5)(10.5,0){1}(17,10){$2$} \put(7,-14){$\vdots$}
\multiframe(0,-20)(10.5,0){1}(17,19){$ $}
\multiframe(0,-30.5)(10.5,0){1}(17,10){$n\!\!-\!\!1$}
\multiframe(0,-41)(10.5,0){1}(17,10){$\partial$}
\end{picture}\normalsize\,.
\ee Substituting this result in (\ref{momo}) proves (\ref{resu1}).

\subsubsection{Second cohomology group}

We now turn to the proof that (\ref{cond2}) implies (\ref{resu2}).
The condition  (\ref{cond2}) reads \be\vspace{1.5cm} \footnotesize
\begin{picture}(30,30)(0,0)
\multiframe(0,10)(10.5,0){1}(17,10){$1$}
\multiframe(17.5,10)(10.5,0){1}(17,10){$n\!\!+\!\!1$}
\multiframe(0,-0.5)(10.5,0){1}(17,10){$2$}
\multiframe(17.5,-0.5)(10.5,0){1}(17,10){$\partial$}
\put(7,-15){$\vdots$} \multiframe(0,-20)(10.5,0){1}(17,19){$ $}
\multiframe(0,-30.5)(10.5,0){1}(17,10){$n$}
\multiframe(0,-41)(10.5,0){1}(17,10){$\partial$}
\end{picture}\normalsize\hspace*{4mm}=~0\,
\ee whose type was already encountered in (\ref{mouche}) above. We
use our previous result (\ref{momotu}) and write \be\vspace{1.5cm}
\footnotesize
\begin{picture}(30,30)(0,0)
\multiframe(0,10)(10.5,0){1}(17,10){$1$}
\multiframe(17.5,10)(10.5,0){1}(17,10){$n\!\!+\!\!1$}
\multiframe(0,-0.5)(10.5,0){1}(17,10){$2$} \put(7,-15){$\vdots$}
\multiframe(0,-20)(10.5,0){1}(17,19){$ $}
\multiframe(0,-30.5)(10.5,0){1}(17,10){$n$}
\multiframe(0,-41)(10.5,0){1}(17,10){$\partial$}
\end{picture}\normalsize
\hspace*{2mm}=\hspace*{2mm} \footnotesize
\begin{picture}(30,30)(0,0)
\multiframe(0,10)(10.5,0){1}(17,10){$1$}
\multiframe(17.5,10)(10.5,0){1}(17,10){$\partial$}
\multiframe(0,-0.5)(10.5,0){1}(17,10){$2$} \put(7,-15){$\vdots$}
\multiframe(0,-20)(10.5,0){1}(17,19){$ $}
\multiframe(0,-30.5)(10.5,0){1}(17,10){$n$}
\multiframe(0,-41)(10.5,0){1}(17,10){$\partial$}
\end{picture}\normalsize
\ee or \be\vspace{1.5cm} \footnotesize \begin{picture}(30,30)(0,0)
\multiframe(0,10)(10.5,0){1}(17,10){$1$}
\multiframe(17.5,10)(10.5,0){1}(17,10){$n\!\!+\!\!1$}
\multiframe(0,-0.5)(10.5,0){1}(17,10){$2$} \put(7,-15){$\vdots$}
\multiframe(0,-20)(10.5,0){1}(17,19){$ $}
\multiframe(0,-30.5)(10.5,0){1}(17,10){$n$}
\multiframe(0,-41)(10.5,0){1}(17,10){$\partial$}
\end{picture}\normalsize
\hspace*{2mm}-\hspace*{2mm} \footnotesize
\begin{picture}(30,30)(0,0)
\multiframe(0,10)(10.5,0){1}(17,10){$1$}
\multiframe(17.5,10)(10.5,0){1}(17,10){$\partial$}
\multiframe(0,-0.5)(10.5,0){1}(17,10){$2$} \put(7,-15){$\vdots$}
\multiframe(0,-20)(10.5,0){1}(17,19){$ $}
\multiframe(0,-30.5)(10.5,0){1}(17,10){$n$}
\multiframe(0,-41)(10.5,0){1}(17,10){$\partial$}
\end{picture}\normalsize~=~0~.
\ee This kind of equation also came before, in (\ref{cond1}), when
i=1. Then we are able to write \be\vspace{1.5cm} \footnotesize
\begin{picture}(30,30)(0,0)
\multiframe(0,10)(10.5,0){1}(17,10){$1$}
\multiframe(17.5,10)(10.5,0){1}(17,10){$n\!\!+\!\!1$}
\multiframe(0,-0.5)(10.5,0){1}(17,10){$2$} \put(7,-15){$\vdots$}
\multiframe(0,-20)(10.5,0){1}(17,19){$ $}
\multiframe(0,-30.5)(10.5,0){1}(17,10){$n$}
\end{picture}\normalsize
\hspace*{2mm}-\hspace*{2mm} \footnotesize
\begin{picture}(30,30)(0,0)
\multiframe(0,10)(10.5,0){1}(17,10){$1$}
\multiframe(17.5,10)(10.5,0){1}(17,10){$\partial$}
\multiframe(0,-0.5)(10.5,0){1}(17,10){$2$} \put(7,-15){$\vdots$}
\multiframe(0,-20)(10.5,0){1}(17,19){$ $}
\multiframe(0,-30.5)(10.5,0){1}(17,10){$n$}
\end{picture}\normalsize~=~
\footnotesize \begin{picture}(30,30)(0,0)
\multiframe(0,10)(10.5,0){1}(17,10){$1$}
\multiframe(17.5,10)(10.5,0){1}(17,10){$n$}
\multiframe(0,-0.5)(10.5,0){1}(17,10){$2$} \put(7,-15){$\vdots$}
\multiframe(0,-20)(10.5,0){1}(17,19){$ $}
\multiframe(0,-30.5)(10.5,0){1}(17,10){$\partial$}
\end{picture}\normalsize
\ee which is the analogue of (\ref{momo}). Equivalently,
\be\vspace{1.5cm} \footnotesize
\begin{picture}(30,30)(0,0)
\multiframe(0,10)(10.5,0){1}(17,10){$1$}
\multiframe(17.5,10)(10.5,0){1}(17,10){$n\!\!+\!\!1$}
\multiframe(0,-0.5)(10.5,0){1}(17,10){$2$} \put(7,-15){$\vdots$}
\multiframe(0,-20)(10.5,0){1}(17,19){$ $}
\multiframe(0,-30.5)(10.5,0){1}(17,10){$n$}
\end{picture}\normalsize
\hspace*{2mm}=\hspace*{2mm} \footnotesize
\begin{picture}(30,30)(0,0)
\multiframe(0,10)(10.5,0){1}(17,10){$1$}
\multiframe(17.5,10)(10.5,0){1}(17,10){$\partial$}
\multiframe(0,-0.5)(10.5,0){1}(17,10){$2$} \put(7,-15){$\vdots$}
\multiframe(0,-20)(10.5,0){1}(17,19){$ $}
\multiframe(0,-30.5)(10.5,0){1}(17,10){$n$}
\end{picture}\normalsize~+~
\footnotesize \begin{picture}(30,30)(0,0)
\multiframe(0,10)(10.5,0){1}(17,10){$1$}
\multiframe(17.5,10)(10.5,0){1}(17,10){$n\!\!+\!\!1$}
\multiframe(0,-0.5)(10.5,0){1}(17,10){$2$} \put(7,-15){$\vdots$}
\multiframe(0,-20)(10.5,0){1}(17,19){$ $}
\multiframe(0,-30.5)(10.5,0){1}(17,10){$\partial$}
\end{picture}\normalsize
\ee which is the desired result.

\subsection{Generalized cohomology in $\Omega^{(*,*)}_{(2)}({\mathbb R}^D)$}

We extend the previous results to the cases where the Young
diagrams have two columns filled by an arbitrary number of cells.
That is to say, we show now that
$H^{(*,*)}_{(*)}(\Omega_{(2)}\big({\mathbb R}^D)\big)\cong 0$. We
will proceed by induction on the number of boxes in the last
(second) column: what we showed in the above subsection
constitutes the ``stage zero" of our induction proof.

We leave for a moment the diagrammatic exposition. For an easier
understanding of the following propositions, we sketch in the
subsection \ref{Diagrammatically} a pictorial translation of the
proof that $H^{(n,l)}_{(1)}(\Omega_2)\cong 0\,,~~0<l<n<D$.
\\

\noindent {\bf Induction hypothesis :} We suppose that the
following results are known : \ba &d^{\{1\}}\m(l_1,l_2)=0\,,~~0<
l_1<D,~ 0 < l_2 < \ell < l_1 &
 \nn\\
&\Rightarrow \m(l_1,l_2) = d^{\{1\}}\n(l_1-1,l_2)\,,& \ea where
the notation $\m(l_1,l_2)$ indicates that
$\m\in\Omega^{(l_1,l_2)}_{(2)}$, similarly
$\n\in\Omega^{(l_1-1,l_2)}_{(2)}$. The integer $\ell$ is fixed.
Also, we suppose that one has the vanishing cohomologies \ba
H^{(l_1,l_2)}_{(1)}(\Omega_{(2)}({\mathbb R}^D))&\cong&
0\,,~~{\rm{and}}
\\
H^{(l_1,l_2)}_{(2)}(\Omega_{(2)}({\mathbb R}^D))&\cong& 0\,, \ea
for $0< l_1<D$ and $0<l_2< \ell< l_1$.
\\

Basically, the induction hypothesis is that one knows the
cohomology of $d^{\{1\}}$ and the generalized cohomology for all
tensors with second column of length strictly smaller than $\ell$.
Now we prove that we have similar results after having put a new
box in the second column, {\it{i.e.}} the second column can have a
length $\ell$. In the induction hypothesis, $0<l_2\leq \ell< l_1$
is therefore substituted for $0<l_2< \ell<l_1$ everywhere. The
case $\ell=l_1$ will be treated separately along the same lines.
This will prove inductively that
$H^{(l_1,l_2)}_{(*)}(\Omega_{(2)}({\mathbb R}^D))\cong 0$ {\it{for
any}} $(l_1,l_2)\in {\mathbb Y}^{(2)}$.

Before starting the proof and for later purposes, we introduce an
order relation in the space $\Omega_{(S)}$ of irreducible tensors
under $GL(D,\mathbb R)$ naturally induced by the lexicographic
order relation (\ref{Ygrading}) for ${\mathbb Y}^{(S)}$.
\\
If $\a(l_1,\ldots,l_S)$ and $\b(l'_1,\ldots,l'_{S'})$ belong to
$\Omega_{(S)}^{(l_1,\ldots,l_S)}$ and
$\Omega_{(S')}^{(l'_1,\ldots,l'_{S'})}$, respectively, then \be
\a(l_1,\ldots,l_S) \ll \b(l'_1,\ldots,l'_{S'}) \ee if and only if
\ba
l_k&=&l'_k\,, ~~1\leq k\leq K\,, ~~{\rm{and}}\nonumber \\
l_{K+1}&\leq& l'_{K+1}\,, \ea where $K$ is an integer satisfying
$1\leq K \leq min(S,S')$. This lexicographical ordering naturally
induces a $\mathbb N$-grading that we call the
{\it{$L\!-$grading}} of $\Omega_{(S)}$.

Turning back to our inductive proof, we first show the
%--------------------------------------------------------
\begin{lemma}
\label{lemma1} \ba &d^{\{1\}}\m(l_1,\ell)=0\,, ~0< l_1<D, ~0<
\ell< l_1\,,&
\label{2col}\\
&{\rm{implies}}~{\rm{that}}& \nonumber \\
&\m(l_1,\ell)=d^{\{1\}}\n(l_1-1,\ell).& \ea
\end{lemma}
%---------------------------------------------------------
\proof{Applying the Poincar\'e lemma on equation (\ref{2col}),
viewing the second column as spectator, yields\footnote{We will
use hated symbols to denote multiforms of $\Omega_{[S]}({\mathbb
R}^D)$, while the unhated tensors belong to $\Omega_{(S)}({\mathbb
R}^D)$.} $\m(l_1,\ell)\simeq d_1 \hat{\n}(l_1-1,\ell)$, where
$\hat{\n}(l_1-1,\ell)\in\O_{[2]}^{l_1-1,\ell}$. Decomposing the
right-hand-side (expressed in terms of multiforms) into irrep. of
$GL(D,\mathbb R)$ gives \ba\m(l_1,\ell)&\simeq&
d^{\{1\}}\n(l_1-1,\ell)+d^{\{2\}}\n(l_1,\ell-1)+d^{\{1\}}\n(l_1,\ell-1)\nn\\
&&+d^{\{2\}}\n(l_1+1,\ell-2)+(\ldots)\,,\ea where the $(\ldots)$
denote tensors of higher L-grading. Because the third and fourth
terms do not belong to $\O_{(2)}^{(l_1,\ell)}$, they must cancel :
\be
d^{\{1\}}\n(l_1,\ell-1)+d^{\{2\}}\n(l_1+1,\ell-2)=0\,.\label{this}\ee
Applying the operator $d^{\{2\}}$ to (\ref{this}) gives
$d^{\{1,2\}}\n(l_1,\ell-1)=0$. The induction hypothesis allow us
to write
$\n(l_1,\ell-1)=d^{\{1\}}\n(l_1-1,\ell-1)+d^{\{2\}}\n(l_1,\ell-2)$.
Plugging back in the decomposition of $\m(l_1,\ell)$, one finds,
after a redefinition of $\n(l_1-1,\ell)$, the result we were
looking for : \be \m(l_1,\ell)=d^{\{1\}}\n(l_1-1,\ell).
\label{resulem2col} \ee}

The case $\ell=l_1$ comes in the same way. The equation \be
d^{\{1\}}\m(l_1,l_1)=0\,,~~0<l_1<D \ee gives, after using standard
Poincar\'e lemma, that $\m(l_1,l_1)\simeq d_1
\hat{\n}(l_1-1,l_1)$,
$\hat{\n}(l_1-1,l_1)\in\O_{[2]}^{l_1-1,l_1}\simeq\O_{[2]}^{l_1,l_1-1}$.
In terms of irrep. of $GL(D,\mathbb R)$, we get $\m(l_1,l_1)\simeq
d^{\{2\}}
\n(l_1,l_1-1)+d^{\{1\}}\n(l_1,l_1-1)+d^{\{2\}}\n(l_1+1,l_1-2)+(\ldots)$
where $(\ldots)$ denote terms of higher L-grading. The sum of the
second and third terms must vanish, and applying $d^{\{2\}}$ gives
$d^{\{1,2\}}\n(l_1,l_1-1)=0$. By our hypothesis of induction we
obtain
$\n(l_1,l_1-1)=d^{\{1\}}\n(l_1-1,l_1-1)+d^{\{2\}}\n(l_1,l_1-2)$.
Here, the result which emerges after substituting the above
equation in the decomposition of $\m(l_1,l_1)$ and field
redefinition, is \be \m(l_1,l_1)=d^{\{1,2\}}\n(l_1-1,l_1-1).
\label{resulem2idemcol} \ee In other words, we got for free the
%--------------------------------------------------------
\begin{proposition}
\label{theoHll1} $H^{(l,l)}_{(1)}(\Omega_{(2)}({\mathbb
R}^D))\cong 0\,,~~~0<l<D.$
\end{proposition}
%--------------------------------------------------------
\noindent Having the Lemma \ref{lemma1} at our disposal, we now
proceed to prove the
%-----------------------------------------------------------------------------
\begin{proposition}
\label{Hnl} $H^{(l_1,\ell)}_{(1)}(\Omega_{(2)}({\mathbb
R}^D))\cong 0\,,~0< l_1<D,~0< \ell< l_1\,.$
\end{proposition}
%-----------------------------------------------------------------------------
\noindent It states that the cocycle conditions \be
d^{\{i\}}\m(l_1,\ell)=0\,,~~i\in\{1,2\},~~0< l_1<D, ~0< \ell<
l_1\, \label{cocycle1}\ee imply that \be
\m(l_1,\ell)=d^{\{1,2\}}\n(l_1-1,\ell-1)\,. \ee \proof{In the case
$i=1$, the conditions (\ref{cocycle1}) give, using Lemma
\ref{lemma1}, that \be \m(l_1,\ell)=d^{\{1\}}\n(l_1-1,\ell).
\label{intermed1} \ee Plugging this in the condition
(\ref{cocycle1}) for $i=2$ yields \be d^{\{1,2\}}\n(l_1-1,\ell)=0.
\label{cocycle12} \ee Using Poincar\'e lemma on the second column,
we have \ba d^{\{1\}}\n(l_1-1,\ell)&\simeq& d_1
\hat{\n}(l_1-1,\ell)\nn\\&\simeq&
d^{\{2\}}\n(l_1,\ell-1)+d^{\{1\}}\n(l_1,\ell-1)\nn\\&&+d^{\{2\}}\n(l_1+1,\ell-2)
+(\ldots)\,,\ea where as before we used the branching rule for
$GL(D,\mathbb R)$ and the $(\ldots)$ correspond to terms of higher
L-grading. The sum of the second and third terms of the
right-hand-side must vanish, so does the action of $d^{\{2\}}$ on
it. As a consequence,
$\n(l_1,\ell-1)=d^{\{1\}}\n(l_1-1,\ell-1)+d^{\{2\}}\n(l_1,\ell-2)$,
hence $d^{\{1\}}\n(l_1-1,\ell)=d^{\{1,2\}}\n(l_1-1,\ell-1)$.
Injecting in (\ref{intermed1}) we finally have \be
\m(l_1,\ell)=d^{\{1,2\}}\n(l_1-1,\ell-1). \ee which proves the
proposition.}

It still remains to show the following vanishing of cohomologies
%----------------------------------------------------------------------------
\begin{proposition}
$H^{(l_1,\ell)}_{(2)}(\Omega_{(2)}({\mathbb R}^D))\cong 0\,,~0<
l_1<D,~0< \ell< l_1\,.$
\end{proposition}
%----------------------------------------------------------------------------
\proof{Actually the cocycle condition was already encountered in
(\ref{cocycle12}).
%The case where $\ell=l_1-1$ in (\ref{cocycle12}) provides the cocycle
%condition for $H^{(l_1,\ell)}_{(2)}(\Omega_{(2)}({\mathbb R}^D))\,,
%~0< l_1<D,~ \ell= l_1$, and we didn't point this out at that time
%because the solution of (\ref{cocycle12}) for $\ell=l_1-1$ in fact proceeds
%exactly along the same lines as for $\ell<l_1-1$.
We can then use the results already obtained in the proof of the
Theorem \ref{Hnl} to write that the cocycle condition \be
d^{\{1,2\}}\m(l_1,\ell)=0\,, \quad 0<l_1<D, ~0<\ell\leq l_1\ee
leads to $d^{\{1\}}\m(l_1,\ell)=d^{\{1,2\}}\m(l_1,\ell-1)$.
Rewriting this equation as
$d^{\{1\}}[\m(l_1,\ell)-d^{\{2\}}\m(l_1,\ell-1)]=0$ and using the
results of Lemma \ref{lemma1} we obtain
$\m(l_1,\ell)-d^{\{2\}}\m(l_1,\ell-1)= d^{\{1\}}\m(l_1-1,\ell)$,
{\it{i.e.}} \be
\m(l_1,\ell)=d^{\{2\}}\m(l_1,\ell-1)+d^{\{1\}}\m(l_1-1,\ell).
\label{cocycle1+2} \ee Had we started with the cocycle condition
$d^{\{1,2\}}\m(l_1,l_1)=0$, $0<l_1<D$, we would have found
$d^{\{1\}}\m(l_1,l_1)=d^{\{1,2\}}\m(l_1,l_1-1)$, then
$\m(l_1,l_1)-d^{\{2\}}\m(l_1,l_1-1)=d^{\{1,2\}}\m(l_1-1,l_1-1)$,
and after a field redefinition, the result \be
\m(l_1,l_1)=d^{\{2\}}\m(l_1,l_1-1). \ee which is the coboundary
condition analogous to (\ref{cocycle1+2}) in the case of maximally
filled tensors in $\O_{2}({\mathbb R}^D)$.}

{\bf{Conclusions}}
\\
Our inductive proof gave us the following results about the
generalized cohomologies of $d^{\{i\}}$, $i\in\{1,2\}$, in the
space $\O_{(2)}$ : \ba H^{(l_1,l_2)}_{(*)}(\Omega_{(2)}({\mathbb
R}^D))\cong 0\,,~\forall (l_1,l_2)\in {\mathbb Y}^{(2)}. \ea
%
%^^^^^^^^^^^^^^^^^^^^^^^^^^^^^^^^^^^^^^^^^^^^^^^^^^^^^^^^^^^^^^^^^^
\subsection{Diagrammatically}\label{Diagrammatically}
%^^^^^^^^^^^^^^^^^^^^^^^^^^^^^^^^^^^^^^^^^^^^^^^^^^^^^^^^^^^^^^^^^^
%

The pictorial translation of the \underline{lemma \ref{lemma1}}
reads \be\vspace{3cm} \footnotesize \begin{picture}(30,30)(0,0)
\multiframe(0,10)(10.5,0){1}(10,10){$ 1$} \put(4,-10){$\vdots$}
\multiframe(0,-20)(10.5,0){1}(10,29.5){$ $}
\multiframe(10.5,10)(10.5,0){1}(10,10){$1 $}
\multiframe(10.5,-9.5)(10.5,0){1}(10,19){$ $}
\multiframe(10.5,-20)(10.5,0){1}(10,10){$l$}
\multiframe(0,-39.5)(10.5,0){1}(10,19){$ $}
\put(4,-34.5){$\vdots$} \put(14,-5){$\vdots$}
\multiframe(0,-49.5)(10.5,0){1}(10,10){$n$}
\multiframe(0,-60)(10.5,0){1}(10,10){$\partial$}
\end{picture}\normalsize
=~0~~~~\Rightarrow~~ \footnotesize \begin{picture}(30,30)(0,0)
\multiframe(0,10)(10.5,0){1}(10,10){$ 1$} \put(4,-10){$\vdots$}
\multiframe(0,-20)(10.5,0){1}(10,29.5){$ $}
\multiframe(10.5,10)(10.5,0){1}(10,10){$1 $}
\multiframe(10.5,-9.5)(10.5,0){1}(10,19){$ $}
\multiframe(10.5,-20)(10.5,0){1}(10,10){$l$}
\multiframe(0,-39.5)(10.5,0){1}(10,19){$ $}
\put(4,-34.5){$\vdots$} \put(14,-5){$\vdots$}
\multiframe(0,-49.5)(10.5,0){1}(10,10){$n$}
\end{picture}\normalsize~=~~~
\footnotesize \begin{picture}(30,30)(0,0)
\multiframe(0,10)(10.5,0){1}(10,10){$ 1$} \put(4,-10){$\vdots$}
\multiframe(0,-20)(10.5,0){1}(10,29.5){$ $}
\multiframe(10.5,10)(10.5,0){1}(10,10){$1 $}
\multiframe(10.5,-9.5)(10.5,0){1}(10,19){$ $}
\multiframe(10.5,-20)(10.5,0){1}(10,10){$l$}
\multiframe(0,-39.5)(10.5,0){1}(10,19){$ $}
\put(4,-34.5){$\vdots$} \put(14,-5){$\vdots$}
\multiframe(0,-49.5)(10.5,0){1}(10,10){$\partial$}
\end{picture}\normalsize~.
\ee For practical purposes, we take $l=2$ to show the lemma
\ref{lemma1} with Young diagrams. Using standard Poincar\'e Lemma,
one has \be\vspace{1cm} \footnotesize
\begin{picture}(30,30)(0,0)
\multiframe(0,20)(10.5,0){1}(10,10){$1$}
\multiframe(0,9.5)(10.5,0){1}(10,10){$2$}
\multiframe(0,-8)(10.5,0){1}(10,17){$ $} \put(4,-5){$\vdots$}
\multiframe(0,-18.5)(10.5,0){1}(10,10){$n$}
\multiframe(0,-29)(10.5,0){1}(10,10){$\partial$}
\multiframe(10.5,20)(10.5,0){1}(10,10){$*$}
\multiframe(10.5,9.5)(10.5,0){1}(10,10){$*$}
\end{picture}\normalsize
~=~0~\Rightarrow~ \footnotesize \begin{picture}(30,30)(0,0)
\multiframe(0,10)(10.5,0){1}(10,10){$1$}
\multiframe(0,-0.5)(10.5,0){1}(10,10){$2$}
\multiframe(0,-18)(10.5,0){1}(10,17){$ $} \put(4,-15){$\vdots$}
\multiframe(0,-28.5)(10.5,0){1}(10,10){$n$}
%\multiframe(0,-29)(10.5,0){1}(10,10){$\partial$}
\multiframe(10.5,10)(10.5,0){1}(10,10){$*$}
\multiframe(10.5,-0.5)(10.5,0){1}(10,10){$*$}
\end{picture}\normalsize\simeq~
\footnotesize \begin{picture}(15,30)(0,0)
\multiframe(0,0)(10.5,0){1}(10,10){$\partial$}
\end{picture}\normalsize
\otimes~\Big(~ \footnotesize \begin{picture}(25,30)(0,0)
\multiframe(0,10)(10.5,0){1}(17,10){$1$}
\multiframe(0,-0.5)(10.5,0){1}(17,10){$2$}
\multiframe(0,-18)(10.5,0){1}(17,17){$ $} \put(6,-15){$\vdots$}
\multiframe(0,-28.5)(10.5,0){1}(17,10){$n\!\!-\!\!1$}
\end{picture}\normalsize
\otimes~ \footnotesize \begin{picture}(15,30)(0,0)
\multiframe(0,5)(10.5,0){1}(10,10){$*$}
\multiframe(0,-5.5)(10.5,0){1}(10,10){$*$}
\end{picture}\normalsize
\Big) \ee i.e. \be\vspace{1.8cm} \footnotesize
\begin{picture}(30,30)(0,0)
\multiframe(0,10)(10.5,0){1}(17,10){$1$}
\multiframe(0,-0.5)(10.5,0){1}(17,10){$2$}
\multiframe(0,-18)(10.5,0){1}(17,17){$ $} \put(7,-15){$\vdots$}
\multiframe(0,-28.5)(10.5,0){1}(17,10){$n\!\!-\!\!1$}
\multiframe(0,-39)(10.5,0){1}(17,10){$n$}
\multiframe(17.5,10)(10.5,0){1}(10,10){$* $}
\multiframe(17.5,-0.5)(10.5,0){1}(10,10){$* $}
\end{picture}\normalsize
\simeq~ \footnotesize \begin{picture}(30,30)(0,0)
\multiframe(0,10)(10.5,0){1}(17,10){$1$}
\multiframe(0,-0.5)(10.5,0){1}(17,10){$2$}
\multiframe(0,-18)(10.5,0){1}(17,17){$ $} \put(7,-15){$\vdots$}
\multiframe(0,-28.5)(10.5,0){1}(17,10){$n\!\!-\!\!1$}
\multiframe(0,-39)(10.5,0){1}(17,10){$\partial$}
\multiframe(17.5,10)(10.5,0){1}(10,10){$* $}
\multiframe(17.5,-0.5)(10.5,0){1}(10,10){$* $}
\end{picture}\normalsize
\oplus \footnotesize \begin{picture}(30,30)(0,0)
\multiframe(0,10)(10.5,0){1}(17,10){$1$}
\multiframe(0,-0.5)(10.5,0){1}(17,10){$2$}
\multiframe(0,-18)(10.5,0){1}(17,17){$ $} \put(7,-15){$\vdots$}
\multiframe(0,-28.5)(10.5,0){1}(17,10){$n\!\!-\!\!1$}
\multiframe(0,-39)(10.5,0){1}(17,10){$* $}
\multiframe(17.5,10)(10.5,0){1}(10,10){$ *$}
\multiframe(17.5,-0.5)(10.5,0){1}(10,10){$\partial$}
\end{picture}\normalsize
\oplus \footnotesize \begin{picture}(30,30)(0,0)
\multiframe(0,20)(10.5,0){1}(17,10){$1$}
\multiframe(0,9.5)(10.5,0){1}(17,10){$2$}
\multiframe(0,-8)(10.5,0){1}(17,17){$ $} \put(7,-5){$\vdots$}
\multiframe(0,-18.5)(10.5,0){1}(17,10){$n\!\!-\!\!1$}
\multiframe(0,-29)(10.5,0){1}(17,10){$*$}
\multiframe(0,-39.5)(10.5,0){1}(17,10){$\partial$}
\multiframe(17.5,20)(10.5,0){1}(10,10){$* $}
\end{picture}\normalsize\oplus
\footnotesize \begin{picture}(30,30)(0,0)
\multiframe(0,20)(10.5,0){1}(17,10){$1$}
\multiframe(0,9.5)(10.5,0){1}(17,10){$2$}
\multiframe(0,-8)(10.5,0){1}(17,17){$ $} \put(7,-5){$\vdots$}
\multiframe(0,-18.5)(10.5,0){1}(17,10){$n\!\!-\!\!1$}
\multiframe(0,-29)(10.5,0){1}(17,10){$* $}
\multiframe(0,-39.5)(10.5,0){1}(17,10){$* $}
\multiframe(17.5,20)(10.5,0){1}(10,10){$ \partial$}
\end{picture}\normalsize
\oplus \footnotesize \begin{picture}(30,30)(0,0)
\multiframe(0,20)(10.5,0){1}(17,10){$1$}
\multiframe(0,9.5)(10.5,0){1}(17,10){$2$}
\multiframe(0,-8)(10.5,0){1}(17,17){$ $} \put(7,-5){$\vdots$}
\multiframe(0,-18.5)(10.5,0){1}(17,10){$n\!\!-\!\!1$}
\multiframe(0,-29)(10.5,0){1}(17,10){$* $}
\multiframe(0,-39.5)(10.5,0){1}(17,10){$* $}
\multiframe(0,-50)(10.5,0){1}(17,10){$ \partial$}
\end{picture}\normalsize~.
\label{debal1} \ee The condition that the sum of the third and
fourth terms of the right-hand-side vanishes, implies due to the
induction hypothesis, that the tensor coming in the second term
writes as \be\vspace{1.5cm} \footnotesize
\begin{picture}(30,30)(0,0)
\multiframe(0,10)(10.5,0){1}(17,10){$1$}
\multiframe(0,-0.5)(10.5,0){1}(17,10){$2$}
\multiframe(0,-18)(10.5,0){1}(17,17){$ $} \put(7,-15){$\vdots$}
\multiframe(0,-28.5)(10.5,0){1}(17,10){$n\!\!-\!\!1$}
\multiframe(0,-39)(10.5,0){1}(17,10){$*$}
\multiframe(17.5,10)(10.5,0){1}(10,10){$*$}
\end{picture}\normalsize
~=~ \footnotesize \begin{picture}(30,30)(0,0)
\multiframe(0,10)(10.5,0){1}(17,10){$1$}
\multiframe(0,-0.5)(10.5,0){1}(17,10){$2$}
\multiframe(0,-18)(10.5,0){1}(17,17){} \put(7,-15){$\vdots$}
\multiframe(0,-28.5)(10.5,0){1}(17,10){$n\!\!-\!\!1$}
\multiframe(0,-39)(10.5,0){1}(17,10){$\partial$}
\multiframe(17.5,10)(10.5,0){1}(10,10){$*$}
\end{picture}\normalsize
~+~ \footnotesize \begin{picture}(30,30)(0,0)
\multiframe(0,10)(10.5,0){1}(17,10){$1$}
\multiframe(0,-0.5)(10.5,0){1}(17,10){$2$}
\multiframe(0,-18)(10.5,0){1}(17,17){$ $} \put(7,-15){$\vdots$}
\multiframe(0,-28.5)(10.5,0){1}(17,10){$n\!\!-\!\!1$}
\multiframe(0,-39)(10.5,0){1}(17,10){$*$}
\multiframe(17.5,10)(10.5,0){1}(10,10){$\partial$}
\end{picture}\normalsize
\ee
which, substituted into (\ref{debal1}), gives \be\vspace{1cm}
\footnotesize \begin{picture}(30,30)(0,0)
\multiframe(0,10)(10.5,0){1}(17,10){$1$}
\multiframe(0,-0.5)(10.5,0){1}(17,10){$2$}
\multiframe(0,-18)(10.5,0){1}(17,17){$ $} \put(7,-15){$\vdots$}
\multiframe(0,-28.5)(10.5,0){1}(17,10){$n\!\!-\!\!1$}
\multiframe(0,-39)(10.5,0){1}(17,10){$n$}
\multiframe(17.5,10)(10.5,0){1}(10,10){$ *$}
\multiframe(17.5,-0.5)(10.5,0){1}(10,10){$* $}
\end{picture}\normalsize
~=~ \footnotesize \begin{picture}(30,30)(0,0)
\multiframe(0,10)(10.5,0){1}(17,10){$1$}
\multiframe(0,-0.5)(10.5,0){1}(17,10){$2$}
\multiframe(0,-18)(10.5,0){1}(17,17){$ $} \put(7,-15){$\vdots$}
\multiframe(0,-28.5)(10.5,0){1}(17,10){$n\!\!-\!\!1$}
\multiframe(0,-39)(10.5,0){1}(17,10){$\partial$}
\multiframe(17.5,10)(10.5,0){1}(10,10){$* $}
\multiframe(17.5,-0.5)(10.5,0){1}(10,10){$ *$}
\end{picture}\normalsize~.
\ee %$\diamond$
\\
The lemma \ref{lemma1} revealed crucial in proving
\underline{$H^{(n,l)}_{(1)}(\Omega_{(2)}({\mathbb R}^D))\cong 0$}.

The equation \be d^{\{1\}} \a(n,l)=0 \ee writes \be\vspace{1cm}
\footnotesize
\begin{picture}(30,30)(0,0) \multiframe(0,35)(10.5,0){1}(10,10){$
1$} \put(4,15){$\vdots$} \multiframe(0,5)(10.5,0){1}(10,29.5){$ $}
\multiframe(10.5,35)(10.5,0){1}(10,10){$ 1$}
\multiframe(10.5,15.5)(10.5,0){1}(10,19){$ $}
\multiframe(10.5,5)(10.5,0){1}(10,10){$l$}
\multiframe(0,-14.5)(10.5,0){1}(10,19){$ $} \put(4,-9.5){$\vdots$}
\put(14,20){$\vdots$} \multiframe(0,-25)(10.5,0){1}(10,10){$n$}
\multiframe(0,-35.5)(10.5,0){1}(10,10){$\partial$}
\end{picture}\normalsize
~=~~0\,. \ee Its solution is, as we just showed pictorially for
$l=2$, \be\vspace{1cm} \footnotesize \begin{picture}(30,30)(0,0)
\multiframe(0,20)(10.5,0){1}(10,10){$1$} \put(4,0){$\vdots$}
\multiframe(0,-10)(10.5,0){1}(10,29.5){$ $}
\multiframe(10.5,20)(10.5,0){1}(10,10){$1$}
\multiframe(10.5,0.5)(10.5,0){1}(10,19){$ $}
\multiframe(10.5,-10)(10.5,0){1}(10,10){$l$}
\multiframe(0,-29.5)(10.5,0){1}(10,19){$ $}
\put(4,-24.5){$\vdots$} \put(14,5){$\vdots$}
\multiframe(0,-40)(10.5,0){1}(10,10){$n$}
\end{picture}\normalsize
~=~~~ \footnotesize \begin{picture}(30,30)(0,0)
\multiframe(0,20)(10.5,0){1}(10,10){$1$} \put(4,0){$\vdots$}
\multiframe(0,-10)(10.5,0){1}(10,29.5){$ $}
\multiframe(10.5,20)(10.5,0){1}(10,10){$1$}
\multiframe(10.5,0.5)(10.5,0){1}(10,19){$ $}
\multiframe(10.5,-10)(10.5,0){1}(10,10){$l$}
\multiframe(0,-29.5)(10.5,0){1}(10,19){$ $}
\put(4,-24.5){$\vdots$} \put(14,5){$\vdots$}
\multiframe(0,-40)(10.5,0){1}(10,10){$\partial$}
\end{picture}\normalsize\,.
\label{inte2} \ee Substituting in the second cocycle condition \be
d^{\{2\}}\a(n,l)=0 \ee yields \be\vspace{1.5cm} \footnotesize
\begin{picture}(30,30)(0,0)
\multiframe(0,25)(10.5,0){1}(17,10){$1$} \put(7,5){$\vdots$}
\multiframe(0,-5)(10.5,0){1}(17,29.5){$ $}
\multiframe(17.5,25)(10.5,0){1}(17,10){$1$}
\multiframe(17.5,5.5)(10.5,0){1}(17,19){$ $}
\multiframe(17.5,-5)(10.5,0){1}(17,10){$l$}
\multiframe(17.5,-15.5)(10.5,0){1}(17,10){$\partial$}
\multiframe(0,-24.5)(10.5,0){1}(17,19){$ $}
\put(7,-19.5){$\vdots$} \put(24,10){$\vdots$}
\multiframe(0,-35)(10.5,0){1}(17,10){$n\!\!-\!\!1$}
\multiframe(0,-45.5)(10.5,0){1}(17,10){$\partial$}
\end{picture}\normalsize
~~~=~~0\,. \label{a(n+d,l+d)} \ee Applying the Poincar\'e lemma on
the second column, viewing the first one as spectator, yields \ba
\footnotesize \begin{picture}(30,30)(0,0)
\multiframe(0,20)(10.5,0){1}(17,10){$1$} \put(7,0){$\vdots$}
\multiframe(0,-10)(10.5,0){1}(17,29.5){$ $}
\multiframe(17.5,20)(10.5,0){1}(17,10){$1$}
\multiframe(17.5,0.5)(10.5,0){1}(17,19){$ $}
\multiframe(17.5,-10)(10.5,0){1}(17,10){$l$}
\multiframe(0,-29.5)(10.5,0){1}(17,19){$ $}
\put(7,-24.5){$\vdots$} \put(24,5){$\vdots$}
\multiframe(0,-40)(10.5,0){1}(17,10){$n\!\!-\!\!1$}
\multiframe(0,-50.5)(10.5,0){1}(17,10){$\partial$}
\end{picture}\normalsize
~~&\simeq&~ \footnotesize \begin{picture}(30,30)(0,0)
\multiframe(0,0)(10.5,0){1}(10,10){$\partial$}
\end{picture}\normalsize\!\!\!\!\!\!\!\!
\otimes\,\,\Big(~~\footnotesize
\begin{picture}(30,30)(0,0)
\multiframe(0,10)(10.5,0){1}(17,10){$1$}
\multiframe(0,-15.5)(10.5,0){1}(17,25){$ $} \put(7,-7.5){$\vdots$}
\multiframe(0,-26)(10.5,0){1}(17,10){$n$}
\end{picture}\normalsize \!\!\!\!\! \otimes \footnotesize
\begin{picture}(30,30)(0,0)
\multiframe(0,10)(10.5,0){1}(17,10){$1$}
\multiframe(0,-10.5)(10.5,0){1}(17,20){$ $} \put(7,-5.5){$\vdots$}
\multiframe(0,-21)(10.5,0){1}(17,10){$l\!\!-\!\!1$}
\end{picture}\normalsize\Big)
\nonumber \\ \nonumber \\ \nonumber \\
~~&\simeq&~ \footnotesize \begin{picture}(30,30)(0,0)
\multiframe(0,20)(10.5,0){1}(17,10){$1$} \put(7,0){$\vdots$}
\multiframe(0,-10)(10.5,0){1}(17,29.5){$ $}
\multiframe(17.5,20)(10.5,0){1}(17,10){$1$}
\multiframe(17.5,0.5)(10.5,0){1}(17,19){$ $}
\multiframe(17.5,-10)(10.5,0){1}(17,10){$l\!\!-\!\!1$}
\multiframe(17.5,-20.5)(10.5,0){1}(17,10){$\partial$}
\multiframe(0,-29.5)(10.5,0){1}(17,19){$ $}
\put(7,-24.5){$\vdots$} \put(24,5){$\vdots$}
\multiframe(0,-40)(10.5,0){1}(17,10){$n$}
\end{picture}\normalsize
~~\oplus \footnotesize \begin{picture}(30,30)(0,0)
\multiframe(0,20)(10.5,0){1}(17,10){$1$} \put(7,0){$\vdots$}
\multiframe(0,-10)(10.5,0){1}(17,29.5){$ $}
\multiframe(17.5,20)(10.5,0){1}(17,10){$1$}
\multiframe(17.5,0.5)(10.5,0){1}(17,19){$ $}
\multiframe(17.5,-10)(10.5,0){1}(17,10){$l\!\!-\!\!1$}
\multiframe(0,-29.5)(10.5,0){1}(17,19){$ $}
\put(7,-24.5){$\vdots$} \put(24,5){$\vdots$}
\multiframe(0,-40)(10.5,0){1}(17,10){$n$}
\multiframe(0,-50.5)(10.5,0){1}(17,10){$\partial$}
\end{picture}\normalsize
~~\oplus\, \footnotesize \begin{picture}(30,30)(0,0)
\multiframe(0,20)(10.5,0){1}(17,10){$1$} \put(7,0){$\vdots$}
\multiframe(0,-10)(10.5,0){1}(17,29.5){$ $}
\multiframe(17.5,20)(10.5,0){1}(17,10){$1$}
\multiframe(17.5,0.5)(10.5,0){1}(17,19){$ $}
\multiframe(17.5,-10)(10.5,0){1}(17,10){$l\!\!-\!\!2$}
\multiframe(17.5,-20.5)(10.5,0){1}(17,10){$\partial$}
\multiframe(0,-29.5)(10.5,0){1}(17,19){$ $}
\put(7,-24.5){$\vdots$} \put(24,5){$\vdots$}
\multiframe(0,-40)(10.5,0){1}(17,10){$n$}
\multiframe(0,-50.5)(10.5,0){1}(17,10){$n\!\!+\!\!1$}
\end{picture}\normalsize
~~\oplus\ldots \label{exp} \ea
\\
\\
\\
where the dots in the above equation correspond to tensors of
higher order L-grading (whose first column has length greater or
equal to $n+2$). The second and third terms must cancel because
they don't have the symmetry of the left-hand-side. Applying
$d^{\{2\}}$ on the sum of the second and third term and using our
hypothesis of reccurence, we obtain \be\vspace{1.5cm}
\footnotesize
\begin{picture}(30,30)(0,0)
\multiframe(0,20)(10.5,0){1}(17,10){$1$} \put(7,-9){$\vdots$}
\multiframe(0,-25)(10.5,0){1}(17,44.5){$ $}
\multiframe(17.5,20)(10.5,0){1}(17,10){$1$}
\multiframe(17.5,3.5)(10.5,0){1}(17,16){$ $}
\multiframe(17.5,-7)(10.5,0){1}(17,10){$l\!\!-\!\!1$}
\multiframe(17.5,-17.5)(10.5,0){1}(17,10){$l$}
\put(24,7){$\vdots$}
\multiframe(0,-35.5)(10.5,0){1}(17,10){$n\!\!-\!\!1$}
\multiframe(0,-46)(10.5,0){1}(17,10){$\partial$}
\end{picture}\normalsize
~~\simeq~ \footnotesize \begin{picture}(30,30)(0,0)
\multiframe(0,20)(10.5,0){1}(17,10){$1$} \put(7,-9){$\vdots$}
\multiframe(0,-25)(10.5,0){1}(17,44.5){$ $}
\multiframe(17.5,20)(10.5,0){1}(17,10){$1$}
\multiframe(17.5,3.5)(10.5,0){1}(17,16){$ $}
\multiframe(17.5,-7)(10.5,0){1}(17,10){$l\!\!-\!\!1$}
\multiframe(17.5,-17.5)(10.5,0){1}(17,10){$\partial$}
\put(24,7){$\vdots$}
\multiframe(0,-35.5)(10.5,0){1}(17,10){$n\!\!-\!\!1$}
\multiframe(0,-46)(10.5,0){1}(17,10){$\partial$}
\end{picture}\normalsize~~~~.
\ee This, substituted back in (\ref{inte2}), gives us the
vanishing of $H^{(n,l)}_{(1)}(\Omega_{(2)}({\mathbb R}^D))$ for
$n\neq D$, $l\neq D$ and $l\neq 0$.

%%%%%%%%%%%%%%%%%%%%%%%%%%%%%%%%%%%%%%%%%%%%%%%%%%%%%%%%%%%%%%%%%%%%%%
\subsection{Generalized Poincar\'e lemma in $\Omega^{(*,\ldots,*)}_{(*)}({\mathbb R}^D)$}\label{section3}
%%%%%%%%%%%%%%%%%%%%%%%%%%%%%%%%%%%%%%%%%%%%%%%%%%%%%%%%%%%%%%%%%%%%%%

Here comes the final inductive proof of \be
H^{(*,\ldots,*)}_{(*)}(\Omega_{(*)}({\mathbb R}^D))\cong 0\,,\ee
for diagrams obeying the assumption of the theorem \ref{GenPoin},
the generalized Poincar\'e lemma.

It is really proved if one has the following inductive
progression:
\begin{proposition}
\label{generalpoincare} Under the assumption that \be
H^{(l_1,\ldots,l_{S-1},l_S)}_{(k)}(\Omega_{(S)}({\mathbb
R}^D))\cong 0 \ee $\forall (l_1,\ldots,l_{S-1})\in {\mathbb
Y}^{(S-1)}$, $\forall k\in\{1,\ldots,S\}$, and $l_S$ fixed such
that $0<l_S<l_{S\!-\!1}$, the following holds : \be
H^{(l_1,\ldots,l_{S-1},l_{S}+1)}_{(k)}(\Omega_{(S)}({\mathbb
R}^D))\cong 0, \ee and \be
H^{(l_1,\ldots,l_{S-1},l_S,1)}_{(l)}(\Omega_{(S+1)}({\mathbb
R}^D))\cong 0\,, \ee $0<l<S+2$.
\end{proposition}
\vspace{.8cm} More explicitely, if the next-coming statements are
fulfilled : \ba &d^{\{I\}}\m(l_1,\ldots,l_{S-1},l_S)=0~~\forall
I\subset\{1,2,\ldots,S\}~\vert ~\# ~I=m&
\nonumber \\
&\Longrightarrow\m(l_1,\ldots,l_{S-1},l_S)=\sum_{J} d^{\{J\}}\n_J&
~~
\nonumber \\
&\forall J\subset\{1,2,\ldots,S\}~\vert~\#~
J=S\!+\!1\!-\!m~~{\rm{and}}~~
d^{\{J\}}\n_J\in\O_{(S)}^{(l_1,\ldots,l_{S-1},l_S)}({\mathbb
R}^D)\,,& \nn\ea it can be showed that the ones below are also
true : \ba &~\bullet
d^{\{I\}}\m(l_1,\ldots,l_{S-1},l_S+1)=0~~\forall
I\subset\{1,2,\ldots,S\}~\vert~\# ~I=m&
\nonumber \\
&\Longrightarrow\m(l_1,\ldots,l_{S-1},l_S+1)=\sum_{J}
d^{\{J\}}\n_J& ~~
\nonumber \\
&\forall J\subset\{1,2,\ldots,S\}~\vert~\#~
J=S\!+\!1\!-\!m~~{\rm{and}}~~
d^{\{J\}}\n_J\in\O_{(S)}^{(l_1,\ldots,l_{S-1},l_S+1)}({\mathbb
R}^D)& \nn\ea and \ba &~\bullet
d^{\{I\}}\m(l_1,\ldots,l_{S-1},l_S,1)=0~~\forall
I\subset\{1,2,\ldots,S,S+1\}~\vert~\# ~I=m&
\nonumber \\
&\Longrightarrow\m(l_1,\ldots,l_{S-1},l_S,1)=\sum_{J}
d^{\{J\}}\n_J& ~~
\nonumber \\
&\forall J\subset\{1,\ldots,S,S+1\}~\vert~\#~
J=S\!+\!2\!-\!m~~{\rm{and}}~~
d^{\{J\}}\n_J\in\O_{(S)}^{(l_1,\ldots,l_S,1)}({\mathbb R}^D)\,.&
\nn\ea
\\
Once this proposition is proved,
$H^{(*,\ldots,*)}_{(*)}(\Omega_{(*)}({\mathbb R}^D))\cong 0$
follows by induction.
%
%""""""""""""""""""""""""""""""""""""""""""""""""""""""""""""""""""""""
\subsection{Plan of the proof of proposition \ref{generalpoincare}}
%""""""""""""""""""""""""""""""""""""""""""""""""""""""""""""""""""""""
%
The plan is a direct generalization of the one we used to prove
that $H^{(l_1,l_2)}_{(k)}(\Omega_{(2)}({\mathbb R}^D))\cong 0$,
$\forall (l_1,l_2)\in {\mathbb Y}^{(2)}$, $l_2\neq 0$, $0<k<3$.
The strategy is to start by proving the
\begin{lemma}
\label{lem1} Knowing that \ba &d^I\m(l_1,\ldots,l_M,\ldots,l_S)=0&
\nonumber \\
&\forall I \subset \{1,2,\ldots,M\}~\vert~\#~I=m, ~0<M<S+1,&
\nonumber \\
&\forall (l_1,\ldots,l_{S-1})\in Y^{(S-1)}~and~ l_S ~fixed ~s.t.~
0<l_S<l_{S\!-\!1}& \nn\ea implies \ba
&\m(l_1,\ldots,l_M,\ldots,l_S)=\sum_{J}d^J\b_J&
\nonumber \\
&\forall J \subset \{1,2,\ldots,M\}~\vert~\#~J=M+1-m,
~d^J\b_J\in\O_{(S)}^{(l_1,\ldots,l_S)}({\mathbb R}^D),& \nn\ea
then one has also that \ba &d^I\m(l_1,\ldots,l_M,\ldots,l_S+1)=0&
\nonumber \\
&\forall I \subset \{1,2,\ldots,M\}~\vert~\#~I=m, ~0<M<S+1& \nn\ea
implies \ba &\m(l_1,\ldots,l_M,\ldots,l_S+1)=\sum_{J}d^J\a_J&
\nonumber \\
&\forall J \subset \{1,2,\ldots,M\}~\vert~\#~J=M+1-m,
~d^J\a_J\in\O_{(S)}^{(l_1,\ldots,l_S+1)}({\mathbb R}^D).& \nn\ea
and that \ba &d^I\m(l_1,\ldots,l_M,\ldots,l_S,1)=0&
\nonumber \\
&\forall I \subset \{1,2,\ldots,M\}~\vert~\#~I=m, ~0<M<S+1& \nn\ea
implies \ba &\m(l_1,\ldots,l_M,\ldots,l_S,1)=\sum_{J}d^J\a_J&
\nonumber \\
&\forall J \subset \{1,2,\ldots,M\}~\vert~\#~J=M+1-m,
~d^J\a_J\in\O_{(S+1)}^{(l_1,\ldots,l_S,1)}({\mathbb R}^D).& \nn\ea
\end{lemma}
Subsequently and with the help of lemma \ref{lem1}, one shows the
\begin{lemma}
\label{lem3} Under proposition \ref{generalpoincare}'s
assumptions, the following vanishing of cohomology groups arises :
\ba H^{(l_1,\ldots,l_S+1)}_{(S)}(\Omega_{(S)}({\mathbb
R}^D))&\cong &0\,,
\nn\\
H^{(l_1,\ldots,l_S,1)}_{(S+1)}(\Omega_{(S+1)}({\mathbb
R}^D))&\cong &0\,.\nn \ea
\end{lemma}
The proof of the proposition \ref{generalpoincare} can directly be
done using the three previous lemma's.

\section{Algebraic proof}

To begin with, we state our hypothesis: we have a tensor
$\a\in\Omega^{(l_1,\ldots,l_S)}_{(S)}({\cal
M})\subset\Omega^{l_1,\ldots,l_S}_{[S]}({\cal M})$ such that \be
\big( \prod_{i \in I} d_i \big) \a = 0 \; \; \; \; \forall I
\subset \{1,2, \dots, S\} \; \,\vert\, \, \# I = m
\label{startprof}\ee with $m \leq S$ a fixed integer such that
$l_{S-m+1}\neq 0$. In other words, our closure condition is weaker
than the one of the generalized cohomology due to proposition
\ref{weaker}. Strictly speaking, we will not proof here the
generalized Poincar\'e lemma, but only what is necessary for all
practical purposes in gauge theories.

The first step of this algebraic proof uses the second theorem of
\cite{Dubois-Violette:2001} which we remind here in the
\begin{lemma}\label{Dubv}
Let $K$ be an arbitrary non-empty subset of $\{1,2, \dots, S\}$.
If the multiform $\a\in\Omega_{[S]}({\cal M})$ is such that all
form degrees are smaller than $D$ and \ba \big( \prod_{i \in I}
d_i \big) \a = 0 \; \; \; \; \forall I \subset K \; \vert \, \# I
= m \nn \ea with $m$ a fixed integer $\leq \# K$, then
\begin{eqnarray}
\a \,\,=\, \,\a_{(m-1)}\,\, +\! \! \! \! \! \! \! \! \!
\sum_{\begin{array}{c}
J \subset K \\
\#J = \#K - m +1
\end{array}}
\! \! \! \! \! \! \! \! \! \big( \prod_{j \in J} d_j \big)
\alpha_J\nn
\end{eqnarray}
where $\a_{(m-1)}$ is a polynomial multiform of degree $\leq m-1$,
that is the coefficients are polynomial functions.
\end{lemma}

With the lemma \ref{Dubv}, we derive from (\ref{startprof}) with
$K=\{1,2, \dots, S\}$ that\be \a= \,\a_{(m-1)}\, + \! \! \! \! \!
\sum_{\begin{array}{c}
J \subset \{1,2, \dots, S\}\\
\#J = S - m +1
\end{array}}
\! \! \! \! \! \big( \prod_{j \in J} d_j \big) \a_J\,,\ee where
$\a_{(m-1)}$ is a polynomial multiform of degree $\leq m-1$. In
consequence one might try to proof the theorem separately for
$\a_{(m-1)}$ and for the second term in the right hand side.
Therefore the proof decomposes in two: we first show that if we
project a multiform $\sum_{J}\! \big( \prod_{j \in J} d_j \big)
\alpha_J$ on the symmetry of $\a$ we get $(\sum_J d^J)$-trivial
terms, that is trivial according to the equivalence relation
(\ref{equivrelatio}). The second part of the proof amounts to show
that $\a_{(m-1)}$ is trivial if its Young diagram is at least
$S-m-1$ columns long. These two results will prove the theorem.

\subsection{Young symmetrization} Let $J$ be a subset of $\{1,2,
\dots, S\}$ with $S-m+1$ elements. A multiform
$\b\in\Omega^{l_1,\ldots,l_S}_{[S]}({\cal M})$ which is equal to
$\big( \prod_{j \in J} d_j \big) \b_J$ is represented by a tensor
product of the form
\begin{center}\footnotesize
\begin{picture}(0,37)(0,-24)
\multiframe(-160,10)(10.5,0){1}(10,10){}
\put(-155.5,-1.5){$\vdots$}\put(-155.5,-13){$\vdots$}\put(-155.5,-24.5){$\vdots$}
\multiframe(-160,-36)(10.5,0){1}(10,10){}
\put(-140,7){$\bigotimes$}

\put(-120,7){$\ldots$} \put(-100,7){$\bigotimes$}

\multiframe(-80,10)(10.5,0){1}(10,10){}
\put(-75.5,-1.5){$\vdots$}\put(-75.5,-13){$\vdots$}
\multiframe(-80,-24.5)(10.5,0){1}(10,10){}
\multiframe(-80,-35)(10.5,0){1}(10,10){$\partial$}\put(-60,7){$\bigotimes$}

\put(-40,7){$\ldots$} \put(-20,7){$\bigotimes$}

\multiframe(0,10)(10.5,0){1}(10,10){}\put(5,-1.5){$\vdots$}
\multiframe(0,-15)(10.5,0){1}(10,10){}
\multiframe(0,-25.5)(10.5,0){1}(10,10){$\partial$}\put(20,7){$\bigotimes$}

\put(40,7){$\ldots$} \put(60,7){$\bigotimes$}

\multiframe(80,10)(10.5,0){1}(10,10){}\put(85,-1.5){$\vdots$}
\multiframe(80,-15)(10.5,0){1}(10,10){}
\end{picture}\normalsize\end{center}
Furthermore $\b_J$ has components with symmetry represented in
terms of Young diagrams as
\begin{center}\footnotesize
\begin{picture}(0,37)(0,-24)
\multiframe(-160,10)(10.5,0){1}(10,10){}
\put(-155.5,-1.5){$\vdots$}\put(-155.5,-13){$\vdots$}\put(-155.5,-24.5){$\vdots$}
\multiframe(-160,-36)(10.5,0){1}(10,10){}
\put(-140,7){$\bigotimes$}

\put(-120,7){$\ldots$} \put(-100,7){$\bigotimes$}

\multiframe(-80,10)(10.5,0){1}(10,10){}
\put(-75.5,-1.5){$\vdots$}\put(-75.5,-13){$\vdots$}
\multiframe(-80,-24.5)(10.5,0){1}(10,10){}\put(-60,7){$\bigotimes$}

\put(-40,7){$\ldots$} \put(-20,7){$\bigotimes$}

\multiframe(0,10)(10.5,0){1}(10,10){}\put(5,-1.5){$\vdots$}
\multiframe(0,-15)(10.5,0){1}(10,10){}\put(20,7){$\bigotimes$}

\put(40,7){$\ldots$} \put(60,7){$\bigotimes$}

\multiframe(80,10)(10.5,0){1}(10,10){}\put(85,-1.5){$\vdots$}
\multiframe(80,-15)(10.5,0){1}(10,10){}
\end{picture}\normalsize\end{center}
Thus the components of the multiform $\b$ transform in the tensor
product of the representation with the previous diagram (symmetry
of $\b_J$) and the completely symmetric representation with
$S-m+1$ boxes (symmetry of $\partial^{S-m+1}$)
\begin{center}\footnotesize
\begin{picture}(0,37)(0,-24)
\multiframe(-160,10)(10.5,0){1}(10,10){}
\put(-155.5,-1.5){$\vdots$}\put(-155.5,-13){$\vdots$}\put(-155.5,-24.5){$\vdots$}
\multiframe(-160,-36)(10.5,0){1}(10,10){}
\put(-140,7){$\bigotimes$}

\put(-120,7){$\ldots$} \put(-100,7){$\bigotimes$}

\multiframe(-80,10)(10.5,0){1}(10,10){}
\put(-75.5,-1.5){$\vdots$}\put(-75.5,-13){$\vdots$}
\multiframe(-80,-24.5)(10.5,0){1}(10,10){}\put(-60,7){$\bigotimes$}

\put(-40,7){$\ldots$} \put(-20,7){$\bigotimes$}

\multiframe(0,10)(10.5,0){1}(10,10){}\put(5,-1.5){$\vdots$}
\multiframe(0,-15)(10.5,0){1}(10,10){}\put(20,7){$\bigotimes$}

\put(40,7){$\ldots$} \put(60,7){$\bigotimes$}

\multiframe(80,10)(10.5,0){1}(10,10){}\put(85,-1.5){$\vdots$}
\multiframe(80,-15)(10.5,0){1}(10,10){}\put(100,7){$\bigotimes$}

\multiframe(120,7)(10.5,0){1}(10,10){$\partial$}\put(140,10){$\ldots$}
\multiframe(160,7)(10.5,0){1}(10,10){$\partial$}

\end{picture}\normalsize\end{center}
With this Young diagram, the $S-m+1$ antisymmetrizations with the
partial derivatives are no longer manifest.

Now one considers more specifically the multiform
$\theta_J\equiv\big( \prod_{j \in J} d_j \big) \alpha_J$ with the
same symmetry properties than $\a$, i.e.
$\theta_J\in\Omega^{(l_1,\ldots,l_S)}_{(S)}({\cal M})$. Thus in
the direct sum of representations obtained from the last tensor
product, we must only keep the ones associated with the Young
diagram $Y_{(l_1,\ldots,l_S)}^{(S)}$. From the lemma
\ref{commutprop} we know that the only possible Young diagrams for
$\a_J$ which gives a non-vanishing contribution to $\theta_J$
belongs the set of diagrams where one has removed the last box in
$S-m+1$ columns such that the final result is still a well-defined
Young diagram. The argument works for all $J$, hence the sum on
all possible $J$ will also obey the rule given in theorem
\ref{GenPoin}.

\subsection{Trivial polynomial irreducible tensors}

The present aim is to show that any polynomial tensor
$\a_{(m-1)}\in\Omega^{(l_1,\ldots,l_S)}_{(S)}({\cal M})$, with at
least $S-m+1$ columns and of degree of polynomiality in $x^\mu$
strictly smaller than $m$, is equal to a sum of $d^J\b_J$ with
$\#J=S-m+1$ and $\b_J\in\Omega_{(S)}({\cal M})$. To proof that, we
will show (in a constructive way) that this is true for any
monomial tensor of degree strictly smaller than $m$ represented by
the Young diagram $Y_{(l_1,\ldots,l_S)}^{(S)}$ with $l_{S-m+1}\neq
0$, which will end the proof of the generalized Poincar\'e lemma.

\subsubsection{Comments on irreducible monomial tensor fields}

Before entering into the core of the proof itself, we should get
more familiar with irreducible monomial tensor fields. Let us
consider an arbitrary monomial tensor field
$M^{(\l_1,\ldots,\l_S),\hat{n}}\in\Omega^{(\l_1,\ldots,\l_S)}_{(S)}({\cal
M})$ of degree of polynomiality equal to $n\leq S$ which reads
explicitly \be
M_{[\mu^1_1\ldots\mu^1_{\l_1}]\ldots[\mu^S_1\ldots\mu^S_{\l_S}]}^{(\l_1,\ldots,\l_S),\hat{n}}=\frac{1}{n!}\,\mu^{(\l_1,\ldots,\l_S)
\otimes
n}_{[\mu^1_1\ldots\mu^1_{\l_1}]\ldots[\mu^S_1\ldots\mu^S_{\l_S}]\,(\nu_1\ldots\nu_n)}\,\,x^{\nu_1}\ldots
x^{\nu_n}\,,\label{defmon}\ee where $\mu^{(\l_1,\ldots,\l_S)
\otimes
n}_{[\mu^1_1\ldots\mu^1_{\l_1}]\ldots[\mu^S_1\ldots\mu^S_{\l_S}]\,(\nu_1\ldots\nu_n)}\in\mathbb
R $. The formula (\ref{defmon}) states that any monomial tensor
$M^{(\l_1,\ldots,\l_S),\hat{n}}$ is characterized by a constant
tensor $\mu^{(\l_1,\ldots,\l_S)\otimes n}$, symmetric in $n$
indices. Conversely, to any constant tensor symmetric in at least
$n$ indices, we can associate a monomial tensor.

If the Young diagram $Y_{(\l_1,\ldots,\l_S)}^{(S)}$ is pictured as
\begin{center}\footnotesize
\begin{picture}(0,37)(0,-24)
\multiframe(-80,10)(10.5,0){1}(10,10){}
\put(-75.5,-1.5){$\vdots$}\put(-75.5,-13){$\vdots$}\put(-75.5,-24.5){$\vdots$}
\multiframe(-80,-37)(10.5,0){1}(10,10){}
\put(-60,7){$\ldots$}\put(-60,-20){$\ldots$}
\multiframe(-40,10)(10.5,0){1}(10,10){}
\put(-35.5,-1.5){$\vdots$}\put(-35.5,-13){$\vdots$}
\multiframe(-40,-24.5)(10.5,0){1}(10,10){}
\put(-20,7){$\ldots$}\put(-20,-15){$\ldots$}
\multiframe(0,10)(10.5,0){1}(10,10){}\put(5,-1.5){$\vdots$}
\multiframe(0,-15)(10.5,0){1}(10,10){}
\end{picture}\normalsize\end{center}
then the constant tensor $\mu$ characterizing
$M^{(\l_1,\ldots,\l_S),\hat{n}}$ is described by the tensor
product diagram $\hat{D}_{(\l_1,\ldots,\l_S)\otimes n}\equiv
Y_{(\l_1,\ldots,\l_S)}^{(S)}\otimes \hat{Y}^S_n$ pictured as
\begin{center}\footnotesize
\begin{picture}(0,37)(0,-24)
\multiframe(-80,10)(10.5,0){1}(10,10){}
\put(-75.5,-1.5){$\vdots$}\put(-75.5,-13){$\vdots$}\put(-75.5,-24.5){$\vdots$}
\multiframe(-80,-37)(10.5,0){1}(10,10){}
\put(-60,7){$\ldots$}\put(-60,-20){$\ldots$}
\multiframe(-40,10)(10.5,0){1}(10,10){}
\put(-35.5,-1.5){$\vdots$}\put(-35.5,-13){$\vdots$}
\multiframe(-40,-24.5)(10.5,0){1}(10,10){}
\put(-20,7){$\ldots$}\put(-20,-15){$\ldots$}
\multiframe(0,10)(10.5,0){1}(10,10){}\put(5,-1.5){$\vdots$}
\multiframe(0,-15)(10.5,0){1}(10,10){} \put(20,7){$\bigotimes$}
\multiframe(40,7)(10.5,0){1}(10,10){$\times$}\put(60,10){$\ldots$}
\multiframe(80,7)(10.5,0){1}(10,10){$\times$}
\end{picture}\normalsize\end{center}
where we used the notation according to which we fill by a cross
symbol $\times$ all the cells in the empty Young diagram
$D_{(\l_1,\ldots,\l_S)\otimes n}\equiv
Y_{(\l_1,\ldots,\l_S)}^{(S)}\otimes Y^S_n$ (associated to
$\mu^{(\l_1,\ldots,\l_S)\otimes n}$) that corresponds to indices
contracted with a $x$ in $M^{(\l_1,\ldots,\l_S),\hat{n}}$. This
explains why the second factor in the tensor product is a row of
$n$ cells, all filled by a cross. In the sequel, the hat will
signify that the diagram is filled with crosses according to the
previous rule.

\noindent\,\,{\bf Notation:} Let us denote the space of monomial
tensor fields on $\cal M$ in the irreducible representation
associated with the Young diagram $Y_{(\l_1,\ldots,\l_S)}^{(S)}$
and of degree $n$ by
$\Omega_{(S)}^{(\l_1,\ldots,\l_S),\hat{n}}({\cal M})$. The space
$\hat{\Omega}^{(\l_1,\ldots,\l_S)\otimes n}$ is defined as the
space of constant tensors in the tensor product representation
$\hat{D}_{(\l_1,\ldots,\l_S)\otimes n}$.

Mathematically speaking, the previous equation (\ref{defmon})
defines a bijective map \be
f:\hat{\Omega}^{(\l_1,\ldots,\l_S)\otimes
n}\rightarrow\Omega_{(S)}^{(\l_1,\ldots,\l_S),\hat{n}}({\cal
M}):\mu^{(\l_1,\ldots,\l_S)\otimes n}\rightarrow
M^{(\l_1,\ldots,\l_S),\hat{n}}\,.\label{biject}\ee

We can decompose the constant tensor
$\m^{(\l_1,\ldots,\l_S)\otimes
n}\in\hat{\Omega}^{(\l_1,\ldots,\l_S)\otimes n}$ according to its
irreducible representations under $GL(D)$:
$\mu^{(\l_1,\ldots,\l_S)\otimes
n}=\mu^{(1),n}+\ldots+\mu^{(r),n}$. This sum corresponds to the
direct sum of Young diagram $\hat{D}_{(\l_1,\ldots,\l_S)\otimes
n}=\hat{Y}_{(1),n}\oplus \ldots\oplus\hat{Y}_{(r),n}$ using
successively the branching rule. But some $\mu^{(i),\hat{n}}$
identically vanish. They are characterized by Young diagrams
$\hat{Y}_{(i),n}$ such that two cells containing a cross are
present in the same column. Let $\mu^J$ be the constant tensors
associated to the Young diagrams $\hat{Y}_{J}^{(S+n)}$ with
$J\subset\{1,\ldots,S+n\}$ (\#J=n) constructed according to the
\emph{selection rule} :
\begin{description}
  \item[1.] One starts from the Young diagram $Y^{(S)}_{(\l_1,\ldots,\l_S)}$.
  \item[2.] One adds $n$ cells containing a cross to this diagram. At each step one can
put a cell either

- at the bottom of a column with no cross or

- in the first row, at the right of the diagram.
  \item[3.] The result must be a well defined Young diagram.
\end{description}
The set $J$ encodes the column whose bottom cell contains a cross.
The selection amounts to require that
$Y^{(S)}_{(\l_1,\ldots,\l_S)}\Subset \hat{Y}_{J}^{(S+n)}$. The
previous construction is nothing else than what is called the
Pieri map in the mathematical litterature.

The monomial tensor $M^{\hat{J}}$ is equal to \be
M^{\hat{J}}_{[\mu^1_1\ldots\mu^1_{\l_1}]\ldots[\mu^S_1\ldots\mu^S_{\l_S}]}=\frac{1}{n!}\,\sum_J\big(\,\mu^J_{[\mu^1_1\ldots\mu^1_{\ell^J_1}]\ldots[\mu^S_1\ldots\mu^S_{\ell^J_S}]\,(\nu_1\ldots\nu_n)}\,\,\prod_{j\in
J}x^{\mu^j_1}\,\big)\,,\label{sometimes}\ee where we introduce
some

\noindent\,\,{\bf Notations:} Let $N$ be a natural number in
${\mathbb N}_0$. For any subset $J\subset \{1,\ldots,N\}$ we
define the numbers $\ell^J_j$ for any $j\in \{1,\ldots,N\}$ by
\be \ell^J_j\equiv\left\{\begin{array}{lll}\l_j\,\quad&\mbox{if}\,\,j\not\in J\,,&\\
\l_j+1\,\quad&\mbox{if}\,\,j\in J\,.&\end{array}\right. \ee Let
$\{Y^{(S+n)}_J\}$ be the set of Young diagrams $Y^{(S+n)}_J$ in
the tensor product $D_{(\l_1,\ldots,\l_S)\otimes n}$ which are
obtained from $Y^{(S)}_{(\l_1,\ldots,\l_S)}$ following the
selection rule. The space of the associated constant tensors
$\mu^J$ is denoted by $\hat{\Omega}_{(S+n)}^J$. There is an
implicit symmetrization in the constant tensor $\mu^J$ due to the
contraction with the $x$'s. The space of monomial tensors
$M^{\hat{J}}$ is denoted by $\Omega_{(S)}^{\hat{J}}({\cal M})$.

Since $\hat{\Omega}^{(\l_1,\ldots,\l_S)\otimes n}=\oplus_J
\hat{\Omega}_{(S+n)}^J$, we can define the restriction of the
bijective map (\ref{biject}) to each irreducible invariant
subspace\be
f^J:\hat{\Omega}_{(S+n)}^J\rightarrow\Omega_{(S)}^{\hat{J}}({\cal
M}):\mu^J\rightarrow M^{\hat{J}}\,.\ee

The partial derivative $\partial_\a$ of the product of x's is
equal to \be\partial_\a\big(\prod_{j\in
J}x^{\mu^j}\big)=\sum_{i\in J}\d_\a^{\mu^i}\big(\prod_{j\in
\overline{J}=J-\{j\}}x^{\mu^j}\big)\,.\ee The polynomial $\partial
M^{\hat{J}}$ is of course of degree $n-1$. Due to the implicit
symmetrization of $\mu^J$, the $n$ terms obtained by the
derivation of $\prod_{j\in J}x^{\mu^j_1}$ in (\ref{sometimes}) are
equal. This gives a factor $n$ which implies that the constant
tensor characterizing $\partial M^{\hat{J}}$ is exactly equal to
$\mu^J$ (with the appropriate symmetrization on the indices
corresponding to cells containing a cross). In terms of diagrams
associated with $\m^J$, the operation of differentiation becomes a
sum on the set of diagrams obtained by removing a cross in one
cell of the Young diagram of $M^{\hat{J}}$.

The operator \be d^I:\Omega^{(\l_1,\ldots,\l_S)}_{(S)}({\cal
M})\rightarrow \Omega^{(\ell^I_1,\ldots,\ell^I_S)}_{(S)}({\cal M})
\ee with $I\subset \{1,\ldots,S\}$ ($\#I=m$). Its action in the
space of monomial tensor fields defines a map
\begin{equation} d^I:\Omega_{(S)}^{\hat{J}}({\cal M})\rightarrow
\Omega_{(S)}^{\hat{K}}({\cal M})
\end{equation}
with $K=J-I$ (\#K=n-m). It induces a map
\begin{equation} \d_I:\hat{\Omega}_{(S+n)}^J\rightarrow
\hat{\Omega}_{(S+n-m)}^K
\end{equation} such that $f^K$ is a morphism between the space of
constant tensors in the representation pictured by
$\hat{Y}^{(S+n)}_J$ and the space of polynomial tensors of order
$n$ pictured by $Y^{(S)}_J$. Indeed, the following square commutes
\ba \xyma{ \hat{\Omega}_{(S+n)}^J \ar[r]^{f^J} \ar[d]^{\d_I} &
\Omega_{(S)}^{\hat{J}}({\cal M})      \ar[d]^{d^I}\\
\hat{\Omega}_{(S+n-m)}^K        \ar[r]^{f^K} &
\Omega_{(S)}^{\hat{K}}({\cal M}) } \label{squarecomp}\ea The Young
diagram $\hat{Y}^{(S+n-m)}_K$ is obtained by the
\emph{superposition rule}:

\begin{description}
  \item[1] One superposes the Young diagram $\hat{Y}^{(S+n)}_J$ with the Young diagram $\tilde{Y}^{(S)}_{(\ell^I_1,\ldots,\ell^I_S)}$ characterizing $d^I$.
  \item[2] One suppresses a cross in the superposed diagram if there is a partial derivative in the same cell
  \item[3] The operator $\d_I$ acts trivially if there exist a cell which contain a partial derivative but no cross or if $Y^{(S)}_{(\ell^I_1,\ldots,\ell^I_S)}$ is not a subdiagram of $Y^{(S+n)}_J$.
\end{description}

\noindent\,\,{\bf Example:} To fix the ideas, let us consider a
specific example: $\m\in\hat{\Omega}^{(1,1)\otimes 2}$. It belongs
to the tensor product representation associated to
\ba\footnotesize
\begin{picture}(0,0)(20,0)
\multiframe(-20,0)(10.5,0){1}(10,10){}
\multiframe(-9.5,0)(10.5,0){1}(10,10){} \put(10.5,0){$\bigotimes$}
\multiframe(30,0)(10.5,0){1}(10,10){$\times$}
\multiframe(40.5,0)(10.5,0){1}(10,10){$\times$}
\end{picture}\normalsize\nn\ea
It decomposes into the sum
$\m=\mu^{(1)}+\mu^{(2)}+\mu^{(3)}+\mu^{(4)}$ where
$\m^{(1)}\in\hat{\Omega}_{(2)}^{(3,1)}$,
$\m^{(2)}\in\hat{\Omega}_{(2)}^{(2,2)}$,
$\m^{(3)}\in\hat{\Omega}_{(2)}^{(2,1,1)}$, and
$\m^{(4)}\in\hat{\Omega}_{(2)}^{(1,1,1,1)}$. Diagramatically, we
have \vspace{0.3cm}\begin{center}\footnotesize
\begin{picture}(0,0)(0,0)
\multiframe(-120,10)(10.5,0){1}(10,10){}
\multiframe(-120,-0.5)(10.5,0){1}(10,10){$\times$}
\multiframe(-120,-11)(10.5,0){1}(10,10){$\times$}
\multiframe(-109.5,10)(10.5,0){1}(10,10){}
\put(-90,10){$\bigotimes$} \multiframe(-70,10)(10.5,0){1}(10,10){}
\multiframe(-70,-0.5)(10.5,0){1}(10,10){$\times$}
\multiframe(-59.5,10)(10.5,0){1}(10,10){}
\multiframe(-59.5,-0.5)(10.5,0){1}(10,10){$\times$}
\put(-40,10){$\bigotimes$} \multiframe(-20,10)(10.5,0){1}(10,10){}
\multiframe(-20,-0.5)(10.5,0){1}(10,10){$\times$}
\multiframe(-9.5,10)(10.5,0){1}(10,10){}
\multiframe(1,10)(10.5,0){1}(10,10){$\times$}
\put(20,10){$\bigotimes$} \multiframe(40,10)(10.5,0){1}(10,10){}
\multiframe(50.5,10)(10.5,0){1}(10,10){}
\multiframe(61,10)(10.5,0){1}(10,10){$\times$}
\multiframe(71.5,10)(10.5,0){1}(10,10){$\times$}
\end{picture}\normalsize\end{center}\vspace{0.3cm}
We have $f(\mu^{(1)})\equiv 0$ since the diagram of $\m^{(1)}$ has
two cells containing a cross which are present in the same column.
Thus we have:
$f(\mu):=M_2=M_2^{\{1,2\}}+M_2^{\{1,3\}}+M_2^{\{3,4\}}$. This
follows effectively the selection rule. Let us now take a look at
$d^{\{1\}}M_2$ where the action of $d^{\{1\}}:=d$ is represented
by the diagram
\begin{center}\footnotesize
\begin{picture}(0,0)(0,0)
\multiframe(0,10)(10.5,0){1}(10,10){}
\multiframe(0,-0.5)(10.5,0){1}(10,10){$\partial$}
\multiframe(10.5,10)(10.5,0){1}(10,10){}
\end{picture}\normalsize\end{center}
In that case, $dM_2^{\{3,4\}}\equiv 0$. For $dM_2=f(\d\m)$, the
constant tensor $\d\m$ falls into the representation
\begin{center}\footnotesize
\begin{picture}(0,0)(0,0)
\multiframe(-30,10)(10.5,0){1}(10,10){}
\multiframe(-30,-0.5)(10.5,0){1}(10,10){}
\multiframe(-19.5,10)(10.5,0){1}(10,10){}
\multiframe(-19.5,-0.5)(10.5,0){1}(10,10){$\times$}
\put(0,10){$\bigotimes$} \multiframe(20,10)(10.5,0){1}(10,10){}
\multiframe(20,-0.5)(10.5,0){1}(10,10){}
\multiframe(30.5,10)(10.5,0){1}(10,10){}
\multiframe(41,10)(10.5,0){1}(10,10){$\times$}
\end{picture}\normalsize\end{center}
The result follows the superposition rules.

With the help of the following lemma, we get closer to the
complete proof of theorem \ref{GenPoin}.
\begin{lemma}Let $\tilde{Y}^{(S)}_{(\ell^I_1,\ldots,\ell^I_S)}$ be the Young diagram representing the map $d^I:\Omega^{\hat{J}}({\cal M})\rightarrow
\Omega^{\hat{K}}({\cal M})$. The corresponding map
$\d_I:\Omega^J\rightarrow \Omega^K$ provides an isomorphism
between $\Omega^J$ and $\Omega^K$ if and only if
\begin{itemize}
\item $Y_{(\ell^I_1,\ldots,\ell^I_S)}^{(S)}$ is a subdiagram of
$Y^{(S+n)}_J$ and
\item Any box in $\tilde{Y}_I$ containing a
partial derivative corresponds to a box containing a cross in
$\hat{Y}^{(S+n)}_J$.
\end{itemize}\label{isomorfism}
\end{lemma}Among other things, the map $\d_I$ is surjective, which
will be extremely useful in the sequel. The proof of the lemma
\ref{isomorfism} amounts to notice that $\d_q$ (i) satisfies the
assumptions of the Schur lemma and (ii) acts non trivially on
$\Omega^J$.

\subsubsection{End of the algebraic proof}

After all these preliminaries, let us come back to our initial
aim. We now want to show that any monomial tensor
$M^{p,\hat{n}}\in\Omega_{(S)}^{(\l_1,\ldots,\l_S),\hat{n}}({\cal
M})$ of polynomiality degree $n\leq m-1$ which is represented by a
Young diagram $Y_{(l_1,\ldots,l_S)}^{(S)}$ with $l_{S-m+1}\neq 0$
can be represented as a direct sum of Young diagrams constructed
by putting a partial derivative in the last cell of $S-m+1$
columns of $Y_{(l_1,\ldots,l_S)}^{(S)}$.

To start with, we know that the maps $f^J$ are bijective, so let
$\m^J\in\hat{\Omega}^{(\l_1,\ldots,\l_S)\otimes n}$ be the inverse
image of the monomial tensor $M^{\hat{J}}\in
\Omega_{(S)}^{\hat{J}}({\cal M})$ by $f^J$, i.e.
$M^{\hat{J}}=f^J(\m^{(l_1,\ldots,l_S)\otimes n})$. These
corresponding Young diagrams obeys the assumption of the following
lemma which makes a link between the coefficients of
$M^{(l_1,\ldots,l_S),\hat{n}}$ and the coefficients of a potential
inverse image by $d^I$ with $\# I=S-m+1$.

\begin{lemma}Let
\begin{description}
\item[-] $Y_{(\ell_1,\ldots,\ell_S)}^{(S)}$ be a Young diagram with at least $S-m+1$ columns ($\ell_{S+m-1}\neq 0$).
\item[-] $I$ be a subset of $\{1,\ldots,S\}$ with $\# I=S-m+1$.
\item[-] $Y_{(\l^I_1,\ldots,\l^I_S)}^{(S)}$ be a Young diagram
constructed from the Young diagram
$Y_{(\ell_1,\ldots,\ell_S)}^{(S)}$ by removing a cell in $S-m+1$
distinct columns characterized by the elements of $I$.
\item[-] $\tilde{Y}^I_{(\ell_1,\ldots,\ell_S)}$ be the Young diagram associated to
$Y_{(\l^I_1,\ldots,\l^I_S)}^{(S)}$ by replacing the $S-m+1$
removed boxes from $Y_{(\ell_1,\ldots,\ell_S)}^{(S)}$ with boxes
containing a partial derivative symbol.
\item[-] $n$ be an integer obeying $n\leq m-1\leq S$.
\item[-] $\hat{D}_{(\ell_1,\ldots,\ell_S)\otimes n}=\oplus_J
\hat{Y}^{(S+n)}_J$ be the decomposition of
$\hat{D}_{(\ell_1,\ldots,\ell_S)\otimes n}$ following the
selection rule.
\item[-] $\hat{D}_{(\l^I_1,\ldots,\l^I_S)\otimes
(S+n-m+1)}=\oplus_K \hat{Y}^{(2S+1+n-m)}_{I,K}$ be the
decomposition of $\hat{D}_{(\l^I_1,\ldots,\l^I_S)\otimes n}$
following the selection rule (one adds $S+n-m+1$ filled cells to
$Y_{(\l^I_1,\ldots,\l^I_S)}^{(S)}$).
\end{description}

First of all, we have
$\{\hat{Y}^{(S+n)}_J\}\subset\{\hat{Y}^{(2S+1+n-m)}_{I,K}\}$.
Secondly, the diagram $\tilde{Y}^I_{(\ell_1,\ldots,\ell_S)}$
defines an application \ba d^I:\Omega_{(S)}^{\hat{J}}({\cal
M})\rightarrow \Omega^{(\ell_1,\ldots,\ell_S)}_{(S)}({\cal
M})\,.\nn\ea Finally, for all $J$ there always exists an $I$ and a
$K$ such that, following the superposition rule, the operator
$d_J^{S-m+1}$ sends $\hat{Y}^{(2S+1+n-m)}_{I,K}$ on
$\hat{Y}^{(S+n)}_J$.\label{squarecompl}
\end{lemma}
\proof{For any empty Young diagram $\hat{Y}^{(S+n)}_J$ there
exists a $K$ and an $I$ such that
$\hat{Y}^{(S+n)}_J=\hat{Y}^{(2S+n-m+1)}_{I,K}$ since the boxes
removed to $Y_{(\ell_1,\ldots,\ell_S)}^{(S)}$ by constructing
$Y_{(\l^I_1,\ldots,\l^I_S)}^{(S)}$ are given back in one of the
diagram of $\hat{D}_{(\l^I_1,\ldots,\l^I_S)\otimes (S+n-m+1)}$ due
to the selection rule. The second point is trivial from the
definition of $\tilde{Y}^I_{(\ell_1,\ldots,\ell_S)}$. The final
point of the lemma \ref{squarecompl} can be understood from the
proof of the first point, working with hatted tensors this time.
We understand from the superposition rule that the removed boxes
correspond to boxes with partial derivatives in
$\tilde{Y}^I_{(\ell_1,\ldots,\ell_S)}$ while all boxes given back
by the selection rule obviously contain a hat.}

To end up, we deduce from lemmas \ref{isomorfism} and
\ref{squarecompl} that, for any given constant tensor $\m^J$, we
have a constant tensor
$\n_J^{I,K}\in\hat{\Omega}_{(2S+1+n-m)}^{I,K}$ such that
$\m^J=\d_I(\n_J^{I,K})$. If we take $\m^J$ to be the inverse image
of $M^{\hat{J}}$, then we have, putting everything together, \ba
d^I\left(f(\n_J^{I,K})\right)
&=& (d^I\circ f^K)(\n^{I,K}_J)\nn\\
&=& (f^J\circ\d^I)(\n^{I,K}_J)\nn\\
&=&f^J(\m^J)=M^{\hat{J}}\nn \ea where we used at the second line
that the square (\ref{squarecomp}) commutes. In conclusion we have
succeeded to proof that any monomial irreducible tensor field of
degree $n\leq m-1$ with at least $S-m+1$ columns is trivial
according to the equivalence relation (\ref{equivrelatio}).
Therefore this is also true for any polynomial tensor of degree
$n\leq m-1$ with at least $S-m+1$ columns.

%%%%%%%%%%%%%%%%%%%%%%%%%%%%%%%%%%%%%%%%%%%%%%%%%%%%%%%%%%%%%%%%
%%%%%%%%%%%%%%%%%%%%%%%%%%%%%%%%%%%%%%%%%%%%%%%%%%%%%%%%%%%%%%%%

\chapter{Proof of the no-go theorem}\label{pmain}

%%%%%%%%%%%%%%%%%%%%%%%%%%%%%%%%%%%%%%%%%%%%%%%%%%%%%%%%%%%%%%%

\section{Cohomology of $\gamma$}

The following lemma completely gives $H(\gamma)$.
\begin{lemma}
The cohomology of $\gamma$ is given by,
\begin{equation}\label{gamma}
H(\gamma,{\cal A})={\cal{I}}\otimes V.
\end{equation}
Here, the algebra ${\cal{I}}$ is the algebra of the local forms
with coefficients that depend only on the variables $F^A_{ijk}$,
the antifields $\phi^*_M$, and all their partial derivatives up to
a finite order (``gauge-invariant" local forms). These variables
are collectively denoted by $\chi$. The algebra $V$ is the
polynomial algebra in the ghosts $\eta^A$ of ghost number two and
their time derivatives.
\end{lemma}
\proof{ The generators of ${\cal{A}}$ can be grouped in three
sets:
\begin{eqnarray}
&&T=\{t^i\}=\{\partial^{}_{{\mu}_1 \ldots
{\mu}_k}F^{A}_{ijk},\partial^{}_{{\mu}_1 \ldots
{\mu}_k}\phi_M^*,\eta^{A(l)},dx^{\mu}\}\\
&&U=\{u^{\alpha}\}=\{\partial^{}_{(i_1 \ldots i_k}A^{A(l)}_{[i)_2
j]_1},\partial^{}_{(i_1 \ldots i_{k-1}}C^{A(l)}_{i_k)}\}\\
&&V=\{v^{\alpha}\}=\{\partial^{}_{i_1 \ldots
i_k}\partial^{}_{[i}C^{A(l)}_{j]},\partial^{}_{i_1 \ldots
i_k}\eta^{A(l)}\}
\end{eqnarray}
($k,l = 0, \cdots$) where $[$ $]$ and $($ $)$ mean respectively
antisymmetrization and symmetrization; the subscript indicates the
order in which the operations are made.

The differential $\gamma$ acts on these three sets in the
following way
\begin{equation}
\gamma T=0,\quad \gamma U=V,\quad \gamma V=0.
\end{equation}
The elements of $U$ and $V$ are in a one-to-one correspondence and
are linearly independent with respect to each other, so they
constitute a manifestly contractible part of the algebra and can
thus be removed from the cohomology.

No element in the algebra of generated by  $T$ is trivial in the
cohomology of $\gamma$, except $0$. Indeed, let us assume the
existence of a local form $F(t^i)\neq 0$ which is $\gamma$-exact,
then
\begin{eqnarray}
F(t^i)=\gamma
G(t^i,u^{\alpha},v^{\alpha})=v^{\alpha}\frac{\partial^L
G}{\partial u^{\alpha}}(t^i,u^{\alpha},v^{\alpha}).
\end{eqnarray}
But this implies that
\begin{equation}
F(t^i)=F(t^i)\mid_{v^{\alpha}=0}=0,
\end{equation}
as announced. }

Note that contrary to what happens in the non-chiral case, the
temporal derivatives of the ghosts $\eta^A$ are non-trivial in
cohomology.  There is thus an infinite number of generators in
ghost number two for $H(\gamma)$, namely, all the $\eta^{A(l)}$'s.
In contrast, in the non-chiral case, one has $\partial_0\eta^A =
\gamma C^A_0$ and so $\partial_0\eta^A$ (and all the subsequent
derivatives) are $\gamma$-exact.  In the chiral case, there is no
$C^A_0$.

Let$\{\omega^I\}$ be a basis of the vector space $V$ of
polynomials in the variables $\eta^A$ and all their time
derivatives. Theorem \ref{gamma} tells us that
\begin{equation}
\gamma \alpha =0, \; \alpha \in \cal{A} \quad\Leftrightarrow
\quad\alpha =\sum_I P_I(\chi)\omega^I+ \gamma\beta.
\end{equation}
Furthermore, because $\omega^I$ is a basis of $V$
\begin{equation}
\sum_I P_I(\chi)\omega^I=\gamma\beta \quad\Rightarrow \quad
P_I(\chi)=0.
\end{equation}

It will be useful in the sequel to choose a special basis $\{
\omega^I \}$. The vector space $V$ of polynomials in the ghosts
$\eta^A$ and their time derivatives splits as the direct sum
$V^{2k}$ of vector spaces with definite pure ghost number $2k$.
The space $V^0$ is one-dimensional and given by the constants. We
may choose $1$ as basis vector for $V^0$, so let us turn to the
less trivial spaces $V^{2k}$ with $k \not= 0$. These spaces are
themselves the direct sums of finite dimensional vector spaces
$V^{2k}_r$ containing the polynomials with exactly $r$ time
derivatives of the $\eta$'s (e.g., $\partial_0 \eta^A \,
\partial_{00} \eta^B$ is in $V^{4}_3$). The following lemma
provides a basis of $V^{2k}$ for $k \not=0$:
\begin{lemma}
Let $V^{2k}$ be the vector space of polynomials in the variables
$\eta^{A(l)}$ with fixed pure ghost number $2 k \neq 0$. $V^{2k}$
is the direct sum
\begin{equation}
V^{2k}=V^{2k}_0\oplus V^{2k}_1 \oplus \ldots,
\end{equation}
where $V^{2k}_m$ is the subspace of $V^{2k}$ containing the
polynomials with exactly $m$ derivatives of $\eta^A$.  One has
dim$V^{2k}_m\leq$ dim$V^{2k}_{m+1}$. There exist a basis of
$V^{2k}_m$
\begin{equation}
\{\omega^{I_m}_{(m)}:\,I_m=1,\ldots,q_m;\, m=0,\ldots\},
\end{equation}
which fulfills
\begin{equation}
\omega^{I_m}_{(m)}=\partial_0\omega^{I_m}_{(m-1)}\quad
(I_m=1,\ldots,q_{m-1}). \label{ploc}
\end{equation}
In other words, the first $q_{m-1}$ basis vectors of $V^{2k}_m$
are directly constructed from the basis vectors of $V^{2k}_{m-1}$
by taking their time derivative $\partial_0$. \label{zap}
\end{lemma}
\proof{We will prove the lemma by induction. For $m=0$, take an
arbitrary basis of $V^{2k}_0$ (space of polynomials in the
undifferentiated ghosts $\eta^A$ of degree $k$). Assume now that a
basis with the required properties exists up to order $m-1$. Let
$\{\omega^I_{(m-1)};\, I=0,\ldots,q_{m-1}\}$ be a basis with those
properties for $V^{2k}_{m-1}$. We want to prove that it is
possible to construct a basis of $V^{2k}_m$ where the first
$q_{m-1}$ basis vectors are the time derivatives of the basis
vectors of $V^{2k}_{m-1}$. We only have to show that the
$\partial_0\omega^I_{(m-1)}$ are linearly independent (because
they can always be completed to form a basis of $V^{2k}_{m}$). In
other words, we must prove that
\begin{equation}
\sum \limits_{I=1}^{q_{m-1}}\lambda_I\partial_0\omega^I_{(m-1)}
=\partial_0(\sum
\limits_{I=1}^{q_{m-1}}\lambda_I\omega^I_{(m-1)})=0 \label{bardaf}
\end{equation}
implies $\lambda_I=0$. But (\ref{bardaf}) is equivalent to
\begin{equation}
\sum \limits_{I=1}^{q_{m-1}}\lambda_I\omega^I_{(m-1)}=K,
\end{equation}
where $K$ is a constant (algebraic Poincar\'e lemma in form degree
0). $K$ must be equal to zero because we are in pure ghost number
$\neq 0$. By hypothesis, the $\omega^I_{(m-1)}$ are linearly
independent, hence the $\lambda_I$ must be all equal to zero,
which ends the proof. }

%%%%%%%%%%%%%%%%%%%%%%%%%%%%%%%%%%%%%%%%%%%%%%%%%%%%%%%%%%%%%%%%%

\section{Cohomology of $\gamma$ modulo $d$ at positive antighost
number} \label{Hgammad}

Let be $a^p$ a local $p$-form of antighost number $k\neq 0$
fulfilling
\begin{equation}
\gamma a^p + db^{p-1}=0. \label{gammad}
\end{equation}
We want to show that if we add to $a^p$ an adequate $d$-trivial
term, the equation (\ref{gammad}) reduces to $\gamma a^p=0$.

{}From (\ref{gammad}), using the algebraic Poincar\'e lemma and
the fact that $\gamma$ is nilpotent and anticommute with $d$, we
can derive the descent equations
\begin{eqnarray}
\gamma a^p &+& db^{p-1}=0\\
\gamma b^{p-1} &+&dc^{p-2}=0\\
&\vdots&\nonumber\\
\gamma e^{q+1}&+&df^q=0 \label{before}\\
\gamma f^{q}&=&0\label{last},
\end{eqnarray}
Indeed, the fact that the antighost number is strictly positive
eliminates the constants. [E.g., from (\ref{gammad}), one derives
$d \gamma b^{p-1} = 0$ and thus $\gamma b^{p-1} +dc^{p-2}=
\hbox{constant}$, but the constant must vanish since it must have
strictly positive antighost number.] We suppose $q<p$, since
otherwise $\gamma a^p =0$, which is the result we want to prove.
The equation (\ref{last}) tells us that $f^q$ is a cocycle of
$\gamma$. It must be non-trivial in $H^q(\gamma)$ because if $f^q
= \gamma g^q$, then (\ref{before}) becomes $\gamma
(e^{q+1}-dg^q)=0$. The redefinition $e^{'q+1}=e^{q+1}-dg^q$ does
not affect the descent equation before (\ref{before}), which means
that the descent stops one step earlier, at $q-1$.

Using lemma \ref{gamma}, we deduce from (\ref{last}) that
\begin{equation}
f^q =\sum_{m,I_m}[\tilde{P}^{(m)}_{I_m}(\chi) +
dx^0\tilde{Q}^{(m)}_{I_m}(\chi)]\omega^{I_m}_{(m)}, \label{bzz}
\end{equation}
where $\tilde{P}^{(m)}_{I_m}$ and $\tilde{Q}^{(m)}_{I_m}$ are
local spatial forms of respective degree $q$ and $q-1$. We take
the basis elements $\omega^{I_m}_{(m)}$ to fulfill the conditions
of lemma \ref{zap}. Differentiating (\ref{bzz}), we find
\begin{eqnarray}
df^q&=&\sum_{m,I_m}\{\tilde{d}\tilde{P}^{(m)}_{I_m}\omega^{I_m}_{(m)}
+\gamma(\tilde{P}^{(m)}_{I_m}\hat{\omega}^{I_m}_{(m)})\nonumber \\
&&+dx^0[(\partial_0\tilde{P}_{I_m}^{(m)}-
\tilde{d}\tilde{Q}_{I_m}^{(m)})\omega^{I_m}_{(m)}
+\tilde{P}^{(m)}_{I_m}\partial_0\omega^{I_m}_{(m)}]\}.
\label{afterbefore}
\end{eqnarray}
The local function $\hat{\omega}^{I_m}_{(m)}$ is defined by
$\tilde{d}\omega^{I_m}_{(m)}=\gamma\hat{\omega}^{I_m}_{(m)}$ ( and
exists thanks to equation (\ref{yep})).

Now, we will show that the component $\tilde{P}^{(m)}_{I_m}$ can
be eliminated from $f^q$ by a trivial redefinition of $f^q$. In
order to satisfy (\ref{before}), the term independent of $dx^0$
and the coefficient of the term linear in $dx^0$ in
(\ref{afterbefore}) must separately be $\gamma$-exact. The second
condition gives explicitly
\begin{equation}
\sum_{m,I_m}[(\partial_0\tilde{P}_{I_m}^{(m)}-\tilde{d}
\tilde{Q}_{I_m}^{(m)})\omega^{I_m}_{(m)}
+\tilde{P}^{(m)}_{I_m}\partial_0\omega^{I_m}_{(m)}]=\gamma\beta,
\label{back2}
\end{equation}
To analyze precisely this equation, we define a degree $T$ by
\begin{equation}
T(\chi)=0,\quad T(\eta^{A(m)})=m.
\end{equation}
In fact, $T$ simply counts the number of time derivative of
$\eta^A$. We can decompose (\ref{back2}) according to the degree
$T$. Let $p$ be the highest degree occurring in $f^q$.  Then, the
highest degree occurring in (\ref{back2}) is $p+1$  and we must
have
\begin{equation}
\sum_{I=1}^{q_p}\tilde{P}^{(p)}_{I}
\partial_0\omega^{I}_{(p)}=\gamma\beta_{p+1}.
\end{equation}
{ } From the proof of the lemma \ref{zap}, we find that
\begin{eqnarray}
\tilde{P}^{(p)}_I=0\quad (I=1,\ldots,q_p)
\end{eqnarray}
because the $\partial_0\omega^{I}_{(p)}$ are linearly independent.
In $T$-degree $p$, (\ref{back2}) gives then
\begin{eqnarray}
\gamma\beta_p&=&-\sum_{I=1}^{q_p}\tilde{d}\tilde{Q}_{I}^{(p)}\omega^{I}_{(p)}
+\sum_{I=1}^{q_{p-1}}\tilde{P}_I^{(p-1)}\partial_0\omega^I_{(p-1)}\\
&=&\sum_{I=1}^{q_{p-1}}(\tilde{P}_I^{(p-1)}\,-
\tilde{d}\tilde{Q}_I^{(p)})\omega^I_{(p)}
-\sum_{I=q_{p-1}+1}^{q_p}\tilde{d}\tilde{Q}_I^{(p)}\omega^I_{(p)},
\end{eqnarray}
where we have used the property (\ref{ploc}) of the basis
$\{\omega^I \}$. This implies that
\begin{equation}
\tilde{P}_I^{(p-1)}=\tilde{d}\tilde{Q}_I^{(p)}\quad
(I=1,\ldots,q_{p-1})
\end{equation}
Inserting this equation in (\ref{bzz}), we find that
$\tilde{P}_I^{(p-1)}$ can be removed from $f^q$ by eliminating a
trivial cocycle of $\gamma$ modulo $d$ and redefining
$\tilde{Q}_I^{(p-1)}$. It only affects $e^{q+1}$ by a $d$-exact
term. Next, the equation (\ref{back2}) at $T$-degree $p-1$ shows
that $\tilde{P}_I^{(p-2)}$ is also $\tilde{d}$-exact and can thus
also be removed. Proceeding in the same way until the order 1 in
$T$, we have proved that all the $\tilde{P}^{(m)}_I$ can be
eliminated from $f^q$.

Looking back at (\ref{back2}) and taking into account that
$\tilde{P}^{(m)}_{I_m}$ can be set equal to zero by the above
argument, we find that
\begin{equation}
\tilde{d}\tilde{Q}^{(m)}_{I_m}=0. \label{tilt}
\end{equation}
Now, we must use the invariant Poincar\'e lemma (invariant means
in the algebra ${\cal{I}}$ of gauge-invariant forms) stating that
\begin{lemma}\label{invP}
Let be $\tilde{P}(\chi)$ a local spatial form of degree $q<5$,
then
\begin{equation}
\tilde{d}\tilde{P}(\chi)=0\Rightarrow
\tilde{P}(\chi)=\tilde{R}(F^{A(l)})+\tilde{d}\tilde{Q}(\chi),
\label{blub}
\end{equation}
where $\tilde{R}(F^{A(l)})$ is a polynomial in the curvature forms
$F^A=\frac{1}{6}F^A_{ijk}dx^i dx^j dx^k$ and all their time
derivatives (with coefficients that may involve $dx^k$, which
takes care of the constant forms).
\end{lemma}
\proof{The set of the generators of the algebra ${\cal{I}}$ is
\begin{equation}
\{\chi\}=\{\partial^{}_{i_1 \ldots
i_k}F^{A(l)}_{ijk},\partial^{}_{i_1 \ldots
i_k}\phi^{*(l)}_M,\eta^{A(l)},dx^{\mu}\} \label{setsetset}
\end{equation}
The 1-form $dx^0$ is not present in our problem since $\tilde{P}$
is a spatial local form (it only involves $dx^k$). Considering $l$
and $A$ as only one label (call it $\alpha$) and forgetting about
$dx^0$, the set (\ref{setsetset}) is the same as the corresponding
set of generators of the algebra ${\cal{I}}$($\equiv H(\gamma)$ in
pureghost number 0) for a system of spatial two-forms $
\{A^\alpha_{ij} \equiv A^A_{ij}, \partial_0 A^A_{ij},
\partial_{00}A^A_{ij}, \cdots \}$ in 5 dimensions. Consequently,
we can simply use the results proved in
\cite{Henneaux:1997,Knaepen:1999} for a system of $p$-forms in any
dimension.} We assumed before that $f^q$ is of degree $q<6$, hence
$\tilde{Q}_I$ is of degree $<5$. Thus, (\ref{tilt}) implies
\begin{equation}
\tilde{Q}^{(m)}_{I_m}=\tilde{d}\tilde{R}^{(m)}_{I_m}, \label{baff}
\end{equation}
where $\tilde{R}^{(m)}_{I_m}$ is a spatial form which only depends
on the variables $\chi$. There is no exterior polynomial in the
curvatures in $\tilde{Q}^{(m)}_{I_m}$ because
$\tilde{Q}^{(m)}_{I_m}$ has strictly positive antighost number. We
can therefore conclude that $f^q$ is trivial in $H^q(\gamma\mid
d)$ and can be eliminated by redefining $e^{q+1}$. The true bottom
is then one step higher. We can proceed in the same way until we
arrive at $\gamma a^{'p}=0$ with $a^{'p}=a^{p}+dg^{p-1}$. This can
be translated into the following lemma
\begin{lemma}\label{gammad'}
Let be a local form $a$ of antighost number $\neq 0$ fulfilling
$\gamma a + db=0$. There exists a local form $c$ such as
$a':=a+dc$ satisfies $\gamma a'=0$.
\end{lemma}

%%%%%%%%%%%%%%%%%%%%%%%%%%%%%%%%%%%%%%%%%%%%%%%%%%%%%%%%%%%%%%%%

\section{Cohomology of $\gamma$ modulo $d$ at zero antighost
number} \label{Hgammadbis}

Now, we want to study $H^{6,0}(\gamma\mid d)$ in pureghost number
0. Let be $a^{(6,0)}\in{\cal{A}}$ of form degree 6, of antighost
and pureghost number 0, and fulfilling $\gamma a^{(6,0)} +
da^{(5,1)}=0$. If $a^{(5,1)}$ is trivial $\gamma$ modulo $d$, this
equation reduces to $\gamma a^{(6,0)}+db^{(5,0)}=0$, which gives
$a^{(6,0)}=f(\partial^{}_{{\mu}_1 \ldots {\mu}_k}F^{A}_{ijk})d^6x$
plus a term trivial in the cohomology of $\gamma$ modulo $d$.

Otherwise, we can derive the non trivial descent equations
\begin{eqnarray}
\gamma a^{(6,0)} &+& da^{(5,1)}=0\label{first}\\
\gamma a^{(5,1)} &+&da^{(4,2)}=0\\
&\vdots&\nonumber\\
\gamma a^{(7-g,g-1)}&+&da^{(6-g,g)}=0 \label{before'}\\
\gamma a^{(6-g,g)}&=&0\label{last'},
\end{eqnarray}
because $pureghost(\gamma a^{(6-i,i)})>0$ eliminates the
constants. If $a^{(6-g,g)}$ is trivial $\gamma$ modulo $d$, the
bottom is really one step higher.

Eq. (\ref{last'}) implies that
\begin{equation}
a^{(6-g,g)}=\sum_I(\tilde{P}^{6-g}_I(\chi)+
dx^0\tilde{Q}^{5-g}_I(\chi))\omega^I+\gamma b^{(6-g,g-1)},
\label{last''}
\end{equation}
where $\tilde{P}^{6-g}_I$ and $\tilde{Q}^{5-g}_I$ are local
spatial forms, the superscript giving the form degree. Because the
pureghost number of $\eta$ is two, $a^{(6-g,g)}$ is non trivial
only for $g$ even. So, three cases are of interest: $g=0,2,4$.

The case $g=0$ corresponds to $\gamma a^{(6,0)} = 0$ and has been
already studied so let us assume $g>0$. The equations
(\ref{before'}) and (\ref{last''}) imply together
\begin{equation}
\sum_I(\partial_0
\tilde{P}^{6-g}_I-\tilde{d}\tilde{Q}^{5-g}_I)\omega^I+
\sum_I\tilde{P}^{6-g}_I\partial_0\omega^I=\gamma\beta.
\label{blob}
\end{equation}
Repeating the same analysis as for the equation (\ref{back2}), we
arrive at the conclusion that $\tilde{P}^{6-g}_I$ is trivial in
the invariant cohomology of $\tilde{d}$ (or vanishes) and can thus
be removed from $a^{(6-g,g)}$ by the addition of trivial terms in
the cohomology of $\gamma$ modulo $d$ and a redefinition of
$\tilde{Q}^{5-g}_I$. The case $g=6$ is then eliminated because in
that case $\tilde{Q}^{5-g}_I$ is not present at all. Hence, there
remains only two cases to examine: $g=2$ and $g=4$.

Once $\tilde{P}^{6-g}_I$ is removed, the equation (\ref{blob})
gives $\tilde{d}\tilde{Q}^{5-g}_I=0$. Using the invariant
Poincar\'e lemma, we find
$\tilde{Q}^{5-g}_I=\tilde{R}^{5-g}_I(F^{A(l)})+
\tilde{d}\tilde{S}^{(4-g,g)}_I(\chi)$. Hence, the form of the
bottom is
\begin{equation}
a^{(6-g,g)}=dx^0\sum_I\tilde{R}^{5-g}_I(F^{A(l)})\omega^I+\gamma
b^{(6-g,g-1)}+dc^{(5-g,g)}.
\end{equation}
But $F^{A(l)}$ is of form degree 3, thus if $g=4$,
$\tilde{R}^{5-g}_I$ must be a constant spatial $1$-form.  In that
instance, the $\omega^I$ must be quadratic in the ghosts
$\eta^{A(l)}$. The lift of such a bottom is obstructed (i.e.,
leads to no $a^{6,0}$) unless it is trivial (see
\cite{Henneaux:1997,Henneaux:1997i,Henneaux:1998,Henneaux:1999,Knaepen:1999}),
so that the case $g=4$ need not be considered. [In the algebra of
$x$-dependent local forms, the argument is simpler: the bottom is
always trivial and removable since it involves a constant 1-form,
which is trivial.]

It only remains to examine the case $g=2$. $\tilde{R}$ must then
be a 3-form. One can take $\tilde{R}$ linear in $F^{A(l)}$. In
that case, the lift gives Chern-Simons terms, which are linear
combinations of $dx^0F^{A(l)}A^{B(m)}$, with
$A^{B(m)}=\frac{1}{2}A^{B(m)}_{ij}dx^idx^j$. Or one can take
$\tilde{R}$ to be a constant 3-form. The corresponding deformation
is linear in the 2-form $A^{A(l)}$ with coefficients that are
constant forms. This second possibility is not $SO(5)$ invariant
and leads to equations of motion that are not Lorentz invariant.
It will not be considered further.

Dropping the latter possibility, all these results can be
summarized in the
\begin{proposition}\label{inter}
The non trivial elements of $H^{6,0}_0(\gamma\mid d,{\cal A})$ are
of two types: (i) those that descend trivially; they are of the
form $f(\partial^{}_{{\mu}_1 \ldots {\mu}_k}B^{A}_{ij})d^6x$; (ii)
those that descend non trivially; they are linear combinations of
the Chern-Simons terms
$\partial^l_0B^{Aij}\partial^m_0A^B_{ij}d^6x$.
\end{proposition}

Note that the kinetic term in the free action is precisely of the
Chern-Simons type (with $l=0$ and $m=1$).

%%%%%%%%%%%%%%%%%%%%%%%%%%%%%%%%%%%%%%%%%%%%%%%%%%%%%%%%%%%%%%%%%%

\section{Invariant cohomology of $\delta$ modulo $\tilde{d}$ in\\
even antifield number}

To pursue the analysis, we  need some results on the cohomology of
the Koszul-Tate differential $\delta$ as well as on its mod-$d$
and mod-$\tilde{d}$ cohomologies.

We can rewrite the action of the Koszul-Tate differential in the
following way
\begin{eqnarray}
\delta A^{A(l)}_{ij}&=&\delta C^{A(l)}_{i}=\delta \eta^{A(l)}=0,\\
\delta A^{*A(l)ij}&=&2\partial_k
F^{A(l)kij}-\epsilon^{ijklm}\partial_k
A^{A(l+1)}_{lm},\\
\delta C^{*A(l)i}&=&\partial_j A^{*A(l)ij},\\
\delta \eta^{*A(l)} &=& \partial_i C^{*A(l)i}.
\end{eqnarray}
If we regard $A$ and $l$ as only one label, these equations
corresponds to an infinite number of coupled non-chiral 2-forms in
5 dimensions.

It is useful to introduce a degree $N$ defined as
\begin{eqnarray}
&& N(\Phi^{*}_M)=1,\quad N(\Phi^M)=0,\\
&& N(\partial_k)=1,\quad N(\partial_0)=0\\
&& N(dx^{\mu})=0.
\end{eqnarray}
$N$ counts the number of spatial derivatives as well as the
antifields (with equal weight given to each). According to this
degree, $\delta$ decomposes as $\delta_0 + \delta_1$. The
differential $\delta_1$ acts exactly in the same way as the
Koszul-Tate differential for a system of free 2-forms in 5
dimensions.

We are now able to prove the
\begin{lemma}\label{delta}
$H_i(\delta,{\cal A})=0$ for $i>0$, where $i$ is the antighost
number, i.e, the cohomology of $\delta$ is empty in antighost
number strictly greater than zero.
\end{lemma}
\proof{From \cite{Henneaux:1997,Knaepen:1999}, we know that
$H_i(\delta_1)=0$. Let be $a\in\cal{A}$ a $\delta$-closed local
function of antighost number $i>0$. We decompose $a$ according to
the degree $N$
\begin{equation}
a=a_1+\ldots+a_m.
\end{equation}
The expansion stops because $a$ is polynomial in the antifields
and the derivatives. Furthermore, $a_0=0$ because $antigh(a)=i>0$.
The equation $\delta a=0$ gives in $N$-degree $m+1$: $\delta_1
a_m=0$. But $H_i(\delta_1)=0$, hence $a_m=\delta_1 b_{m-1}$. We
can define an $a'$ as being
\begin{equation}
a'=a-\delta b_{m-1}=a_1+\ldots+a_{m-2}+a'_{m-1},
\end{equation}
with $a'_{m-1}=a_{m-1}-\delta_0 b_{m-1}$. We can proceed in the
same way as before with $a'$, whose component of higher $N$-degree
is of degree less than $m$. We will then find a new $a'$ of
highest degree less than $m-1$, and so on, each time lowering the
$N$-degree. After a finite number of steps, we arrive at
$a^{'}=a^{'}_1=a-\delta b$. Then, $\delta a=0$ implies $\delta_1
a'_1=0$. Hence, $a'_1=\delta_1 b_0=\delta b_0$ because
$\delta_0\Phi^M=0$. In conclusion $a=\delta b$, with
$b=b_0+\ldots+b_{m-1}$. }

Of course, this lemma is really a consequence of general known
results on the cohomology of the Koszul-Tate differential.  It
simply confirms, in a sense, that we have correctly taken into
account all gauge symmetries and reducibility identities in
constructing the antifield spectrum.

The cohomological space $H^{5,inv}_k(\delta\mid\tilde{d})$ is
defined as $H^{5}_k(\delta\mid\tilde{d},\widetilde{\cal
A}\cap{\cal I})$ and is called the local, spatial,
\emph{invariant} Koszul-Tate cohomology since we work in the space
of local spatial forms that belongs to ${\cal{I}}$, i.e. that are
invariant. We want to compute it for $k$ even and $\neq 0$. To do
this, we will proceed as in the proof of lemma \ref{delta}. We
first prove the requested result for $\delta_1$; we then use
``cohomological perturbation" techniques to extend the result to
$\delta$.
\begin{lemma}\label{delta1}
For $k=2,4,6,\ldots$
\begin{equation}
H_k^{5,inv}(\delta_1\mid\tilde{d})=0.
\end{equation}
\end{lemma}
Again, this result is simply a particular case of more general
results, which were previously known, but for completeness, we
prove it here. \proof{Firstly, the theorem 9.1 of
\cite{Barnich:1995i} says that for a linear gauge theory of
reducibility order $p$ in $n$ dimensions $H^n_k(\delta\mid d)=0$
for $k>p+2$. A system of Abelian spatial 2-forms in 5 dimensions
is a linear gauge theory of reducibility order 1 (see subsection
\ref{degreesf}), thus, we can state that
$H^5_k(\delta_1\mid\tilde{d})=0$ for $k>3$.

Secondly, the theorem 7.4 of \cite{Henneaux:1997} gives here :
$H^5_2(\delta_1\mid\tilde{d})=0$.

Finally, the theorem 10.1 of \cite{Henneaux:1997} says that for a
system of space-time p-form gauge fields of the same degree
$H^n_k(\delta\mid d)\cong H^{n,inv}_k(\delta\mid d)$ for $k>0$.
For the system under consideration here, this can be translated
into: $H^5_k(\delta_1\mid \tilde{d})\cong H^{5,inv}_k(\delta_1\mid
\tilde{d})$ for $k>0$. Putting all these results together
completes the proof. } Let be $a^5(\chi)$ a local spatial $5$-form
in ${\cal{I}}$ of strictly positive and even antighost number,
satisfying
\begin{equation}
\delta a^5(\chi) + \tilde{d} b^4(\chi)=0.\label{modd}
\end{equation}
We can decompose $a^5$ and $b^4$ according to the degree $N$
\begin{eqnarray}
&& a^5=a^5_1+\ldots+a^5_n,\\
&& b^4=b^4_1+\ldots+b^4_m.
\end{eqnarray}
$a^5_0=0$ and $b^4_0=0$ because $a^5$ and $b^4$ are of antighost
number $>0$. We can always suppose $m\leq n$ because if $m>n$,
(\ref{modd}) gives in $N$-degree $m+1$: $\tilde{d} b^4_m=0$. Using
the invariant Poincar\'e lemma, this yields $b^4_m=\tilde{d}
c^3_{m-1}$. Hence, $b^4_m$ only contributes to $b^4$ by a
$\tilde{d}$-trivial term which can be eliminated. Proceeding in
the same way until $m=n$, we arrive at the equation
\begin{equation}
\delta_1 a^5_n(\chi) + \tilde{d} b^4_n(\chi)=0. \label{keykey}
\end{equation}
It has already been noticed above that the algebra ${\cal{I}}$
without dependence on $dx^0$ is the same as for a system of
spatial 2-forms. We can thus use the lemma \ref{delta1} in
(\ref{keykey}) to find that
\begin{eqnarray}
a^5_n(\chi)=\delta_1 e^5_{n-1}(\chi) + \tilde{d} f^4_{n-1}(\chi).
\end{eqnarray}
Therefore, $a^{'5}=a^5-\delta e^5_{n-1} - \tilde{d} f^4_{n-1}$
satisfies the same properties as $a^5$, except that its component
of highest $N$-degree is of degree $<D$. We can now apply the same
reasoning as before to $a^{'5}$, and so on, until we arrive at
\begin{equation}
a^{'5}=a^{'5}_1=a^5-\delta (\sum_{i=1}^{n-1}e^5_i) - \tilde{d}
(\sum_{i=1}^{n-1}f^4_i)
\end{equation}
This leads to
\begin{equation}
a^{'5}_1=\delta_1 e^5_0(\chi) + \tilde{d} f^4_0(\chi).
\end{equation}
But $\delta_1 e^5_0=\delta e^5_0$ because $\delta_0\Phi^M=0$.
Eventually, we have $a^5=\delta e^5(\chi) + \tilde{d} f^4(\chi)$,
with $e^5=\sum\limits_{i=0}^{n-1}e^5_i$ and
$f^4=\sum\limits_{i=0}^{n-1}f^4_i$. This gives the awaited lemma:
\begin{lemma}\label{deltadinv}
For $k=2,4,\ldots$
\begin{equation}
H_k^{5,inv}(\delta\mid\tilde{d})=0.
\end{equation}
\end{lemma}

%%%%%%%%%%%%%%%%%%%%%%%%%%%%%%%%%%%%%%%%%%%%%%%%%%%%%%%%%%%%%%%%%%%

\section{Decomposition of the Wess-Zumino equation}

We now have all the necessary tools to solve the Wess-Zumino
consistency condition that controls the consistent deformations
(to first-order) of the action,
\begin{equation}
sa^6 + db^5 = 0, \label{WZ}
\end{equation}
where $a^6$ and $b^5$ are local forms of respective form degrees 6
and 5, and ghost number 0 and 1. These forms are defined up to the
following allowed redefinitions
\begin{eqnarray}
&& a^6\rightarrow a^6+sf^6+dg^5\label{redef1}\\
&& b^5\rightarrow b^5+sg^5+dh^4, \label{redef2}
\end{eqnarray}
which preserve (\ref{WZ}). We can decompose $a^6$ and $b^5$
according to antighost number, which gives
\begin{eqnarray}
a^6&=&a^6_0+ \ldots +a^6_k,\\
b^5&=&b^5_0+\ldots+b^5_q,
\end{eqnarray}
with $a^6_k\neq 0$.

We suppose $k>0$ and we will show that $a^6_k$ can be eliminated
if we redefine $a^6$ in an appropriate way. In antighost number
$k$, the equation (\ref{WZ}) just reads
\begin{equation}
\gamma a^6_k + db^5_k=0. \label{Hgamma}
\end{equation}
We can always assume $k\geq q$ because if $q>k$, the equation
(\ref{WZ}) gives in highest antighost number $db^5_q=0$. Using the
algebraic Poincar\'e lemma, we find that $b^5_q=dc^4_q$. Hence, we
can remove the component $b^5_q$ up to a $d$-trivial redefinition
of $b^5$.

{ } From the lemmas \ref{gamma} and \ref{gammad'}, we know that
Eq. (\ref{Hgamma}) implies
\begin{equation}
a^6_k=\sum_I P_I(\chi)\omega^I+\gamma f^6_k + dg^5_k.
\end{equation}
The $\gamma$ modulo $d$ trivial part of $a^6_k$ can be eliminated
by redefining $a^6$ in the following way
\begin{equation}
a^6\rightarrow a^6-sf^6_k-dg^5_k.
\end{equation}
We notice that $H^{6,0}_k(\gamma)$ is non trivial only in even
antighost number $k$ (because $\eta$ is of pureghost number 2).
This implies that we can assume $k$ to be even.

The Wess-Zumino consistency condition in antighost number $k-1$ is
\begin{equation}
\gamma a^6_{k-1}+\delta a^6_k+db^5_{k-1}=0. \label{WZ2}
\end{equation}
The term $b^5_{k-1}$ is invariant because (\ref{WZ2}) implies
$d(\gamma b^5_{k-1})=0$. Therefore, the algebraic Poincar\'e lemma
gives $\gamma b^5_{k-1} + dc^4_{k-1}=0$ because $k>1$.

From the lemma \ref{gammad'} we know that we can suppose $\gamma
b^5_{k-1}=0$ without affecting $a^6$. Furthermore, if
$b^5_{k-1}=\gamma c^5_{k-1}$ we can eliminate $b^5_{k-1}$ by
redefining $b^5$ in the following way: $b^5\rightarrow
b^5-sc^5_{k-1}$, which does not modify $a^6_k$.

Therefore, we can assume
\begin{eqnarray}
a^6_k&=&\sum_Idx^0\tilde{P}^5_I\omega^I,\label{dring1}\\
b^5_{k-1}&=&\sum_I(\tilde{Q}^5_I+dx^0\tilde{R}^4_I)\omega^I.
\label{dring2}
\end{eqnarray}
The $\tilde{P}^5_I$, $\tilde{Q}^5_I$, and $\tilde{R}^4_I$ are
local spatial forms belonging to ${\cal{I}}$.

Inserting (\ref{dring1}) and (\ref{dring2}) in (\ref{WZ2}), we
find
\begin{eqnarray}
\gamma a^6_{k-1} &=&\sum_I\{-\tilde{d}\tilde{Q}^5_I\omega^I
-\gamma[(\tilde{Q}^5_I+dx^0\tilde{R}^4_I)\hat{\omega}^I]\\
&&+dx^0[(\delta\tilde{P}^5_I+\tilde{d}\tilde{R}^4_I
-\partial_0\tilde{Q}^5_I)\omega^I
-\tilde{Q}^5_I\partial_0\omega^I]\},
\end{eqnarray}
with $\tilde{d}\omega^I=\gamma\hat{\omega}^I$. This implies that
\begin{equation}
\sum_I[(\delta\tilde{P}^5_I+\tilde{d}\tilde{R}^4_I
-\partial_0\tilde{Q}^5_I)\omega^I
-\tilde{Q}^5_I\partial_0\omega^I]=\gamma\beta.
\end{equation}
If we analyse this equation in the same way as the equation
(\ref{back2}), we can prove that
$\tilde{Q}^5_I=\delta\tilde{P}^5_I+\tilde{d}\tilde{R}^4_I$ (or
simply vanishes). Inserting these equations in (\ref{dring2}), we
find that $b^5_{k-1}$ is of the form
\begin{equation}
b^5_{k-1}=\delta c^5_{k}+de^4_{k-1}+\gamma f^5_{k-1}
+dx^0\sum_I\tilde{R}^{'4}_I(\chi)\omega^I,
\end{equation}
where $c^5_{k}$ and $e^4_{k-1}$ belong to $H(\gamma)$. In
conclusion, we can eliminate $\tilde{Q}^5_I$ from $b^5_{k-1}$ by
redefining $a^6$ and $b^5$ in the following way
\begin{eqnarray}
&& a^6\rightarrow a^6-d(c^5_k+f^5_{k-1}),\\
&& b^5\rightarrow b^5-s(c^5_k+f^5_{k-1})-de^4_{k-1},
\end{eqnarray}
which does not affect the condition $\gamma a^6_k=0$, because
$\gamma c^5_{k}=0$.

Therefore, we can finally assume
\begin{equation}
a^6_k=\sum_Idx^0\tilde{P}^5_I(\chi)\omega^I,\quad b^5_{k-1}=\sum_I
dx^0\tilde{R}^4_I(\chi)\omega^I.
\end{equation}
The equation (\ref{WZ2}) becomes
\begin{equation}
\gamma a^{'}_{k-1}+dx^0\sum_I(\delta
\tilde{P}^5_I(\chi)+\tilde{d}\tilde{R}^4_I(\chi))\omega^I=0,
\end{equation}
which implies that $\delta
\tilde{P}^5_I(\chi)+\tilde{d}\tilde{R}^4_I(\chi)=0$. We know that
we are in even antighost number, thus we can use the lemma
\ref{deltadinv} to find that $\tilde{P}^5_I=\delta
\tilde{S}^5_I(\chi)+\tilde{d}\tilde{T}^4_I(\chi)$. Hence,
\begin{equation}
a^6_k = sf^6_{k+1}+dg^5_k+\gamma h^6_k,
\end{equation}
where we have defined
\begin{eqnarray}
&& f^6_{k+1}=-dx^0\sum_I\tilde{S}^5_I\omega^I,\quad
g^5_k=-dx^0\sum_I
\tilde{T}^4_I\omega^I,\\
&& h^6_k=dx^0\sum_I \tilde{T}^4_I\hat{\omega}^I,\quad
\tilde{d}\omega^I=\gamma\hat{\omega}^I.
\end{eqnarray}
Thus $a^6_k$ can be completely eliminated by redefining $a^6$ as
\begin{equation}
a^{'6}=a^6-s(f^6_{k+1}+ h^6_k)-dg^5_k,
\end{equation}
which only affects the components of antighost number $<k$.
Repeating the argument at lower antighost numbers enables one to
remove successively $a_{k-1}$, $a_{k-2}$, ..., up to $a_1$. This
completes the proof of the proposition \ref{noantif}.

A direct consequence of the proposition \ref{noantif} is the
\begin{corollary}$H^{6,0}(s \mid d)\cong H^{6,0}(\gamma \mid d)$\end{corollary}
Indeed, for antifield-independent local forms, the cocycle
condition $H^{6,0}(s \mid d)$ reduces to the cocycle condition for
$H^{6,0}(\gamma \mid d)$. Furthermore, for vanishing antifield
number $\gamma$-exact (mod-$d$) solutions are also $s$-exact
(mod-$d$). Thus, we are led to consider $H^{6,0}(\gamma \mid d)$.
This cohomology is given by the proposition \ref{inter}. [The
terms in that cohomology that vanish on-shell are trivial in the
$s$-cohomology.] Thus, the only consistent deformations of the
free action for a system of Abelian chiral $2$-forms are either
functions of the curvatures or of the Chern-Simons type. In both
cases, the integrated deformations are off-shell gauge invariant
and yield no modification of the gauge transformations.

%%%%%%%%%%%%%%Bibliography%%%%%%%%%%%%%%%%%%%%%%%%%

\end{document}